\documentclass[11pt]{article}
\usepackage{amsmath,amsthm,epsfig,amsfonts,color, bm, graphicx,graphicx,fancyhdr,mathrsfs}
\usepackage{rotating}
\allowdisplaybreaks
\usepackage{comment}
\usepackage{booktabs}
\usepackage{setspace}
\usepackage{array}
\usepackage{amsmath}
\usepackage{amssymb}
\usepackage{xurl} 
\usepackage[hidelinks]{hyperref}

\usepackage{indentfirst}
\usepackage{enumerate}
\usepackage{comment}
\usepackage{dsfont}
\usepackage{natbib}
\usepackage[misc]{ifsym}
\usepackage[toc,page]{appendix}
\usepackage{multirow}
\usepackage{float}
\usepackage{todonotes}
\usepackage{ulem}  

\usepackage{booktabs}
\usepackage{tabularx}
\usepackage{graphicx}
\usepackage{subcaption} 
\usepackage[labelsep=period]{caption}

\usepackage[table]{xcolor}
\usepackage{colortbl}

\definecolor{lowgreen}{RGB}{120,200,120}
\definecolor{highred}{RGB}{220,80,80}

\newcommand{\HeatPct}[2]{\cellcolor{highred!#2!lowgreen}{#1}}

\usepackage{siunitx}
\newtheorem{remark}{{Remark}}

\def\marginnote#1{\setbox0=\vtop{\hsize4pc
\small\raggedright\noindent\baselineskip9pt \rightskip=0.5pc plus
1.5pc #1}\leavevmode \vadjust{\dimen0=\dp0
\kern-\ht0\hbox{\kern-4.00pc\box0}\kern-\dimen0}}
\def\lboxit#1{\vbox{\hrule\hbox{\vrule\kern6pt
\vbox{\kern6pt#1\kern6pt}\kern6pt\vrule}\hrule}}

\usepackage{xr}

\begin{document}
\thispagestyle{empty}
\title{Quantifying Social Inflation in Liability Insurance with Advanced Statistical Methods}
 
\author{
Tsz Chai Fung\thanks{Maurice R. Greenberg School of Risk Science, Georgia State University.}~~~~~Lie Ma\thanks{Brighthouse Financial, Charlotte, North Carolina.}~~~~~Liang Peng$^\star$~~~~~Fang Yang$^\star$}

\date{\today }
	\maketitle
	\begin{quote}
\begin{center}
\textbf{Abstract}
\end{center}

 \smallskip
\noindent 

Social inflation, which is the rise in liability claim costs beyond general economic inflation, has become a major concern for insurers and reinsurers, yet it is difficult to quantify because litigation outcomes are heavy-tailed and the mix of cases reaching verdict versus settlement changes over time. Using a large database of US jury verdicts and settlements, we develop case-mix-adjusted social inflation measures through multiple channels that matter to reinsurers: plaintiff win rates (a frequency-type channel), settlement propensity (a frequency-type channel), and verdict/settlement severity. The approach combines rolling-window logistic regression for probabilities and quantile (value-at-risk) regression for severities, with uncertainty quantified via a random-weighted bootstrap. We find statistically significant relative increases in plaintiff win probability of approximately 20\%–30\% from 2009 to 2024, alongside a statistically significant relative decline in settlement probability of more than 10\% over the same period. The dominant channel is verdict severity: Even after controlling for explanatory variables, verdict awards show a sharp rise after 2020, increasing by more than 100\% from 2020 to 2024, whereas settlement amounts show limited and often statistically insignificant inflation. Therefore, inflation in total amounts payable to plaintiffs closely tracks verdict severity. Social inflation is more pronounced in corporate-defendant and uninsured-defendant cases and in states without tort caps or third-party litigation funding regulation. In addition, we find that social inflation has impacts not only on ``nuclear verdicts'' but also, in a similar manner, on moderate losses.

\end{quote}

\section*{Executive Summary}
Social inflation, which is the increase in liability claim costs beyond what can be explained by general economic inflation, has become a material concern for insurers and reinsurers. We decompose this claim-cost growth into changes in the likelihood that plaintiffs receive payments and, more importantly, changes in the size of those payments when they occur. Quantifying social inflation is difficult because litigation outcomes are heavy-tailed (a few outcomes can be extremely large) and because the mix of cases reaching verdict versus settlement changes over time. As a result, simple trend measures can be misleading: An apparent increase in outcomes can reflect changing case composition rather than a true shift in the underlying litigation environment.

This paper addresses that measurement problem using VerdictSearch, a large US database of jury verdicts and settlements maintained by ALM. We analyze 74,188 cases with verdict or settlement dates from 2009 to 2024 (with award and settlement amounts adjusted to December 2024 dollars using the consumer price index). The dataset provides ground-up outcomes (verdict/settlement results and amounts payable to plaintiffs) together with a rich set of case characteristics, including factual features (state, injury/death indicators, case-type categories, party structure such as corporate/insured defendants, and plaintiff demographics) and strategic features (trial length, jury deliberation length, and attorney/expert involvement). This structure allows us to separate changes driven by evolving case mix from changes that remain after holding observable case characteristics fixed.

A central descriptive finding is that the VerdictSearch sample changes markedly over time. Case counts decline substantially, especially after 2020, consistent with pandemic disruptions and the longer-run ``vanishing trial'' phenomenon. At the same time, the cases that do reach verdict or settlement become more complex and severe on average (e.g., more injuries, experts, and attorneys), implying strong selection effects. Because severe and complex cases are more likely to be litigated fully and reported, any unadjusted time trend in outcomes can overstate social inflation by conflating a true shift in outcomes with a shift in what types of cases are observed. The data also show pronounced heavy tails: Verdict awards' 95th percentiles are typically an order of magnitude above medians, and totals can be dominated by a handful of extreme cases. These features motivate a distribution-robust approach rather than an average-based index.

Methodologically, the paper develops a family of social inflation indices for multiple channels that matter to insurance and reinsurance: (1) plaintiff win probability conditional on reaching verdict, (2) settlement probability, (3) verdict award severity conditional on a plaintiff win, (4) settlement severity conditional on settlement, and (5) total amount payable to the plaintiff. Severity channels are modeled using quantile (value-at-risk) regression to accommodate heavy tails, and probabilities are modeled using logistic regression. To stabilize estimation given declining annual sample sizes, the regressions use a rolling-window design for covariate effects. Uncertainty is reported using a random-weighted bootstrap, which quantifies estimation (parameter) uncertainty.

The paper reports three closely related index specifications throughout the results. A ``naive'' specification applies no covariate adjustments and therefore mixes true temporal change with evolving case mix. A ``headline'' specification controls for factual explanatory variables (case characteristics that are largely determined at the outset of litigation and not easily altered by litigation tactics) to adjust for evolving case composition; this is the paper's advocated primary measure of social inflation because it aligns with the definition of social inflation as inflation beyond economic inflation after controlling for observable case characteristics. A ``residual'' specification additionally conditions on strategic litigation variables (e.g., trial/jury deliberation length and attorney/expert involvement), which can change with litigation decisions; the residual specification is best interpreted as a decomposition tool that shows how much of the headline trend is statistically associated with observable changes in litigation intensity, rather than a measure to replace the headline index.

At a high level, the empirical results show that social inflation is present in multiple channels, but the dominant channel is verdict severity, particularly after 2020. Plaintiff win probability conditional on verdict increases over time, while settlement probability declines, indicating that cases become less likely to settle and more likely to proceed to verdict. Inflation in settlement severity, once adjusted for case mix and other variables, is generally smaller and often statistically insignificant compared to inflation in verdict severity. As a result, social inflation in total amounts payable to plaintiffs tends to closely track verdict severity rather than settlement severity. The paper also reports meaningful heterogeneity: Social inflation is more pronounced in corporate-defendant and uninsured-defendant cases and in states without tort caps or third-party litigation funding regulation. Finally, when comparing severity inflation across quantiles, the paper finds little statistical evidence that the upper tail inflates systematically faster than more moderate outcomes after adjustment, suggesting a broad upward shift in verdict severity rather than a phenomenon confined to extreme awards alone.

Table \ref{tab:exec} presents a compact high-level summary of total plaintiff payments that illustrates why case-mix adjustment is essential and why the headline index is the recommended measure. In the table, ASIR is the annual social inflation rate (percentage change from year to year) and CSII is the cumulative social inflation index (normalized to 100 in 2009). The naive series grows extremely rapidly over time, reaching a CSII above 1,000 by 2024, while the headline (case-mix-adjusted) series is materially lower, reaching a CSII of 317.1 by 2024. This gap indicates that failing to adjust for evolving case composition can produce substantial overstatement of social inflation. The residual series is lower than the headline series in later years (CSII 230.3 by 2024), indicating that observed litigation-intensity proxies explain part (but not all) of the case-mix-adjusted trend. The paper advocates the headline series as the primary social inflation index because it adjusts for evolving case mix using factual (largely exogenous) case characteristics while still capturing outcome growth beyond general economic inflation; the residual series is interpreted as an attribution view that conditions on endogenous strategic variables to assess how measured litigation intensity relates to the headline trend.

\setcounter{table}{0}
\renewcommand{\thetable}{ES.\arabic{table}}

\begin{table}[!htbp]
\centering
\small
\caption{High-level social inflation summary for total verdict/settlement amount. ``Naive'' uses no adjustments; ``Headline'' controls for evolving case-mix; ``Residual'' additionally removes the effects of evolving litigation strategies.}
\label{tab:exec}
\begin{tabular}{lrrrrrr}
\toprule
& \multicolumn{2}{c}{\textbf{Naive}} & \multicolumn{2}{c}{\textbf{Headline}} & \multicolumn{2}{c}{\textbf{Residual}} \\
\cmidrule(lr){2-3}\cmidrule(lr){4-5}\cmidrule(lr){6-7}
Year & ASIR (\%) & CSII & ASIR (\%) & CSII & ASIR (\%) & CSII \\
\midrule
2010 &  17.0 &  117.0 &  14.4 &  114.4 &  16.7 &  116.7 \\
2011 &  15.8 &  135.5 &  --4.8 &  108.9 &  --7.4 &  108.1 \\
2012 &   3.7 &  140.5 &  12.3 &  122.3 &  12.1 &  121.2 \\
2013 & --34.4 &   92.1 & --33.2 &   81.7 & --27.3 &   88.1 \\
2014 &  24.1 &  114.3 &   0.4 &   82.0 &  --3.2 &   85.3 \\
2015 &  --3.5 &  110.4 &  --5.2 &   77.7 &   6.6 &   90.9 \\
2016 &  18.8 &  131.1 &   2.8 &   79.9 &   0.3 &   91.2 \\
2017 &  16.3 &  152.5 &  --2.3 &   78.0 &  --5.2 &   86.5 \\
2018 &  57.7 &  240.5 &  18.7 &   92.7 &  16.1 &  100.4 \\
2019 &  --2.9 &  233.6 &   0.8 &   93.4 &   3.2 &  103.6 \\
2020 &   5.1 &  245.4 &   2.4 &   95.6 &   4.7 &  108.5 \\
2021 &  25.1 &  307.0 &  45.7 &  139.3 &  13.3 &  122.9 \\
2022 &  25.6 &  385.6 &  51.5 &  211.1 &  45.4 &  178.7 \\
2023 &  56.7 &  604.1 &  10.3 &  232.8 &   4.6 &  186.9 \\
2024 &  67.5 & 1,012.0&  36.2 &  317.1 &  23.2 &  230.3 \\
\bottomrule
\end{tabular}
\end{table}

Finally, the paper is explicit about limitations and how future work could strengthen the connection to insurer-paid losses. VerdictSearch provides ground-up litigation outcomes; it does not directly observe the insurer-paid amounts after deductibles, limits, attachments, and reinsurance structures, nor does it provide exposure-based claim counts or unreported below-retention ``near-misses.'' Accordingly, the indices should be interpreted as measures of social inflation in litigated ground-up outcomes, which are a leading indicator of liability loss-cost pressure but may require additional modeling and data to translate into insurer-paid claim inflation. The paper discusses potential extensions using claims or statutory data and notes practical challenges, including missing policy-feature information and selection effects in insurer-paid records.

\section{Introduction}\label{sec:intro}
\subsection{Definition and scope}\label{sec:intro:def}
The term \textit{social inflation} was used in the 1970s by Warren Buffett to describe ``a broadening definition by society and juries of what is covered by insurance policies'' (Buffett 1978). However, quantifying social inflation has been challenging because the term is often defined too vaguely (Mackeprang 2020). It is therefore critical to clearly define the concept so that we can collect appropriate data and conduct statistical analyses that provide meaningful insight into how to construct a social inflation index.

Social inflation is commonly defined as the portion of insurance losses that exceeds what can be explained by general economic inflation. Accordingly, claim cost increases beyond economic inflation are categorized as social inflation (NAIC 2025). The Insurance Information Institute (2022) offers a closely related definition, characterizing social inflation as the impact of rising litigation costs on insurance payouts.

Although there is no universally agreed-upon definition of social inflation, commonly cited components include the growth of ``nuclear verdicts" and rising claim costs driven by factors such as increased litigation, broader interpretations of liability, more plaintiff-friendly legal outcomes, and larger jury awards (Bergen 2020; Dickinson and Sutton 2020; Djazayeri 2020; Moorcraft 2020; Oh 2022). In this paper, we adopt the definition in Djazayeri (2020): Social inflation refers to the trend of rising insurance costs due to increased litigation and more plaintiff-friendly outcomes, reflected in settlement and verdict payments that rise faster than general economic inflation and beyond what can be explained by observable case or claim characteristics. At present, social inflation is most prominently observed in the United States. However, it is widely expected that other countries, such as Australia, Canada, and the United Kingdom, will experience similar pressures in the near future. 

\subsection{Evidence and results in the literature}\label{sec:intro:lit}
With these definitions in mind, we see that empirical evidence from claims, litigation, and verdict data increasingly supports the presence of social inflation. Social inflation has attracted heightened attention in the insurance and reinsurance industries because liability claim costs have risen rapidly, often far outpacing broad economic indicators such as the Consumer Price Index (CPI). In recent years, the United States has experienced a sharp increase in both litigation activity and exceptionally large jury awards. 
Since 2016, annual tort costs, including damages, settlements, and legal fees, have grown by about 7\% per year, exceeding \$500 billion in 2022 (McKnight and Hinton 2024). Since 2015, jury awards exceeding \$10 million have tripled, whereas cases with awards above \$100 million have increased by roughly 30\% (Travelers 2023). This pattern is commonly described as social inflation: growth in claim severity that cannot be fully explained by standard economic drivers. Consistent with this view, Boerlin et al. (2024) report that social inflation in the United States increased by an average of 5.4\% annually from 2017 to 2022 and peaked at 7\% in 2023 for the first time in two decades, compared with average annual economic inflation of 3.7\% over the same period. The study also contrasts prior US episodes of social inflation in the 1980s and 2000s with the current episode beginning in the mid-2010s and attributes the recent trend in part to outsized court verdicts, particularly in personal injury cases.

Traditionally, insurers have adjusted policy prices for inflation using standard macroeconomic indicators. In recent years, however, industry research suggests that liability claim costs are increasingly being driven upward by noneconomic social and legal forces, often grouped under the label \textit{social inflation} (Lynch and Moore 2023; Boerlin et al. 2024). Lynch and Moore (2023), in a study supported by the Insurance Information Institute and the Casualty Actuarial Society (CAS), estimate that US commercial auto liability claim payouts from 2012 to 2021 were approximately \$30 billion higher than would have been expected based on standard macroeconomic indicators, due in part to social inflation. They also identify legal system abuse and third-party litigation funding (TPLF) as two key contributors to claim cost growth beyond general inflation. Boerlin et al. (2024) further report that social inflation has become a primary driver of growth in US liability claims; largely due to an increase in large court verdicts, it has raised US liability claims by 57\% over the past decade.

Rising claim costs not only create a riskier and more expensive operating environment for businesses but also pose significant challenges for insurers. In particular, large-dollar verdicts can have substantial adverse consequences for firms, including constraints on innovation, elevated out-of-pocket claim-related costs, higher insurance premiums, and, in severe cases, financial distress or bankruptcy.

Complementing the empirical evidence, Dickinson and Sutton (2020) identify four structural shifts that are intensifying social inflation: changes in juror perspectives, the expansion of litigation funding, a more sophisticated plaintiff’s bar, and the normalization of nuclear verdicts. Together, these forces contribute to higher liability claim frequency and severity. As a result, reinsurers, who typically absorb the highest layers of loss, are increasingly concerned about both the magnitude and the unpredictability of such claims.

Understanding social inflation in a rigorous, quantitative manner is critical for several reasons. First, insurers and reinsurers rely on accurate predictive models of claim severity to price policies and set reserves appropriately. When large claims consistently exceed traditional economic forecasts, it signals the presence of deeper systemic shifts. Second, the impact of social inflation is not uniform; extreme jury verdicts and nuclear settlements may disproportionately affect claims at the highest layers of insurance coverage. Consequently, reinsurers, who cover high-severity portions of claims, face potentially underrecognized exposures, which will eventually heighten their solvency risk and in general adversely affect broader market stability. Third, quantitatively measuring portions of claim inflation and attributing them to various social or legal factors will help reinsurers to better understand and prepare for their most important source of risk exposures, thereby enabling better risk management practices.

Researchers and industry professionals have begun exploring social inflation’s drivers and consequences. For
instance, Oh (2022) reports that social inflation accounts for nearly two-thirds of the roughly 30\% increase in insurance premiums since 2018. The analysis further suggests that shifts in jury sentiment materially influence verdicts and settlements, with significant implications for insurers that are directly exposed to these legal outcomes. Several industry reports (Dixon et al. 2024; Francis 2024; Wellington 2023) reveal not only that verdict sizes are escalating but also that insurer strategies such as curtailing coverage or raising premiums create knock-on effects for insured entities (e.g., businesses adjusting product prices or halting certain high-liability operations). More specifically, Dixon et al.\ (2024) document that since 2010, (1) plaintiff win rates in cases that reach a verdict have increased, (2) inflation-adjusted trial awards have risen, and (3) the share of very large awards (i.e., \$5 million or more) has grown. Furthermore, academic analysis (Frees et al. 2014, chapter 12.3) points to the importance of deeper statistical methods, such as
factor analysis and principal components, to isolate latent societal or legal factors that co-move with rising claim severity.

While prior studies have documented the presence and potential drivers of social inflation, several methodological limitations constrain what can be inferred for pricing and risk management. First, much of the existing evidence is based on summary statistics or aggregate time trends from the full population of observed claims or verdicts, without explicitly accounting for how the underlying case mix evolves over time. As our VerdictSearch data illustrate, the litigation environment is characterized by pronounced selection effects (e.g., vanishing trials and post-pandemic disruptions), shifting case compositions across jurisdictions and liability types, and increasing case complexity; these changes can mechanically generate apparent inflation in outcomes even when the underlying legal environment is unchanged. A small number of recent studies have begun to incorporate regression-based adjustments (e.g., Oh 2022), but there remains room to more fully capture the distributional nature of social inflation in heavy-tailed losses. In particular, linear or mean-focused regressions do not directly address how the tail of the severity distribution evolves, nor whether the most extreme outcomes respond differently to shifts in societal and legal norms. Relatedly, much of the literature does not cleanly separate frequency-related channels (e.g., changes in plaintiff win or settlement probability) from severity-related channels (changes in award or settlement size). Finally, uncertainty quantification is often limited, despite the fact that litigation outcomes are far more variable and data-scarce than the large-scale price baskets used to estimate CPI. For reinsurers managing tail risk, parameter and model uncertainty can be material and should be formally reported.

\subsection{Statistical approach and our contributions}\label{sec:intro:contribution}
To address these gaps in the research, this paper uses VerdictSearch, a leading database of US jury verdicts and settlements that enables a direct, ground-up view of amounts payable to plaintiffs and the case characteristics surrounding those outcomes. We first document clear time variation in case composition and demonstrate that both factual case characteristics and strategic litigation variables are strongly associated with verdict/settlement outcomes and amounts; together, these findings imply that naive trend-based measures can overstate social inflation by conflating genuine shifts with evolving case mix and litigation behavior. Building on this evidence, we develop case-mix-adjusted measures of social inflation---annual social inflation rates (ASIRs) and cumulative social inflation indices (CSIIs)---for plaintiff win probability, settlement probability, verdict award severity, settlement severity, and total plaintiff payments, using rolling-window logistic and quantile regression (value-at-risk, or VaR) frameworks and reporting estimation (parameter) uncertainty via a random-weighted bootstrap (i.e., the uncertainty conditional on model specification). Because VerdictSearch contains outcomes for litigated cases rather than exposure-based claim counts, the plaintiff-winning and settlement-propensity channels mentioned above should be interpreted as litigation-outcome (conditional) probabilities that inform only one component of the claim frequency trend and do not capture the underlying accident/claim arrival process. Empirically, we find statistically significant social inflation in plaintiff win probability (by which we mean increases in plaintiff win probability that contribute to loss inflation through a frequency-type channel) and, more prominently, in verdict award severity (especially post-2020), whereas social inflation in settlement severity is generally not as substantial after adjustment to allow for other variables. 

\begin{table}[!htbp]
\centering
\small
\caption{High-level social inflation summary for total verdict/settlement amount. Values are extracted from Figure \ref{fig:idx_sev_t1_0}. ``Naive'' uses no adjustments; ``Headline'' controls for evolving case mix; ``Residual'' additionally removes the effects of evolving litigation strategies. CSII is indexed to 100 in 2009.}
\label{tab:highlevel_totalpay_fig49}
\begin{tabular}{lrrrrrr}
\toprule
& \multicolumn{2}{c}{\textbf{Naive}} & \multicolumn{2}{c}{\textbf{Headline}} & \multicolumn{2}{c}{\textbf{Residual}} \\
\cmidrule(lr){2-3}\cmidrule(lr){4-5}\cmidrule(lr){6-7}
Year & ASIR (\%) & CSII & ASIR (\%) & CSII & ASIR (\%) & CSII \\
\midrule
2010 &  17.0 &  117.0 &  14.4 &  114.4 &  16.7 &  116.7 \\
2011 &  15.8 &  135.5 &  --4.8&  108.9 &  --7.4&  108.1 \\
2012 &   3.7 &  140.5 &  12.3 &  122.3 &  12.1 &  121.2 \\
2013 & --34.4&   92.1 & --33.2&   81.7 & --27.3&   88.1 \\
2014 &  24.1 &  114.3 &   0.4 &   82.0 &  --3.2&   85.3 \\
2015 &  --3.5&  110.4 &  --5.2&   77.7 &   6.6 &   90.9 \\
2016 &  18.8 &  131.1 &   2.8 &   79.9 &   0.3 &   91.2 \\
2017 &  16.3 &  152.5 &  --2.3&   78.0 &  --5.2&   86.5 \\
2018 &  57.7 &  240.5 &  18.7 &   92.7 &  16.1 &  100.4 \\
2019 &  --2.9&  233.6 &   0.8 &   93.4 &   3.2 &  103.6 \\
2020 &   5.1 &  245.4 &   2.4 &   95.6 &   4.7 &  108.5 \\
2021 &  25.1 &  307.0 &  45.7 &  139.3 &  13.3 &  122.9 \\
2022 &  25.6 &  385.6 &  51.5 &  211.1 &  45.4 &  178.7 \\
2023 &  56.7 &  604.1 &  10.3 &  232.8 &   4.6 &  186.9 \\
2024 &  67.5 & 1,012.0&  36.2 &  317.1 &  23.2 &  230.3 \\
\bottomrule
\end{tabular}
\end{table}

Table \ref{tab:highlevel_totalpay_fig49} summarizes the high-level social inflation numbers for total verdict/settlement amounts obtained in our study. The ``naive'' series (no adjustments) produces much larger cumulative inflation than the case-mix-adjusted series, showing that unadjusted trends can substantially overstate social inflation when the mix of litigated cases changes over time. We therefore advocate for the ``headline'' series as the main social inflation index, because it adjusts for observable case-mix differences while still reflecting inflation beyond general economic inflation. The ``residual'' series additionally removes the effects of evolving litigation strategies and is best interpreted as a decomposition tool: The gap between the headline and residual series indicates how much of the headline trend is associated with changes in observable litigation intensity, while the residual trend reflects social inflation that cannot be explained by those strategic effects.

The headline social inflation results above are based on the full VerdictSearch dataset, which spans various types of liability cases. For practitioners interested in a particular insurance line of business (LoB), we also produce line-specific indices based on proxy LoB groupings constructed from VerdictSearch case-type labels: motor vehicle cases (auto liability), general liability, professional liability, and “others.” We also document meaningful heterogeneity across case environments: Social inflation is more pronounced in cases with corporate defendants, in cases with uninsured defendants, and in states without tort caps or TPLF regulation, broadly echoing prevailing industry hypotheses. Additional detailed social inflation summary tables, reported by these proxy LoB groupings and by other key case characteristics (including defendant type, state regulatory environment, and jury versus bench trials), are provided in Tables \ref{apx:tab:summary_corp} through \ref{apx:tab:summary_trial} in Appendix \ref{apx:table}. 

Finally, we test the “nuclear verdict” hypothesis by comparing adjusted severity inflation across quantiles and find little statistical evidence that the upper tail inflates systematically faster than more moderate outcomes, suggesting that social inflation behaves more like a broad upward shift of the conditional severity distribution than a phenomenon confined to extreme awards.

The remainder of the paper is organized as follows. Section \ref{sec:data} describes the VerdictSearch data and presents summary statistics for the response and explanatory variables. Section \ref{sec:prelim} reports preliminary cross-sectional analyses using logistic and quantile regressions to characterize how case characteristics and litigation strategies relate to outcomes and severities. Section \ref{sec:method} develops the methodology for constructing ASIRs and CSIIs (with uncertainty quantification) across probability and severity channels. Section \ref{sec:result} applies these methods to estimate and compare social inflation dynamics over time and across key subsamples and legal environments. Section \ref{sec:diff} formalizes and evaluates differential social inflation across risk levels to assess whether severity inflation is concentrated in the tail. Finally, Section \ref{sec:discuss} concludes with several avenues for future research directions.

\section{Data} \label{sec:data}

\subsection{Data description} \label{sec:data:desc}

In this project, we primarily focus on data from VerdictSearch, a leading subscription-based database of jury verdicts and settlements maintained by ALM (formerly American Lawyer Media). VerdictSearch collects detailed post-trial and settlement cases from courts across the United States, drawing both on attorney-submitted reports and on ALM’s network of professional editors and reporters, who standardize key information on parties, claims, case types, court locations, liability findings, and damage details. As ALM explained to us over email, “VerdictSearch strives to report as many jury verdicts, decisions and settlements as possible. Although many cases are submitted by attorneys, we also rely on a diligent team of editors and reporters who scour docket lists, cultivate relationships with law firms, and search the Internet and news sources, including the ALM family of legal publications.” This hybrid collection process, together with ALM’s long-standing role as a national legal publisher and its consolidation of multiple regional verdict reporters, makes VerdictSearch one of the most comprehensive and systematically curated sources of US civil trial and settlement outcomes information available to researchers and thus a credible basis for our analysis of social inflation.

From VerdictSearch, we purchased records on 139,575 civil verdict and settlement events with resolution dates spanning from 2002 through early 2025. For the main analysis, we restrict our attention to the 74,188 cases that reached a verdict or settlement between 2009 and 2024. We exclude matters resolved in 2025 because the reporting for that calendar year is still incomplete in our extract. We also do not use cases prior to 2009, as the contemporary notion of social inflation and the associated surge in concern over rapidly escalating liability claim costs and so-called nuclear verdicts becomes salient only in the 2010s. Focusing on the 2009–2024 period therefore aligns the sample window with the era in which social inflation has been most prominently discussed by insurers, reinsurers, and practitioners, while still providing a sufficiently long panel of outcomes to study medium-run trends.

The raw VerdictSearch extract is organized as a set of linked data files that together provide detailed information on each case. The case details file contains one record per case and includes the unique case name and identifier; the case end date (i.e., verdict or settlement date); court name and location; parties involved; case type(s); unstructured textual summaries of the case facts, injuries, and verdict results; jury and trial information; the final outcome (plaintiff win, defendant win, or settlement); and the associated award amount. The injuries file lists, for each case, all injuries or deaths recorded, including the specific injury types for each injured person as well as that person’s age and gender. The insurance file captures the names of insurance carrier(s) and accompanying descriptive text wherever insurers are reported as involved. The plaintiffs and defendants files provide person- or entity-level information on each side of the case: For plaintiffs, this includes name, age, gender, marital status, and related demographics; for defendants, it includes each defendant’s name and an indicator for whether the defendant is a corporate entity. The attorneys file records each attorney’s name, firm information, and role (whether they represent the plaintiff or the defendant), while the experts file similarly lists each expert’s name, title, profile, area(s) of expertise, and whether the expert is retained by the plaintiff or by the defense.

By carefully cleaning and merging these files at the case level, and by extracting as much relevant information as possible from both the structured fields and the unstructured textual components, we constructed a cleaned, analysis-ready dataset. The key variables used in our empirical work are summarized in Tables \ref{tab:summary-stats_response} to \ref{tab:summary-stats_strategic} and can be broadly grouped into three categories: response variables, factual explanatory variables, and strategic explanatory variables. 

\textbf{Response variables} capture the final resolution of the case and form the main objects for measuring social inflation. In particular, a categorical case outcome variable indicates whether the case ends in a plaintiff win (“P”), defendant win (“D”), or settlement (“S”), and a continuous award amount variable contains the total award or settlement amount payable to the plaintiff(s), which is typically zero when the case outcome is “D” and is usually positive when the case outcome is either “P” or “S.” To ensure comparability of dollar amounts over time, we adjust all award and settlement figures using monthly CPI data obtained directly from the US Bureau of Labor Statistics, so that all award values are expressed in December 2024 dollars.

The remaining variables serve as explanatory factors that help account for heterogeneity in case outcomes and allow us to separate changes in the underlying case mix from changes attributable to social inflation. We distinguish between factual explanatory variables and strategic explanatory variables. \textbf{Factual explanatory variables} consist of case characteristics that are largely determined at the outset of litigation and reflect objective features that are difficult to alter through litigation strategy once the dispute is in court. These include the state in which the litigation takes place, the numbers of plaintiffs and defendants, plaintiff demographics (age, gender, marital status, parental status), the number and categorical composition of injuries or deaths, the number and categorical composition of case type(s),\footnote{VerdictSearch classifies cases using case-type descriptors that do not generally map one-to-one to statutory insurance lines. For line-oriented interpretation, we therefore construct LoB proxy subsets using transparent rules. The motor vehicle case-type category (a proxy for the auto line) provides a relatively clean mapping. General liability and professional liability are proxied using text-based filters on VerdictSearch case-type descriptors: General  liability includes premises- and product-liability-type matters (e.g., premises accidents and slips/falls; product liability/breach of warranty; and related categories such as wrongful death), while professional liability includes malpractice and professional-services-type matters (e.g., medical/dental/nursing home negligence, legal malpractice, and other professional negligence categories). Cases not falling into any of the categories above are classified as “others.” These subsets are intended as practitioner-relevant approximations, and we emphasize that they are not definitive statutory LoB classifications.} the calendar year of resolution, indicators for whether defendants are individuals or corporate entities, and indicators for whether the defendants are insured. 

In contrast, \textbf{strategic explanatory variables} are those that can be directly or indirectly shaped by human decisions and litigation tactics and are therefore informative about how evolving strategies may contribute to social inflation. These include the length of the trial phase, the length of jury deliberations, the number and composition of experts retained by each side, and the number of attorneys representing plaintiffs and defendants, respectively. Note that for some continuous variables recorded as proportions in Tables \ref{tab:summary-stats_factual} and \ref{tab:summary-stats_strategic} (e.g., plaintiff gender), the sample means across categories do not sum to 1 because a subset of cases have missing or indeterminate classifications (e.g., gender may not be disclosed if the plaintiff is a corporate plaintiff). For example, sample means of \texttt{NUM\_P\_GEN\_M} and \texttt{NUM\_P\_GEN\_F} are 0.4580 and 0.4250, respectively, so their sum is less than 1 due to these unclassified observations. Specifically, the plaintiff demographic fields (age group, gender, and marital status) can be categorized as ``unspecified.'' The category ``unspecified'' reflects cases in which these demographics are not recorded or not meaningfully defined at the case level, most commonly when the plaintiff is a nonindividual entity (e.g., a corporation or an estate/trust/representative in wrongful-death-type cases) and, more generally, when demographic details are missing or inconsistently reported.

\begin{table}[!h]
\centering
\caption{Summary statistics of response variables.}
\label{tab:summary-stats_response}
\resizebox{\columnwidth}{!}{
\begin{tabular}{llllrr}
\toprule
\multicolumn{1}{c}{Variable name} & \multicolumn{1}{c}{Variable code}    & \multicolumn{1}{c}{Description}                          & \multicolumn{1}{c}{Type} & \multicolumn{1}{c}{Mean} & \multicolumn{1}{c}{Max} \\ \midrule
Case outcome                      & \texttt{OUTCOME\_P} & Indicator that plaintiff wins                            & Categorical              & 0.4375                   & 1                       \\
  & \texttt{OUTCOME\_D} & Indicator that defendant wins                            &                          & 0.2996                   & 1                       \\
  & \texttt{OUTCOME\_S} & Indicator that case results in settlement                &                          & 0.2629                   & 1                       \\ \hline
Award amount                      & \texttt{AWARD.log}  & Log of award amount (CPI adjusted) & Continuous               & \multicolumn{1}{c}{--}   & 26.56                   \\ \bottomrule
\end{tabular}
}
\end{table}

\begin{table}[!h]
\centering
\caption{Summary statistics of factual explanatory variables.}
\label{tab:summary-stats_factual}
\resizebox{\columnwidth}{!}{
\begin{tabular}{llllrr}
\toprule
\multicolumn{1}{c}{Variable name} & \multicolumn{1}{c}{Variable code}    & \multicolumn{1}{c}{Description}                          & \multicolumn{1}{c}{Type} & \multicolumn{1}{c}{Mean} & \multicolumn{1}{c}{Max} \\ \midrule
State                             & \texttt{STATE\_TX}                           & Texas                                                        & Categorical              & 0.1621                   & 1                       \\
  & \texttt{STATE\_CA}                           & California                                                   &                          & 0.1441                   & 1                       \\
  & \texttt{STATE\_NY}                           & New York                                                     &                          & 0.1368                   & 1                       \\
  & \texttt{STATE\_FL}                           & Florida                                                      &                          & 0.0788                   & 1                       \\
  & \texttt{STATE\_PA}                           & Pennsylvania                                                 &                          & 0.0702                   & 1                       \\
  & \texttt{STATE\_NJ}                           & New Jersey                                                   &                          & 0.0700                   & 1                       \\
  & \texttt{STATE\_OH}                           & Ohio                                                         &                          & 0.0424                   & 1                       \\
  & \texttt{STATE\_MI}                           & Michigan                                                     &                          & 0.0394                   & 1                       \\
  & \texttt{STATE\_GA}                           & Georgia                                                      &                          & 0.0411                   & 1                       \\
  & \texttt{STATE\_IL}                           & Illinois                                                     &                          & 0.0404                   & 1                       \\
  & \texttt{STATE\_NEWENG}                       & New England states (Connecticut, Maine, etc.) &                          & 0.0530                   & 1                       \\
  & \texttt{STATE\_CAROLI}                       & North or South Carolina                                      &                          & 0.0381                   & 1                       \\
  & \texttt{STATE\_DCMETR}                       & DC metro (DC, Maryland, [West] Virginia)&                          & 0.0612                   & 1                       \\ \hline
Plaintiff count                   & \texttt{NUM\_P.log}                          & Log of number of plaintiffs& Continuous               & 0.2492                   & 5.53                    \\ \hline
Defendant count                   & \texttt{NUM\_D.log}                          & Log of number of defendants& Continuous               & 0.5170                   & 4.76                    \\ \hline
Plaintiff's age                   & \texttt{NUM\_P\_AGE\_0}                      & Proportion of plaintiffs ages 0--15 years& Continuous             & 0.0315                   & 1                       \\
  & \texttt{NUM\_P\_AGE\_16}                     & Proportion of plaintiffs ages 16--29 years&                        & 0.1190                   & 1                       \\
  & \texttt{NUM\_P\_AGE\_30}                     & Proportion of plaintiffs ages 30--39 years&                          & 0.1332                   & 1                       \\
  & \texttt{NUM\_P\_AGE\_40}                     & Proportion of plaintiffs ages 40--49 years&                          & 0.1745                   & 1                       \\
  & \texttt{NUM\_P\_AGE\_50}                     & Proportion of plaintiffs ages 50--59 years&                          & 0.1540                   & 1                       \\
  & \texttt{NUM\_P\_AGE\_60}                     & Proportion of plaintiffs ages 60--69 years&                          & 0.0788                   & 1                       \\
  & \texttt{NUM\_P\_AGE\_70}                     & Proportion of plaintiffs ages 70+ years&                          & 0.0453                   & 1                       \\ \hline
Plaintiff's gender                & \texttt{NUM\_P\_GEN\_M}                      & Proportion of male plaintiffs                                & Continuous             & 0.4580                   & 1                       \\
  & \texttt{NUM\_P\_GEN\_F}                      & Proportion of female plaintiffs                              &                        & 0.4250                   & 1                       \\ \hline
Plaintiff's marital status& \texttt{NUM\_P\_MAR\_M}                      & Proportion of married plaintiffs                             & Continuous             & 0.2339                   & 1                       \\
  & \texttt{NUM\_P\_MAR\_S}                      & Proportion of single plaintiffs                              &                      & 0.1906                   & 1                       \\ \hline
Plaintiff's parental status& \texttt{NUM\_P\_CHD\_Y}                      & Proportion of plaintiffs with children                       & Continuous             & 0.2471                   & 1                       \\
  & \texttt{NUM\_P\_CHD\_N}                      & Proportion of plaintiffs without children                    &                          & 0.1069                   & 1                       \\ \hline
Injury indicator                  & \texttt{(INJURY\_NUM\textgreater{}0)}        & Indicator for injuries involved                              & Binary                   & 0.7089                   & 1                       \\ \hline
Injury count                      & \texttt{log(pmax(INJURY\_NUM,1))}            & Log of number of injuries& Continuous               & 0.0801                   & 3.33                    \\ \hline
Death indicator                   & \texttt{(INJURY\_NUM\_DEATH\textgreater{}0)} & Indicator for deaths involved                                & Binary                   & 0.0652                   & 1                       \\ \hline
Injury type                       & \texttt{INJURY\_NUM\_TYPE\_ARM}              & Proportion of injuries involving arms                        & Continuous            & 0.0930                   & 1                       \\
  & \texttt{INJURY\_NUM\_TYPE\_BACK}             & Proportion of injuries involving back                        &                          & 0.2498                   & 1                       \\
  & \texttt{INJURY\_NUM\_TYPE\_BODY}             & Proportion of injuries involving body                        &                          & 0.1549                   & 1                       \\
  & \texttt{INJURY\_NUM\_TYPE\_HEAD}             & Proportion of injuries involving head                        &                          & 0.1527                   & 1                       \\
  & \texttt{INJURY\_NUM\_TYPE\_INT}              & Proportion of injuries involving internal organs             &                          & 0.1839                   & 1                       \\
  & \texttt{INJURY\_NUM\_TYPE\_LEG}              & Proportion of injuries involving legs                        &                          & 0.1784                   & 1                       \\
  & \texttt{INJURY\_NUM\_TYPE\_MENT}             & Proportion of injuries involving mental damages              &                          & 0.1388                   & 1                       \\
  & \texttt{INJURY\_NUM\_TYPE\_NECK}             & Proportion of injuries involving neck                        &                          & 0.2936                   & 1                       \\
  & \texttt{INJURY\_NUM\_TYPE\_SURG}             & Proportion of injuries requiring surgery                     &                          & 0.1713                   & 1                       \\ \hline
Case type count                   & \texttt{CASETYPE\_NUM.log}                   & Log of number of case types involved& Continuous               & 1.0133                   & 2.89                    \\ \hline
Case type category                & \texttt{CASETYPE\_NUM\_MOTOR}                & Indicator for case involving motor liability                 & Binary                   & 0.4229                   & 1                       \\
  & \texttt{CASETYPE\_NUM\_GLIAB}                & Indicator for case involving general liability               &                          & 0.1805                   & 1                       \\
  & \texttt{CASETYPE\_NUM\_PLIAB}                & Indicator for case involving professional liability          &                          & 0.0864                   & 1                       \\ \hline
Corporate defendant               & \texttt{IDX\_D\_CORP}                        & Indicator for corporate defendants involved                  & Binary                   & 0.5165                   & 1                       \\ \hline
Insured defendant                 & \texttt{IDX\_INS\_CORP}                      & Indicator for insured defendants involved                    & Binary                   & 0.4982                   & 1                       \\ \hline
Year                              & \texttt{YEAR}                                & Calender year of case end date ($2009,\ldots,2024$)          & Categorical              & --                       & --                     \\ \bottomrule
\end{tabular}
}
\end{table}

\begin{table}[!h]
\centering
\caption{Summary statistics of strategic explanatory variables.}
\label{tab:summary-stats_strategic}
\resizebox{\columnwidth}{!}{
\begin{tabular}{llllrr}
\toprule
\multicolumn{1}{c}{Variable name} & \multicolumn{1}{c}{Variable code}    & \multicolumn{1}{c}{Description}                          & \multicolumn{1}{c}{Type} & \multicolumn{1}{c}{Mean} & \multicolumn{1}{c}{Max} \\ \midrule
Trial length                      & \texttt{TRIAL\_LEN\_1D}             & Indicator for trial length of 1 day                                & Categorical              & 0.0813                   & 1                       \\
  & \texttt{TRIAL\_LEN\_2D}             & Indicator for trial length of 2 days                               &                          & 0.1043                   & 1                       \\
  & \texttt{TRIAL\_LEN\_3D}             & Indicator for trial length of 3 days                               &                          & 0.0949                   & 1                       \\
  & \texttt{TRIAL\_LEN\_4D}             & Indicator for trial length $\geq$4 days and \textless{}1 week      &                          & 0.0657                   & 1                       \\
  & \texttt{TRIAL\_LEN\_1W}             & Indicator for trial length $\geq$1 week and \textless{}2 weeks     &                          & 0.1297                   & 1                       \\
  & \texttt{TRIAL\_LEN\_2W}             & Indicator for trial length $\geq$2 weeks and \textless{}1 month    &                          & 0.0630                   & 1                       \\
  & \texttt{TRIAL\_LEN\_1M}             & Indicator for trial length $\geq$1 month                           &                          & 0.0121                   & 1                       \\ \hline
Jury deliberation length& \texttt{JURY\_LEN\_1I}              & Indicator for jury deliberation \textless{}1 hour                  & Categorical              & 0.0773                   & 1                       \\
  & \texttt{JURY\_LEN\_1H}              & Indicator for jury deliberation $\geq$1 hour and \textless{}2 hours&                          & 0.1075                   & 1                       \\
  & \texttt{JURY\_LEN\_2H}              & Indicator for jury deliberation $\geq$2 hours and \textless{}3 hours&                          & 0.1086                   & 1                       \\
  & \texttt{JURY\_LEN\_3H}              & Indicator for jury deliberation $\geq$3 hours and \textless{}4 hours&                          & 0.0713                   & 1                       \\
  & \texttt{JURY\_LEN\_4H}              & Indicator for jury deliberation $\geq$4 hours and \textless{}1 day&                          & 0.0876                   & 1                       \\
  & \texttt{JURY\_LEN\_1D}              & Indicator for jury deliberation $\geq$1 day and \textless{}2 days&                          & 0.0362                   & 1                       \\
  & \texttt{JURY\_LEN\_2D}              & Indicator for jury deliberation $\geq$2 days                       &                          & 0.0248                   & 1                       \\ \hline
Plaintiff  with expert            & \texttt{(EXPERT\_P\textgreater{}0)} & Indicator for plaintiff experts involved                           & Binary                   & 0.4885                   & 1                       \\ \hline
Plaintiff expert \#               & \texttt{log(pmax(EXPERT\_P,1))}     & Log of number of plaintiff experts involved                        & Continuous               & 0.4454                   & 4.86                    \\ \hline
Plaintiff expert type             & \texttt{EXPERT\_D\_ME}              & Proportion of plaintiff experts specializing in medical& Continuous             & 0.3509                   & 1                       \\
  & \texttt{EXPERT\_D\_BU}              & Proportion of plaintiff experts specializing in business&                          & 0.0512                   & 1                       \\
  & \texttt{EXPERT\_D\_TE}              & Proportion of plaintiff experts specializing in technical&                          & 0.0317                   & 1                       \\
  & \texttt{EXPERT\_D\_AC}              & Proportion of plaintiff experts specializing in accident&                          & 0.0212                   & 1                       \\ \hline
Defendant with expert             & \texttt{(EXPERT\_D\textgreater{}0)} & Indicator for defendant experts involved                           & Binary                   & 0.3898                   & 1                       \\ \hline
Defendant expert \#               & \texttt{log(pmax(EXPERT\_D,1))}     & Log of number of defendant experts involved                        & Continuous               & 0.3157                   & 4.93                    \\ \hline
Defendant expert type             & \texttt{EXPERT\_D\_ME}              & Proportion of defendant experts specializing in medical& Continuous            & 0.2801                   & 1                       \\
  & \texttt{EXPERT\_D\_BU}              & Proportion of defendant experts specializing in business&                          & 0.0247                   & 1                       \\
  & \texttt{EXPERT\_D\_TE}              & Proportion of defendant experts specializing in technical&                          & 0.0304                   & 1                       \\
  & \texttt{EXPERT\_D\_AC}              & Proportion of defendant experts specializing in accident&                          & 0.0226                   & 1                       \\ \hline
Plaintiff attorney \#             & \texttt{NUM\_P\_ATT.log}            & Log of number of plaintiff attorneys                               & Continuous               & 0.2991                   & 3.91                    \\ \hline
Defendant attorney \#             & \texttt{NUM\_D\_ATT.log}            & Log of number of defendant attorneys                               & Continuous               & 0.3722                   & 3.78                   \\ \bottomrule
\end{tabular}
}
\end{table}

\subsection{Summary statistics} \label{sec:data:stat}
\begin{figure}[!h]
  \centering
  \begin{subfigure}[t]{0.48\textwidth}
    \centering
    \includegraphics[width=\linewidth]{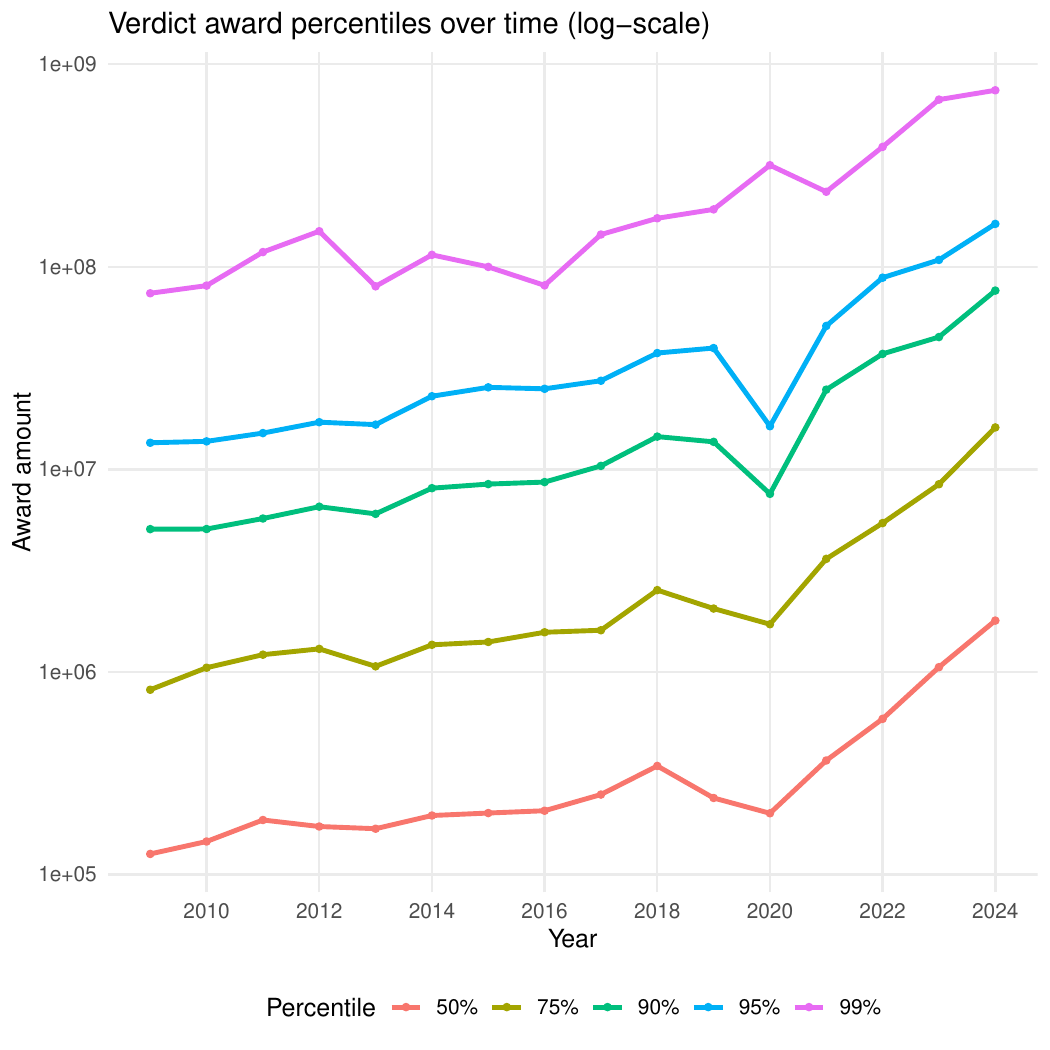}
    \label{fig:award1}
  \end{subfigure}
  \hfill
  \begin{subfigure}[t]{0.48\textwidth}
    \centering
    \includegraphics[width=\linewidth]{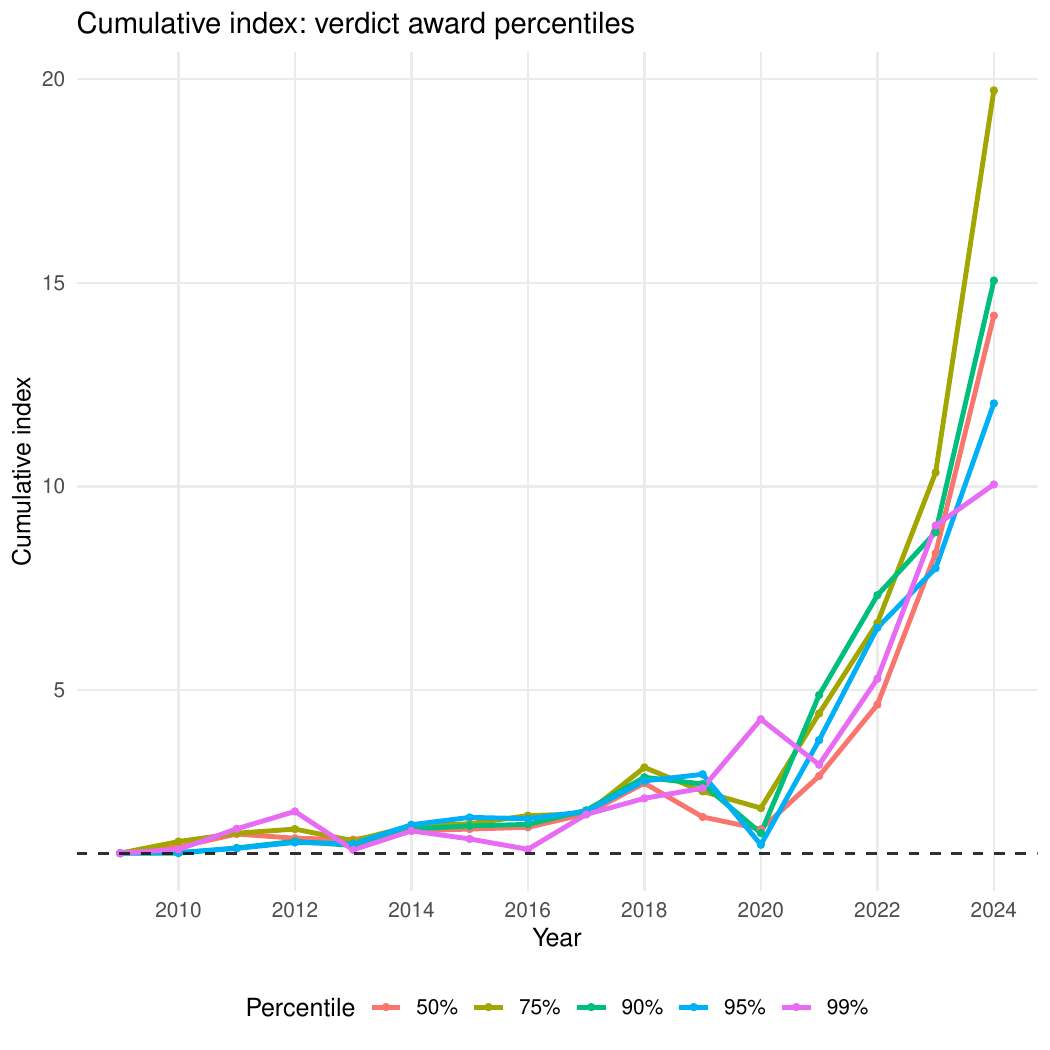}
    \label{fig:award1-index}
  \end{subfigure}
  \caption{Verdict award percentiles: levels (\textit{left}) and cumulative index (\textit{right}).}
  \label{fig:award-byside}
\end{figure}

\begin{figure}[!h]
  \centering
  \begin{subfigure}[t]{0.48\textwidth}
    \centering
    \includegraphics[width=\linewidth]{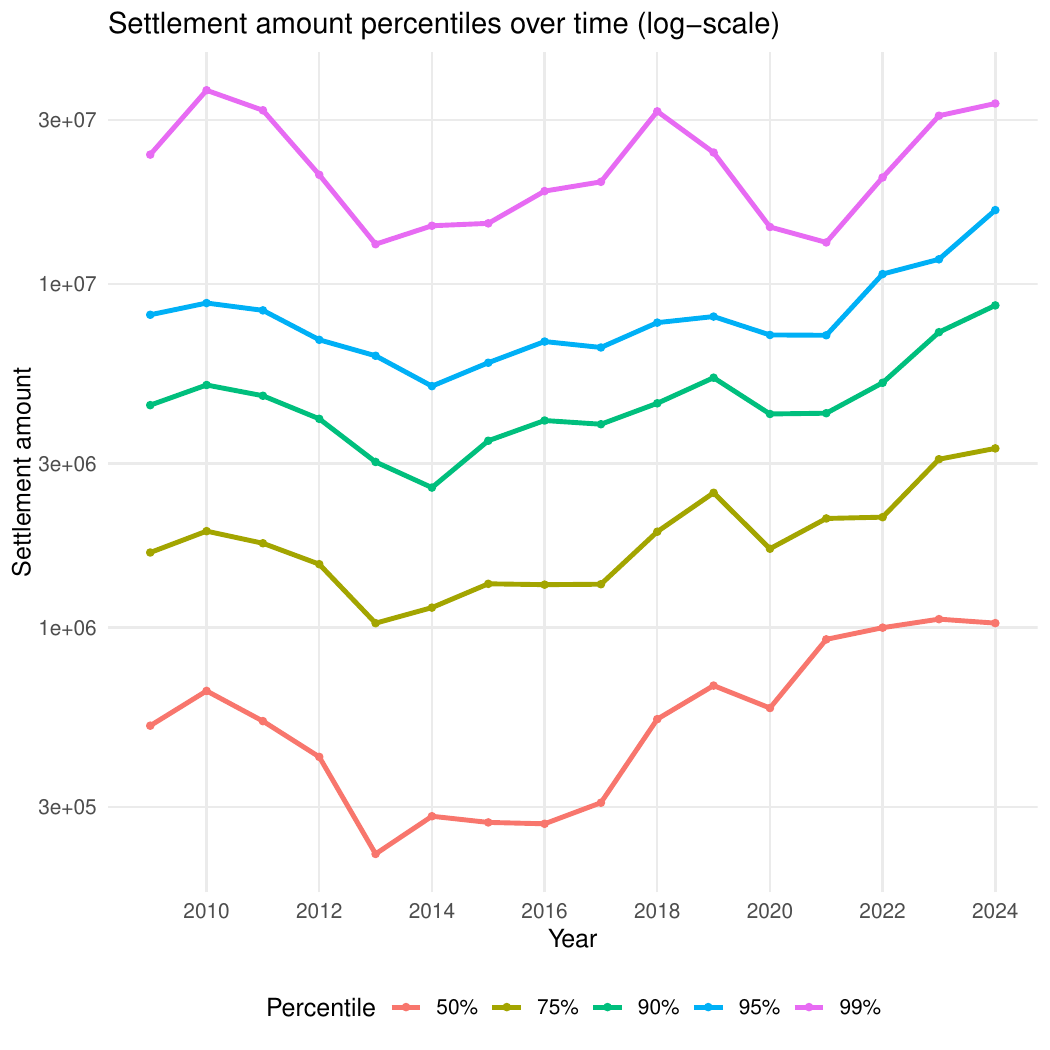}
    \label{fig:award2}
  \end{subfigure}
  \hfill
  \begin{subfigure}[t]{0.48\textwidth}
    \centering
    \includegraphics[width=\linewidth]{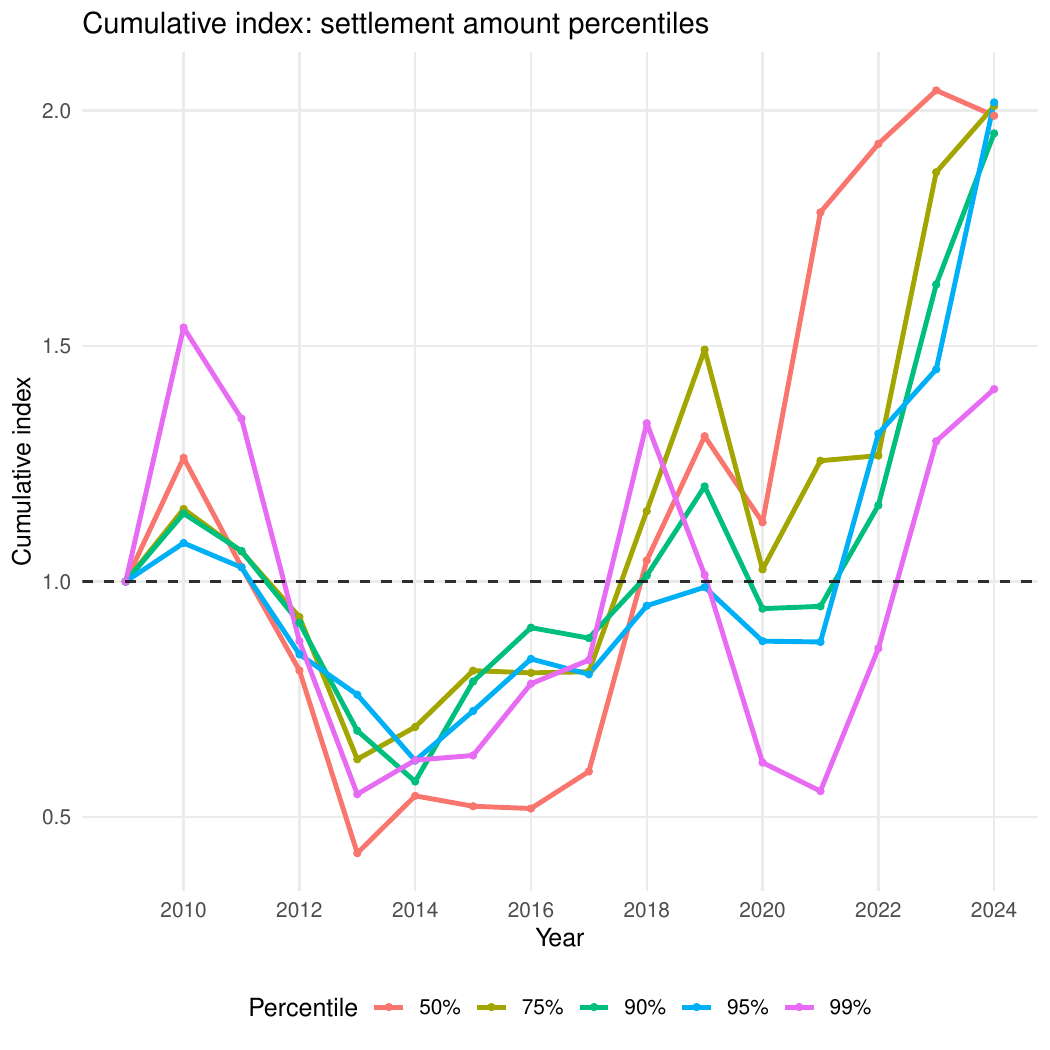}
    \label{fig:award2-index}
  \end{subfigure}
  \caption{Settlement amount percentiles: levels (\textit{left}) and cumulative index (\textit{right}).}
  \label{fig:settlement -byside}
\end{figure}

\begin{figure}[!h]
  \centering
  \begin{subfigure}[t]{0.48\textwidth}
    \centering
    \includegraphics[width=\linewidth]{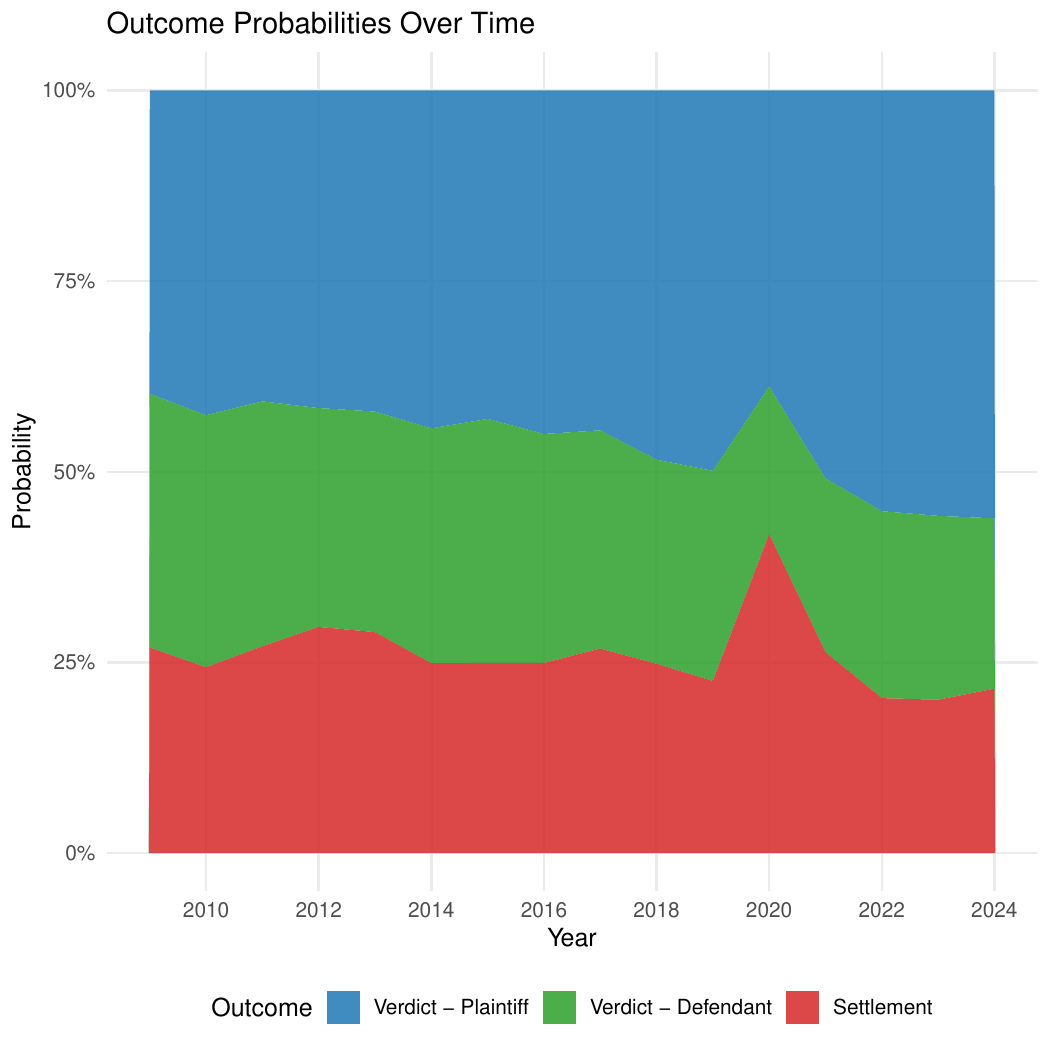}
    \label{fig:award3}
  \end{subfigure}
  \hfill
  \begin{subfigure}[t]{0.48\textwidth}
    \centering
    \includegraphics[width=\linewidth]{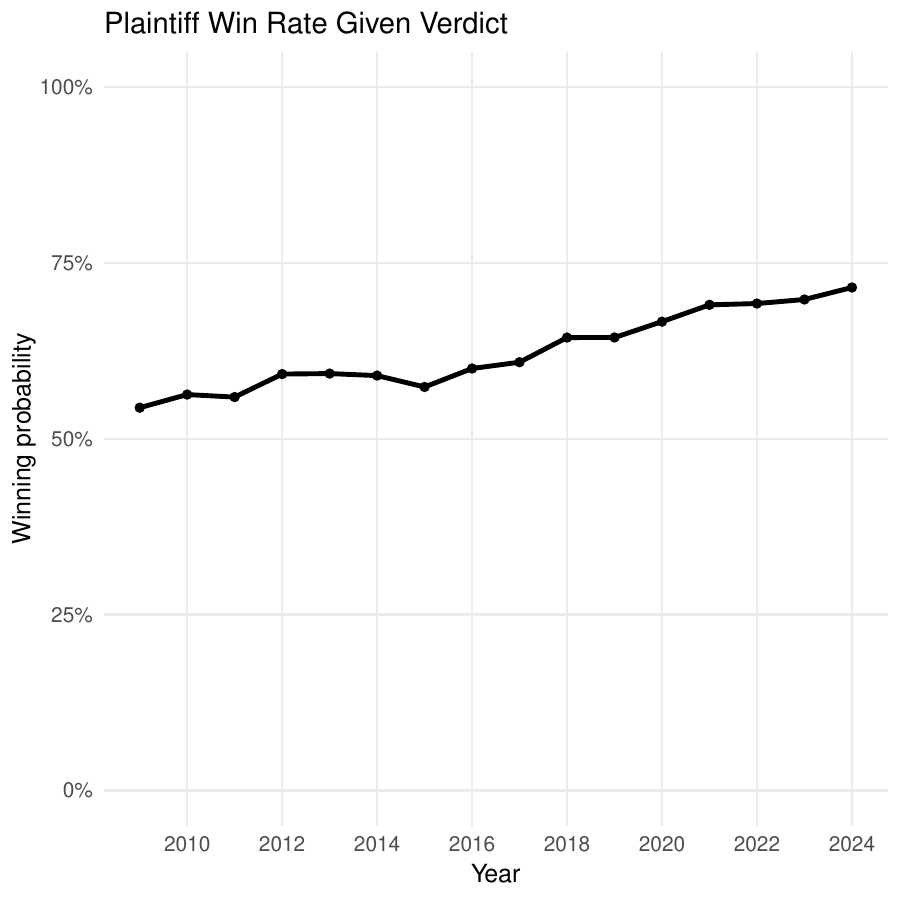}
    \label{fig:award3-index}
  \end{subfigure}
  \caption{Outcome probabilities (\textit{left}) and plaintiff win probability (\textit{right}).}
  \label{fig:outcome -byside}
\end{figure}

\begin{figure}[!h]
\centering
\includegraphics[scale=0.5]{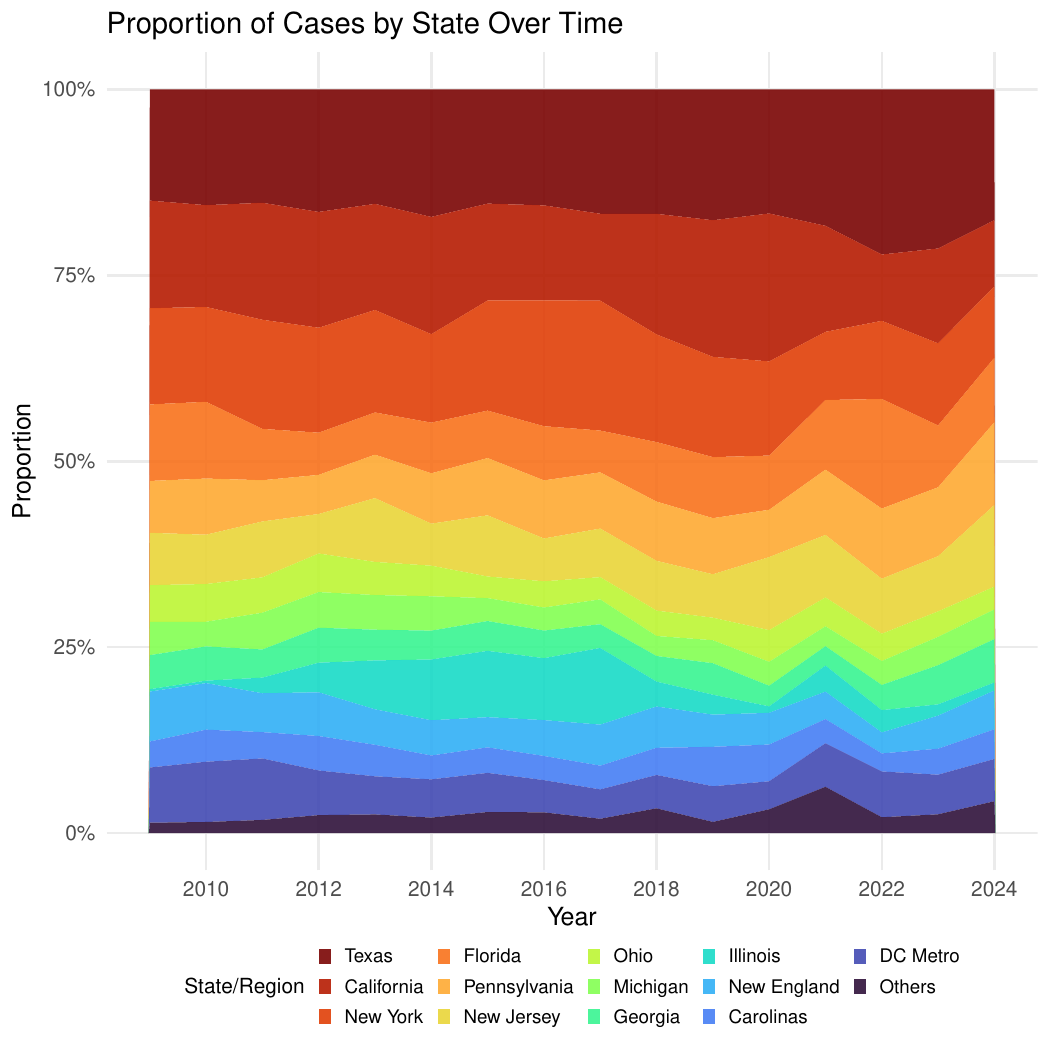}
\caption{Proportion of cases by state.}
\label{fig:case by state}
\end{figure}

\begin{figure}[!h]
\centering
\includegraphics[width=0.5\textwidth]{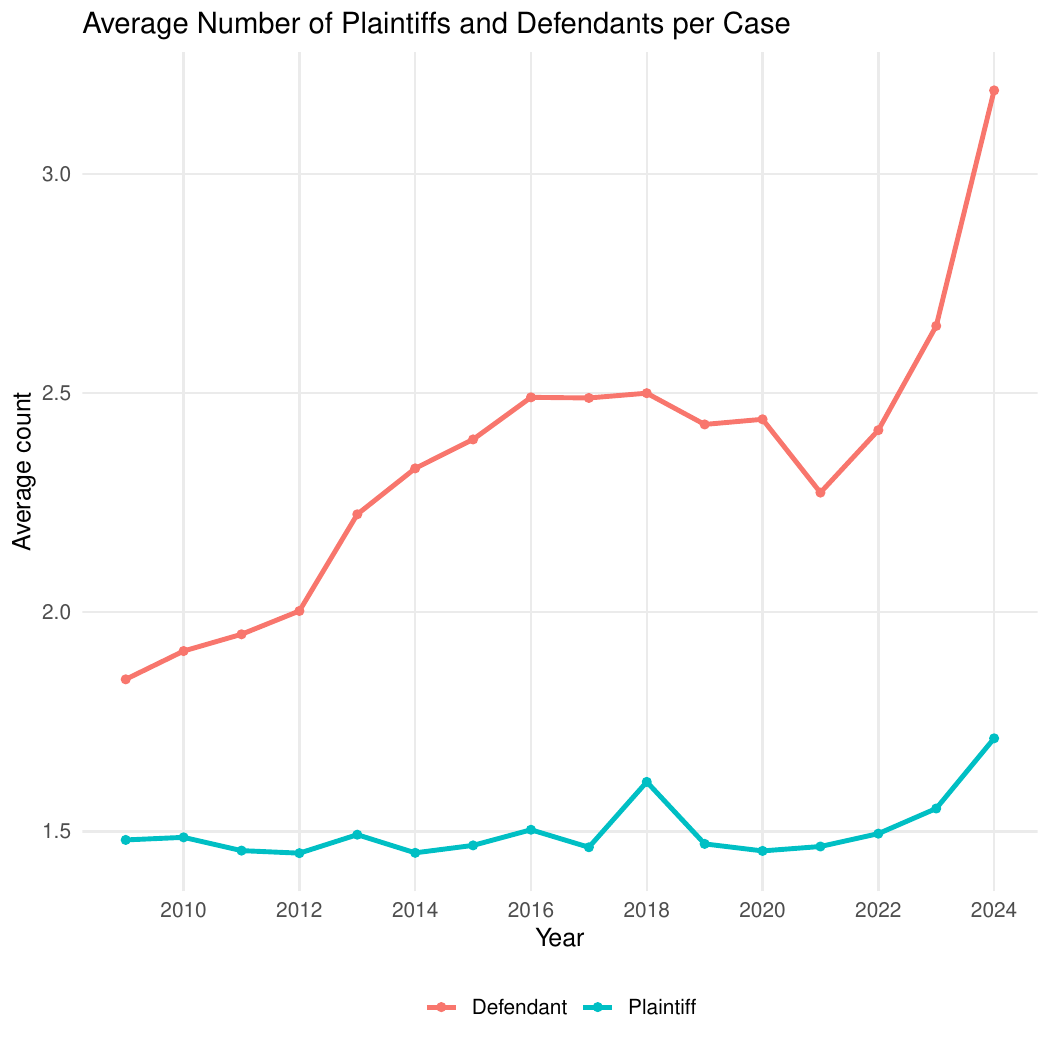}
\caption{Average number of plaintiffs and defendants per case.}
\label{fig:P and D per case}
\end{figure}

\begin{figure}[!h]
  \centering
  \begin{subfigure}[t]{0.48\textwidth}
    \centering
    \includegraphics[width=\linewidth]{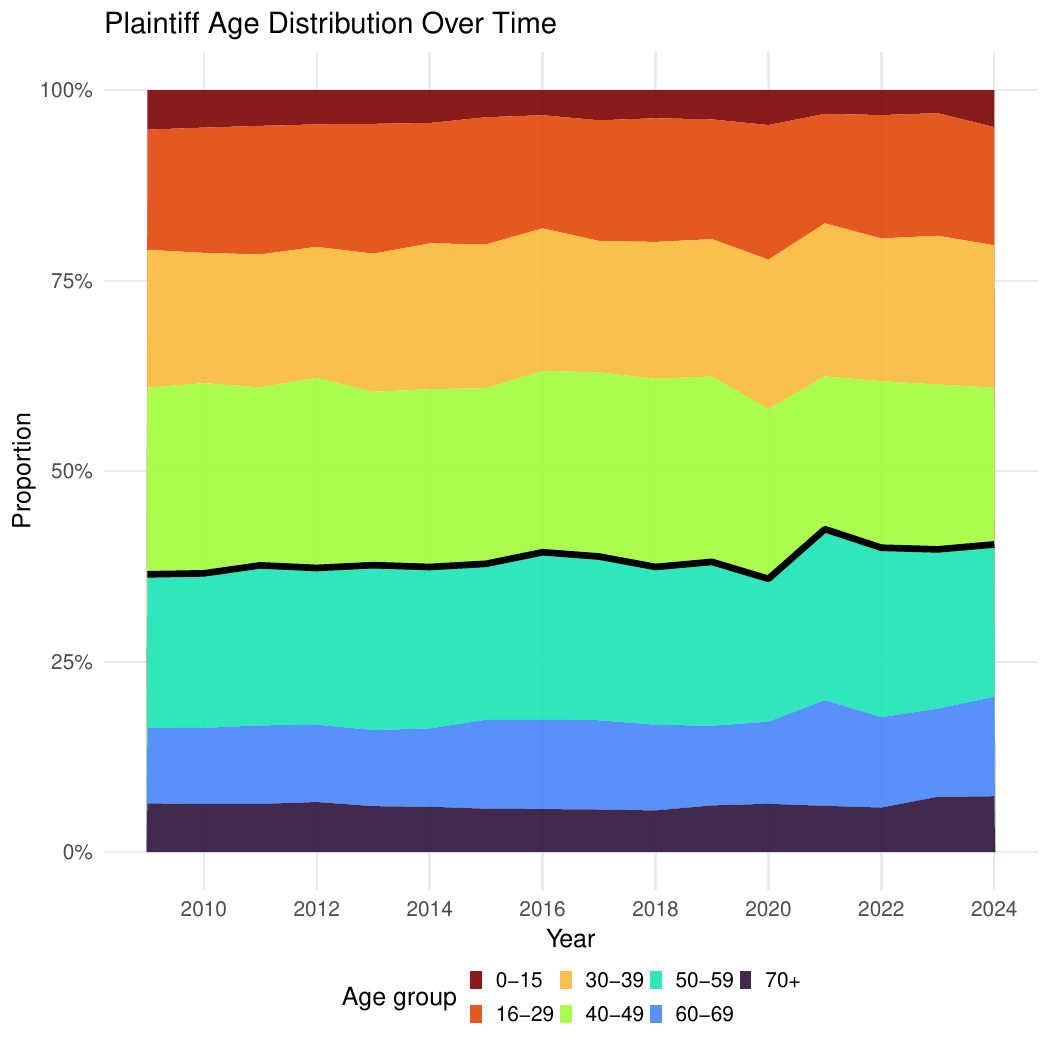}
    \caption{Plaintiff age distribution over time. Black line: \% of plaintiffs $\geq50$ years old.}
    \label{fig:age}
  \end{subfigure}\hfill
  \begin{subfigure}[t]{0.48\textwidth}
    \centering
    \includegraphics[width=\linewidth]{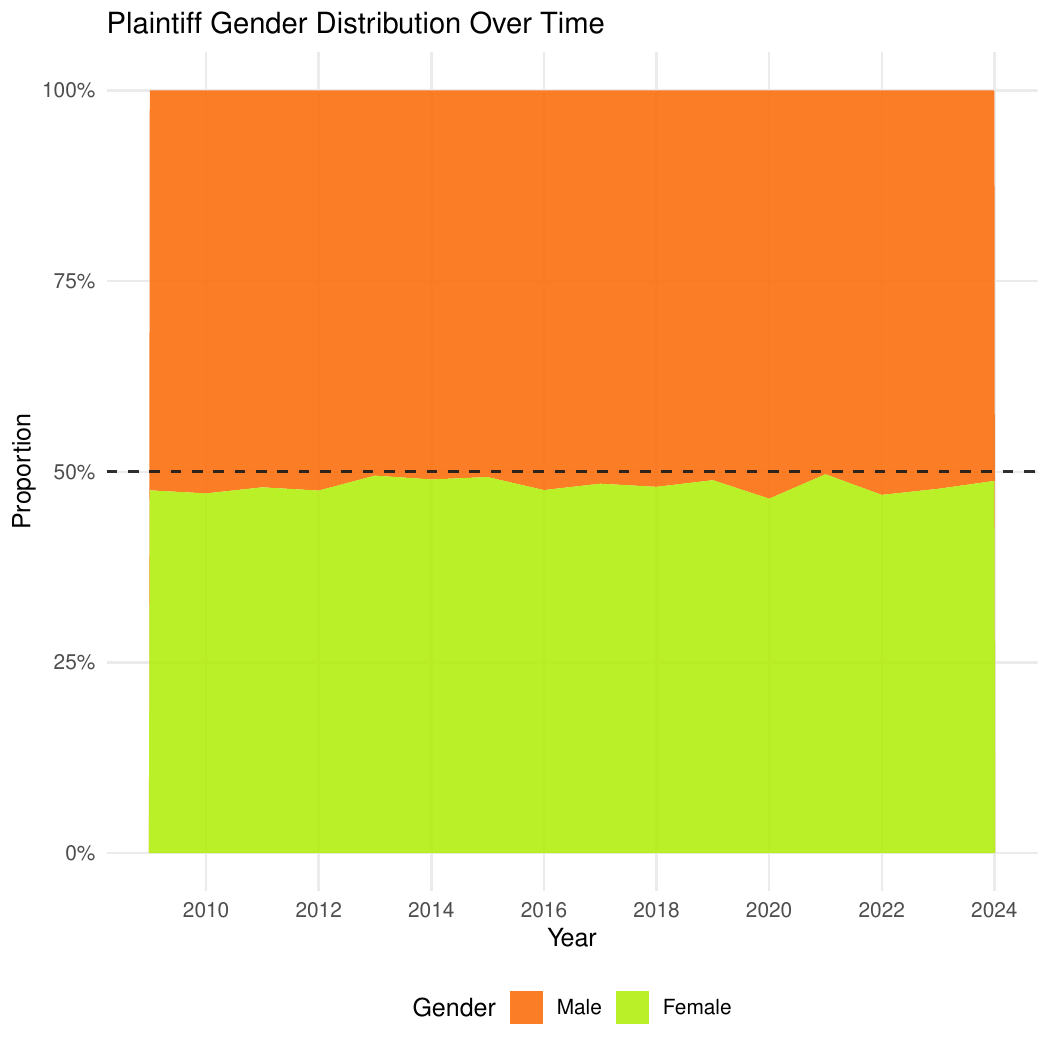}
    \caption{Plaintiff gender distribution.}
    \label{fig:gender}
  \end{subfigure}

  \vspace{0.75em}

  \begin{subfigure}[t]{0.48\textwidth}
    \centering
    \includegraphics[width=\linewidth]{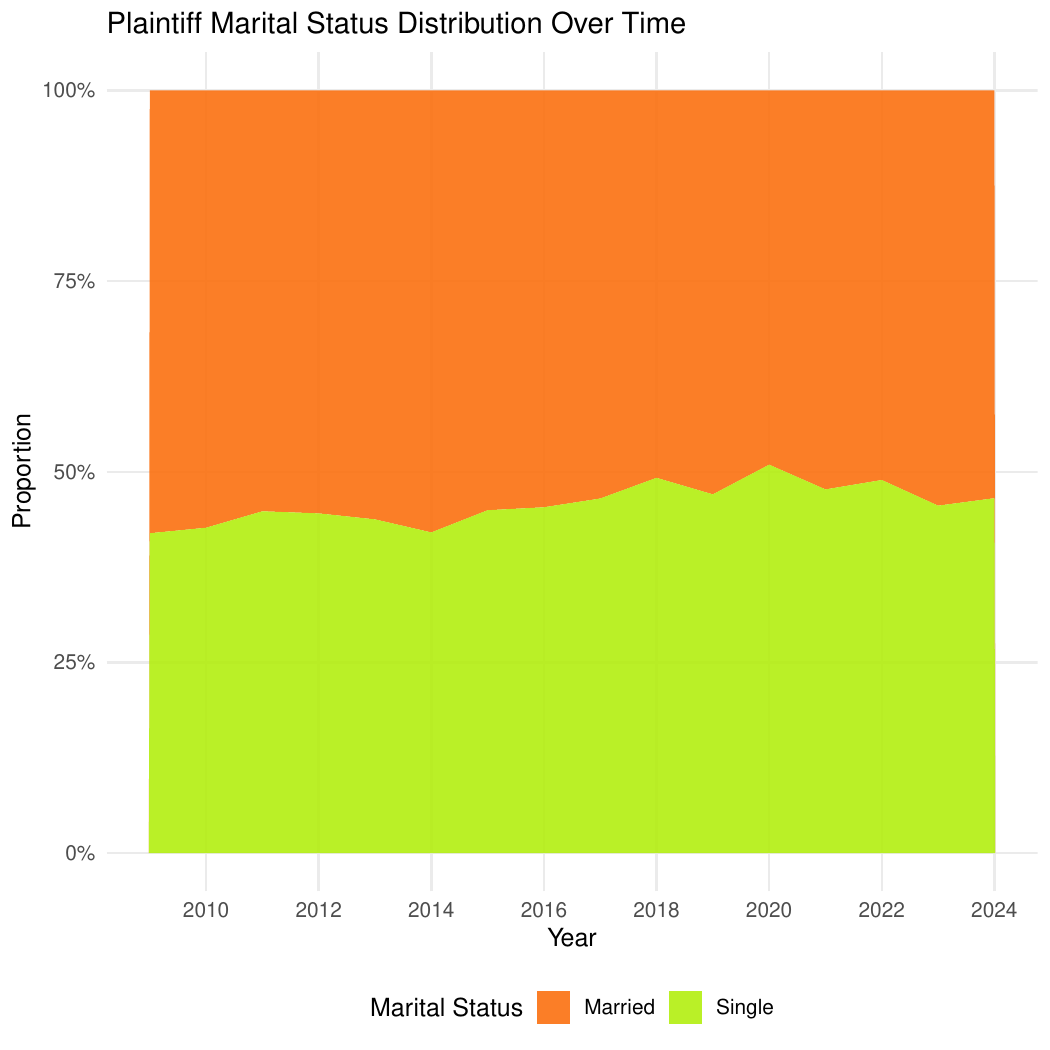}
    \caption{Plaintiff marital status distribution.}
    \label{fig:marital}
  \end{subfigure}\hfill
  \begin{subfigure}[t]{0.48\textwidth}
    \centering
    \includegraphics[width=\linewidth]{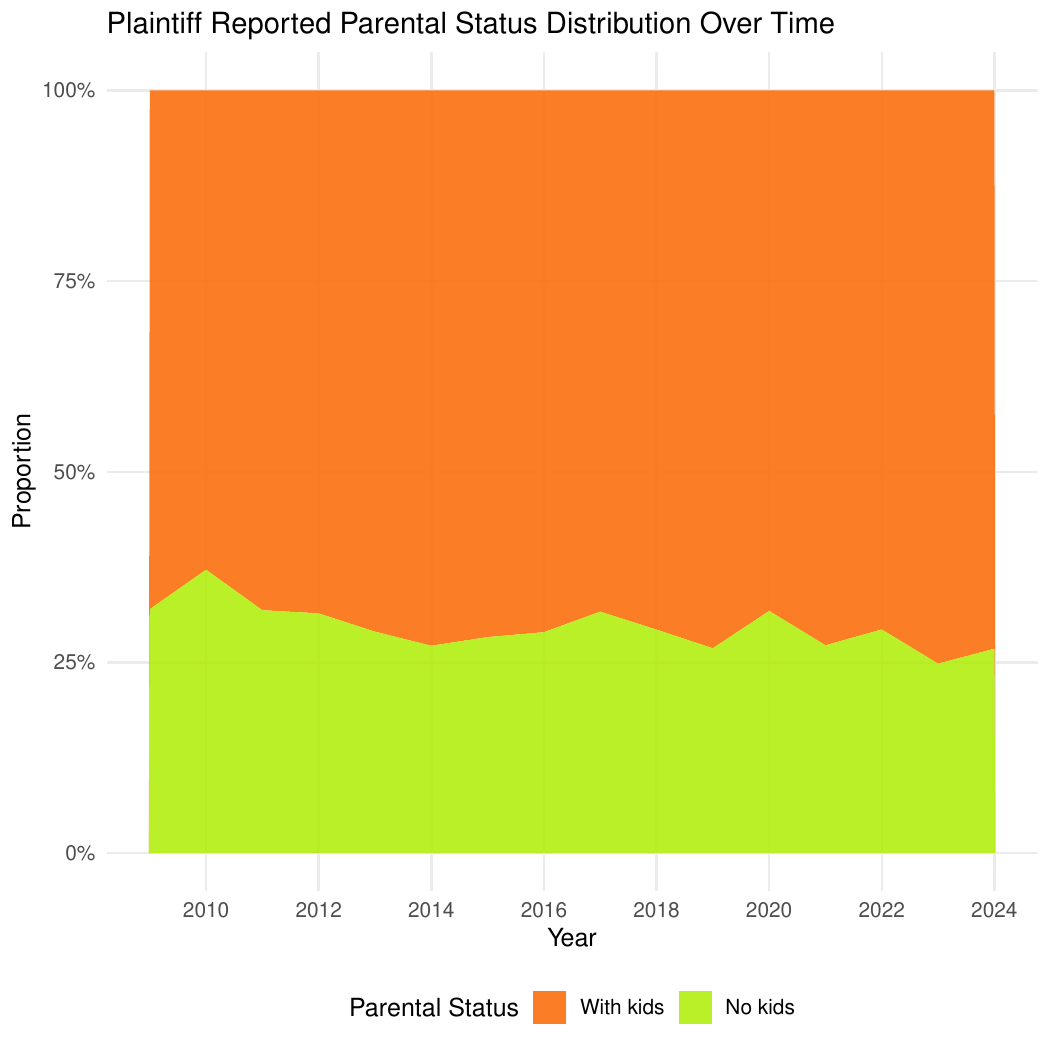}
    \caption{Plaintiff reported parental status distribution.}
    \label{fig:parental}
  \end{subfigure}

  \caption{Plaintiff demographics over time.}
  \label{fig:plaintiff_demographics_2x2}
\end{figure}

\begin{figure}[!h]
  \centering
  \begin{subfigure}[t]{0.48\textwidth}
    \centering
    \includegraphics[width=\linewidth]{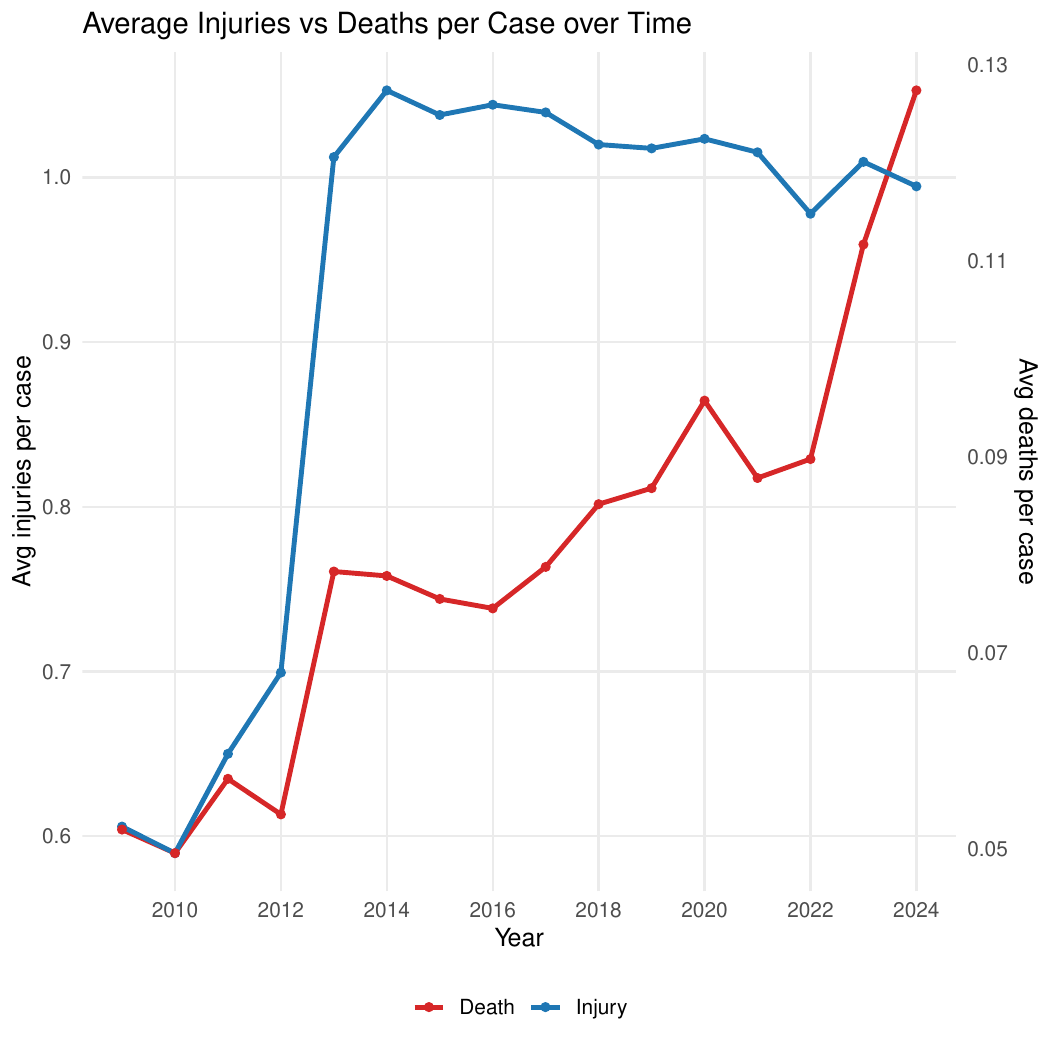}
    \label{fig:Average Injuries}
  \end{subfigure}
  \hfill
  \begin{subfigure}[t]{0.48\textwidth}
    \centering
    \includegraphics[width=\linewidth]{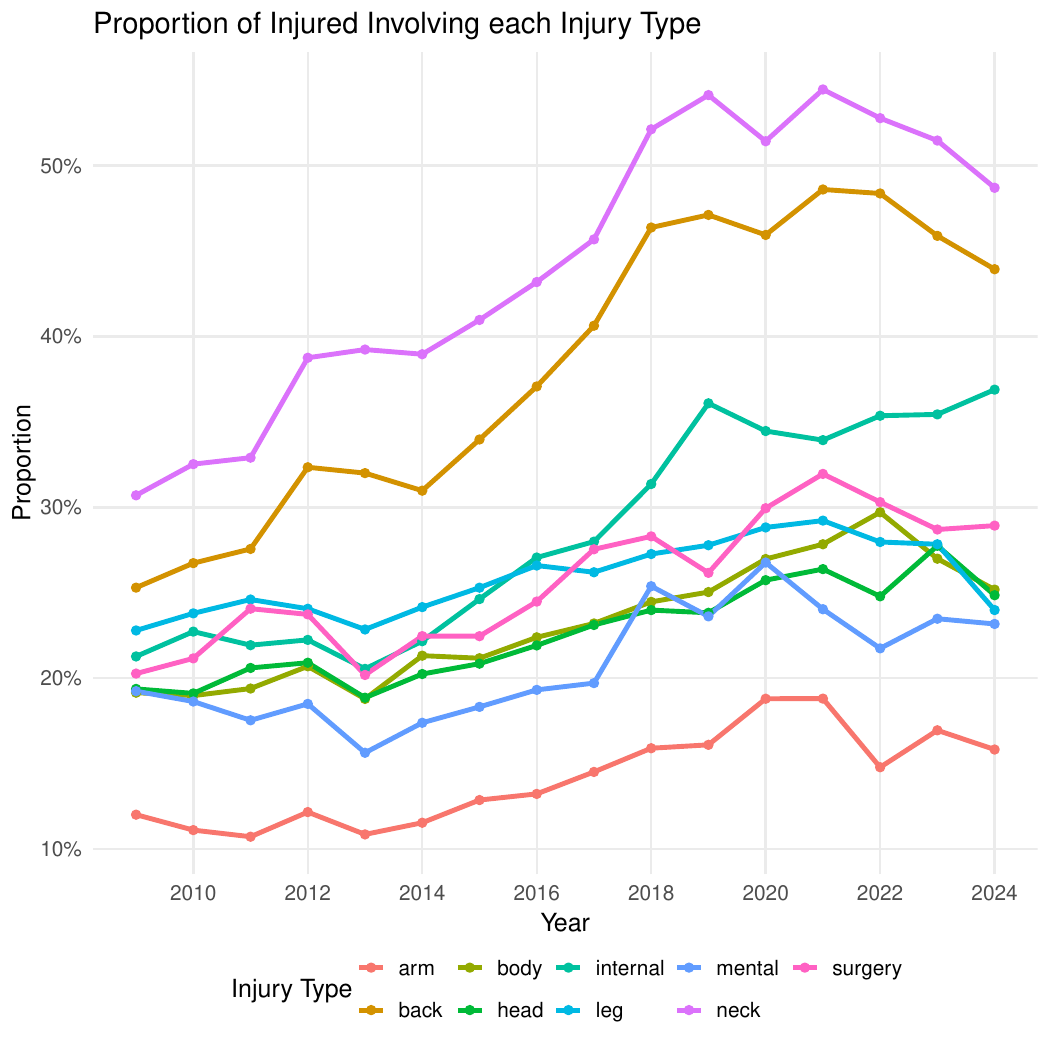}
    \label{fig:injury type}
  \end{subfigure}
  \caption{Average injuries vs. deaths (\textit{left}) and proportion of injury type (\textit{right}).}
  \label{fig:injury-byside}
 \end{figure}

\begin{figure}[!h]
  \centering
  \begin{subfigure}[t]{0.48\textwidth}
    \centering
    \includegraphics[width=\linewidth]{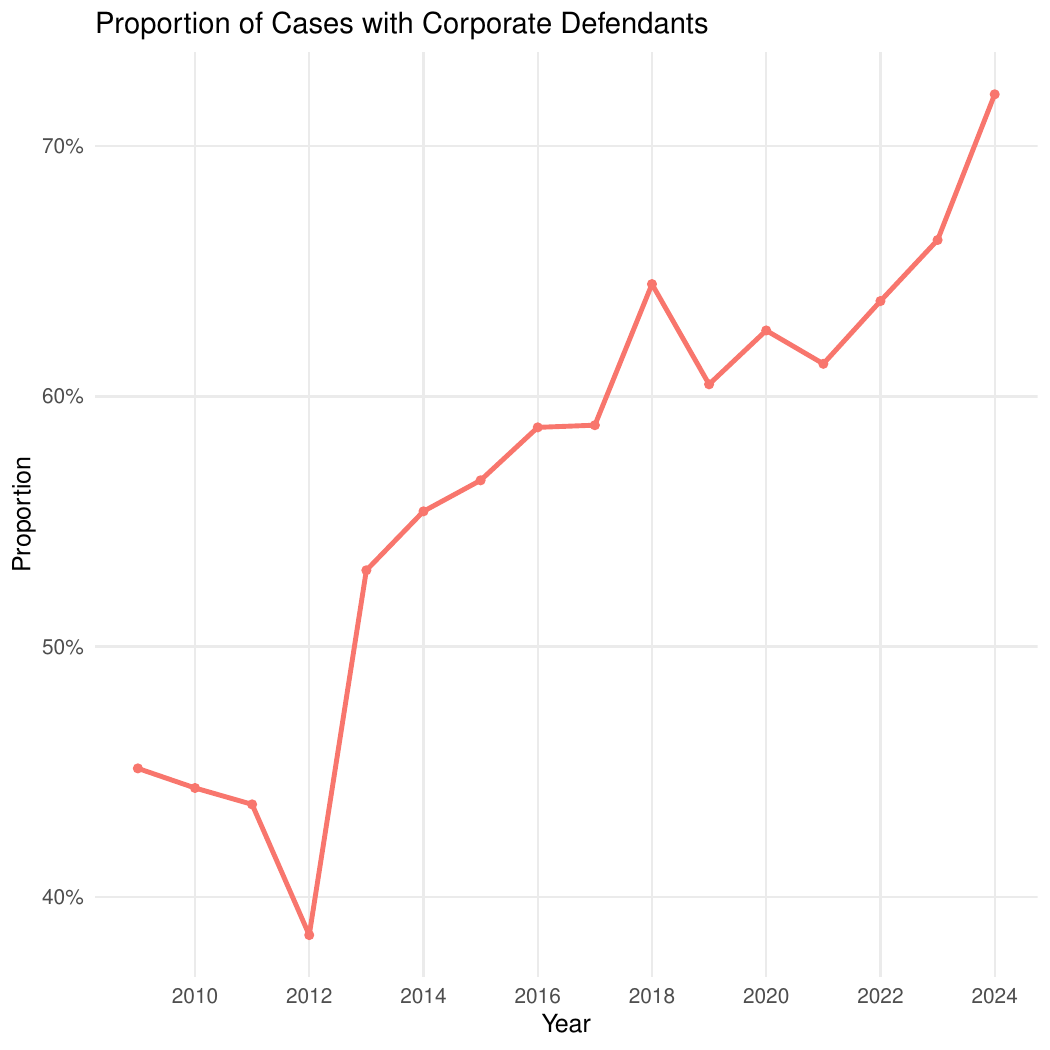}
    \label{fig:Corporate}
  \end{subfigure}
  \hfill
  \begin{subfigure}[t]{0.48\textwidth}
    \centering
    \includegraphics[width=\linewidth]{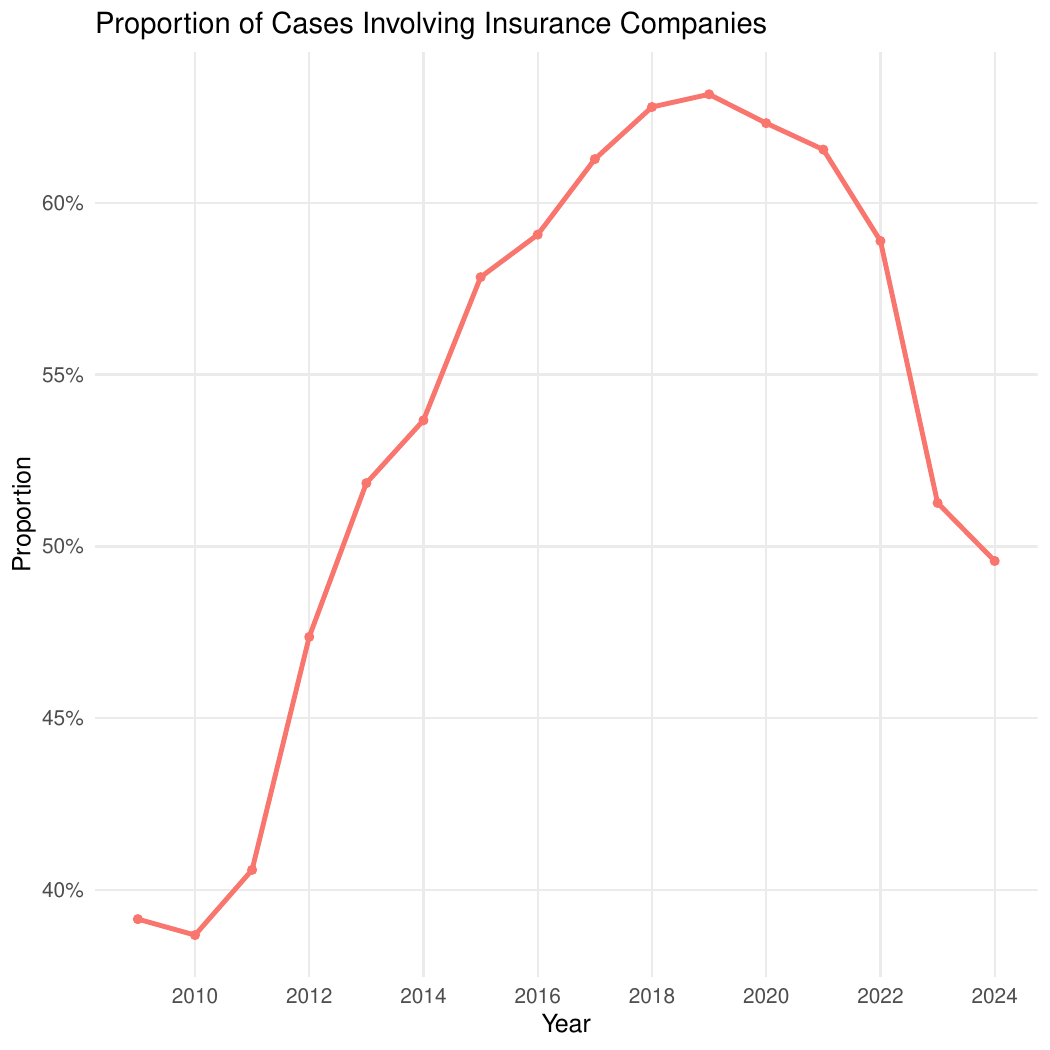}
    \label{fig: Insurance}
  \end{subfigure}
  \caption{Proportion of cases with corporate defendants (\textit{left}) and insured defendants (\textit{right}).}
  \label{fig:Proportion of Cases-byside}
\end{figure}

\begin{figure}[!h]
  \centering
  \begin{subfigure}[t]{0.48\textwidth}
    \centering
    \includegraphics[width=\linewidth]{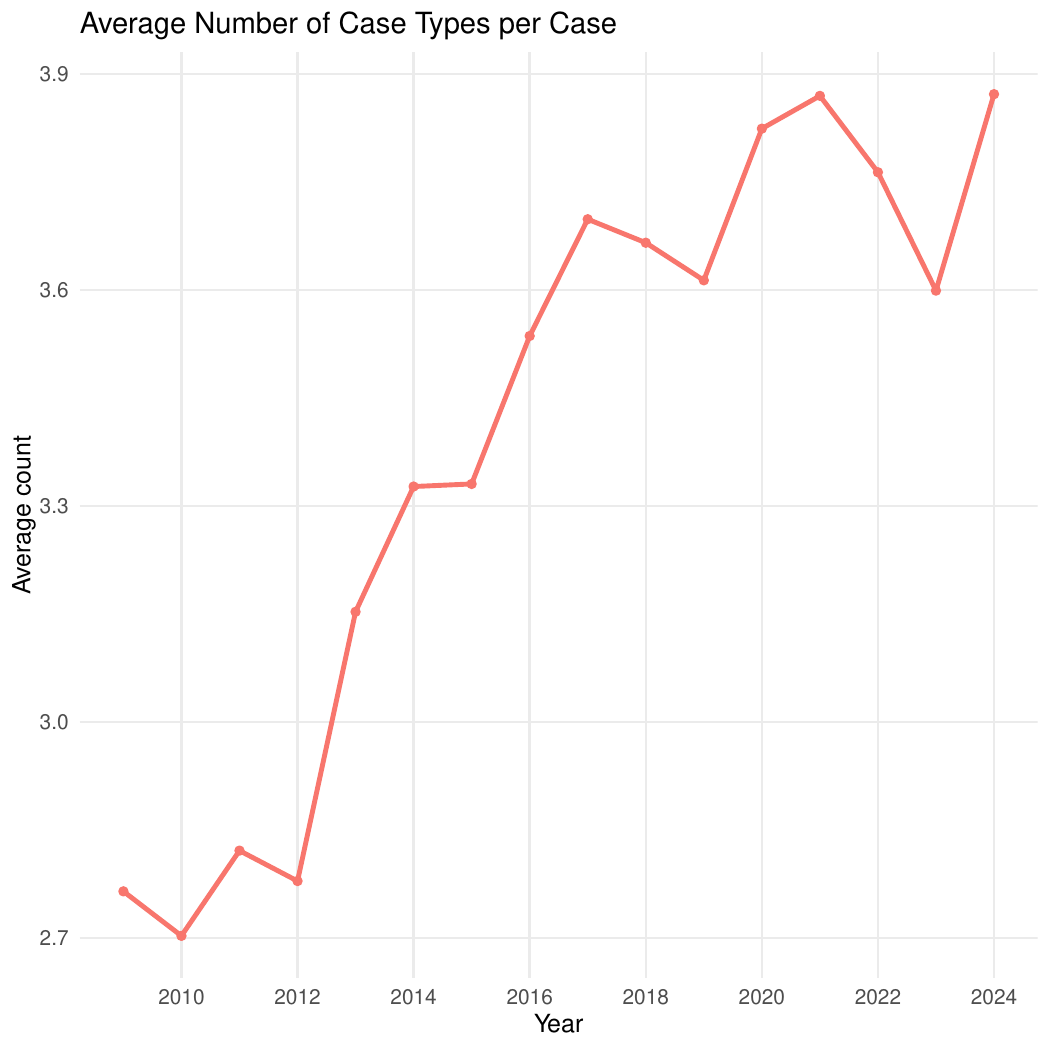}
    \label{fig:Number of Case Types}
  \end{subfigure}
  \hfill
  \begin{subfigure}[t]{0.48\textwidth}
    \centering
    \includegraphics[width=\linewidth]{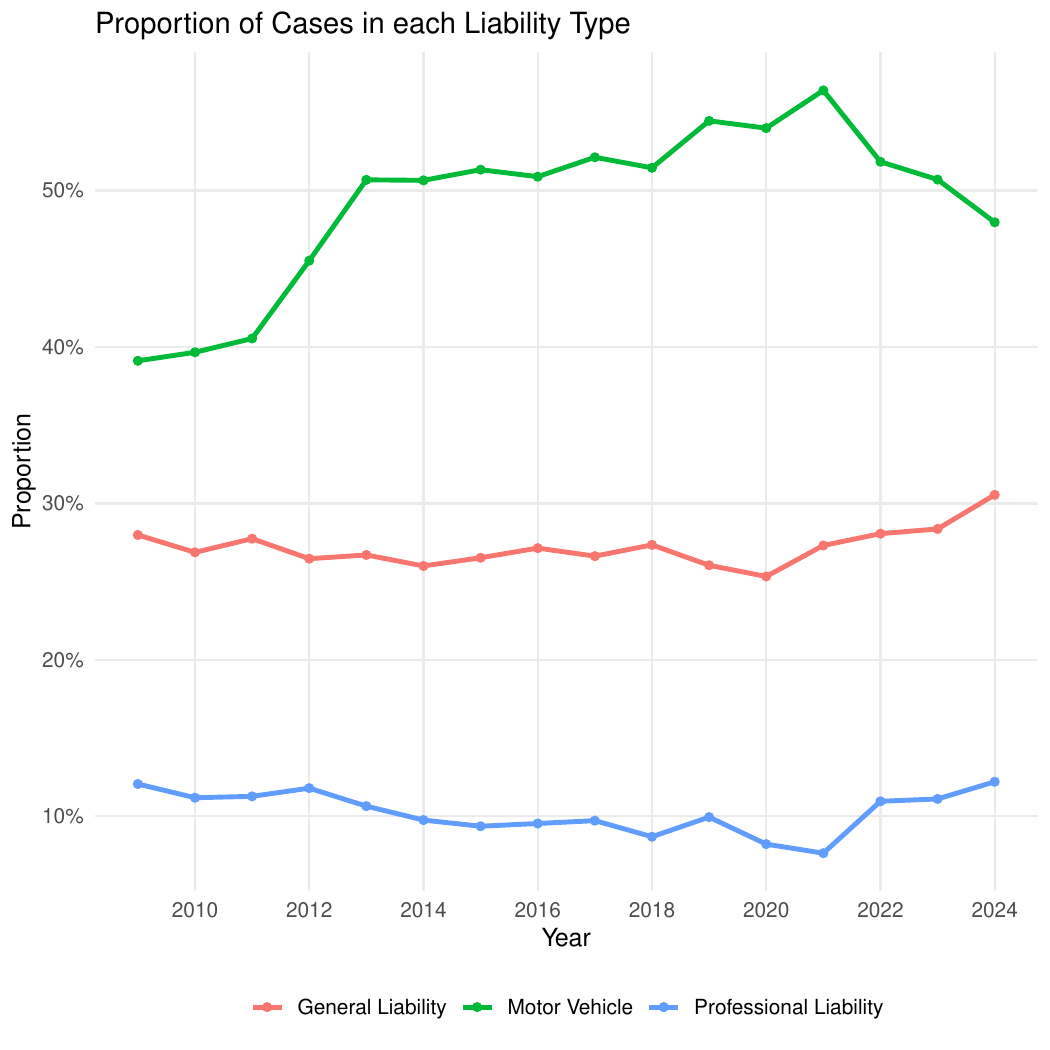}
    \label{fig: Liability Type}
  \end{subfigure}
  \caption{Average number of case types per case (\textit{left}) and proportion of cases in each liability category (\textit{right}).}
  \label{fig:Cases types-byside}
\end{figure}

\begin{figure}[!h]
  \centering
  \begin{subfigure}[t]{0.48\textwidth}
    \centering
    \includegraphics[width=\linewidth]{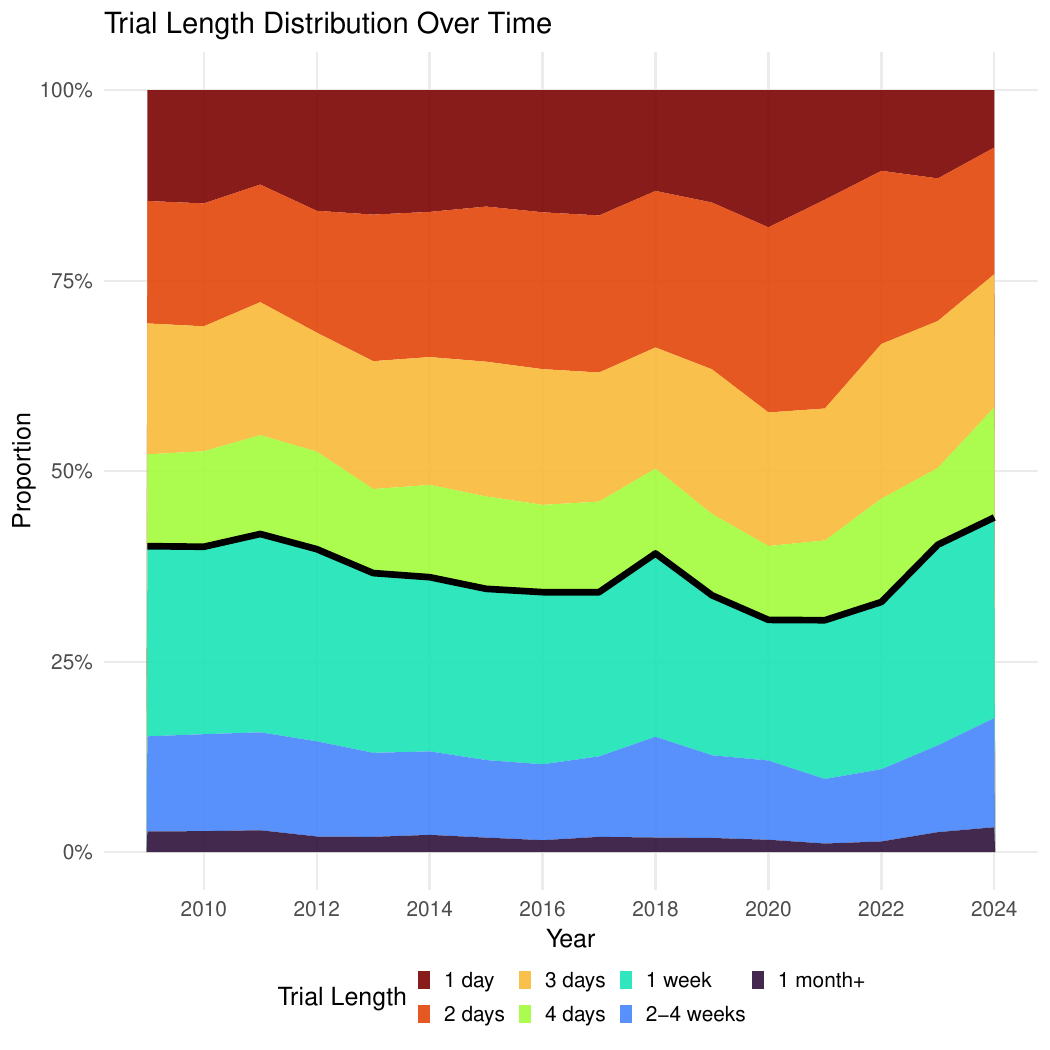}
    \label{fig:Trial}
  \end{subfigure}
  \hfill
  \begin{subfigure}[t]{0.48\textwidth}
    \centering
    \includegraphics[width=\linewidth]{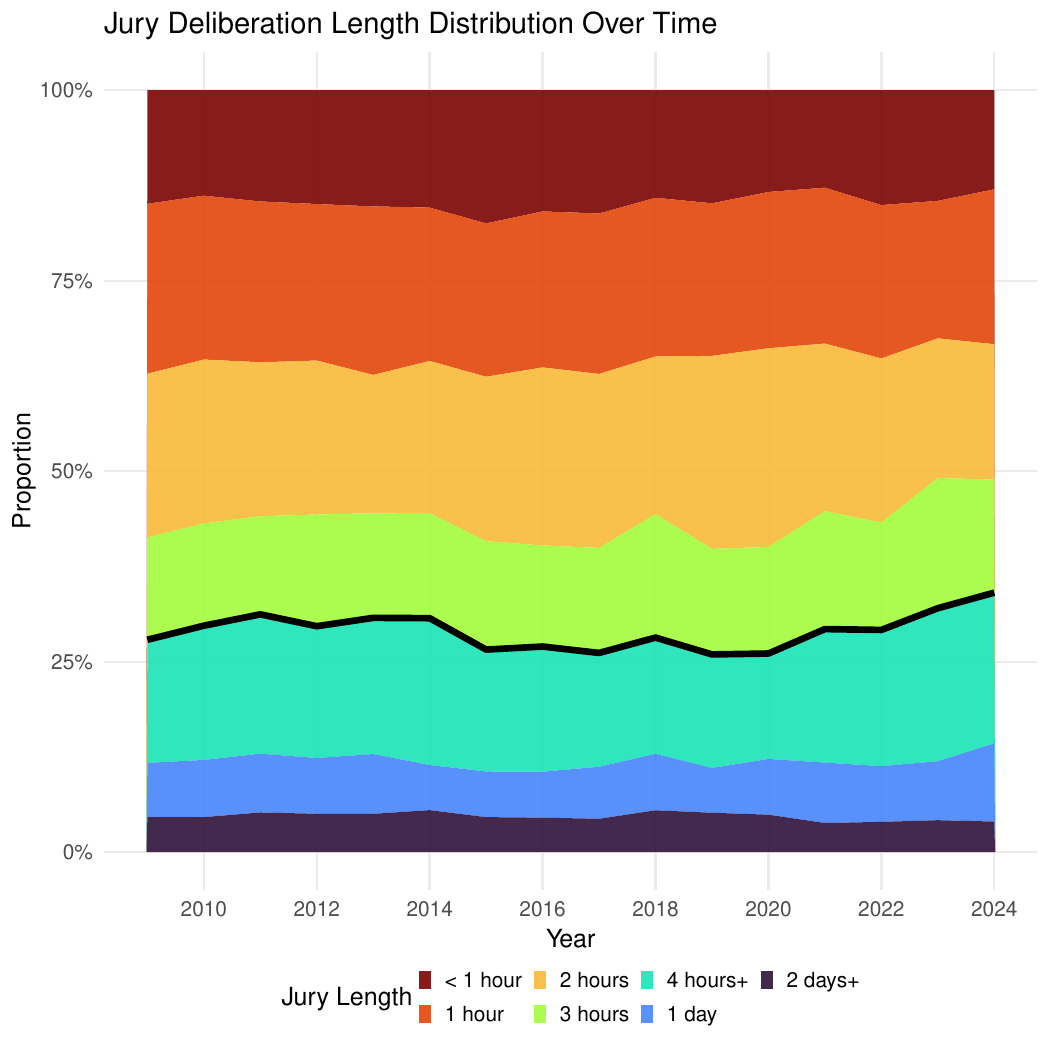}
    \label{fig: Jury}
  \end{subfigure}
  \caption{Composition of cases by trial length (\textit{left}; black line represents the \% of cases with trial length $\geq1$ week) and jury deliberation length (\textit{right}; black line represents the \% of cases with jury deliberation length $\geq4$ hours).}
  \label{fig:Trail and Jury-byside}
\end{figure}

\begin{figure}[!h]
  \centering
  \begin{subfigure}[t]{0.48\textwidth}
    \centering
    \includegraphics[width=\linewidth]{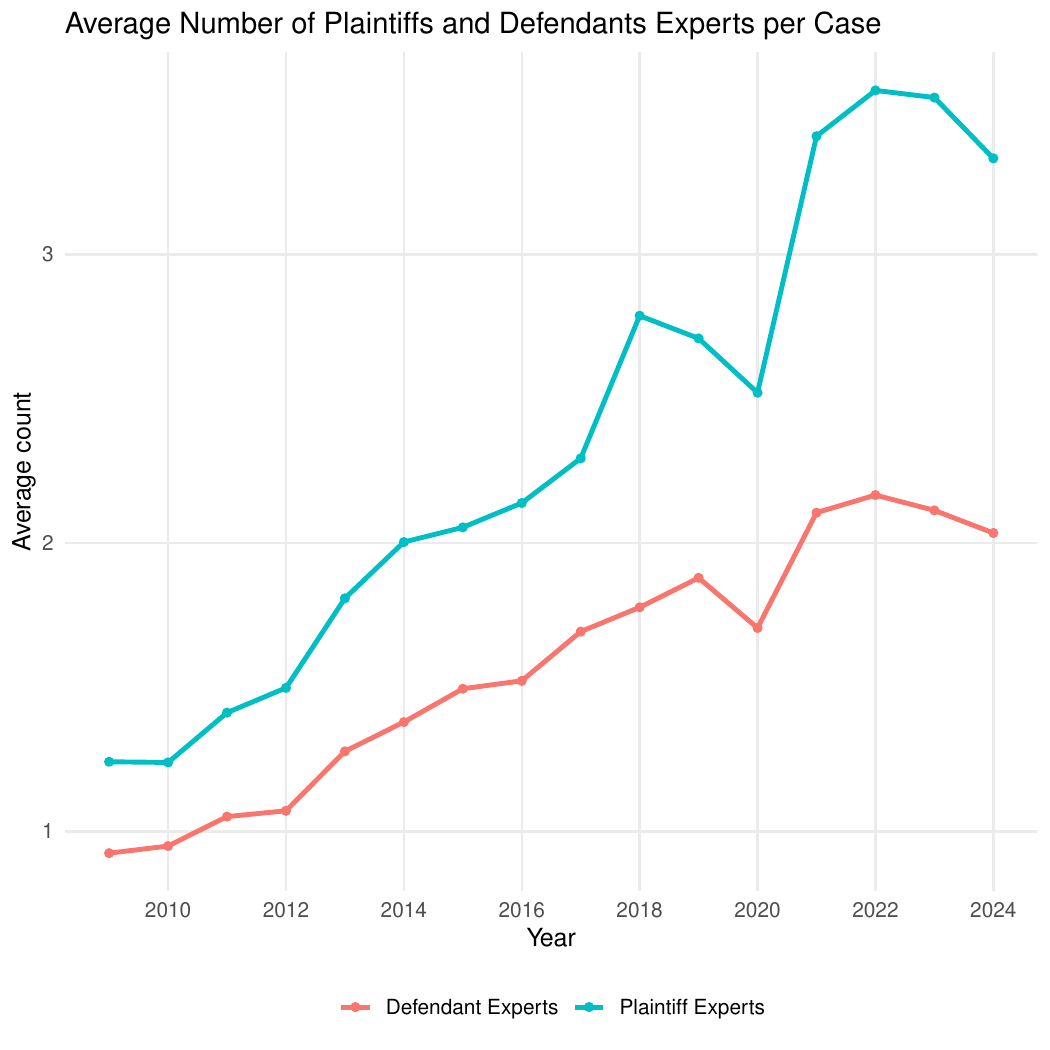}
    \label{fig:Experts}
  \end{subfigure}
  \hfill
  \begin{subfigure}[t]{0.48\textwidth}
    \centering
    \includegraphics[width=\linewidth]{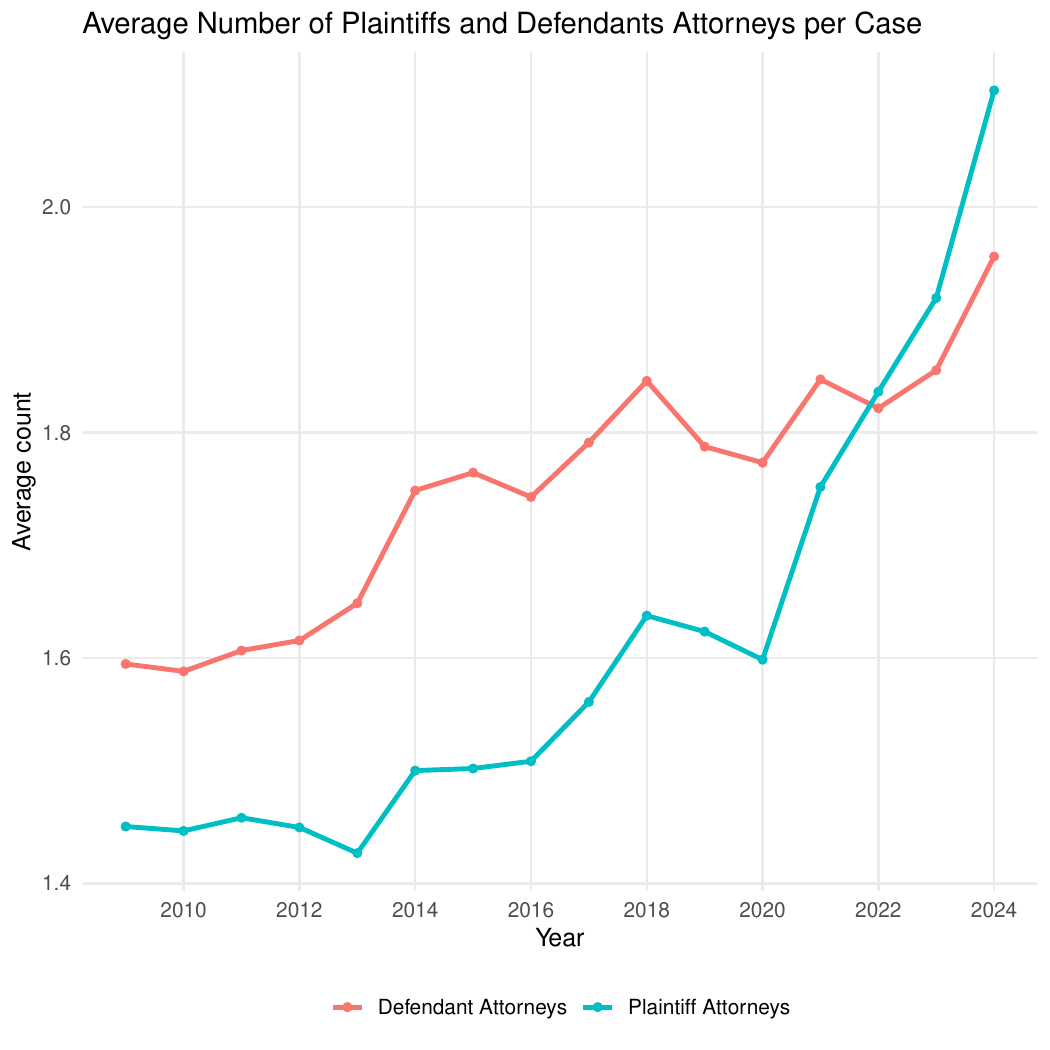}
    \label{fig: Attorneys}
  \end{subfigure}
  \caption{Average number of plaintiff and defendant experts (\textit{left}) and attorneys per case (\textit{right}).}
  \label{fig:Experts and Attorneys -byside}
\end{figure}

\begin{figure}[!h]
  \centering
  \begin{subfigure}[t]{0.48\textwidth}
    \centering
    \includegraphics[width=\linewidth]{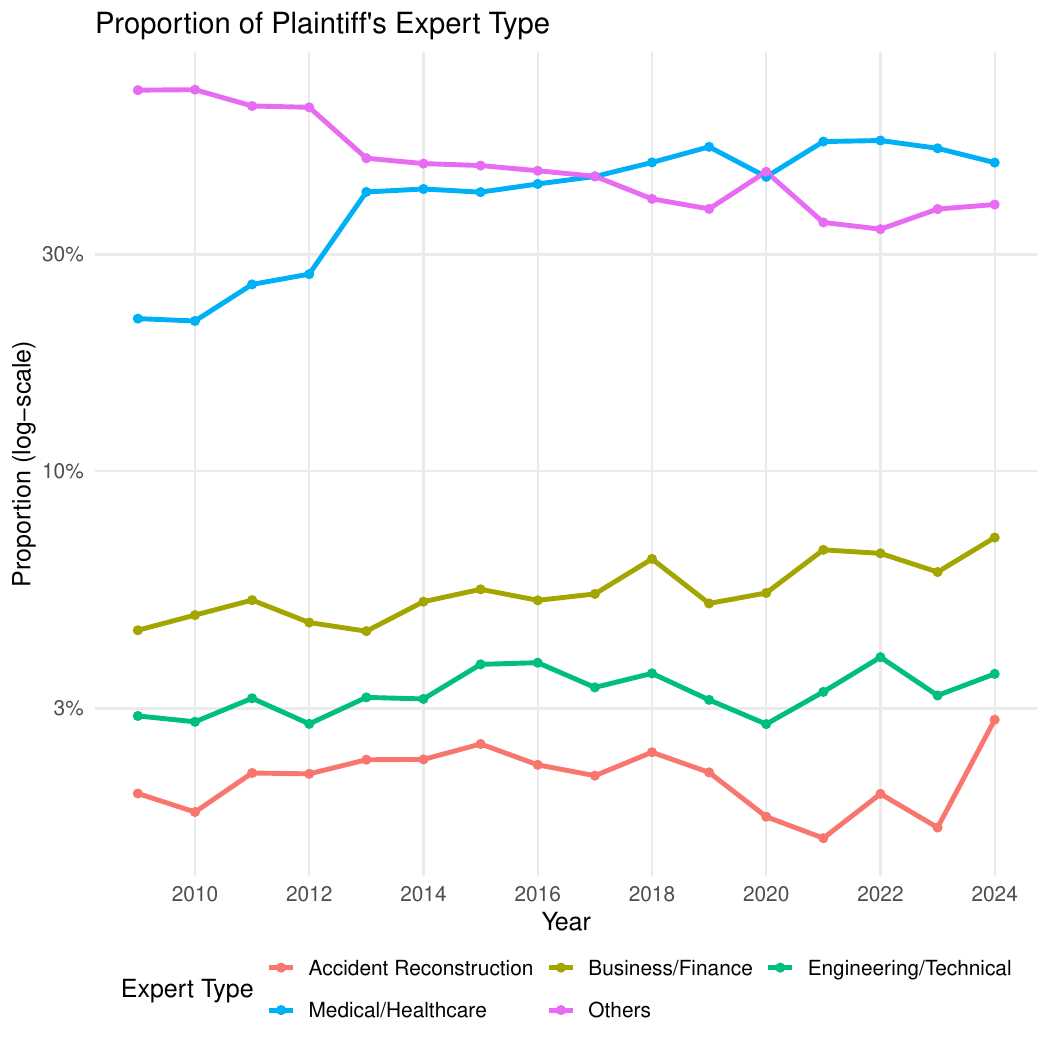}
    \label{fig:P Expert Type}
  \end{subfigure}
  \hfill
  \begin{subfigure}[t]{0.48\textwidth}
    \centering
    \includegraphics[width=\linewidth]{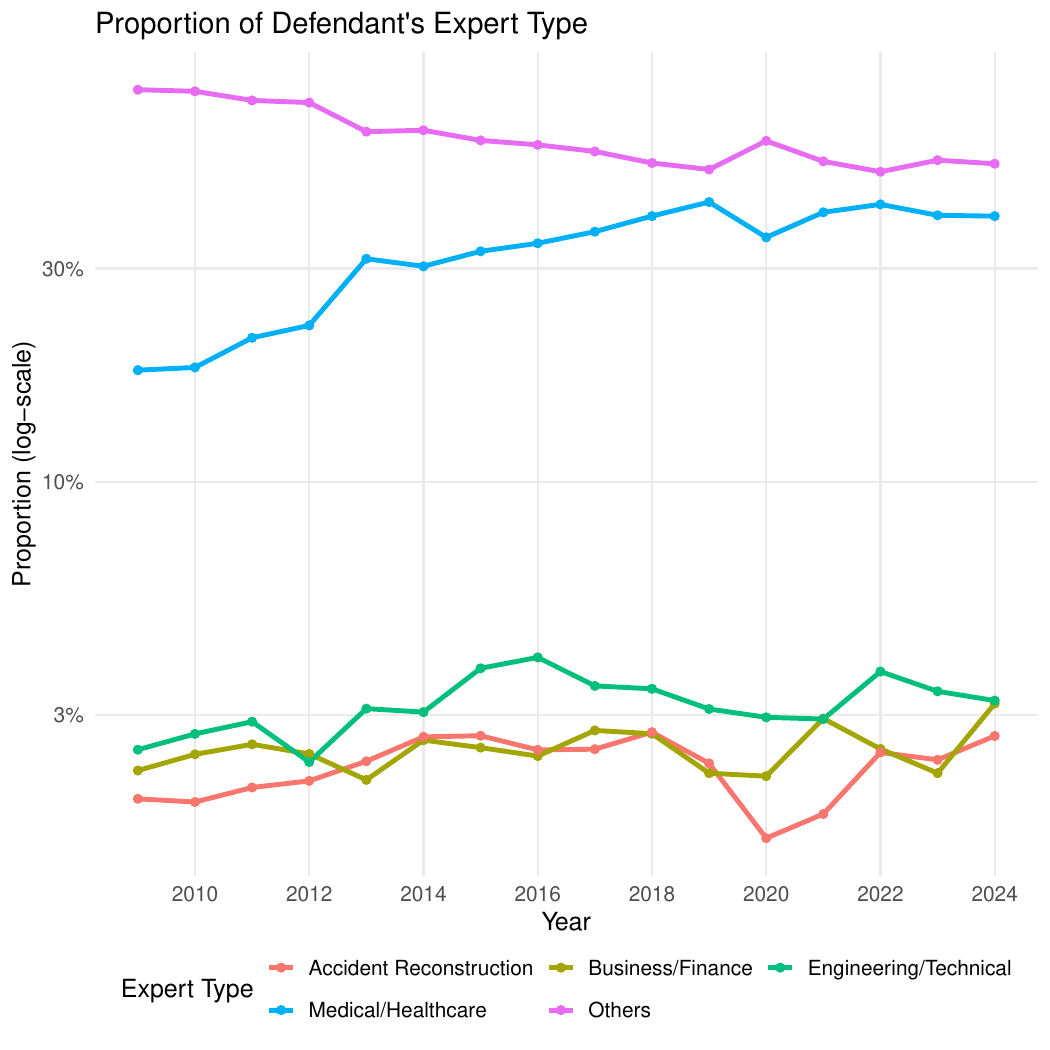}
    \label{fig: D Expert Type}
  \end{subfigure}
  \caption{Plaintiff (\textit{left}) and defendant (\textit{right}) expert-type composition over time.}
  \label{fig:expert type -byside}
\end{figure}

This subsection summarizes the main descriptive patterns in our data and highlights features that are particularly relevant for measuring social inflation. We first examine how the volume of reported cases and the distribution of plaintiff awards and settlements evolve over time (Table \ref{tab:cases-by-year}), then consider the composition of case outcomes (Table \ref{tab:outcome-by-party}), and finally describe temporal changes in key factual and strategic covariates using Figures \ref{fig:award-byside} to \ref{fig:expert type -byside}. Together, these summary statistics provide initial evidence of both selection effects and evolving litigation strategies, and motivate the more formal modeling in later sections.

Table \ref{tab:cases-by-year} reports, by calendar year from 2009 to 2024, the number of cases in our sample and key quantiles of inflation-adjusted plaintiff verdict awards and settlement amounts. The number of VerdictSearch cases in our working sample declines markedly over time, falling from roughly 10,000 cases in 2009 to around 3,000--5,000 cases per year in the late 2010s, and then to only 1,300--1,800 cases per year after 2020. The sharpest drop coincides with the onset of the COVID-19 pandemic, when court closures and the suspension or postponement of jury trials caused substantial case backlogs and pushed many disputes toward negotiated resolution or delay. More broadly, the sustained decline is consistent with the well-documented “vanishing trial” phenomenon in the United States: Civil jury trials are costly, time-consuming, and risky for both sides, and courts increasingly encourage or mandate alternative dispute resolution (ADR) mechanisms such as mediation and arbitration. As a result, a growing share of civil disputes are resolved outside of trial. A further contributing factor is that VerdictSearch is an editorial product rather than a census: Coverage depends on attorney submissions and the efforts of ALM editors and reporters, so capture rates may shift with staffing, jurisdictional emphasis, and submission practices. At the same time, ALM has maintained a consistent collection strategy and explicitly strives to gather as many verdicts, decisions, and settlements as possible, which mitigates (but does not eliminate) the risk that sampling changes alone drive the observed decline.

The shrinking number of reported verdicts has important implications for selection. As trials become rarer, the cases that do proceed to verdict are increasingly likely to be the most severe, complex, and high-stakes disputes; routine, lower-value, or predictable cases are disproportionately filtered out through early settlement or ADR. Consequently, the VerdictSearch sample may become progressively more concentrated in high-stakes cases over time. If one were to track average award or settlement amounts over calendar years without accounting for this changing case mix, the resulting trend would tend to overstate social inflation because it conflates genuine shifts in the underlying loss severity distribution with a selection effect that drops lower-severity cases from the trial population. This observation motivates our later emphasis on controlling for case complexity and other explanatory variables when estimating social inflation.

Table \ref{tab:cases-by-year} also illustrates the extreme heavy-tailed nature of both plaintiff awards and settlements. For verdicts, the 95th percentile is an order of magnitude larger than the median in most years; for example, in 2009 the median plaintiff award was about \$0.13 million, while the 95th percentile was roughly \$13.5 million, with all dollar amounts converted to December 2024 dollars using monthly CPI for All Urban Consumers (CPI-U) to adjust for general economic inflation. Similar, though somewhat less pronounced, patterns appear for settlements. Total annual plaintiff award volumes fluctuate dramatically, driven by a small number of extraordinarily large cases. The most striking example is 2021, when a Texas jury returned a \$301 billion verdict (approximately \$342.1 billion in December 2024 dollars), the largest personal injury award in US history; this single case generates the enormous spike in total plaintiff awards in that year. Overall, the heavy-tailedness of plaintiff verdict awards is noticeably more severe than that of settlements, suggesting that extreme awards are predominantly a verdict phenomenon rather than a settlement phenomenon. For the purpose of measuring social inflation, such heavy tails make traditional average-based indices unsuitable: Just as a consumer price index would be unreliable if a few items could occasionally cost billions of dollars, simple averages of awards are highly unstable and can be dominated by a handful of extreme observations. A central challenge, therefore, is to construct social inflation measures that reflect changes in the loss distribution while remaining robust to these outliers.

Table \ref{tab:outcome-by-party} summarizes the distribution of case outcomes in our sample (from 2009 to 2024): 43.7\% of cases end in a plaintiff verdict (P), 30.0\% in a defendant verdict (D), and 26.3\% in settlement (S). All three outcomes appear in substantial numbers, and none dominates the sample. This balanced composition highlights the importance of modeling how the probabilities of P, D, and S evolve over time and across covariate profiles and of understanding how shifts in the mix of outcomes (rather than changes in award levels alone) contribute to social inflation.

\begin{table}[!h]
\centering
\caption{Number of verdict cases and award quantiles by year (in millions). Cell shading is column-wise: Green indicates smaller values and red indicates larger values within the same column (years 2009--2024).}
\label{tab:cases-by-year}
\resizebox{0.75\columnwidth}{!}{
\begin{tabular}{@{}l r r r r r r r@{}}
\toprule
\textbf{Year} & \textbf{Count} &
\multicolumn{3}{c}{\textbf{Plaintiff award}} &
\multicolumn{3}{c}{\textbf{Settlement amount}}\\
\cmidrule(lr){3-5}\cmidrule(lr){6-8}
 &  & \textbf{50\%} & \textbf{95\%} & \textbf{Total} & \textbf{50\%} & \textbf{95\%} & \textbf{Total} \\
\midrule
2009 & \HeatPct{10,025}{100} & \HeatPct{0.13}{0}  & \HeatPct{13.50}{0}   & \HeatPct{19,130.33}{20}  & \HeatPct{0.52}{40} & \HeatPct{8.14}{67}  & \HeatPct{15,009.77}{100} \\
2010 & \HeatPct{9,282}{93}   & \HeatPct{0.15}{7}  & \HeatPct{13.80}{7}   & \HeatPct{19,979.62}{33}  & \HeatPct{0.65}{67} & \HeatPct{8.80}{80}  & \HeatPct{7,938.85}{93}  \\
2011 & \HeatPct{8,190}{87}   & \HeatPct{0.19}{27} & \HeatPct{15.10}{13}  & \HeatPct{235,397.38}{93} & \HeatPct{0.53}{47} & \HeatPct{8.38}{73}  & \HeatPct{6,703.12}{87}  \\
2012 & \HeatPct{6,783}{80}   & \HeatPct{0.17}{17} & \HeatPct{17.10}{33}  & \HeatPct{21,656.93}{40}  & \HeatPct{0.42}{33} & \HeatPct{6.88}{33}  & \HeatPct{4,333.59}{80}  \\
2013 & \HeatPct{6,539}{73}   & \HeatPct{0.17}{17} & \HeatPct{16.60}{27}  & \HeatPct{17,466.65}{13}  & \HeatPct{0.22}{0}  & \HeatPct{6.18}{13}  & \HeatPct{2,299.69}{53}  \\
2014 & \HeatPct{5,485}{67}   & \HeatPct{0.20}{40} & \HeatPct{23.00}{40}  & \HeatPct{59,838.27}{87}  & \HeatPct{0.28}{20} & \HeatPct{5.04}{0}   & \HeatPct{1,695.44}{27}  \\
2015 & \HeatPct{5,192}{60}   & \HeatPct{0.20}{40} & \HeatPct{25.40}{53}  & \HeatPct{13,786.49}{7}   & \HeatPct{0.27}{10} & \HeatPct{5.90}{7}   & \HeatPct{2,221.12}{47}  \\
2016 & \HeatPct{4,709}{53}   & \HeatPct{0.21}{53} & \HeatPct{25.00}{47}  & \HeatPct{24,633.94}{53}  & \HeatPct{0.27}{10} & \HeatPct{6.80}{27}  & \HeatPct{2,112.11}{40}  \\
2017 & \HeatPct{4,075}{47}   & \HeatPct{0.25}{67} & \HeatPct{27.40}{60}  & \HeatPct{22,931.14}{47}  & \HeatPct{0.31}{27} & \HeatPct{6.54}{20}  & \HeatPct{2,330.58}{60}  \\
2018 & \HeatPct{2,900}{33}   & \HeatPct{0.34}{73} & \HeatPct{37.50}{67}  & \HeatPct{19,698.51}{27}  & \HeatPct{0.54}{53} & \HeatPct{7.72}{53}  & \HeatPct{2,635.35}{73}  \\
2019 & \HeatPct{3,168}{40}   & \HeatPct{0.24}{60} & \HeatPct{39.70}{73}  & \HeatPct{30,404.74}{67}  & \HeatPct{0.68}{73} & \HeatPct{8.04}{60}  & \HeatPct{1,730.55}{33}  \\
2020 & \HeatPct{1,619}{20}   & \HeatPct{0.20}{40} & \HeatPct{16.30}{20}  & \HeatPct{8,008.00}{0}    & \HeatPct{0.58}{60} & \HeatPct{7.11}{47}  & \HeatPct{1,116.40}{20}  \\
2021 & \HeatPct{1,571}{7}    & \HeatPct{0.37}{80} & \HeatPct{51.10}{80}  & \HeatPct{354,766.08}{100}& \HeatPct{0.92}{80} & \HeatPct{7.09}{40}  & \HeatPct{783.32}{0}     \\
2022 & \HeatPct{1,771}{27}   & \HeatPct{0.59}{87} & \HeatPct{88.40}{87}  & \HeatPct{31,015.33}{73}  & \HeatPct{1.00}{87} & \HeatPct{10.69}{87} & \HeatPct{932.13}{7}     \\
2023 & \HeatPct{1,576}{13}   & \HeatPct{1.06}{93} & \HeatPct{108.00}{93} & \HeatPct{26,981.86}{60}  & \HeatPct{1.06}{100}& \HeatPct{11.80}{93} & \HeatPct{1,106.70}{13}  \\
2024 & \HeatPct{1,303}{0}    & \HeatPct{1.79}{100}& \HeatPct{163.00}{100}& \HeatPct{43,217.10}{80}  & \HeatPct{1.03}{93} & \HeatPct{16.41}{100}& \HeatPct{2,393.42}{67}  \\
\midrule
\textbf{Total} & \textbf{74,188} &
\multicolumn{2}{r}{\textbf{—}} & \textbf{968,882.87} &
\multicolumn{2}{r}{\textbf{—}} & \textbf{52,851.14} \\
\bottomrule
\end{tabular}
}
\end{table}

\begin{table}[!h]
\centering
\caption{Verdict outcome summary (for all cases from 2009 to 2024).}
\label{tab:outcome-by-party}
\resizebox{0.38\columnwidth}{!}{
\begin{tabular}{@{}lrr@{}}
\toprule
\textbf{Outcome} & \textbf{Count} & \textbf{Percent} \\
\midrule
Plaintiff (P)  & \num{32456} & \SI{43.7}{\percent} \\
Defendant (D)    & \num{22228} & \SI{30.0}{\percent} \\
Settlement (S) & \num{19504} & \SI{26.3}{\percent} \\
\midrule
\textbf{Total} & \textbf{\num{74188}} & \textbf{\SI{100}{\percent}} \\
\bottomrule
\end{tabular}
}
\end{table}

Figures \ref{fig:award-byside} to \ref{fig:outcome -byside} provide a more detailed view of outcome sizes and probabilities over time. Figure \ref{fig:award-byside} plots, for each calendar year, several percentiles (50th, 75th, 90th, 95th, 99th) of inflation-adjusted plaintiff verdict awards on a log scale (left panel) and the corresponding cumulative index series normalized to 1 in 2009 (right panel). All percentiles show a clear upward trajectory, especially after 2015 and again after the COVID-19 disruption in 2020, providing prima facie evidence that verdict awards have been rising faster than general price inflation. The cumulative index curves for different percentiles are relatively parallel, indicating that the increase is not confined to the extreme tail: Small, moderate, and large verdicts all drift upward at similar relative rates. This pattern suggests that social inflation operates across the entire loss distribution rather than being driven solely by a small number of nuclear verdicts. Nevertheless, given the selection concerns noted above, these descriptive trends must be interpreted cautiously and will be revisited in models that control for case characteristics.

Figure \ref{fig:settlement -byside} presents analogous plots for settlement amounts. Settlement percentiles also trend upward over time, but the slopes are much flatter than those for verdicts, and the post-2015 and post-2020 accelerations are considerably weaker. The cumulative indices for settlements remain closer to 1 over the sample period and diverge less across percentiles. This contrast between Figures \ref{fig:award-byside} and \ref{fig:settlement -byside} is consistent with the narrative that juries have become more sympathetic to plaintiffs and more willing to award large damages, while settlement dynamics are more constrained by negotiation, insurer oversight, and the mutual desire to avoid trial risk. In other words, verdict amounts appear to be the primary channel through which social inflation manifests in our data, with settlements playing a more muted role.

Figure \ref{fig:outcome -byside} examines the composition of case outcomes over time (left panel) and the plaintiff win rate conditional on a verdict (right panel). The left panel shows that the probability of settlement exhibits a gentle downward trend from 2009 onward, with a pronounced spike in 2020 when settlements temporarily become more common, consistent with courts pushing parties toward settlement during pandemic-induced backlogs and trial suspensions. After 2020, settlement shares generally resume their decline. The right panel reveals a steady increase in the plaintiff win rate provided that the case reaches a verdict, rising from the mid-50\% range to around 70\% by 2024. These patterns suggest that social inflation is reflected not only in higher award amounts conditional on plaintiff victory, but also in a growing propensity for plaintiffs to win when cases do go to verdict. They may also indicate that plaintiffs become less willing to settle over time, especially in an environment where jury awards are perceived to be increasingly generous.

Figures \ref{fig:case by state} and \ref{fig:P and D per case} describe how the geographic and party structures of cases evolve. Although VerdictSearch records cases from all 50 states plus the District of Columbia, Puerto Rico, and the Virgin Islands, Figure \ref{fig:case by state} shows that the bulk of our observations come from a handful of large, densely populated jurisdictions, notably Texas, California, New York, Florida, Pennsylvania, New Jersey, and the broader DC metro and New England regions. Over time, the share of cases reported in Texas rises, while the share in California declines. This compositional movement likely reflects both real differences in litigation activity, e.g., Texas’s reputation as an increasingly important venue for large verdicts, and changes in ALM’s coverage emphasis or attorney submission behavior. Figure \ref{fig:P and D per case} shows that the average number of defendants per case increases steadily over the period, especially after 2020, while the average number of plaintiffs rises more modestly. This suggests that cases reaching VerdictSearch increasingly involve multiple defendants, consistent with either enhanced plaintiff incentives to sue more entities in order to maximize potential recovery or selection effects whereby more complex, multiparty disputes are more likely to proceed to full litigation and be reported.

Plaintiff demographics, by contrast, are relatively stable. Figure \ref{fig:plaintiff_demographics_2x2} plots the composition of plaintiffs by age group, gender, marital status, and parental status over time. Across the sample from 2009 to 2024, these distributions show no pronounced shifts: The age mix remains centered on working-age adults, the gender split stays close to balanced, and the proportions of married/single plaintiffs and those with/without children fluctuate only mildly. This stability may still be useful information for our later analysis because it implies that major demographic shifts are unlikely to be responsible for the observed trends in outcomes and award levels.

Figure \ref{fig:injury-byside} focuses on injuries. The left panel shows that the average number of reported injuries per case rises over time, while the average number of deaths per case increases more sharply, especially in the post-pandemic years. The right panel shows that back and neck injuries are consistently the most prevalent injury types and that the proportion of cases involving surgery and internal injuries has also crept upward. These patterns suggest that cases in the VerdictSearch sample have become more severe and medically complex over time and that fatal cases represent a growing share of the docket. These observations are again consistent with selection toward more serious disputes.

Figure \ref{fig:Proportion of Cases-byside} documents trends in the share of corporate and insured defendants. The proportion of cases involving corporate defendants has been increasing steadily since 2012, while the proportion involving insured defendants rises from 2009 to around 2019 and then declines through 2024. The upward drift in corporate-defendant involvement is consistent with selection into litigation of higher-stakes, higher-limit disputes and with plaintiffs’ incentives to target deep-pocketed corporate entities, potentially leveraging perceived anti-corporate sentiment to obtain larger awards.

Case complexity, as proxied by the number and types of liability categories, is summarized in Figure \ref{fig:Cases types-byside}. The left panel shows a clear upward trend in the average number of case types coded per case, indicating that disputes increasingly involve multiple liability lines (e.g., motor vehicle plus general liability, or general plus professional liability). The right panel shows the shares of cases involving motor, general, and professional liability; motor cases remain the largest category throughout. The rising number of case types per dispute reinforces the picture of growing complexity among litigated cases.

Figure \ref{fig:Trail and Jury-byside} presents the distribution of trial length and jury deliberation length over time. Both trial and deliberation durations lengthen gradually, with a more pronounced stretching of the trial-length distribution after 2020. Longer trials and deliberations are consistent with more complex fact patterns, with larger stakes that justify extensive evidentiary presentation, and potentially with court backlogs that force trials to be spread out over longer calendar periods. They may also reflect a strategic response, especially on the plaintiff side, of assembling more comprehensive narratives and evidentiary records to support higher damages claims.

Figures \ref{fig:Experts and Attorneys -byside} and \ref{fig:expert type -byside} illustrate how expert and attorney involvement has evolved. Figure \ref{fig:Experts and Attorneys -byside} shows that the average numbers of experts and attorneys per case has increased steadily for both plaintiffs and defendants, but with faster growth on the plaintiff side: Plaintiffs add experts and attorneys at a higher rate than defendants, particularly in the later years. Figure \ref{fig:expert type -byside} decomposes expert participation by type, revealing that medical/healthcare experts dominate on both sides and that their share has grown over time. Business/finance experts, while less common, are used more frequently by plaintiffs than by defendants, consistent with plaintiffs’ need to quantify their economic losses and justify large compensatory and punitive awards. These patterns point to increasingly resource-intensive litigation and suggest that plaintiffs, in particular, have become more aggressive in deploying experts and legal representation to maximize potential compensation.

Taken together, these descriptive patterns yield four main insights that guide the remainder of the paper. First, there is strong evidence of selection effects: Even with ALM's efforts to maintain consistent data collection practices, exogenous factors beyond ALM's control, such as changes in the litigation environment and in what cases are most likely to be publicly reported and captured by third-party databases, lead the VerdictSearch sample to become increasingly concentrated in severe, complex, and high-profile cases over time, as reflected in the declining case counts; the rising numbers of plaintiffs, defendants, injuries, deaths, experts, and attorneys per case; and the growing involvement of corporate defendants. Any attempt to measure social inflation that ignores this evolving case mix is likely to overstate the underlying trend. Second, the summary statistics document a clear escalation in plaintiff-side litigation intensity: Plaintiffs hire more attorneys and experts, pursue longer trials and deliberations, and appear increasingly willing to avoid settlement in favor of verdicts, all of which may contribute directly to social inflation. Third, extreme verdict amounts make average-based indices highly unstable and ill-suited for social inflation measurement; robust, distribution-based approaches are required to avoid undue influence from a few nuclear verdicts. Finally, social inflation appears to operate across the full spectrum of cases: Plaintiff win probabilities rise, settlement probabilities fall, and verdict amounts increase in a broadly parallel fashion across quantiles. While these patterns suggest that social inflation is not confined to the most spectacular outliers, rigorous conclusions require models that explicitly control for selection and covariate effects, which we develop in the subsequent sections.

\section{Preliminary analysis} \label{sec:prelim}
Section \ref{sec:data} documented how outcomes and case characteristics in the VerdictSearch data have evolved over time, highlighting patterns such as rising plaintiff win rates, increasing award levels, and growing case complexity. In this section, we turn from descriptive trends to a more granular, cross-sectional analysis of how individual explanatory variables listed in Tables \ref{tab:summary-stats_factual} and \ref{tab:summary-stats_strategic} are associated with four key response variables: (1) the probability that the plaintiff wins provided that the case proceeds to a verdict, (2) the probability that the case resolves via settlement, (3) the size of the verdict award conditional on a plaintiff win, and (4) the size of the settlement amount conditional on settlement. Using standard logistic regression for the binary outcomes and quantile regression for the (log) award and settlement amounts, we estimate how each factual and strategic covariate affects these responses after controlling for the full set of other variables. Together with the descriptive evidence discussed in Section \ref{sec:data}, these variable-level effects illustrate how shifts in case mix and litigation strategies can mechanically generate apparent social inflation if not properly accounted for. For example, if cases with corporate defendants tend to produce larger verdicts and settlements, and the share of cases involving corporate defendants rises over time, then any naive trend in average award sizes will embed both genuine social inflation and a composition shift toward inherently higher-stakes disputes, emphasizing the need for case-mix adjustment in our subsequent social inflation measures.

\subsection{Variable effects on plaintiff win probability} \label{sec:prelim:prob_p}
\begin{figure}[!h]
  \centering
  \begin{subfigure}[t]{0.48\textwidth}
    \centering
    \includegraphics[width=\linewidth]{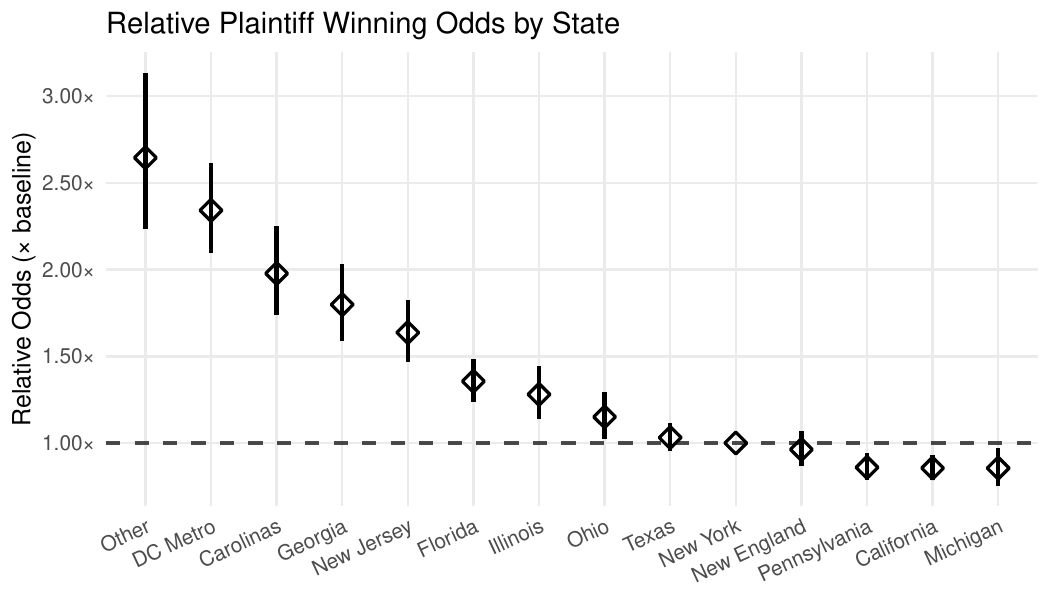}
  \end{subfigure}\hfill
  \begin{subfigure}[t]{0.48\textwidth}
    \centering
    \includegraphics[width=\linewidth]{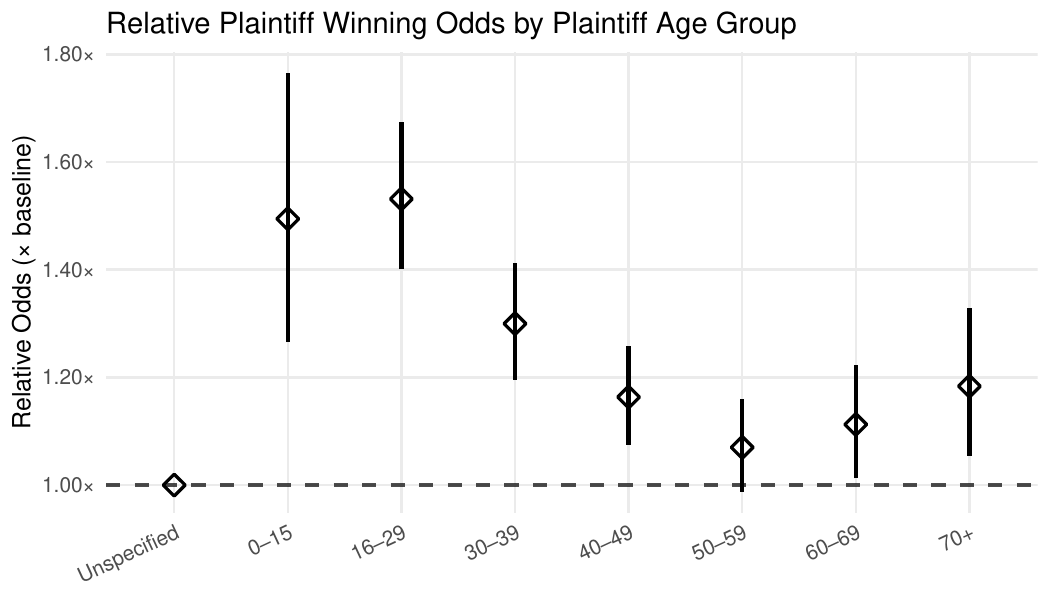}
  \end{subfigure}\hfill
    \begin{subfigure}[t]{0.48\textwidth}
    \centering
    \includegraphics[width=\linewidth]{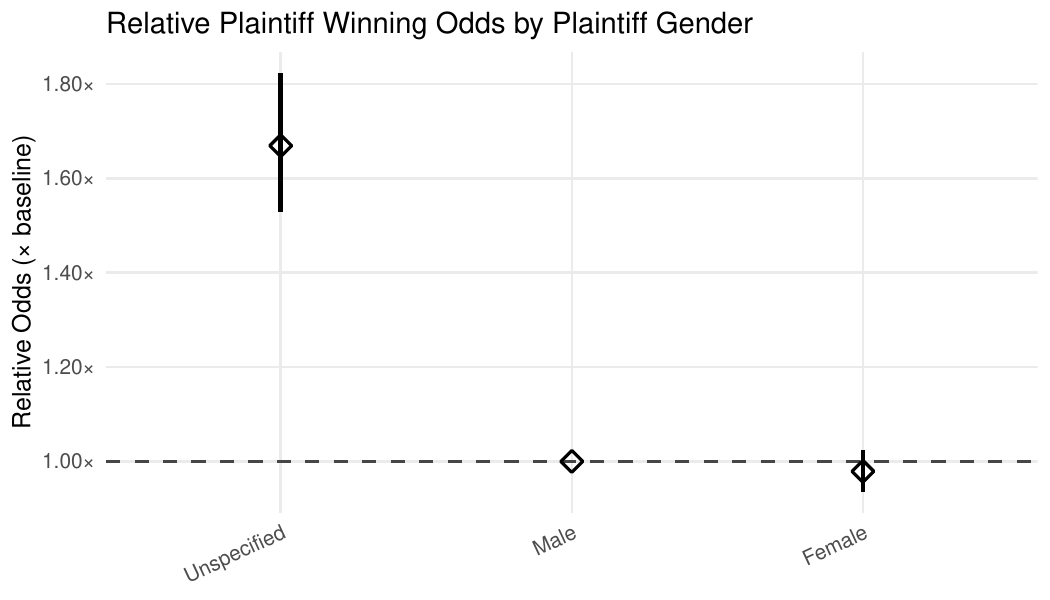}
  \end{subfigure}\hfill
  \begin{subfigure}[t]{0.48\textwidth}
    \centering
    \includegraphics[width=\linewidth]{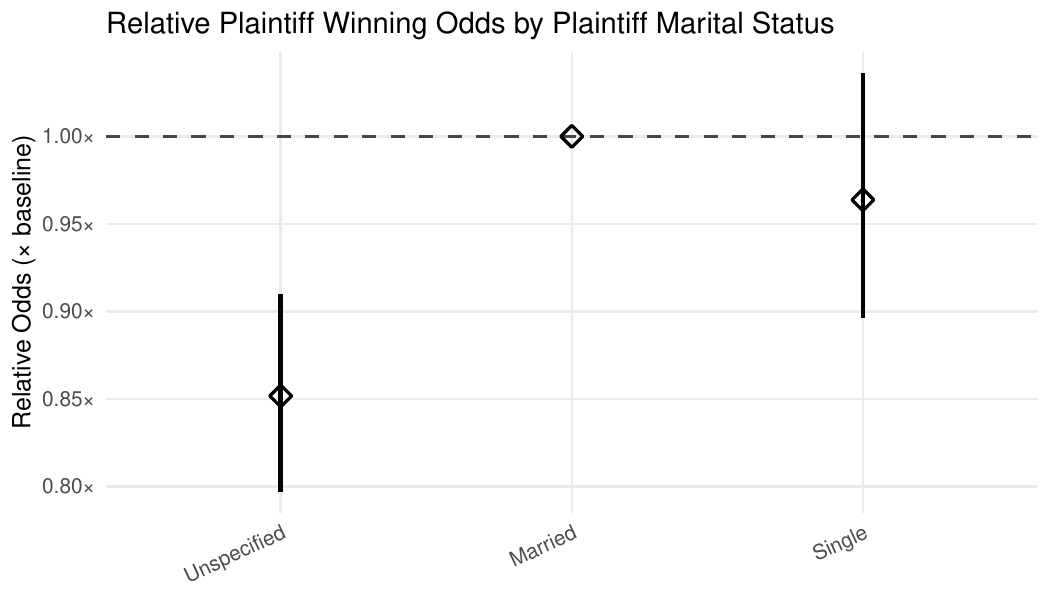}
  \end{subfigure}\hfill
  \begin{subfigure}[t]{0.48\textwidth}
    \centering
    \includegraphics[width=\linewidth]{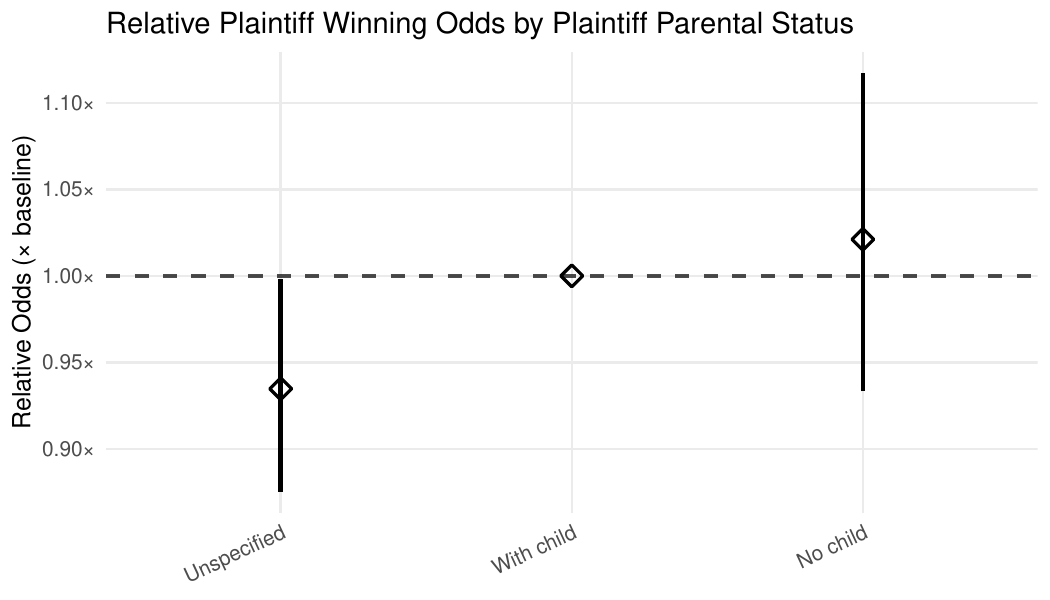}
  \end{subfigure}\hfill
  \begin{subfigure}[t]{0.48\textwidth}
    \centering
    \includegraphics[width=\linewidth]{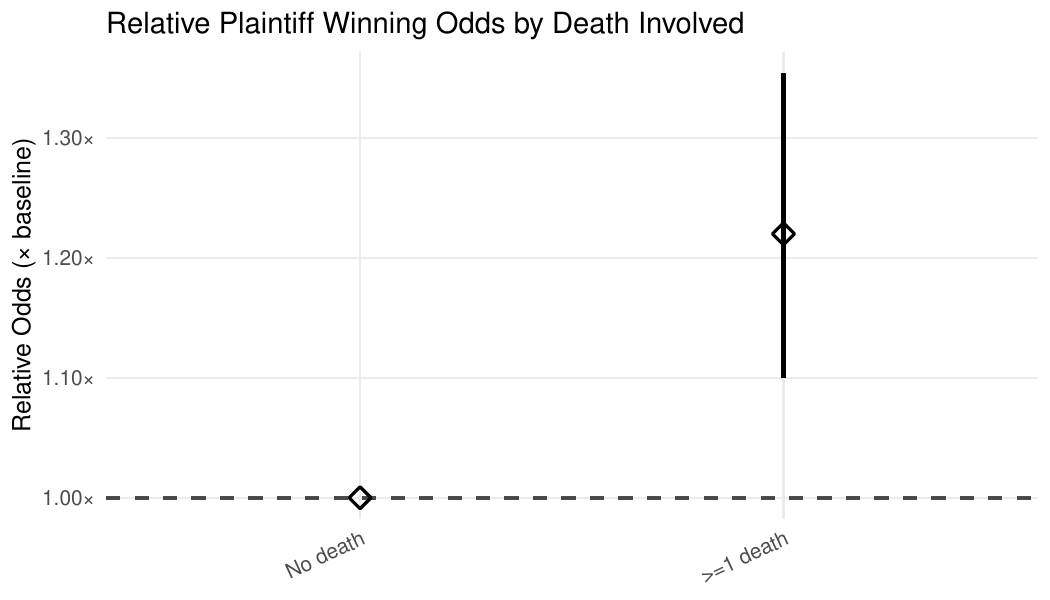}
  \end{subfigure}\hfill
  \begin{subfigure}[t]{0.48\textwidth}
    \centering
    \includegraphics[width=\linewidth]{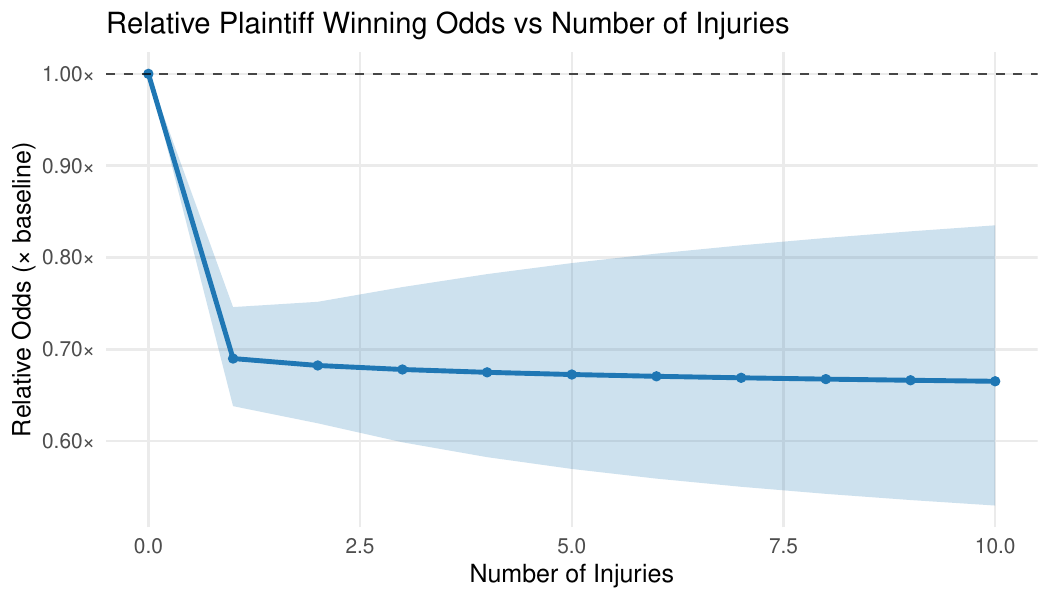}
  \end{subfigure}\hfill
  \begin{subfigure}[t]{0.48\textwidth}
    \centering
    \includegraphics[width=\linewidth]{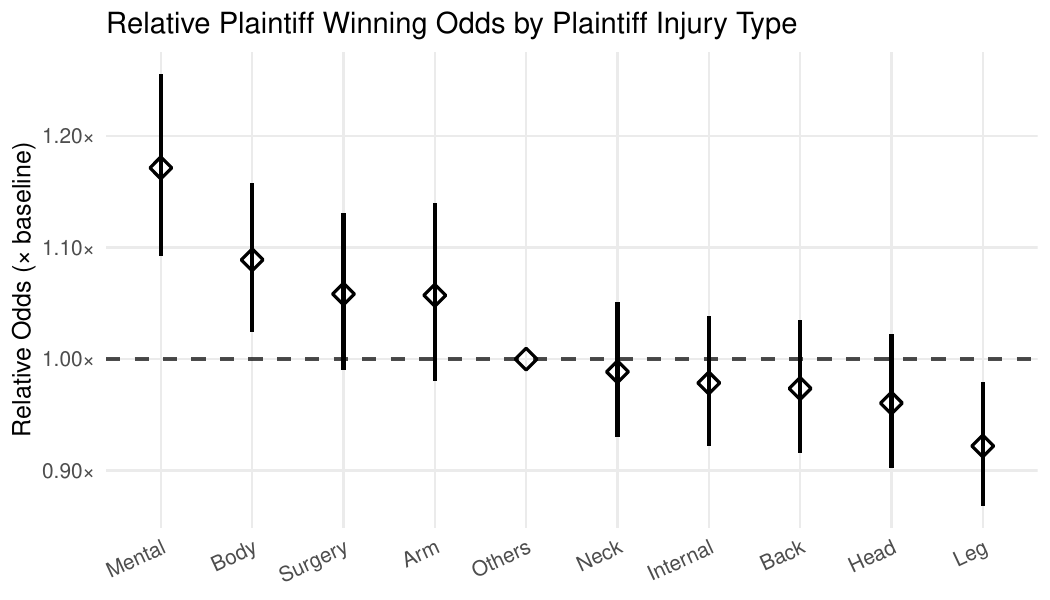}
  \end{subfigure}\hfill
  \caption{Relative plaintiff winning odds vs. variables, with 95\% confidence intervals (part 1).}
  \label{fig:cov_prob_p_fig1}
\end{figure}

\begin{figure}[!h]
  \centering
    \begin{subfigure}[t]{0.48\textwidth}
    \centering
    \includegraphics[width=\linewidth]{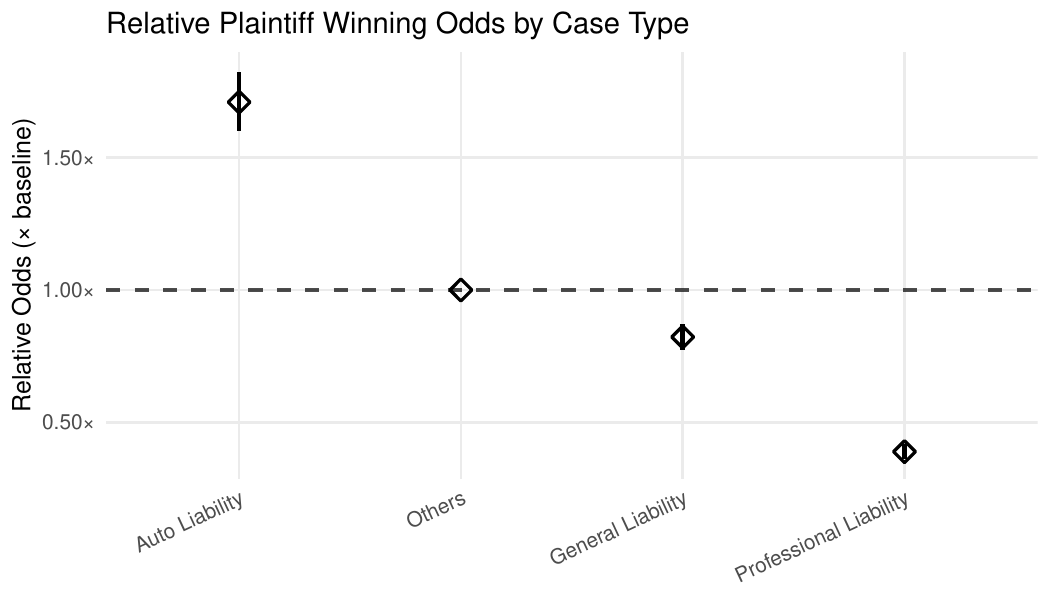}
  \end{subfigure}\hfill
  \begin{subfigure}[t]{0.48\textwidth}
    \centering
    \includegraphics[width=\linewidth]{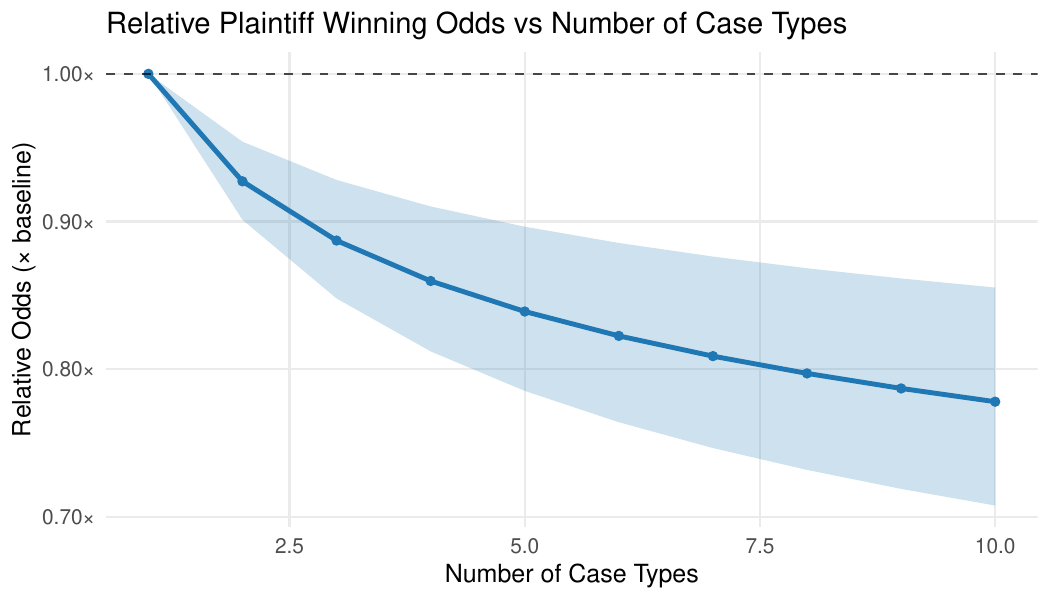}
  \end{subfigure}\hfill
  \begin{subfigure}[t]{0.48\textwidth}
    \centering
    \includegraphics[width=\linewidth]{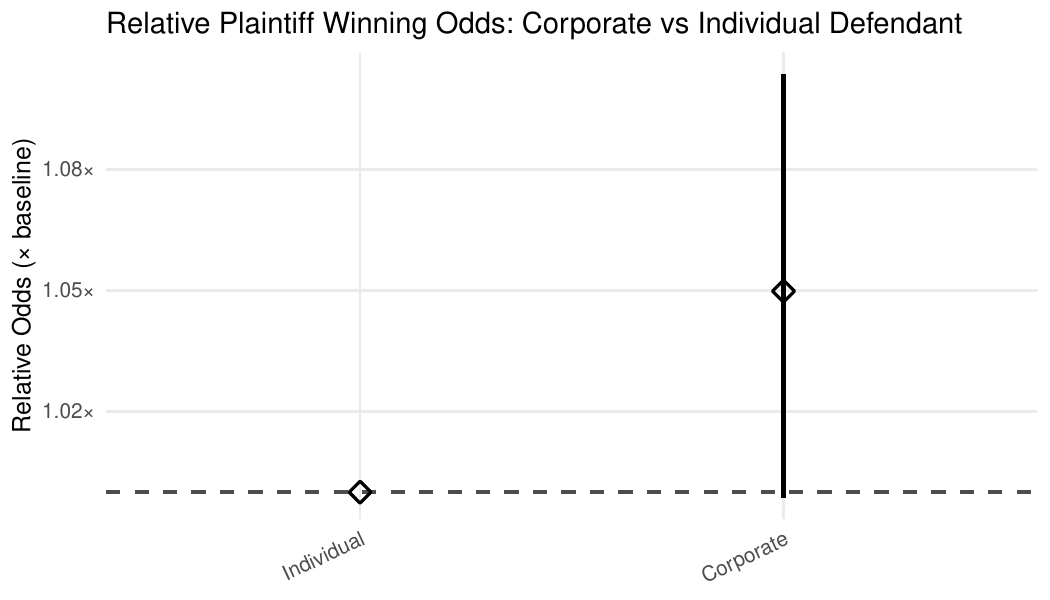}
  \end{subfigure}\hfill
  \begin{subfigure}[t]{0.48\textwidth}
    \centering
    \includegraphics[width=\linewidth]{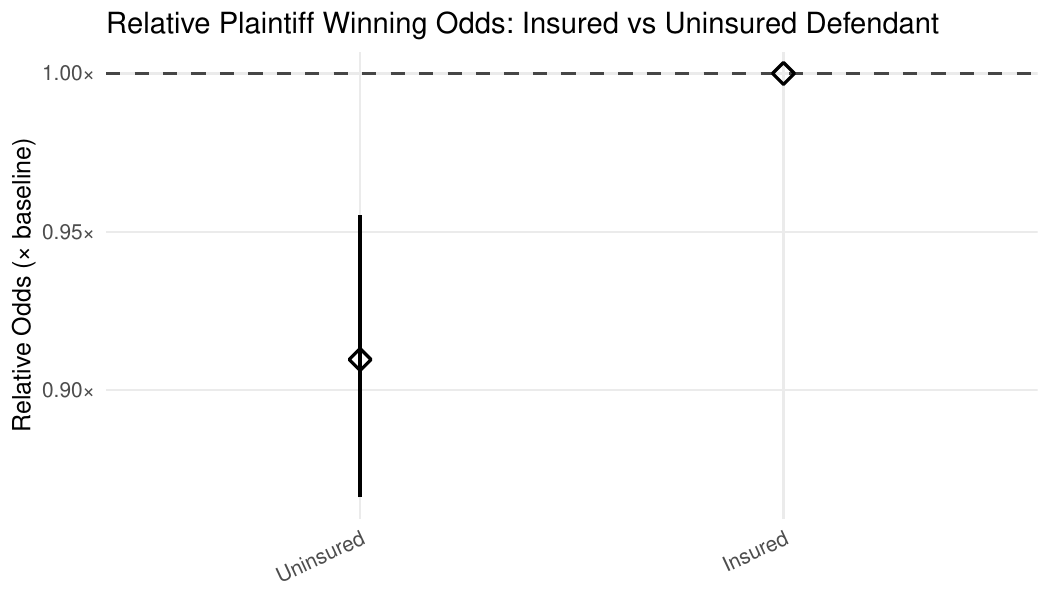}
  \end{subfigure}\hfill
    \begin{subfigure}[t]{0.48\textwidth}
    \centering
    \includegraphics[width=\linewidth]{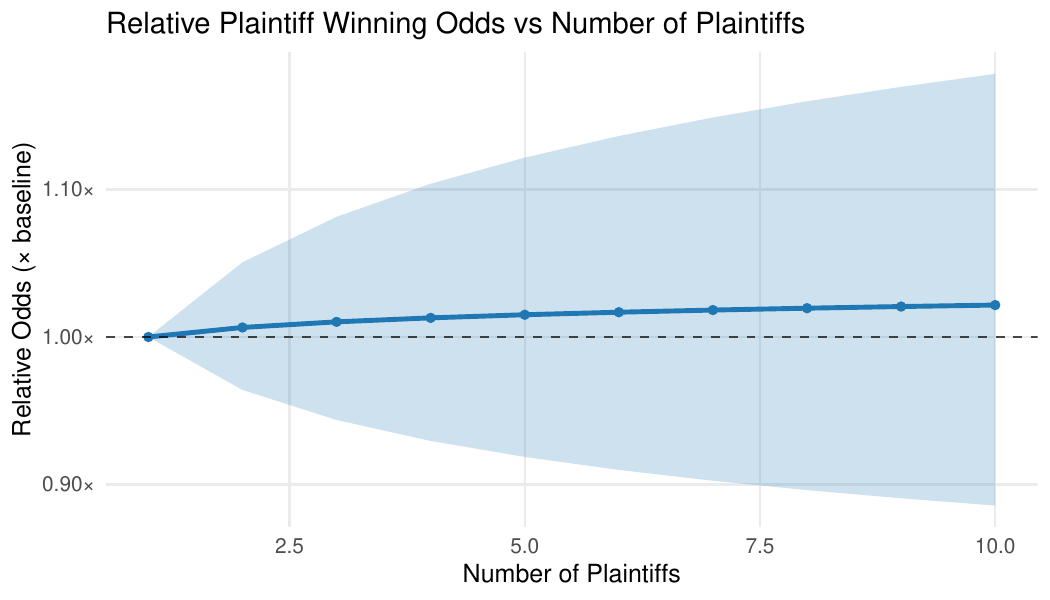}
  \end{subfigure}\hfill
  \begin{subfigure}[t]{0.48\textwidth}
    \centering
    \includegraphics[width=\linewidth]{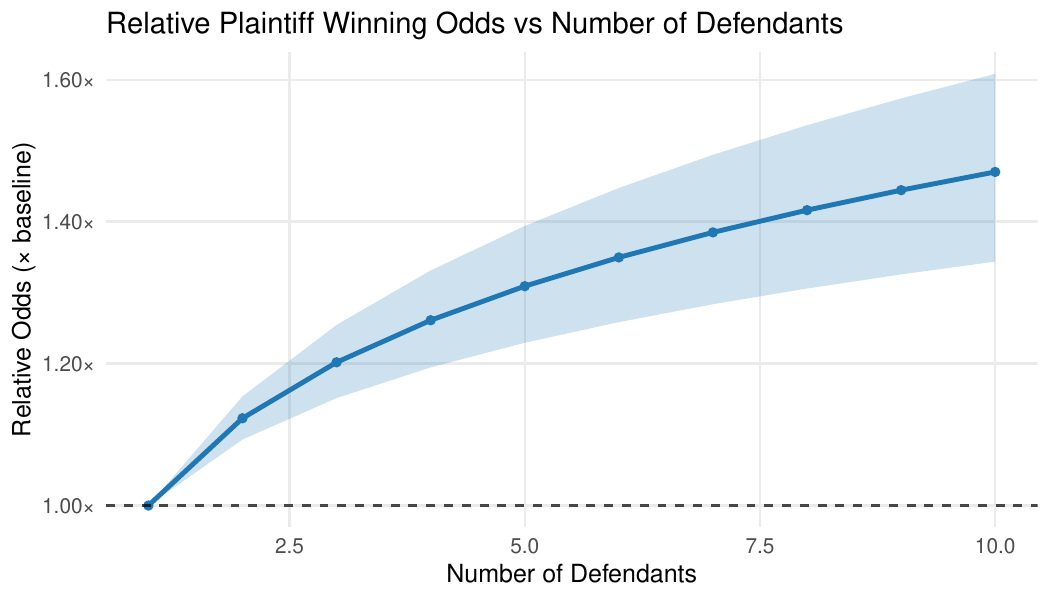}
  \end{subfigure}\hfill
  \begin{subfigure}[t]{0.48\textwidth}
    \centering
    \includegraphics[width=\linewidth]{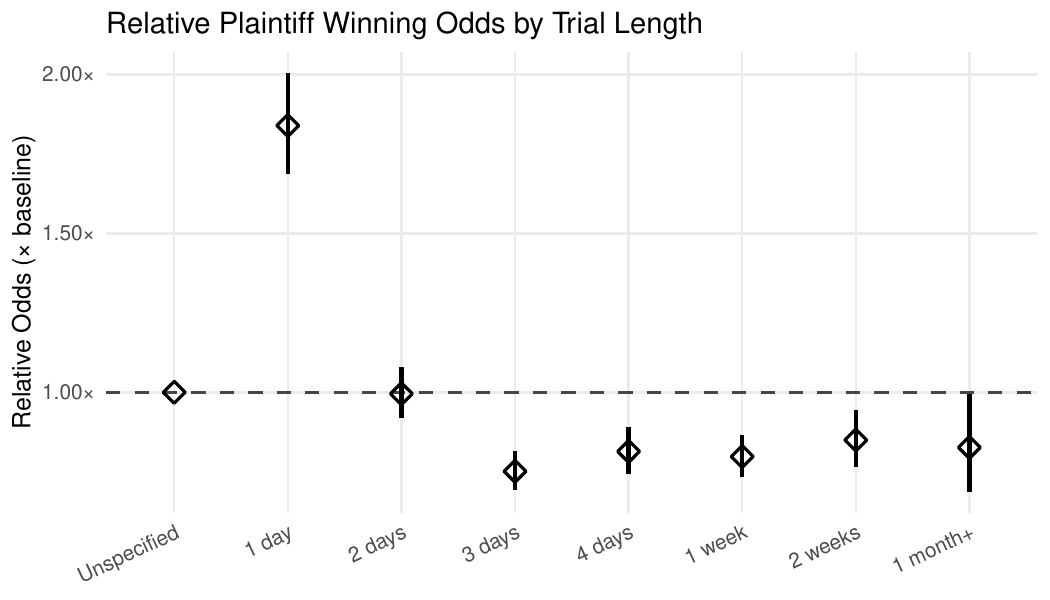}
  \end{subfigure}\hfill
  \begin{subfigure}[t]{0.48\textwidth}
    \centering
    \includegraphics[width=\linewidth]{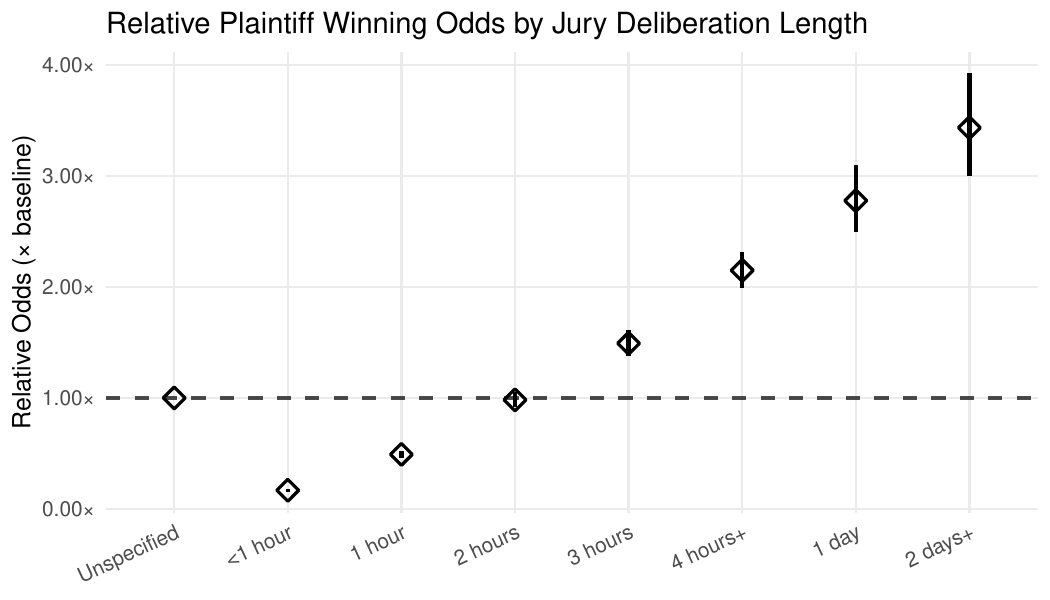}
  \end{subfigure}\hfill
  \caption{Relative plaintiff winning odds vs. variables, with 95\% confidence intervals (part 2).}
  \label{fig:cov_prob_p_fig2}
\end{figure}

\begin{figure}[!h]
  \centering
    \begin{subfigure}[t]{0.48\textwidth}
    \centering
    \includegraphics[width=\linewidth]{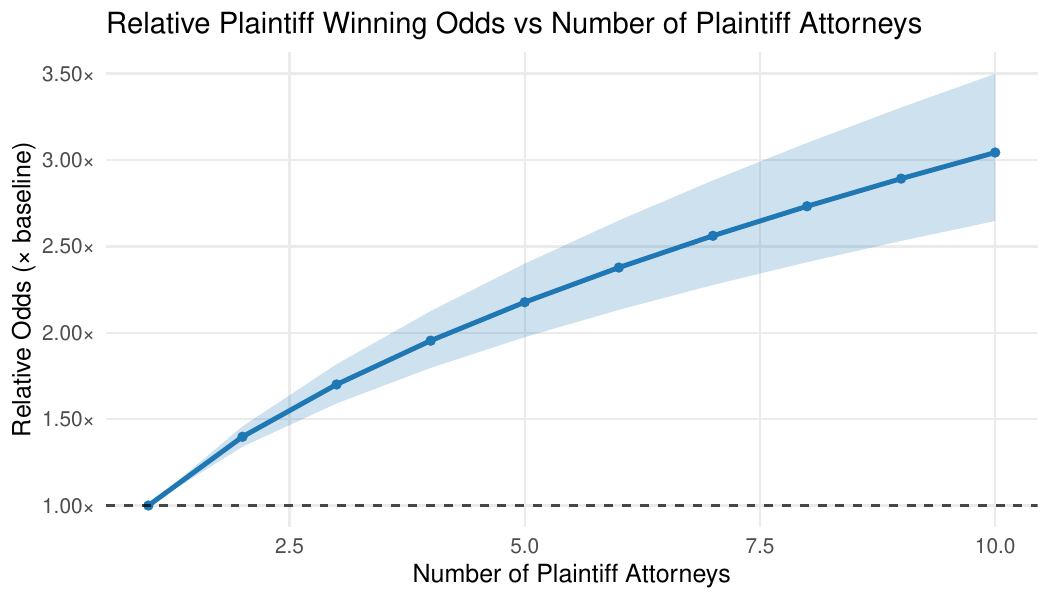}
  \end{subfigure}\hfill
  \begin{subfigure}[t]{0.48\textwidth}
    \centering
    \includegraphics[width=\linewidth]{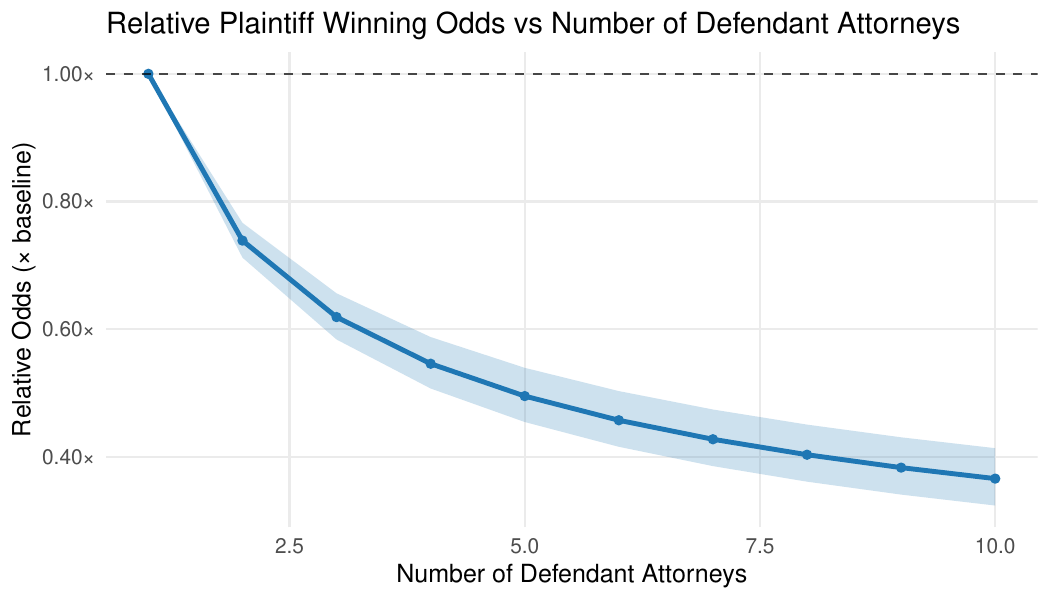}
  \end{subfigure}\hfill
  \begin{subfigure}[t]{0.48\textwidth}
    \centering
    \includegraphics[width=\linewidth]{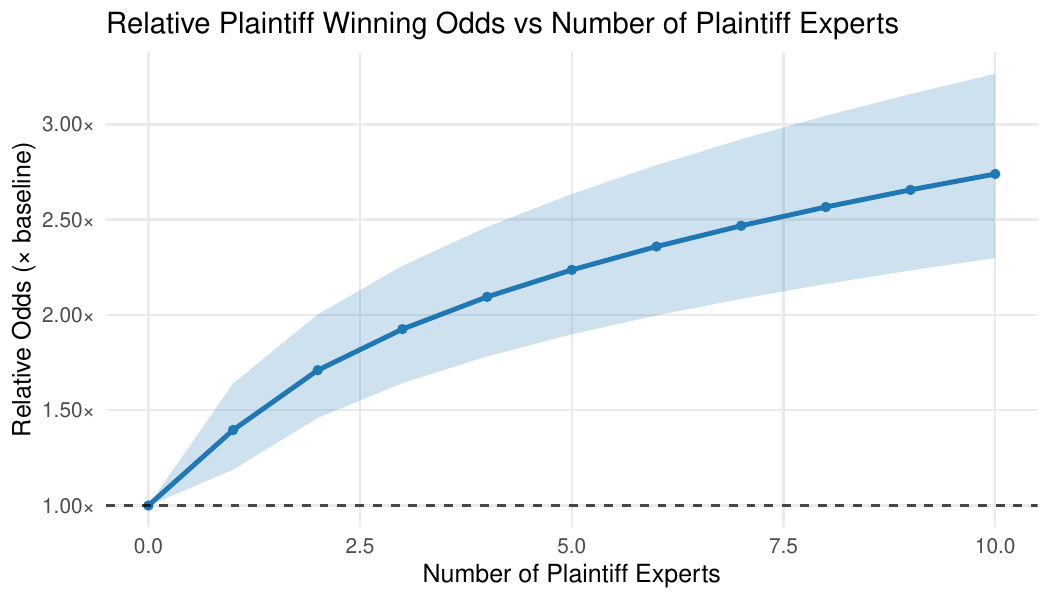}
  \end{subfigure}\hfill
  \begin{subfigure}[t]{0.48\textwidth}
    \centering
    \includegraphics[width=\linewidth]{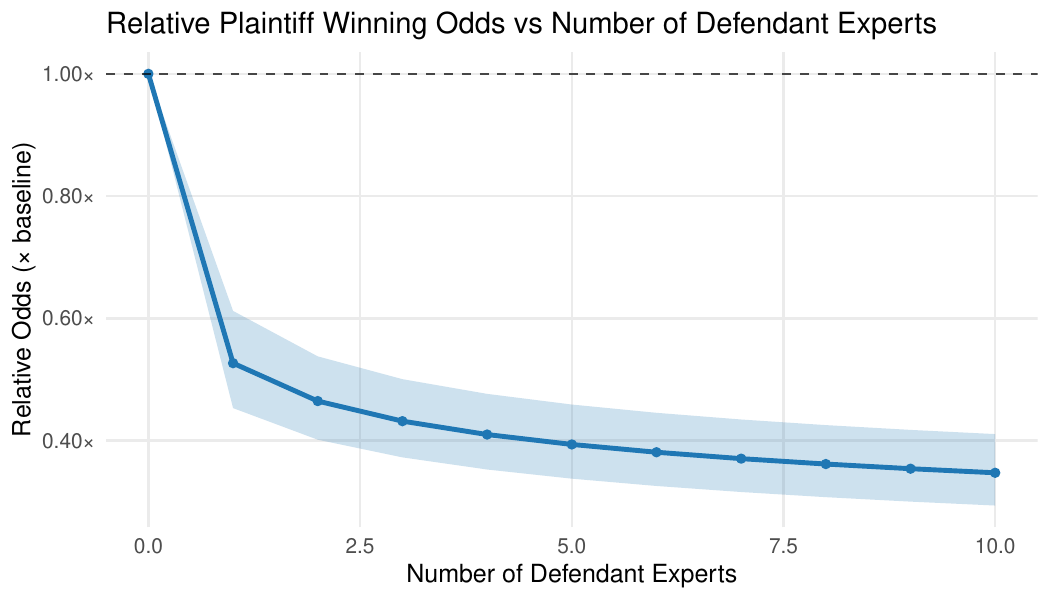}
  \end{subfigure}\hfill
  \begin{subfigure}[t]{0.48\textwidth}
    \centering
    \includegraphics[width=\linewidth]{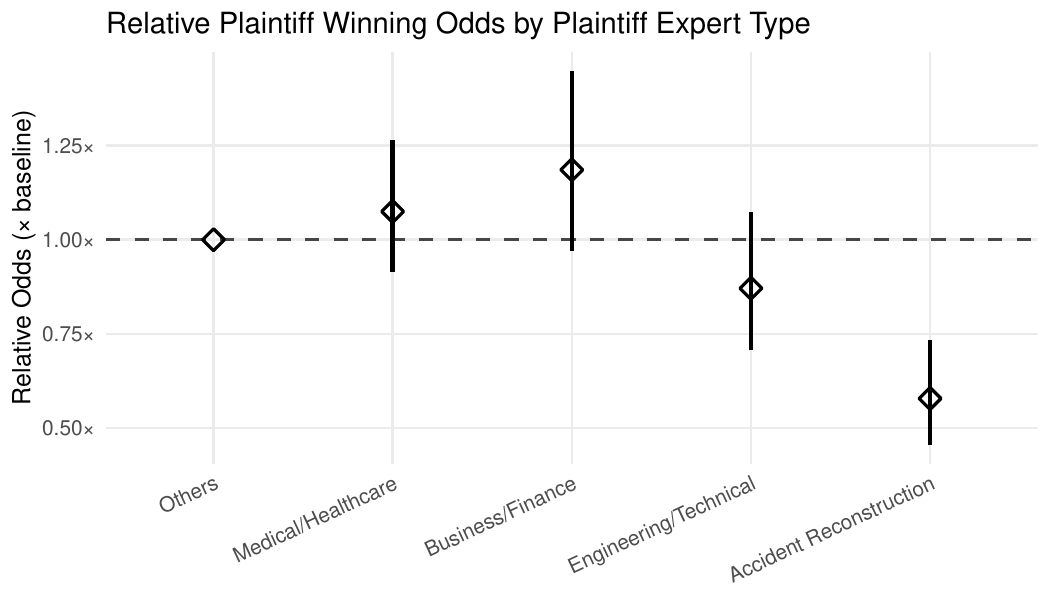}
  \end{subfigure}\hfill
  \begin{subfigure}[t]{0.48\textwidth}
    \centering
    \includegraphics[width=\linewidth]{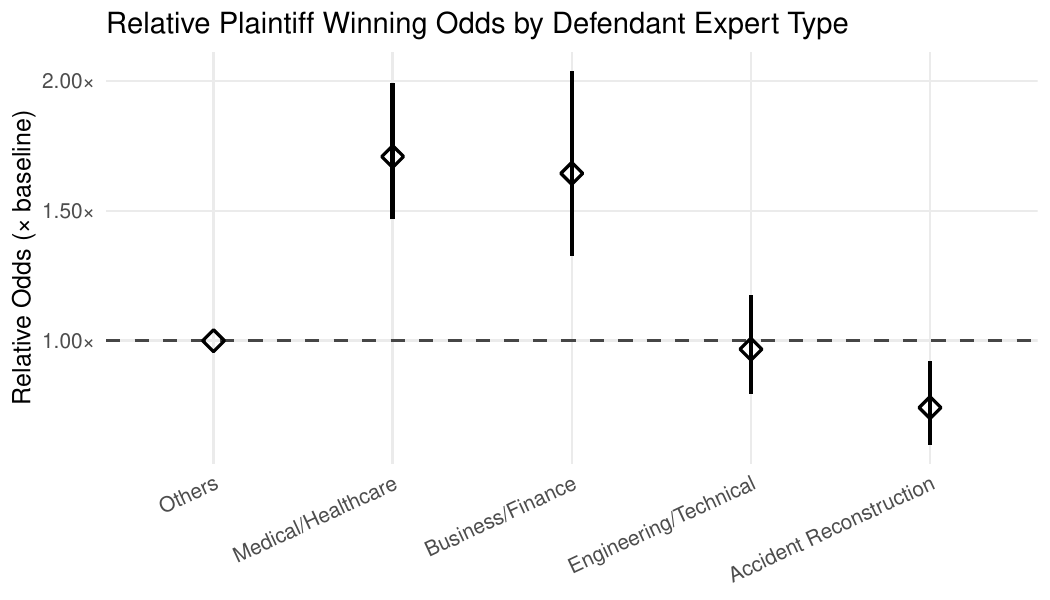}
  \end{subfigure}\hfill
  \caption{Relative plaintiff winning odds vs. variables, with 95\% confidence intervals (part 3).}
  \label{fig:cov_prob_p_fig3}
\end{figure}

We first examine how plaintiff victory probabilities vary with case characteristics by fitting a logistic regression model to all cases that end in a verdict (for the plaintiff or defense), excluding settlements. The dependent variable is an indicator equal to 1 if the plaintiff wins and 0 if the defendant wins. Logistic regression models the log-odds of plaintiff success as a linear function of the explanatory variables in Tables \ref{tab:summary-stats_factual} and \ref{tab:summary-stats_strategic}. Intuitively, the odds of plaintiff victory are the ratio of the probability that the plaintiff wins to the probability that the plaintiff loses; larger odds correspond to a higher win probability. The model is estimated using the standard \texttt{glm()} function with a binomial logit link in \texttt{R}, which returns a coefficient and standard error for each covariate. Exponentiating a coefficient gives the multiplicative effect of that covariate on the odds of plaintiff victory, holding all other variables constant. We report these effects as relative plaintiff winning odds, together with 95\% confidence intervals, in Figures \ref{fig:cov_prob_p_fig1} to \ref{fig:cov_prob_p_fig3}. Values above 1 indicate that the covariate increases the odds that the plaintiff wins; values below 1 indicate a decrease.

The figures reveal substantial geographic heterogeneity in plaintiff success. Relative to the baseline state, the odds that the plaintiff wins differ by a factor of roughly 2 to 3 across states. Cases in Pennsylvania, California, and Michigan exhibit significantly lower plaintiff winning odds, while cases arising in the DC Metro region (District of Columbia, Maryland, and Virginia), the Carolinas, Georgia, Florida, and New Jersey tend to be more favorable to plaintiffs. The ``other'' category, which aggregates less populous Midwest and Mountain states, shows the highest plaintiff winning odds, even after conditioning on injuries, case type, and other controls. These patterns are consistent with long-standing practitioner perceptions of more plaintiff-friendly venues in certain jurisdictions and more defense-oriented environments in others, and hence the importance of controlling for geography cannot be disregarded when comparing plaintiff success over time.

Turning to plaintiff demographics, age emerges as an important predictor of success, whereas gender, marital status, and parental status do not. Cases involving younger plaintiffs, particularly children, teenagers, and young adults (under 30 years), are significantly more likely to result in plaintiff victories than cases involving middle-aged or older plaintiffs. Interestingly, observations coded as ``unspecified'' age have the lowest plaintiff winning odds of all. In VerdictSearch, ``unspecified'' demographics often arise when the plaintiff is not a natural person (e.g., a corporate plaintiff or an estate in a wrongful-death case) or when demographic fields are missing or inconsistently recorded. Thus, the reduced win odds for the ``unspecified'' group likely reflect these underlying case settings and data-availability patterns rather than the effect of age per se. In contrast, the estimated effects of gender, marital status, and parental status are close to 1 and generally statistically insignificant: Male and female plaintiffs, married and single plaintiffs, and plaintiffs with and without children have very similar win probabilities once other factors are held fixed.

Injury and fatality variables also play a role. Controlling for other covariates, cases involving at least one death exhibit roughly 20\%--25\% higher odds of plaintiff victory than otherwise comparable nonfatal cases, and this effect is statistically significant. At the same time, cases in which no injuries are recorded, such as certain property liability or contract disputes cases, have relatively high plaintiff winning odds. However, conditional on injuries being present, the total number of injuries per case has only a modest incremental effect on plaintiff success; our estimates do not indicate a statistically significant pattern. Injury type matters more: Plaintiff winning odds vary moderately across injury categories, with cases involving mental or psychological harm exhibiting the highest success odds, and cases centered on leg injuries exhibiting the lowest. The gap between these extremes is on the order of 20\%–30\%, suggesting that certain injury narratives (e.g., mental trauma, internal injuries, or surgery) may resonate more with fact finders than others. 

Case categories and party structure further shape plaintiff success. Motor vehicle liability cases have much higher plaintiff winning odds than general liability and, especially, professional liability cases. Conditional on the same covariate profile, the odds that the plaintiff prevails in a motor vehicle case are more than 3 times those in a professional liability case. This aligns with the idea that jurors may find negligence or fault easier to establish in straightforward automobile accidents than in complex professional malpractice disputes. Cases with a larger number of coded case types, indicative of multiple types of liability involved, tend to be less favorable to plaintiffs; the relative odds decline steadily as the number of case types increases, consistent with the view that more complex, multi-issue disputes create more avenues for defense arguments and greater uncertainty. With respect to defendants, we find that cases involving at least one corporate defendant have slightly higher plaintiff winning odds than cases with only individual defendants, but the estimated effect is small (a relative odds ratio of about 1.05) and only marginally statistically significant. Cases involving insured defendants show a more clearly positive effect: The presence of insurance coverage is associated with moderately higher plaintiff winning odds.

Trial and jury deliberation durations exhibit an interesting asymmetry. Very short trials (one day) are strongly associated with a higher plaintiff success rate. This pattern suggests that simple, straightforward disputes that can be tried in a day tend to favor plaintiffs, perhaps because liability is relatively clear and the defendant has limited room to generate doubt. By contrast, long trials are often complex and may give defendants more scope to raise competing narratives. However, longer jury deliberations are positively and strongly associated with plaintiff wins. Compared to cases decided within an hour, cases in which the jury deliberates for a full day or multiple days exhibit substantially higher plaintiff winning odds, with relative odds ratios that can exceed 3. Combined, these results suggest that plaintiffs fare best in cases that are simple enough to be tried quickly but sufficiently contested to prompt extended jury discussion rather than immediate dismissal of the plaintiff’s claims.

Finally, the models highlight the impact of expert testimony and legal representation. The number of plaintiff attorneys and experts is positively associated with plaintiff success. Adding more plaintiff attorneys or experts steadily raises the odds of a plaintiff win, with diminishing but still meaningful gains at higher counts. In contrast, additional defense attorneys and experts are associated with lower plaintiff winning odds; as the defense team grows, the likelihood that the plaintiff prevails declines. These patterns are associational and admit multiple interpretations. One possibility is a strategic-resources channel: Plaintiffs may improve their prospects by deploying more legal manpower and specialized expertise, while defendants may reduce plaintiff success by expanding their own representation. An alternative interpretation is selection or reverse causality: Cases with inherently stronger merits may attract more attorneys and experts on the side that expects an advantage. Accordingly, we interpret these coefficients as evidence that litigation resources are strongly associated with outcomes, while causal effects would require additional experimental variation. Expert specialization also matters. The presence of medical and business/finance experts on the plaintiff side typically coincides with higher plaintiff winning odds, especially in cases involving bodily injury and economic loss. On the defense side, medical and business experts appear to mitigate some of this advantage, while accident reconstruction experts tend to be associated with lower plaintiff winning odds, consistent with their role in contesting liability in motor and premises cases.


\subsection{Variable effects on settlement probability} \label{sec:prelim:prob_s}
\begin{figure}[H]
  \centering
  \begin{subfigure}[t]{0.48\textwidth}
    \centering
    \includegraphics[width=\linewidth]{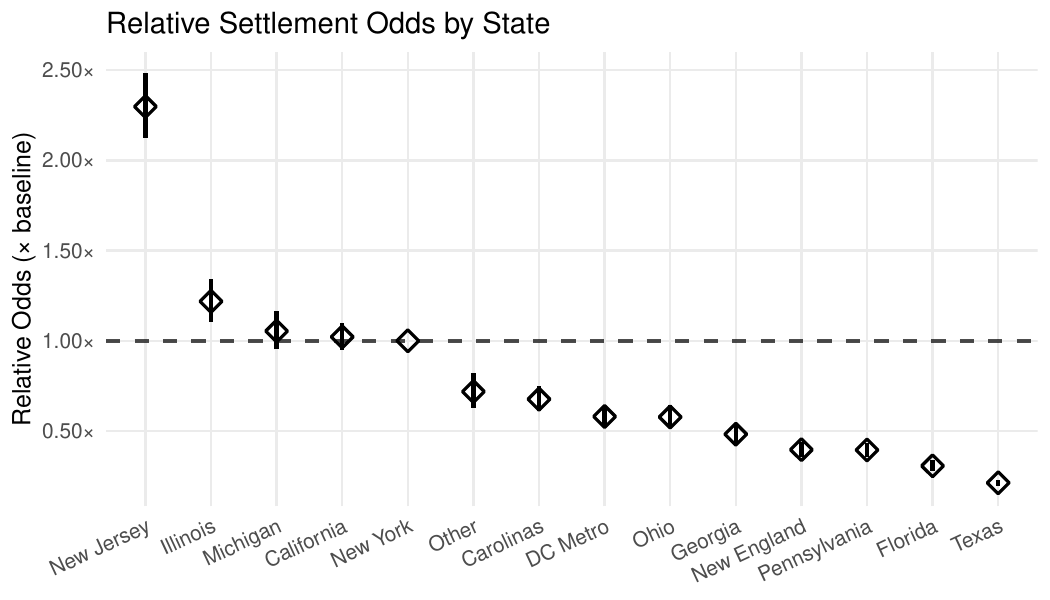}
  \end{subfigure}\hfill
  \begin{subfigure}[t]{0.48\textwidth}
    \centering
    \includegraphics[width=\linewidth]{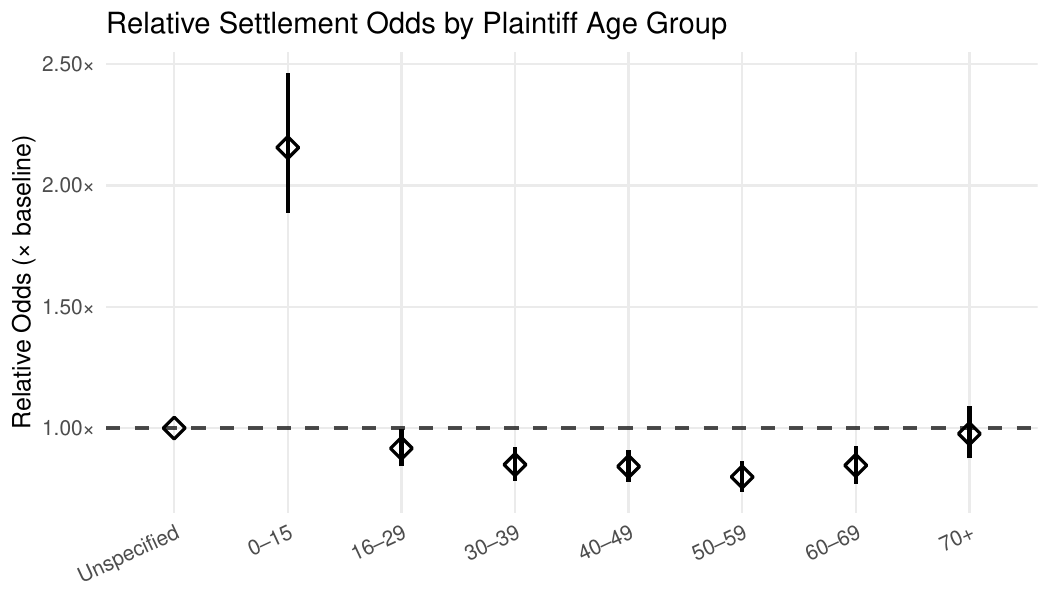}
  \end{subfigure}\hfill
    \begin{subfigure}[t]{0.48\textwidth}
    \centering
    \includegraphics[width=\linewidth]{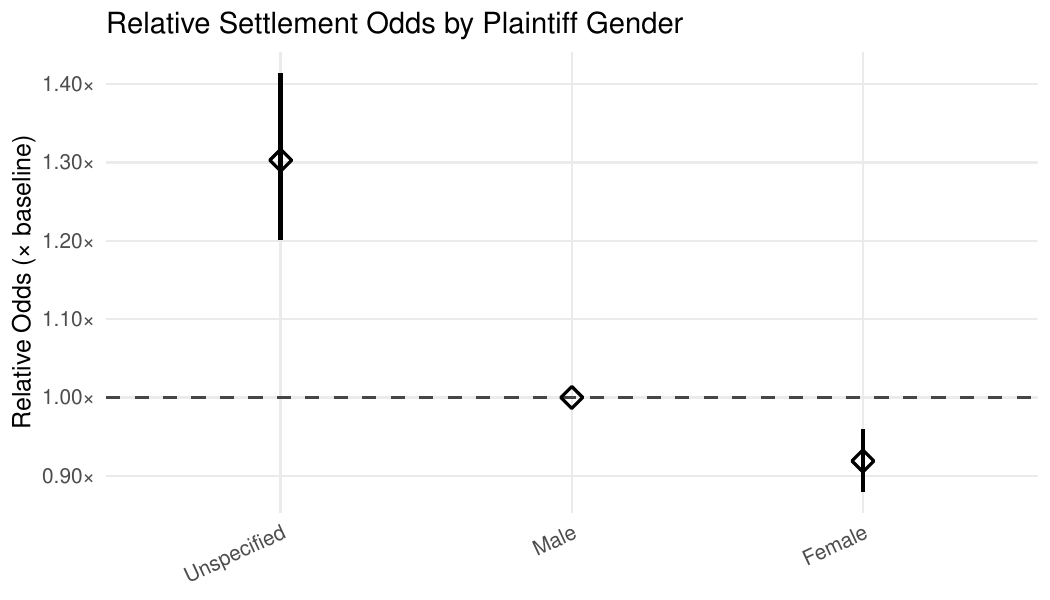}
  \end{subfigure}\hfill
  \begin{subfigure}[t]{0.48\textwidth}
    \centering
    \includegraphics[width=\linewidth]{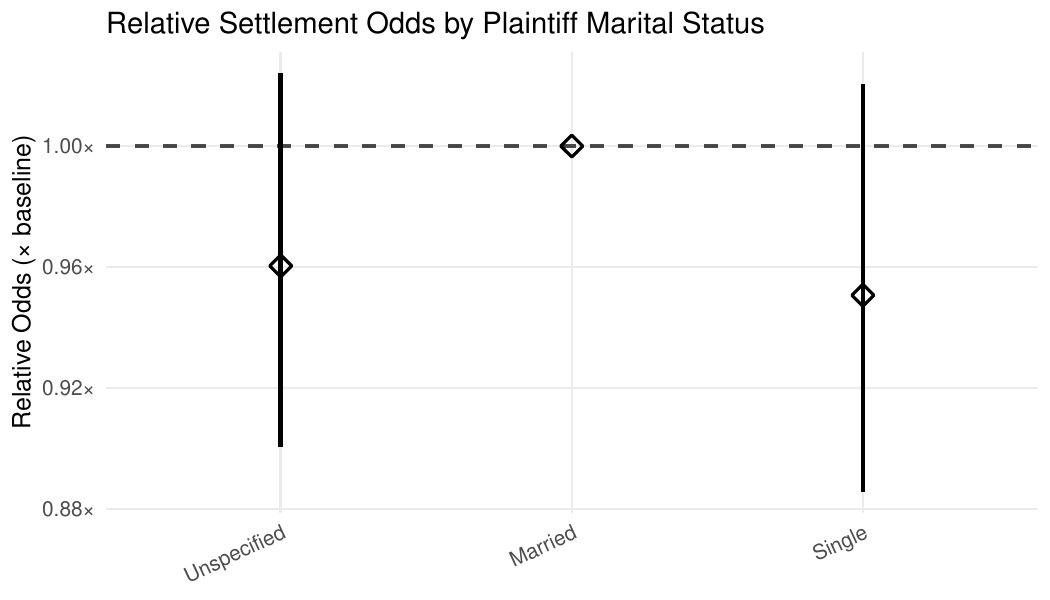}
  \end{subfigure}\hfill
  \begin{subfigure}[t]{0.48\textwidth}
    \centering
    \includegraphics[width=\linewidth]{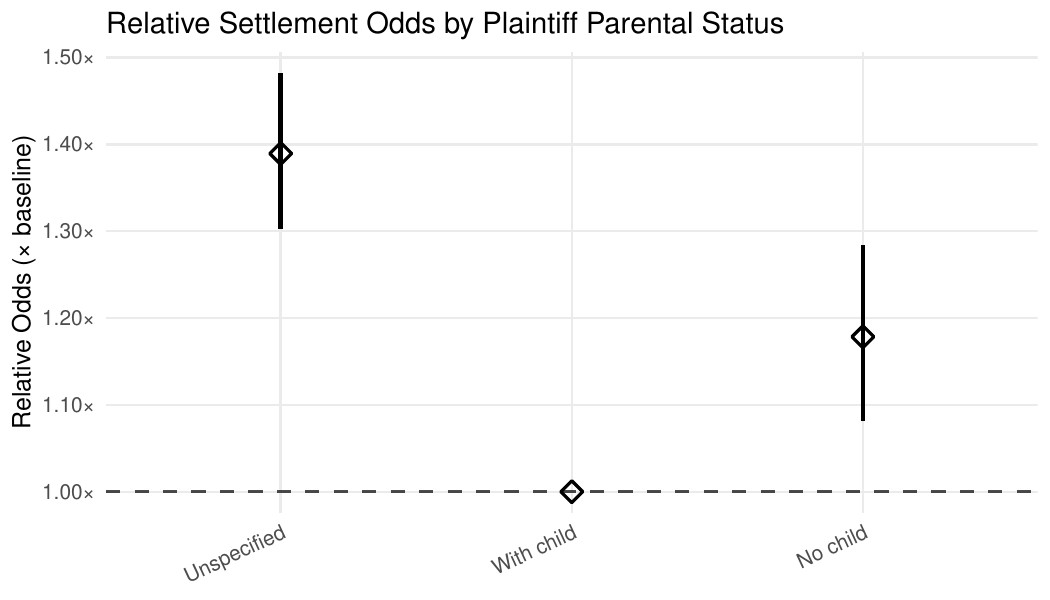}
  \end{subfigure}\hfill
  \begin{subfigure}[t]{0.48\textwidth}
    \centering
    \includegraphics[width=\linewidth]{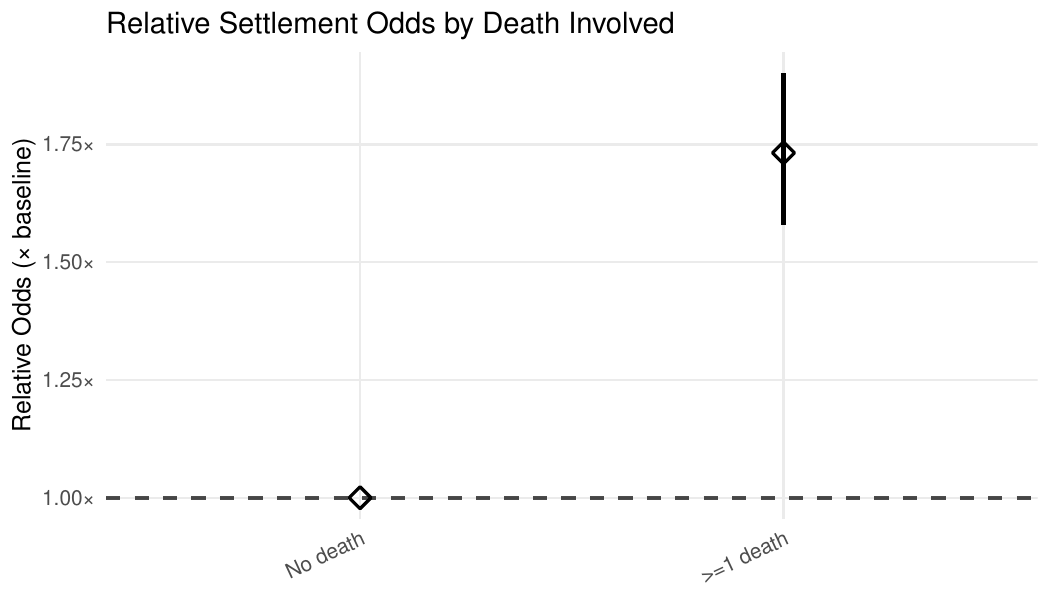}
  \end{subfigure}\hfill
  \begin{subfigure}[t]{0.48\textwidth}
    \centering
    \includegraphics[width=\linewidth]{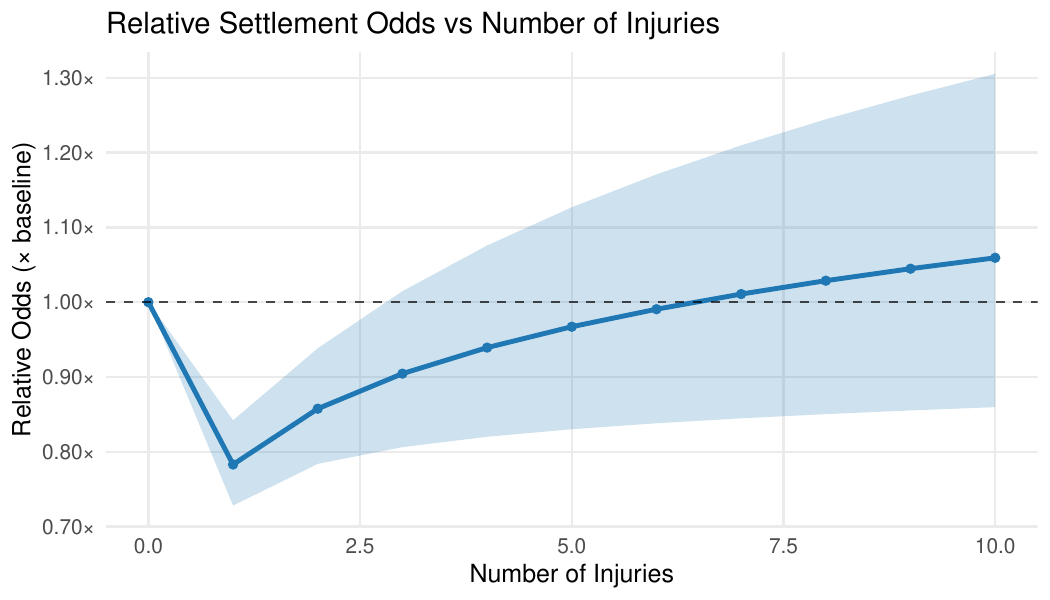}
  \end{subfigure}\hfill
  \begin{subfigure}[t]{0.48\textwidth}
    \centering
    \includegraphics[width=\linewidth]{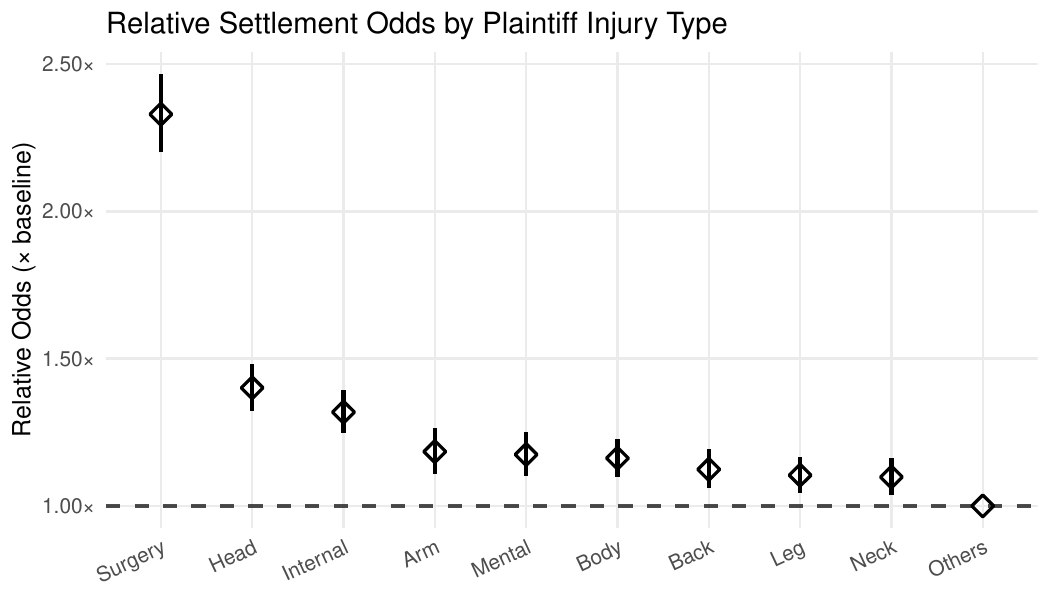}
  \end{subfigure}\hfill
  \begin{subfigure}[t]{0.48\textwidth}
    \centering
    \includegraphics[width=\linewidth]{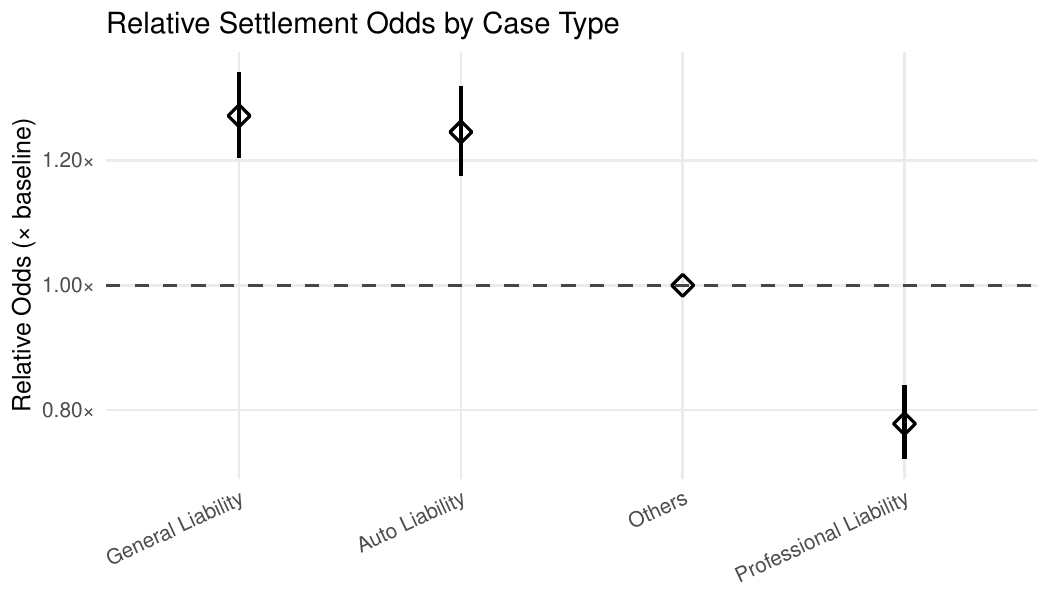}
  \end{subfigure}\hfill
  \begin{subfigure}[t]{0.48\textwidth}
    \centering
    \includegraphics[width=\linewidth]{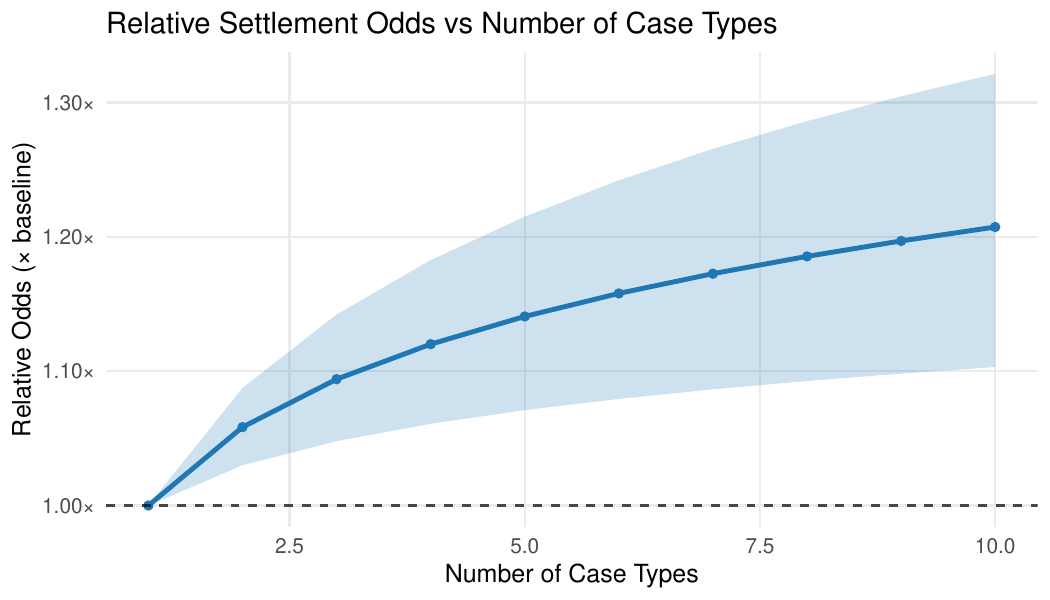}
  \end{subfigure}\hfill
  \caption{Relative settlement odds vs. variables, with 95\% confidence intervals (part 1).}
  \label{fig:cov_prob_s_fig1}
\end{figure}

\begin{figure}[H]
  \centering
  \begin{subfigure}[t]{0.48\textwidth}
    \centering
    \includegraphics[width=\linewidth]{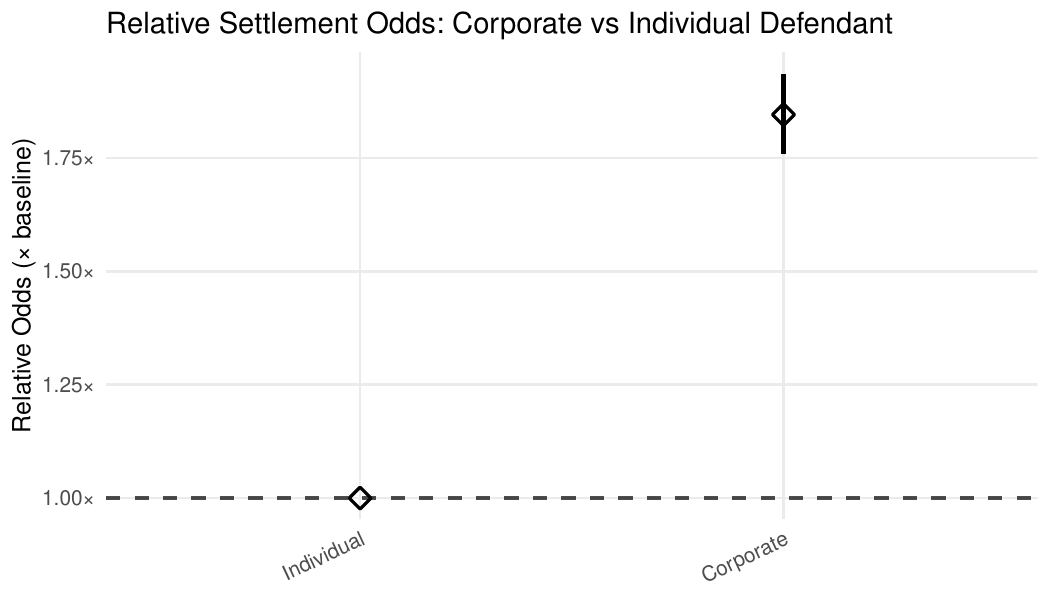}
  \end{subfigure}\hfill
  \begin{subfigure}[t]{0.48\textwidth}
    \centering
    \includegraphics[width=\linewidth]{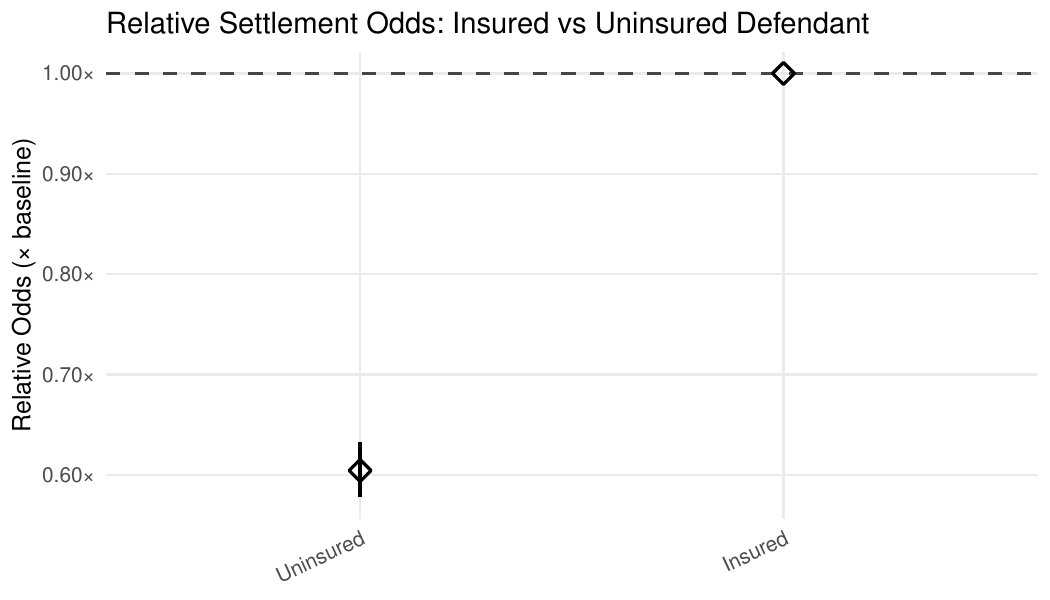}
  \end{subfigure}\hfill
    \begin{subfigure}[t]{0.48\textwidth}
    \centering
    \includegraphics[width=\linewidth]{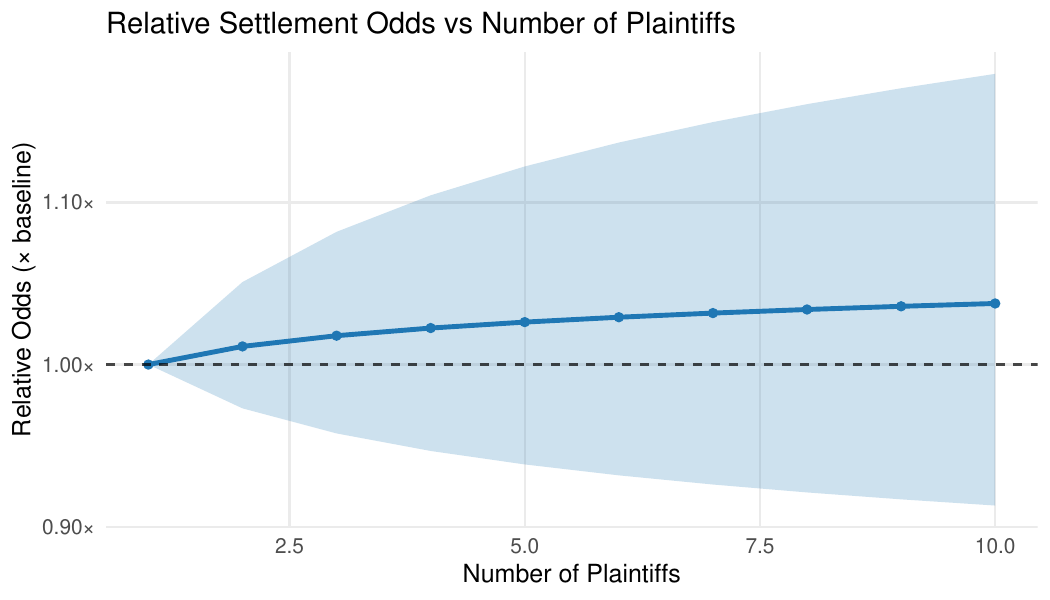}
  \end{subfigure}\hfill
  \begin{subfigure}[t]{0.48\textwidth}
    \centering
    \includegraphics[width=\linewidth]{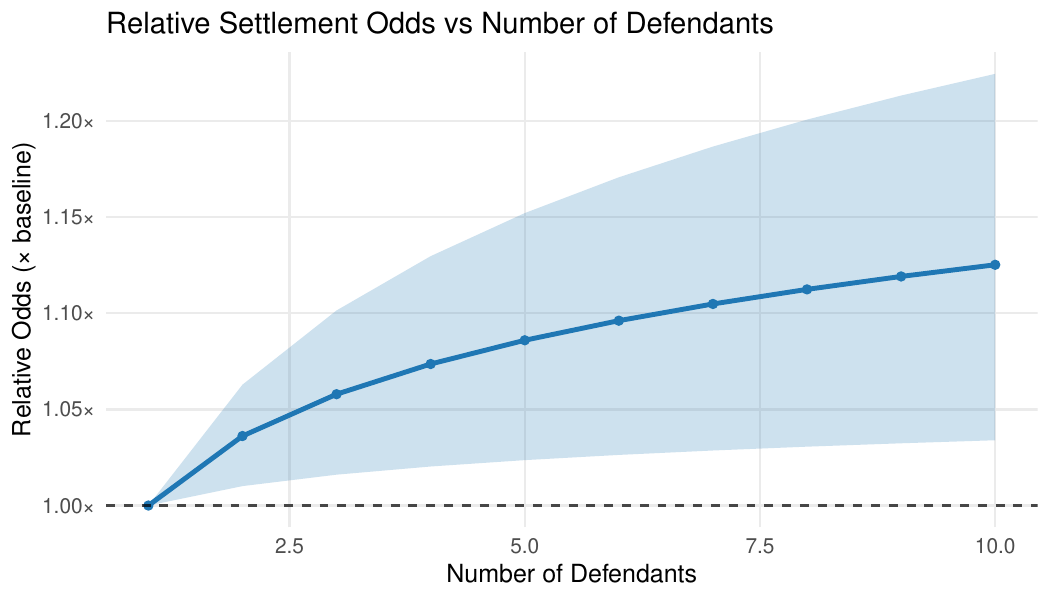}
  \end{subfigure}\hfill
  \begin{subfigure}[t]{0.48\textwidth}
    \centering
    \includegraphics[width=\linewidth]{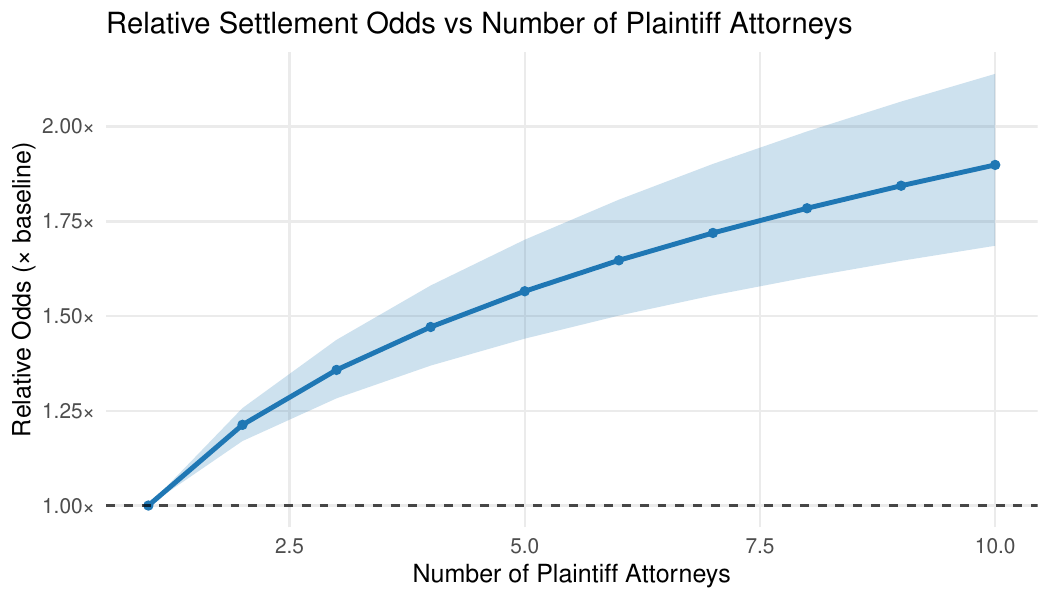}
  \end{subfigure}\hfill
  \begin{subfigure}[t]{0.48\textwidth}
    \centering
    \includegraphics[width=\linewidth]{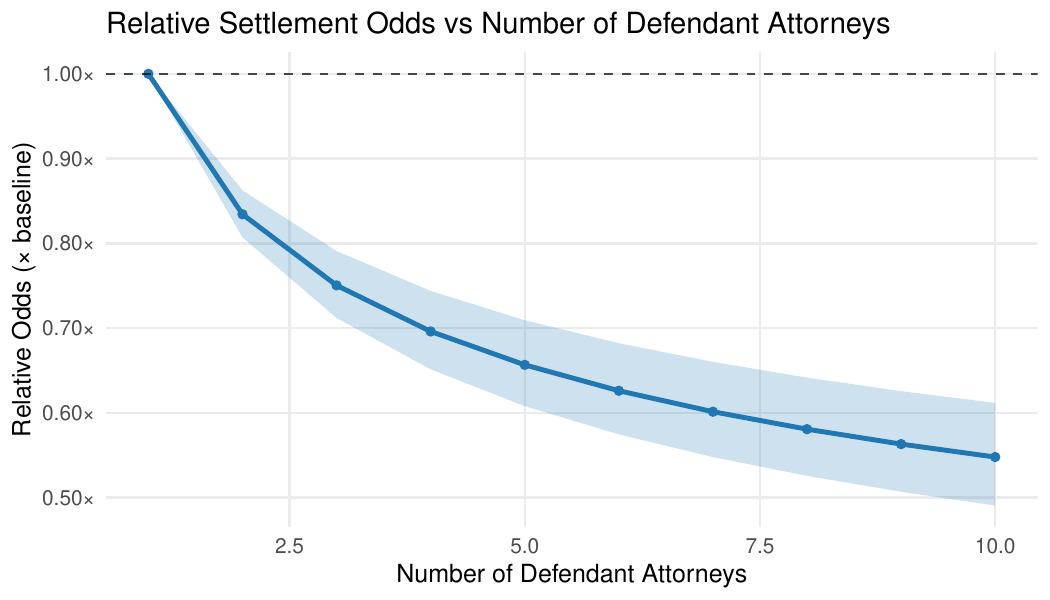}
  \end{subfigure}\hfill
  \begin{subfigure}[t]{0.48\textwidth}
    \centering
    \includegraphics[width=\linewidth]{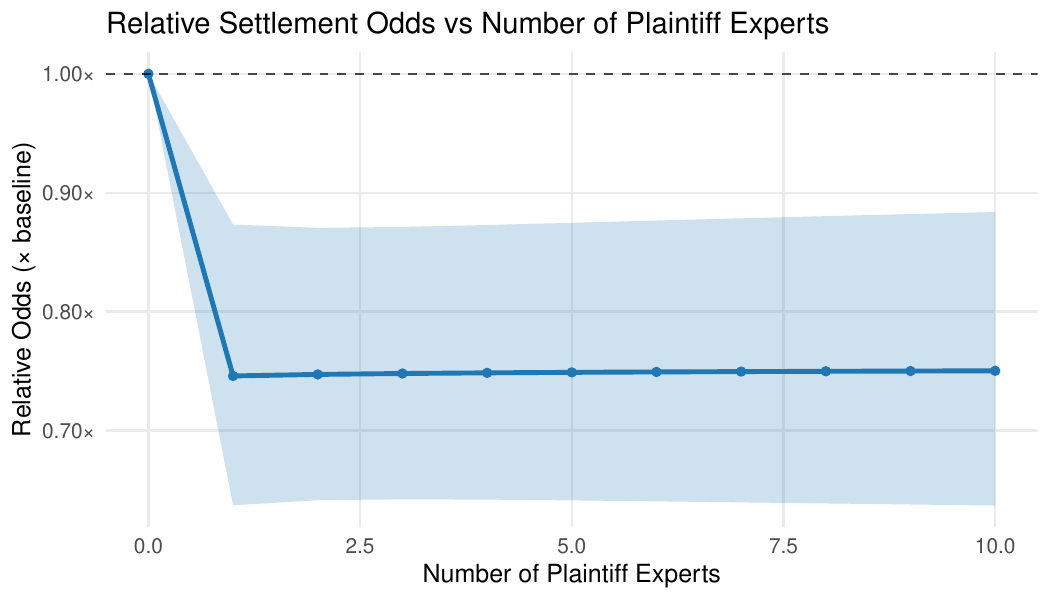}
  \end{subfigure}\hfill
  \begin{subfigure}[t]{0.48\textwidth}
    \centering
    \includegraphics[width=\linewidth]{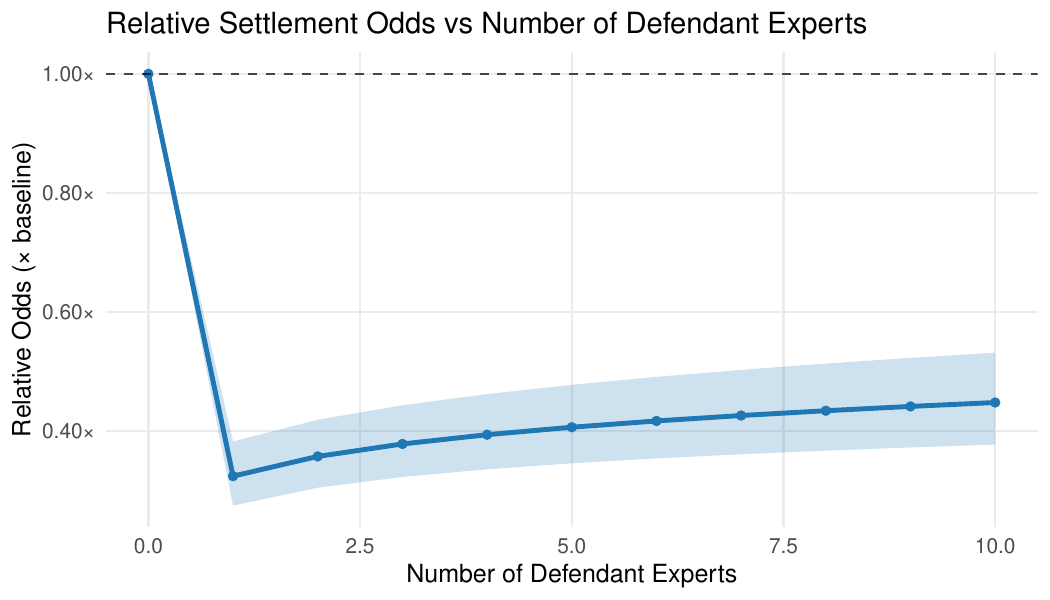}
  \end{subfigure}\hfill
  \begin{subfigure}[t]{0.48\textwidth}
    \centering
    \includegraphics[width=\linewidth]{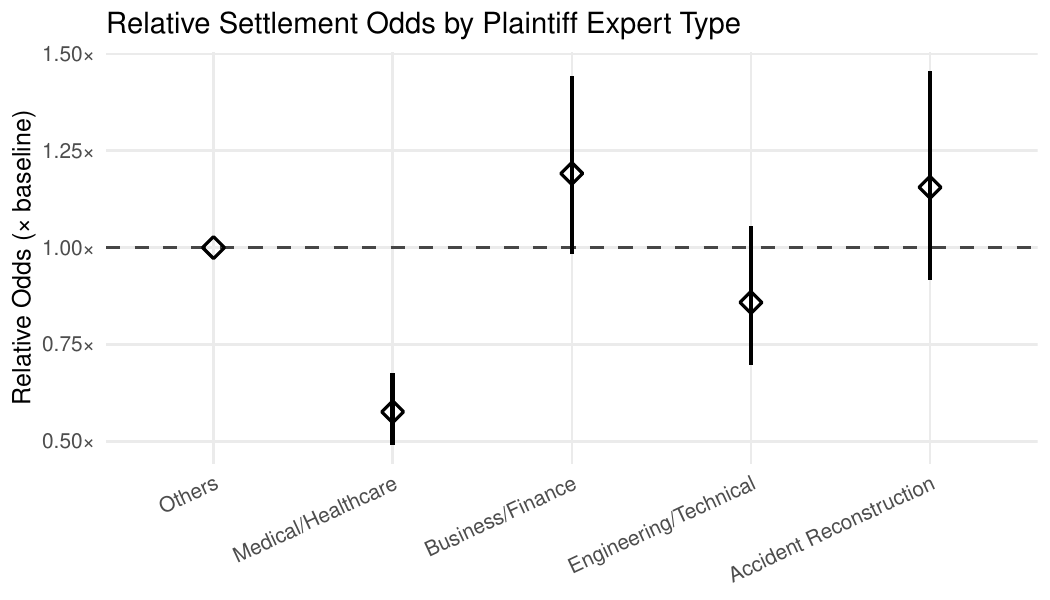}
  \end{subfigure}\hfill
  \begin{subfigure}[t]{0.48\textwidth}
    \centering
    \includegraphics[width=\linewidth]{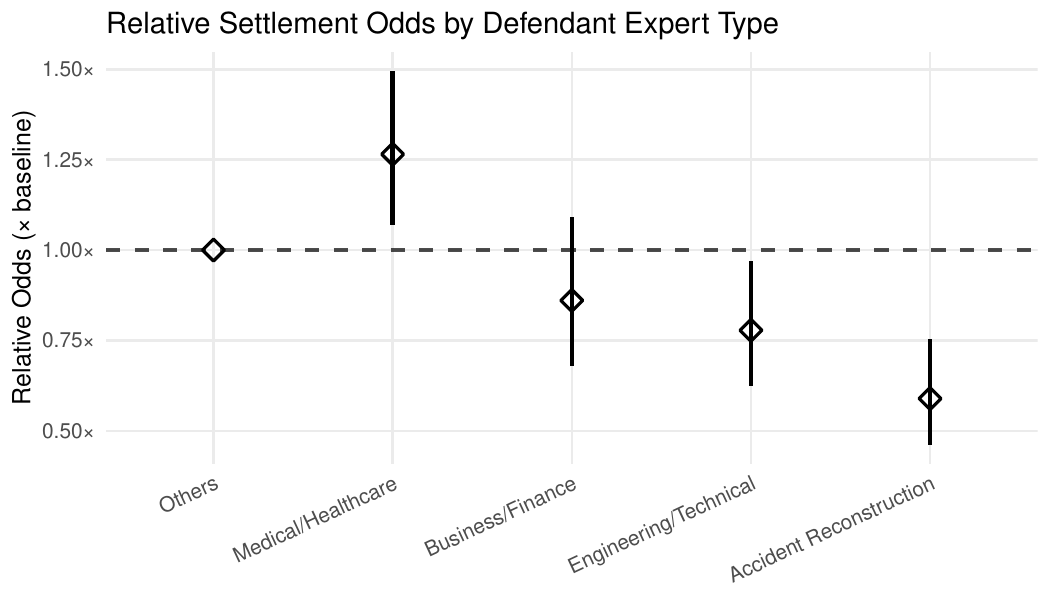}
  \end{subfigure}\hfill
  \caption{Relative settlement odds vs. variables, with 95\% confidence intervals (part 2).}
  \label{fig:cov_prob_s_fig2}
\end{figure}

We next analyze the factors determining whether a case resolves by settlement rather than verdict. To do so, we fit a second logistic regression where the response variable is an indicator equal to 1 if the case ends in a settlement and 0 if it ends in either a plaintiff or defense verdict. As in Section \ref{sec:prelim:prob_p}, the explanatory variables are the full set of factual and strategic covariates from Tables \ref{tab:summary-stats_factual} and \ref{tab:summary-stats_strategic}. We report exponentiated coefficients estimated from \texttt{glm()} as relative settlement odds in Figures \ref{fig:cov_prob_s_fig1} and \ref{fig:cov_prob_s_fig2}.

Settlement probabilities also display pronounced geographic variation. Figure  \ref{fig:cov_prob_s_fig1} shows that cases in New Jersey have the highest settlement odds among the major jurisdictions, with relative odds about three times those in more verdict-oriented states like Texas and Ohio. These differences may reflect both local legal cultures and institutional factors, such as court congestion, the availability of mediation programs, and historical practice patterns regarding pretrial resolution.

Plaintiff age again plays a notable role. Cases involving children and teenagers are significantly more likely to settle than those involving older plaintiffs. This is consistent with the idea that trials can be particularly emotionally draining and retraumatizing for underage victims, while settlement may help protect children's well-being. In contrast, gender, marital status, and parental status exert more modest effects on settlement behavior. Cases with female plaintiffs are slightly less likely to settle than those with male plaintiffs, and cases in which plaintiffs have no children are somewhat more likely to settle than cases involving parents.

With respect to injuries and deaths, the presence of at least one fatality significantly increases the settlement probability, with 75\% higher settlement odds, consistent with the notion that both sides may prefer to avoid the uncertainty of trying very high-stakes wrongful death claims. Injury type is also associated with settlement likelihood: Surgery, head injuries, and internal injuries, which typically signal greater severity and medical complexity, are linked to higher settlement odds. Cases involving either many injuries or no injuries are generally more likely to settle than cases involving exactly one injury.

Case categories and party structure show somewhat similar patterns to those observed for plaintiff victory. General and motor liability cases tend to have the highest settlement probabilities. Cases involving corporate defendants have higher settlement odds than those with only individual defendants, while cases with insured defendants tend to be more likely to settle than cases with uninsured defendants. With respect to expert involvement, the presence of plaintiff and/or defense experts is associated with lower settlement odds. However, the involvement of additional plaintiff or defense experts does not materially impact the settlement odds. 

\subsection{Variable effects on verdict amount} \label{sec:prelim:sev_p}
\begin{figure}[H]
  \centering
  \begin{subfigure}[t]{0.48\textwidth}
    \centering
    \includegraphics[width=\linewidth]{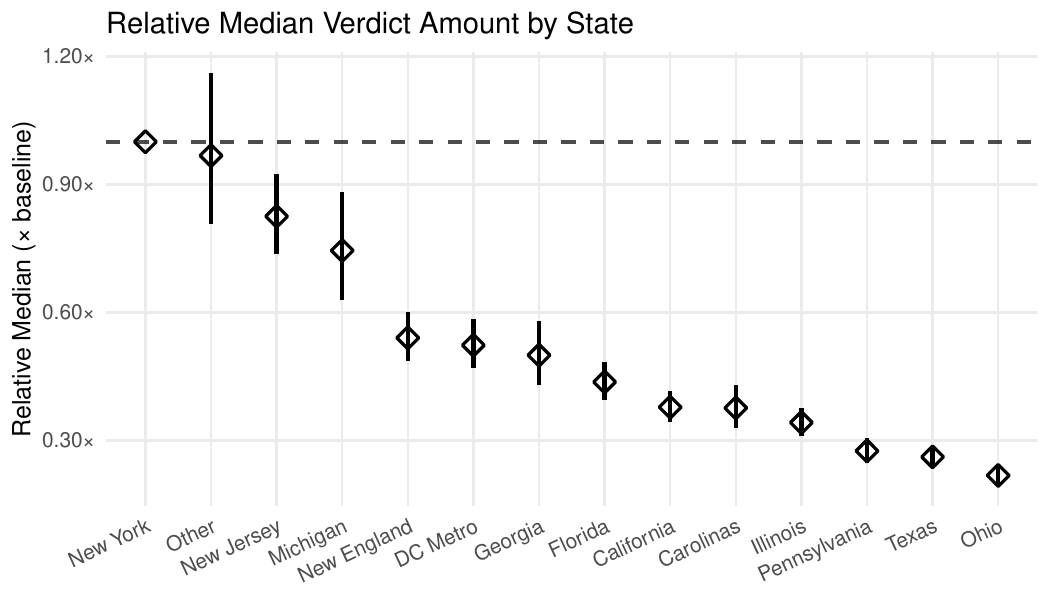}
  \end{subfigure}\hfill
  \begin{subfigure}[t]{0.48\textwidth}
    \centering
    \includegraphics[width=\linewidth]{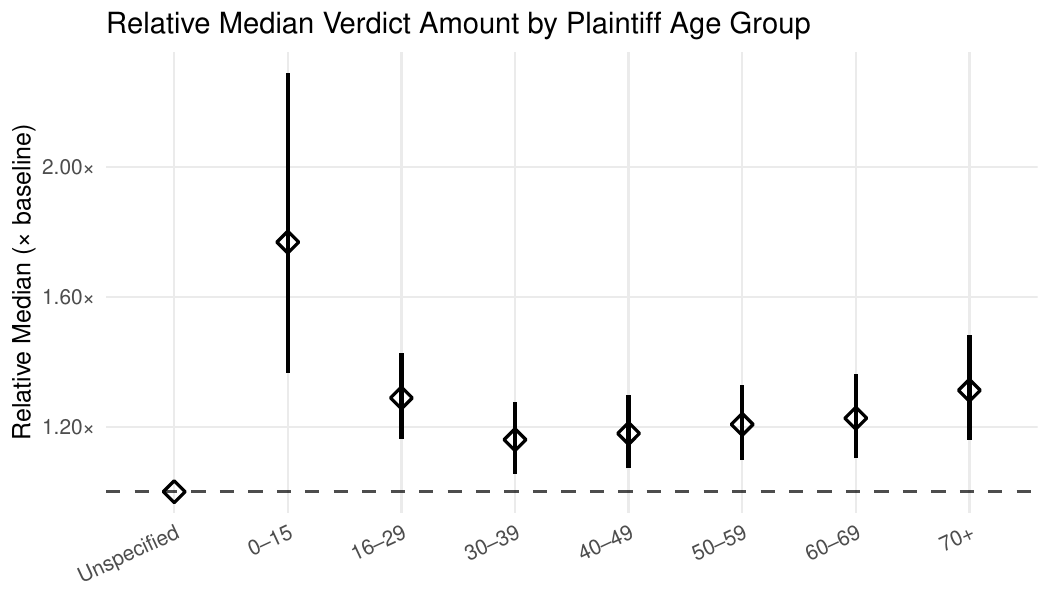}
  \end{subfigure}\hfill
    \begin{subfigure}[t]{0.48\textwidth}
    \centering
    \includegraphics[width=\linewidth]{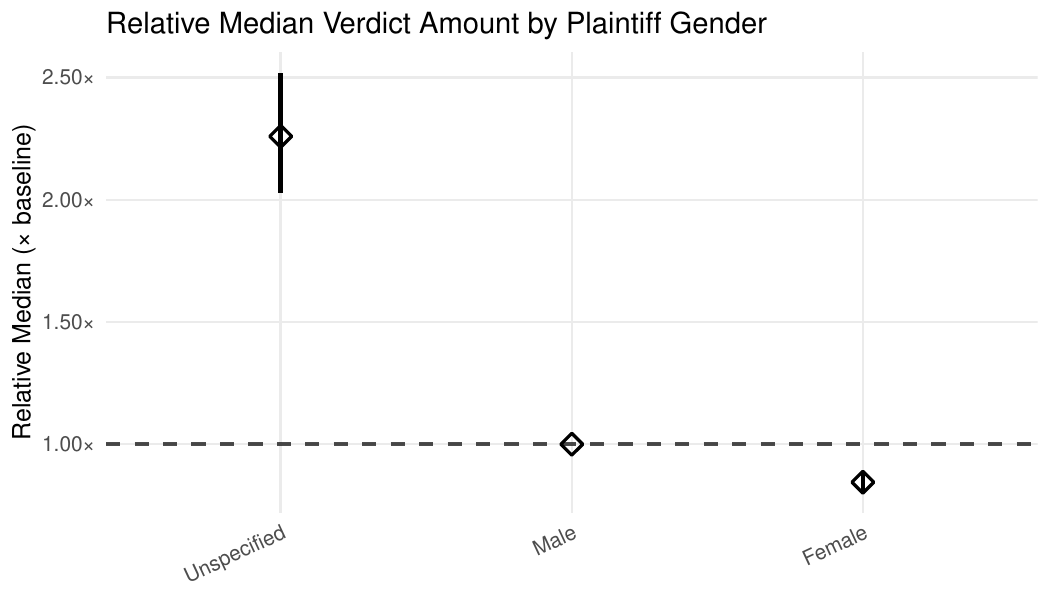}
  \end{subfigure}\hfill
  \begin{subfigure}[t]{0.48\textwidth}
    \centering
    \includegraphics[width=\linewidth]{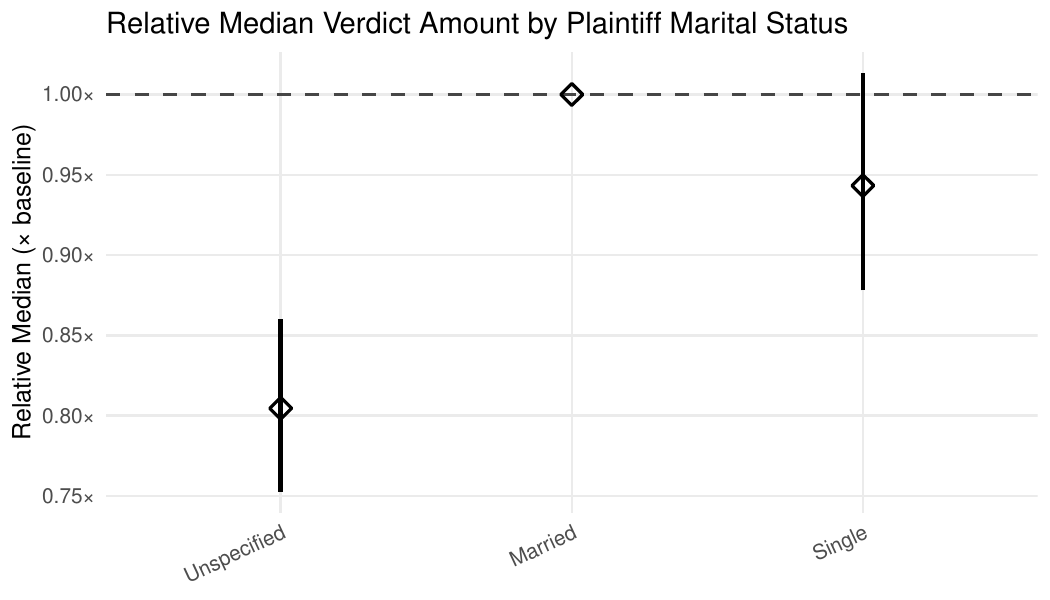}
  \end{subfigure}\hfill
  \begin{subfigure}[t]{0.48\textwidth}
    \centering
    \includegraphics[width=\linewidth]{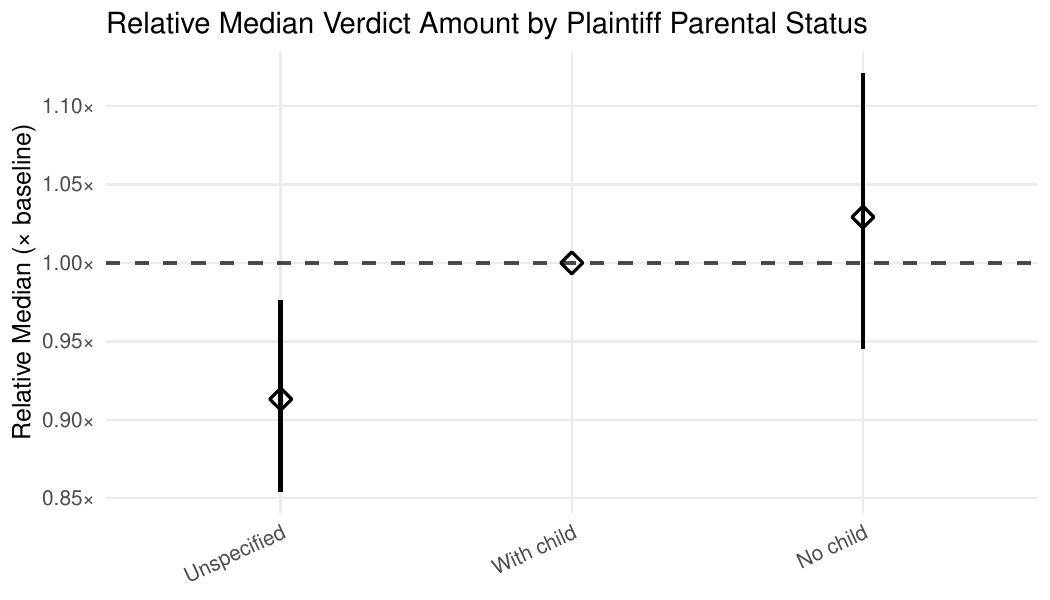}
  \end{subfigure}\hfill
  \begin{subfigure}[t]{0.48\textwidth}
    \centering
    \includegraphics[width=\linewidth]{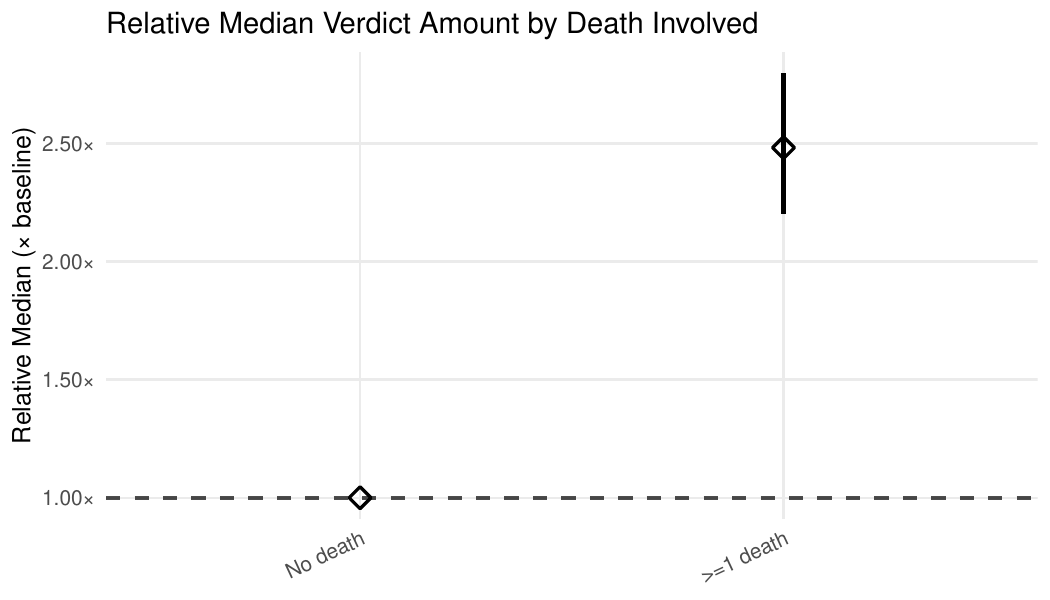}
  \end{subfigure}\hfill
  \begin{subfigure}[t]{0.48\textwidth}
    \centering
    \includegraphics[width=\linewidth]{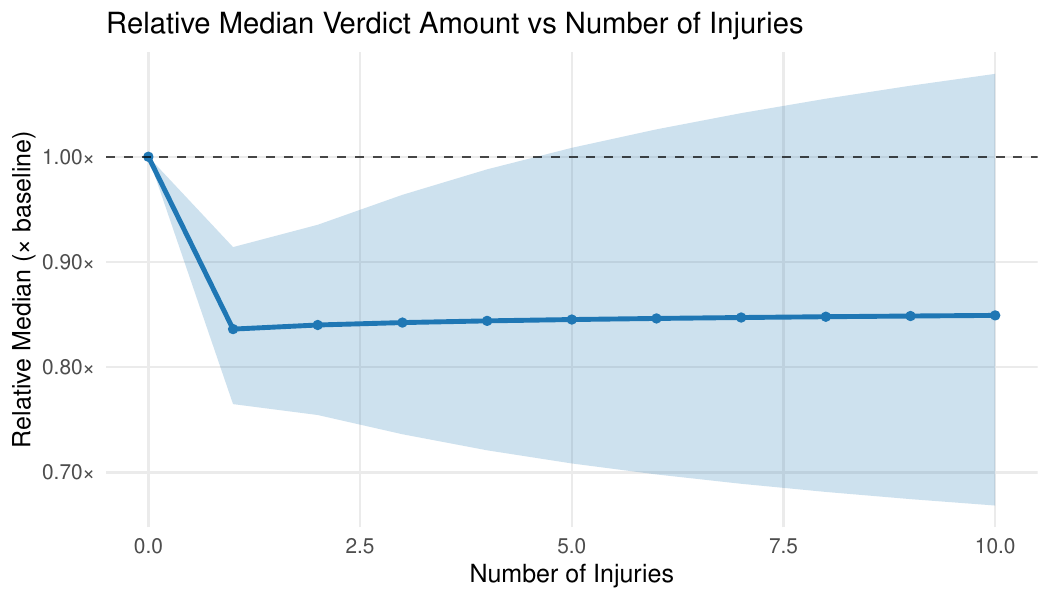}
  \end{subfigure}\hfill
  \begin{subfigure}[t]{0.48\textwidth}
    \centering
    \includegraphics[width=\linewidth]{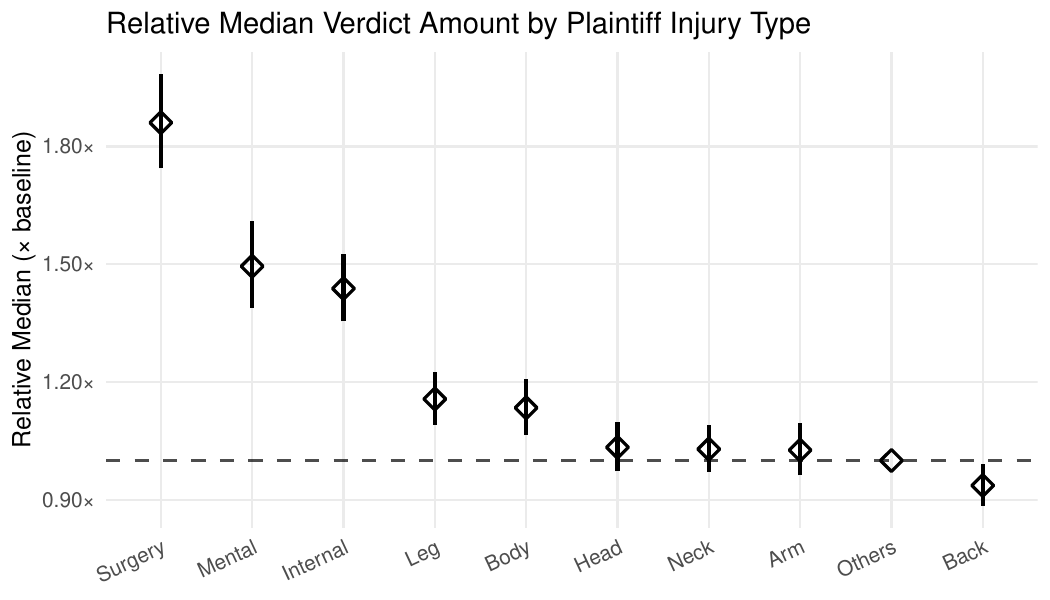}
  \end{subfigure}\hfill
  \caption{Relative median verdict amount vs. variables, with 95\% confidence intervals (part 1).}
  \label{fig:cov_sev_p_fig1}
\end{figure}

\begin{figure}[H]
  \centering
    \begin{subfigure}[t]{0.48\textwidth}
    \centering
    \includegraphics[width=\linewidth]{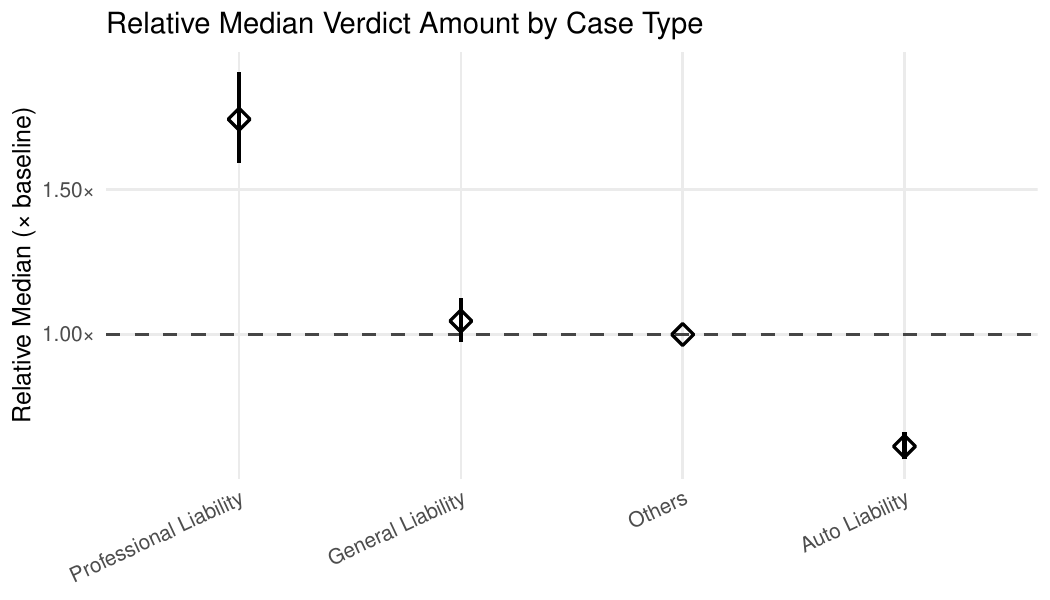}
  \end{subfigure}\hfill
  \begin{subfigure}[t]{0.48\textwidth}
    \centering
    \includegraphics[width=\linewidth]{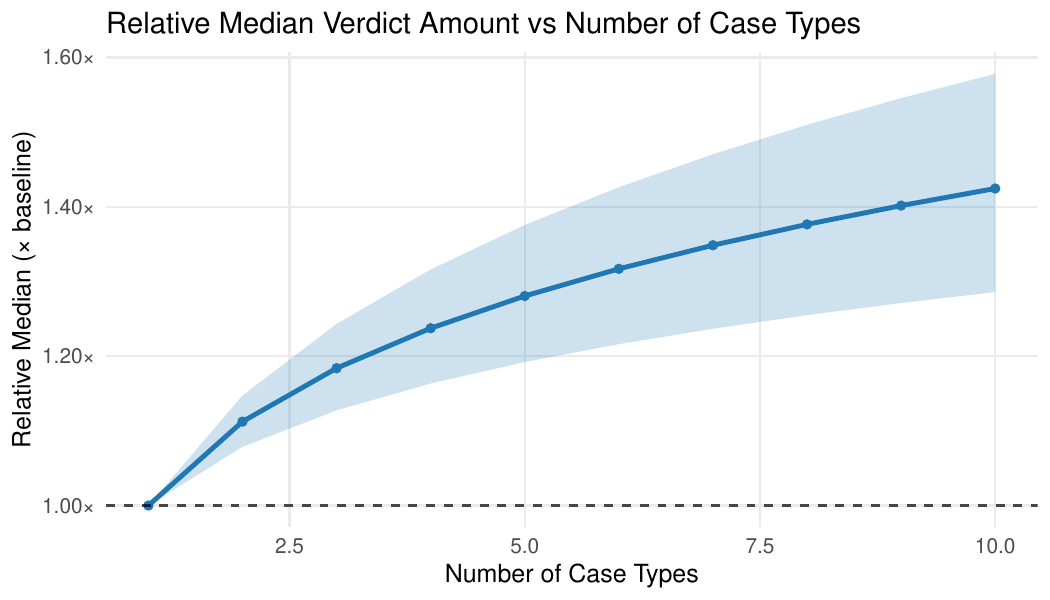}
  \end{subfigure}\hfill
  \begin{subfigure}[t]{0.48\textwidth}
    \centering
    \includegraphics[width=\linewidth]{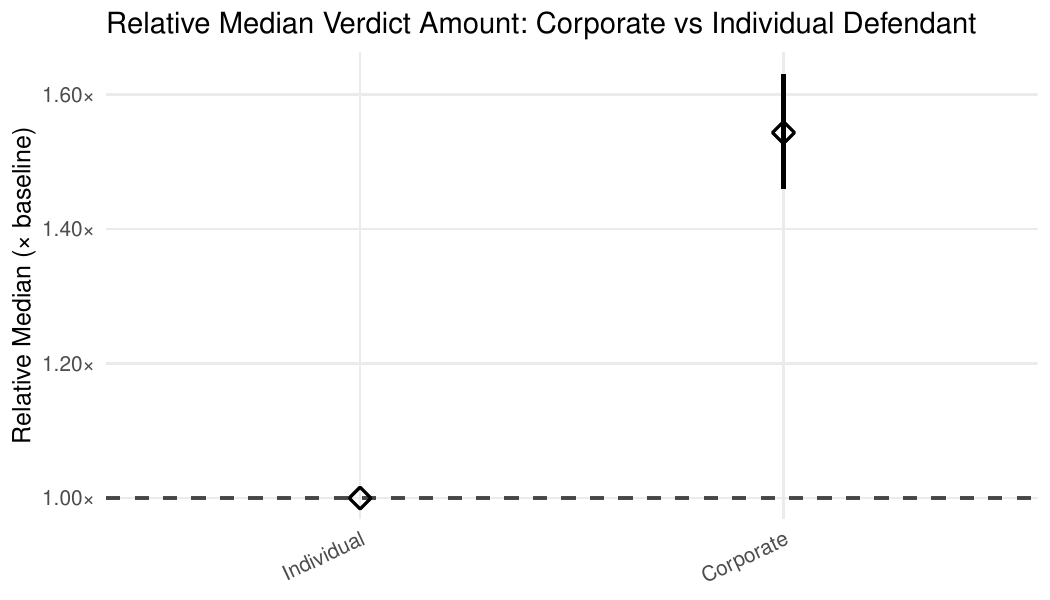}
  \end{subfigure}\hfill
  \begin{subfigure}[t]{0.48\textwidth}
    \centering
    \includegraphics[width=\linewidth]{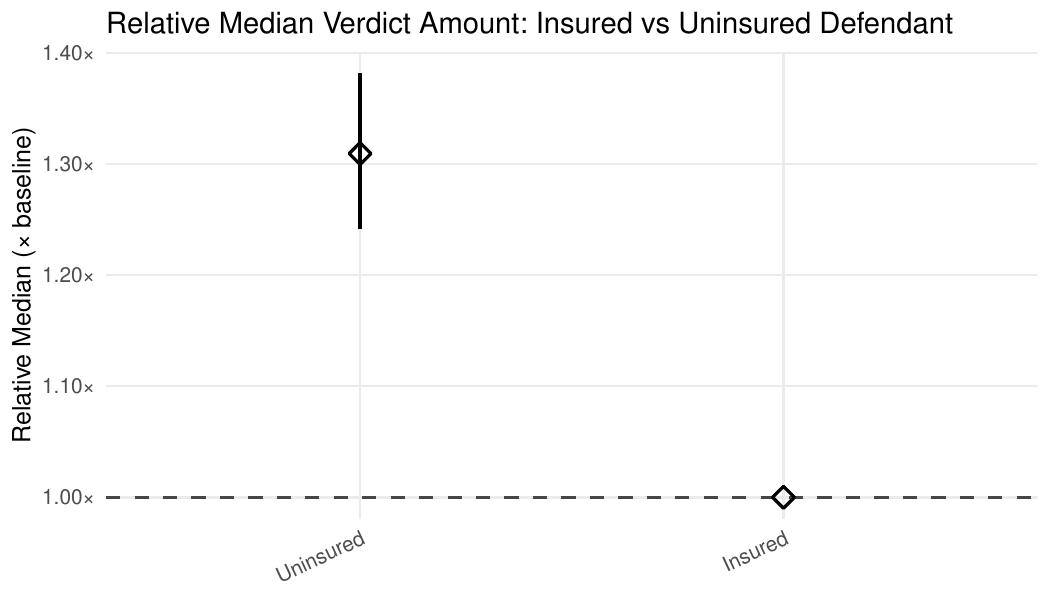}
  \end{subfigure}\hfill
    \begin{subfigure}[t]{0.48\textwidth}
    \centering
    \includegraphics[width=\linewidth]{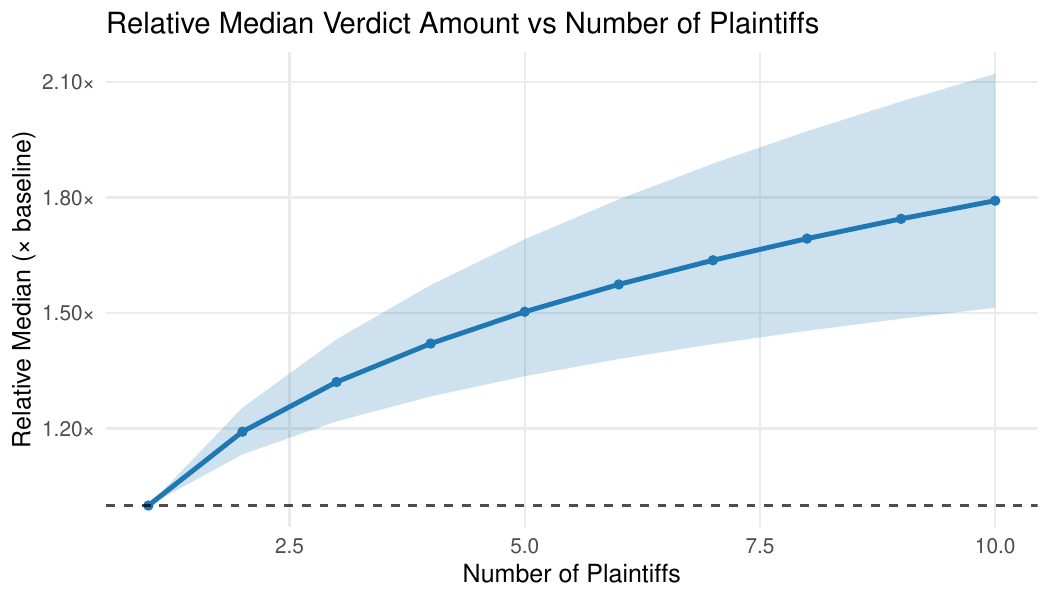}
  \end{subfigure}\hfill
  \begin{subfigure}[t]{0.48\textwidth}
    \centering
    \includegraphics[width=\linewidth]{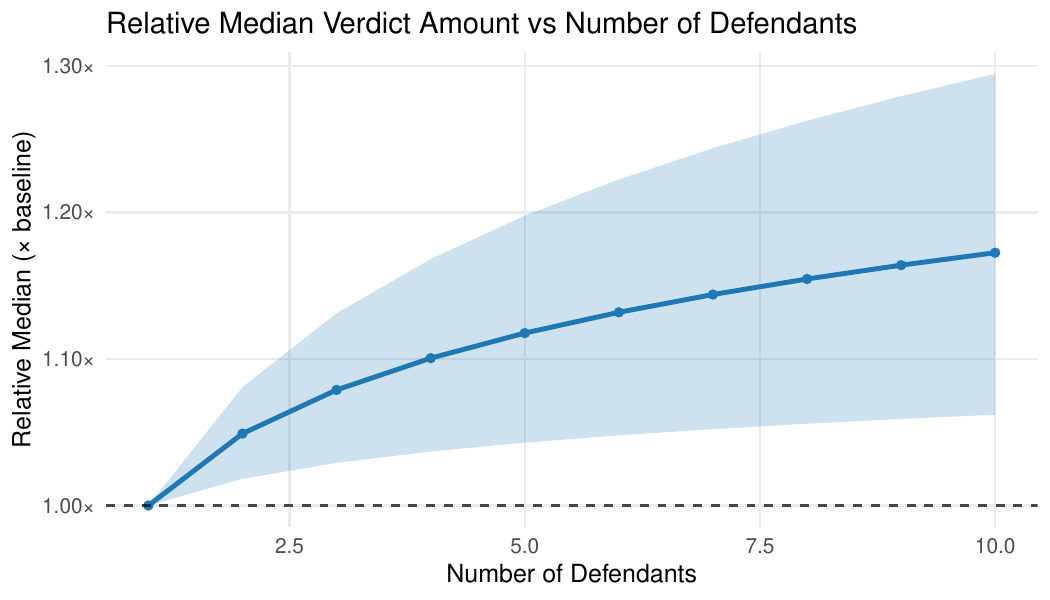}
  \end{subfigure}\hfill
  \begin{subfigure}[t]{0.48\textwidth}
    \centering
    \includegraphics[width=\linewidth]{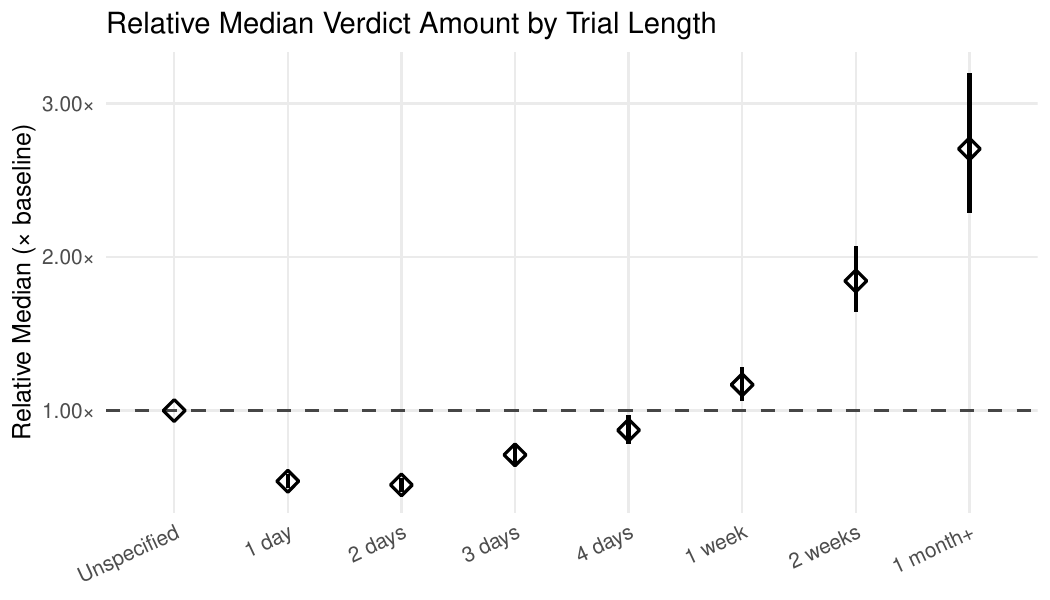}
  \end{subfigure}\hfill
  \begin{subfigure}[t]{0.48\textwidth}
    \centering
    \includegraphics[width=\linewidth]{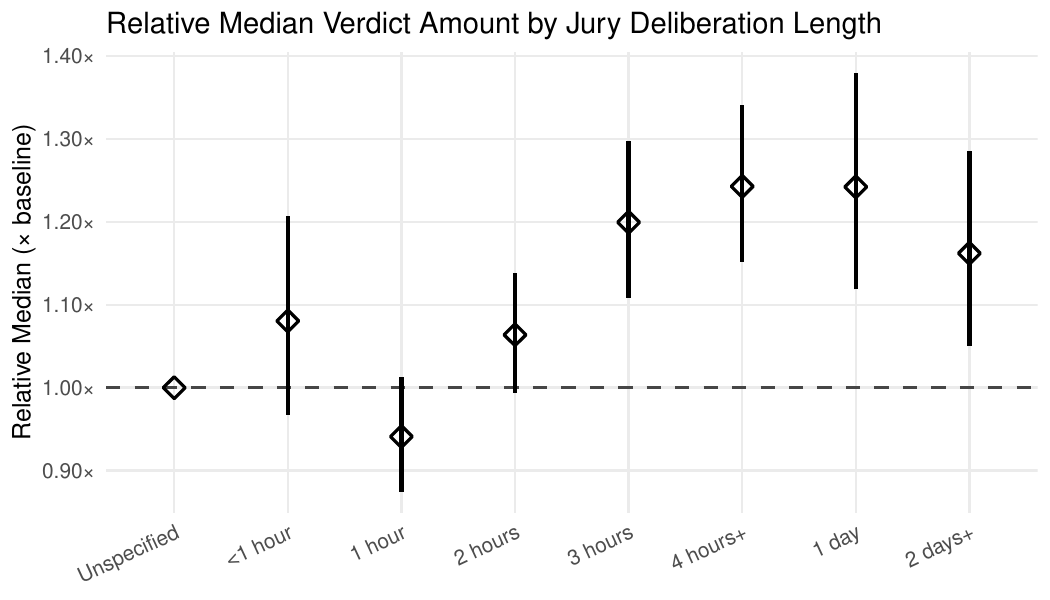}
  \end{subfigure}\hfill
  \caption{Relative median verdict amount vs. variables, with 95\% confidence intervals (part 2).}
  \label{fig:cov_sev_p_fig2}
\end{figure}

\begin{figure}[H]
  \centering
    \begin{subfigure}[t]{0.48\textwidth}
    \centering
    \includegraphics[width=\linewidth]{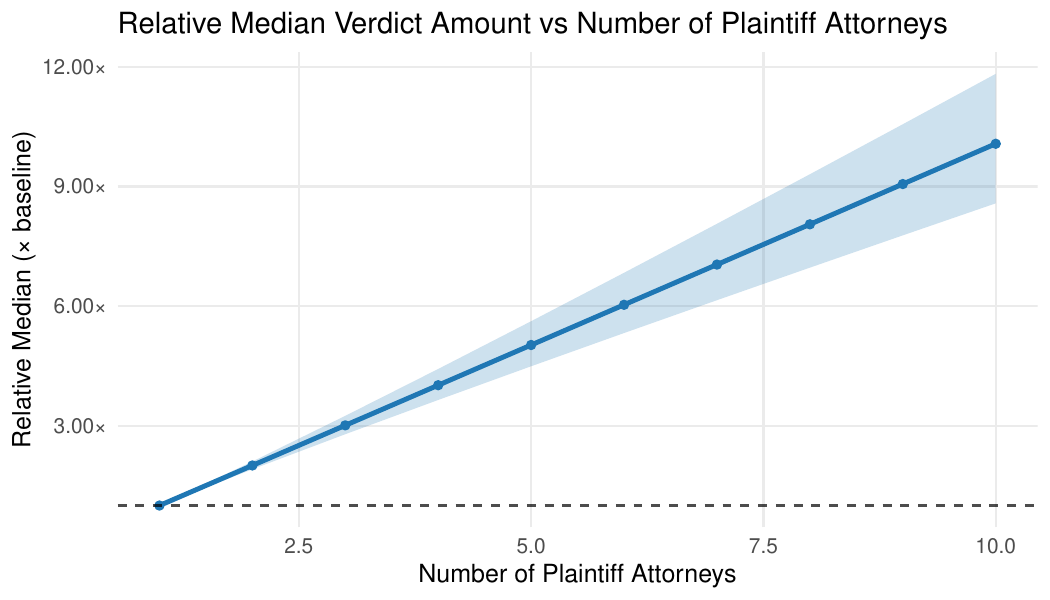}
  \end{subfigure}\hfill
  \begin{subfigure}[t]{0.48\textwidth}
    \centering
    \includegraphics[width=\linewidth]{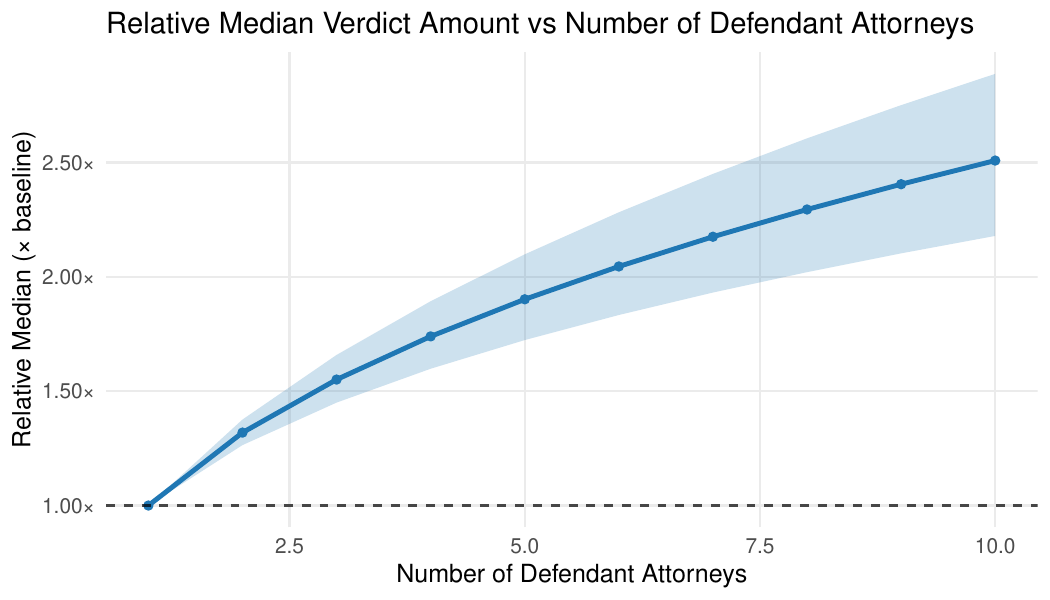}
  \end{subfigure}\hfill
  \begin{subfigure}[t]{0.48\textwidth}
    \centering
    \includegraphics[width=\linewidth]{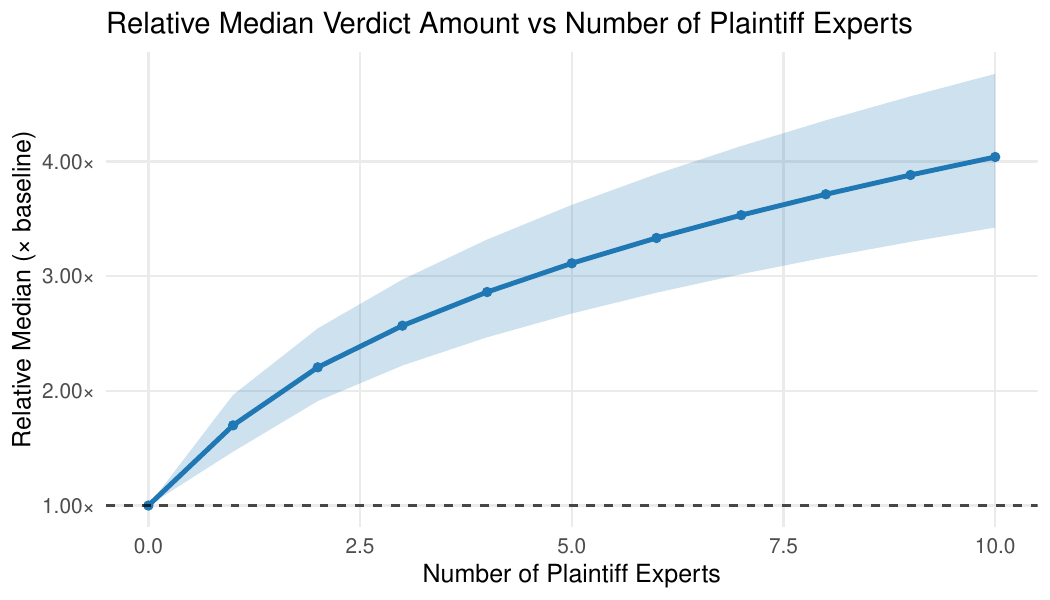}
  \end{subfigure}\hfill
  \begin{subfigure}[t]{0.48\textwidth}
    \centering
    \includegraphics[width=\linewidth]{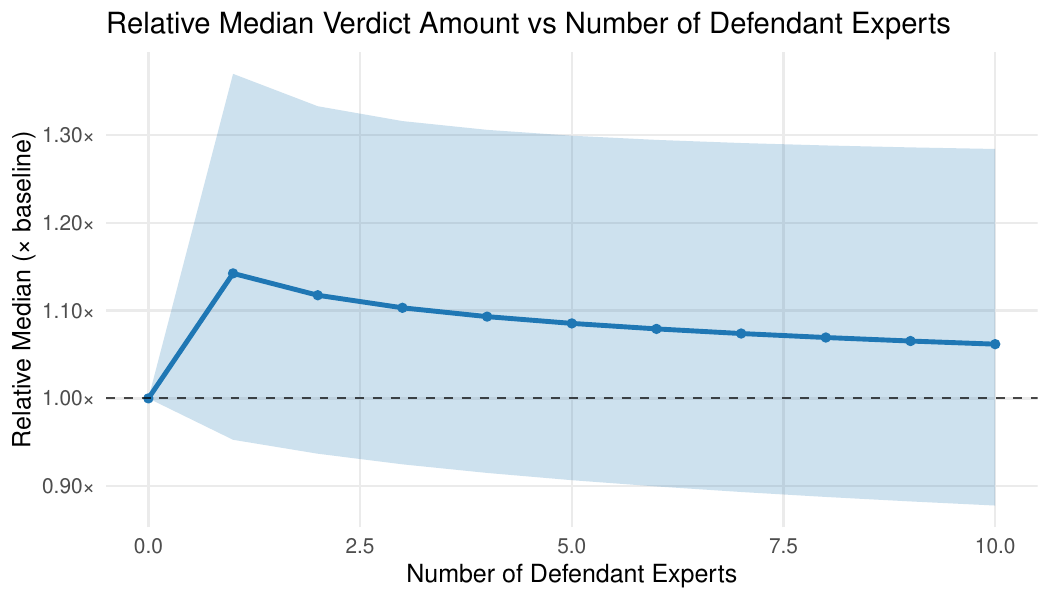}
  \end{subfigure}\hfill
  \begin{subfigure}[t]{0.48\textwidth}
    \centering
    \includegraphics[width=\linewidth]{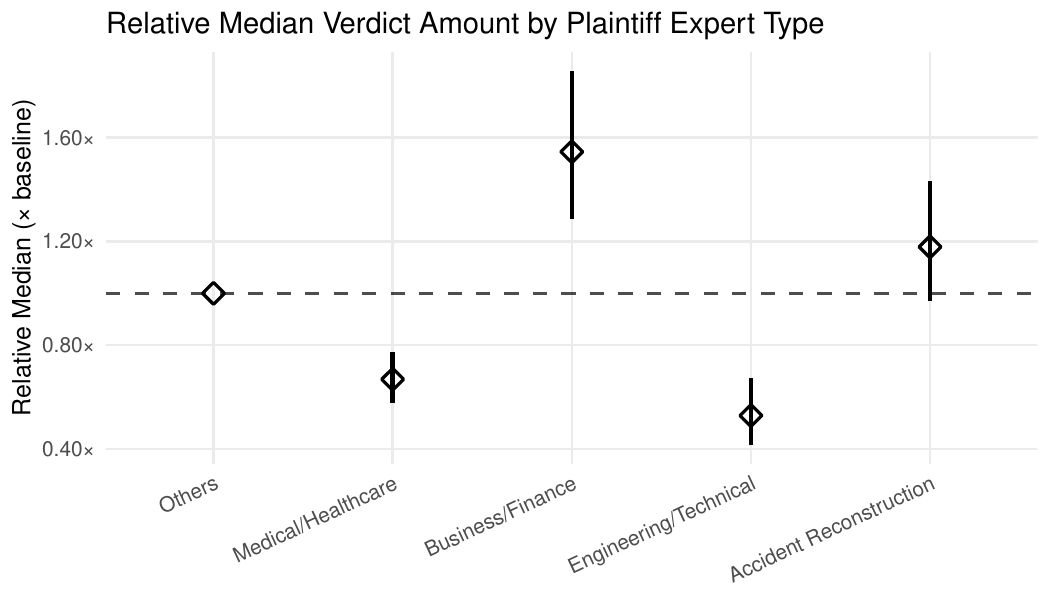}
  \end{subfigure}\hfill
  \begin{subfigure}[t]{0.48\textwidth}
    \centering
    \includegraphics[width=\linewidth]{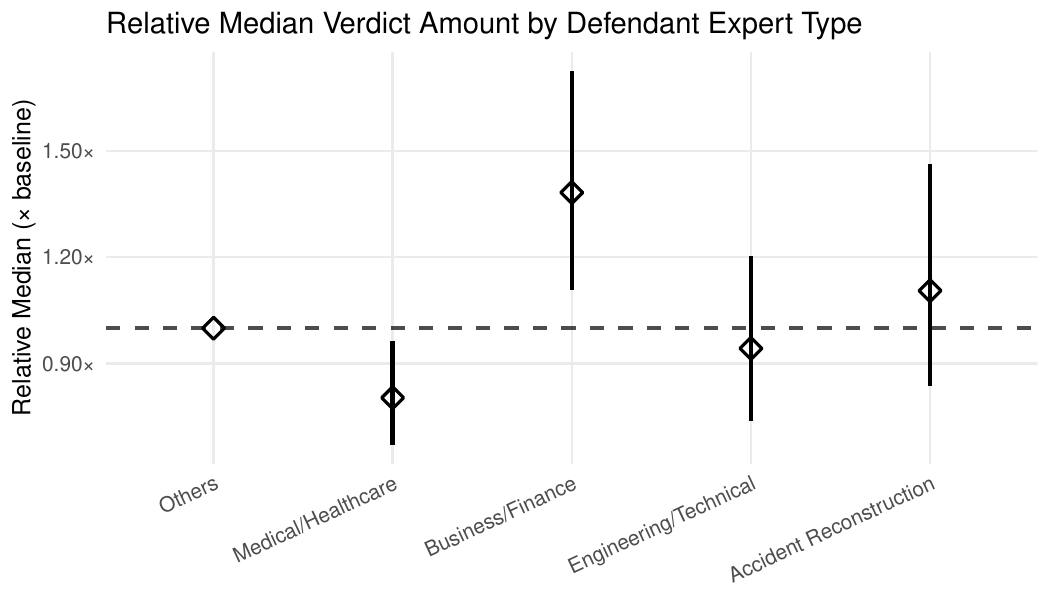}
  \end{subfigure}\hfill
  \caption{Relative median verdict amount vs. variables, with 95\% confidence intervals (part 3).}
  \label{fig:cov_sev_p_fig3}
\end{figure}

We now turn to the size of verdict awards, conditional on plaintiff success. Because the distribution of verdict amounts is highly skewed and heavy-tailed, we model the median (rather than the mean) of the log award amount using quantile regression. Specifically, we estimate a quantile regression of the median log verdict award on the full set of factual and strategic variables, restricting the sample to cases in which the plaintiff wins and receives a positive award. Quantile regression is a flexible, distribution-free method: It does not impose any parametric distributional assumption on award amounts but instead assumes that a given conditional quantile (here, the median) of the log award is a linear function of the covariates. We implement the model using the \texttt{rq()} function from the \texttt{quantreg} package in \texttt{R}. Exponentiating the estimated coefficients yields multiplicative effects on the conditional median award, reported as relative median verdict amounts in Figures \ref{fig:cov_sev_p_fig1} to \ref{fig:cov_sev_p_fig3}, together with 95\% confidence intervals.

As with win probabilities and settlement odds, we find large geographic differences in median verdict awards. Cases in New York have the highest conditional median verdict amounts, roughly four times those in low-award jurisdictions such as Texas and Ohio. New Jersey, Michigan, the DC Metro region, New England, and California also exhibit elevated median awards relative to the baseline, while some Southern and Midwestern states show lower medians. These patterns reflect both local tort environments and venue-specific norms regarding damages. They also imply that any aggregate measure of social inflation must carefully separate jurisdictional composition shifts from genuine changes in award levels.

Plaintiff demographics have weaker but still noticeable effects on verdict magnitudes. Cases involving very young plaintiffs (birth to 15 years of age) tend to yield higher median verdict amounts than those involving older plaintiffs. This is consistent with the greater lifetime economic losses and noneconomic damages that can be claimed for children and teenagers. By contrast, verdict medians are relatively flat across adult age groups. Male plaintiffs obtain somewhat higher median awards than female plaintiffs, but the difference is moderate. The impacts of marital and parental status on the verdict award are insignificant.

Injury and death variables, unsurprisingly, exert a strong influence on award size. Fatal cases are associated with substantially larger median verdict amounts than nonfatal cases---approximately 1.5 times larger---reflecting both economic losses such as lost income and financial support and noneconomic components of wrongful death damages. Among nonfatal injuries, median awards are highest for surgery, mental injuries, and internal injuries and lowest for arm, back, and neck injuries. The relative medians across injury types can differ by a factor of approximately 1.5 to 2, meaning that cases involving surgery and severe internal or mental injuries receive median awards about 50\% to 100\% higher than those associated with less severe injury categories. This pattern indicates that injury type functions as a coarse proxy for injury severity and long-term impact on damages.

Case categories and party structure are strongly associated with award magnitude. Professional liability cases, which often involve malpractice in medical, legal, or other high-stakes professional services, produce the largest median verdicts, followed by general liability cases, with motor vehicle cases generally yielding smaller verdict amounts. Each additional case type coded for a dispute is associated with a noticeable increase in the median verdict amount, consistent with more complex cases involving multiple theories of liability and overlapping damages. Cases involving corporate defendants have substantially higher median verdicts than those with only individual defendants. This pattern aligns with the intuition that corporate defendants are more often involved in large-stakes litigation and that juries may perceive deeper pockets and greater responsibility in such cases. The presence of insured defendants, however, is associated with lower median awards, reflecting possible legal efforts and strategies by insurance companies to lower the plaintiff award amounts.

Procedural and strategic variables also contribute to award size. Longer trials are associated with much larger median verdict amounts: Cases that last a week or longer result in median awards that are multiples higher than those associated with one-day trials. Extended jury deliberations are similarly linked to higher median awards: Cases in which juries deliberate for a full day or multiple days tend to yield significantly larger median awards than those decided within two hours. These relationships likely reflect higher intrinsic stakes for large cases that justify longer trials and deliberations.

Finally, the intensity and specialization of legal and expert resources are strongly related to award size. The median verdict amount increases steeply with the number of plaintiff attorneys and, to a lesser yet still highly significant extent, with the number of defense attorneys, consistent with greater resources being deployed in larger cases. Adding plaintiff experts is associated with substantial increases in the median award, up to several-fold as the number of experts rises from one to 10, while the number of defense experts does not show a significant association. With respect to expert types, cases in which plaintiffs or defendants retain business/finance experts tend to have considerably higher median verdicts, reflecting the fact that high-stakes cases often require testimony from experts specializing in business areas.


\subsection{Variable effects on settlement amount} \label{sec:prelim:sev_s}
\begin{figure}[H]
  \centering
  \begin{subfigure}[t]{0.48\textwidth}
    \centering
    \includegraphics[width=\linewidth]{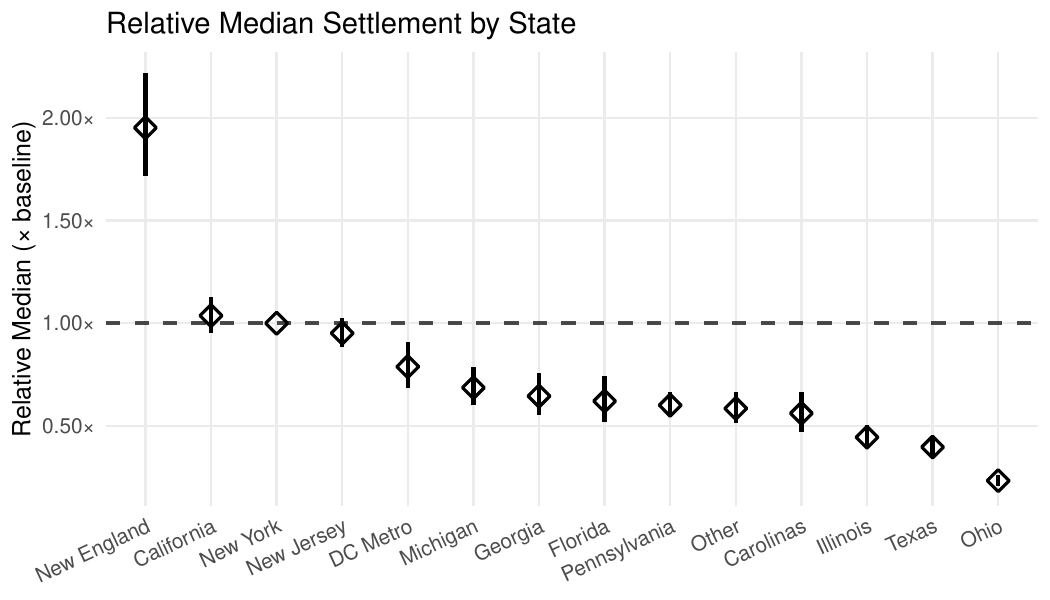}
  \end{subfigure}\hfill
  \begin{subfigure}[t]{0.48\textwidth}
    \centering
    \includegraphics[width=\linewidth]{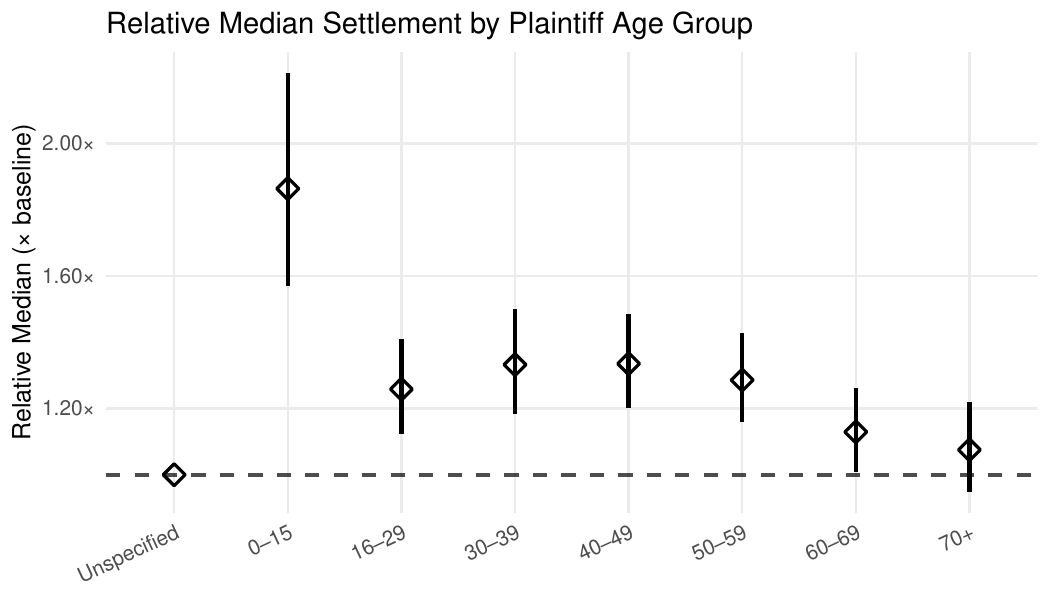}
  \end{subfigure}\hfill
    \begin{subfigure}[t]{0.48\textwidth}
    \centering
    \includegraphics[width=\linewidth]{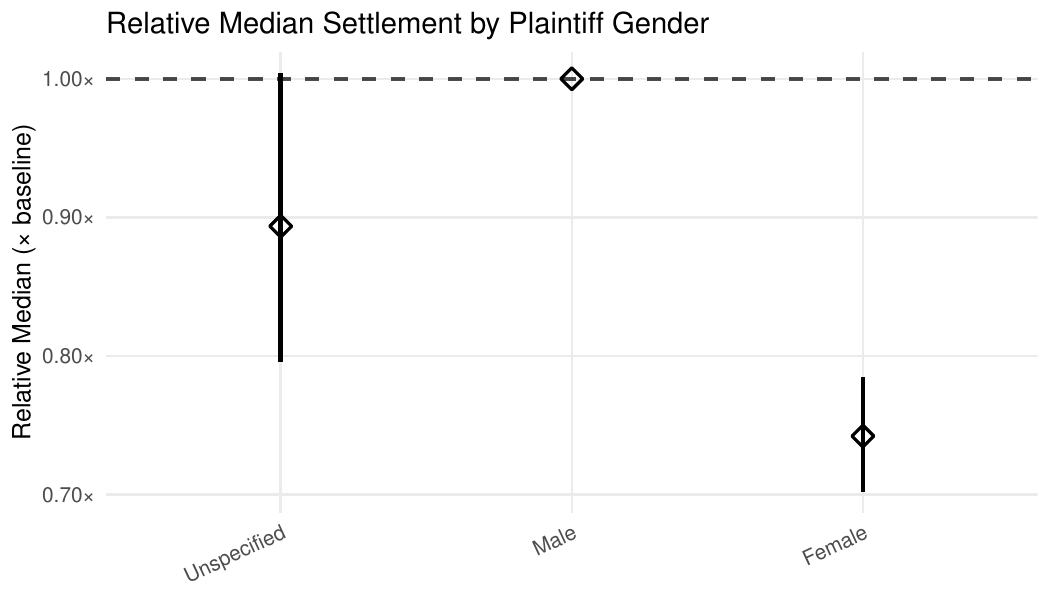}
  \end{subfigure}\hfill
  \begin{subfigure}[t]{0.48\textwidth}
    \centering
    \includegraphics[width=\linewidth]{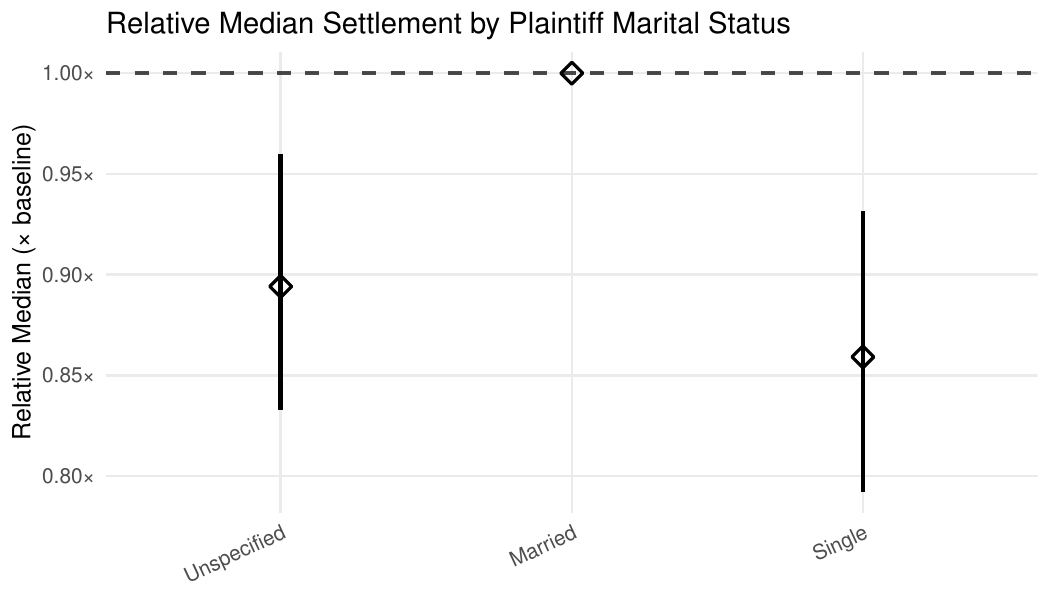}
  \end{subfigure}\hfill
  \begin{subfigure}[t]{0.48\textwidth}
    \centering
    \includegraphics[width=\linewidth]{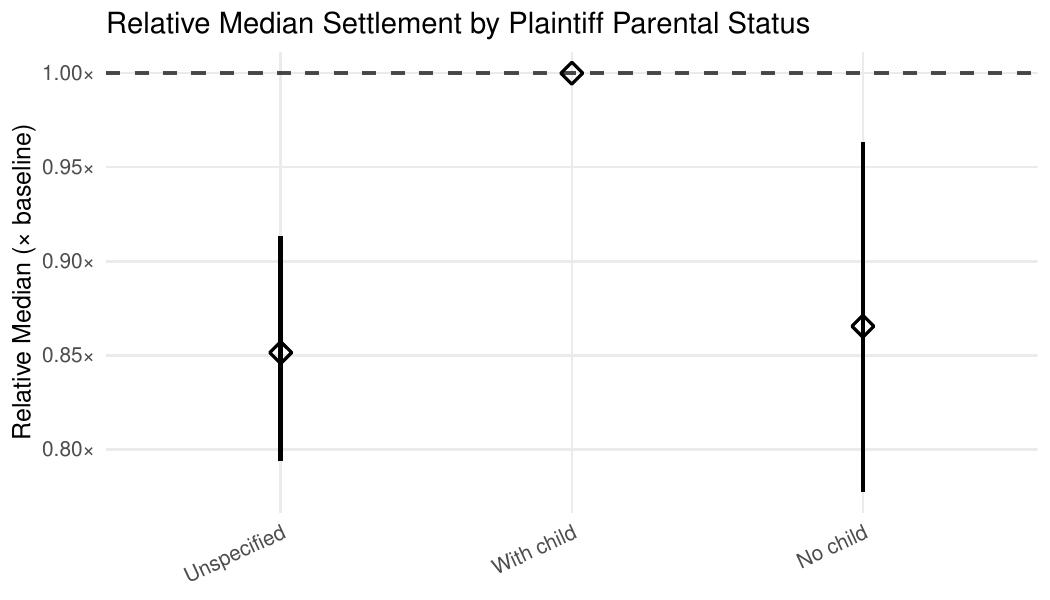}
  \end{subfigure}\hfill
  \begin{subfigure}[t]{0.48\textwidth}
    \centering
    \includegraphics[width=\linewidth]{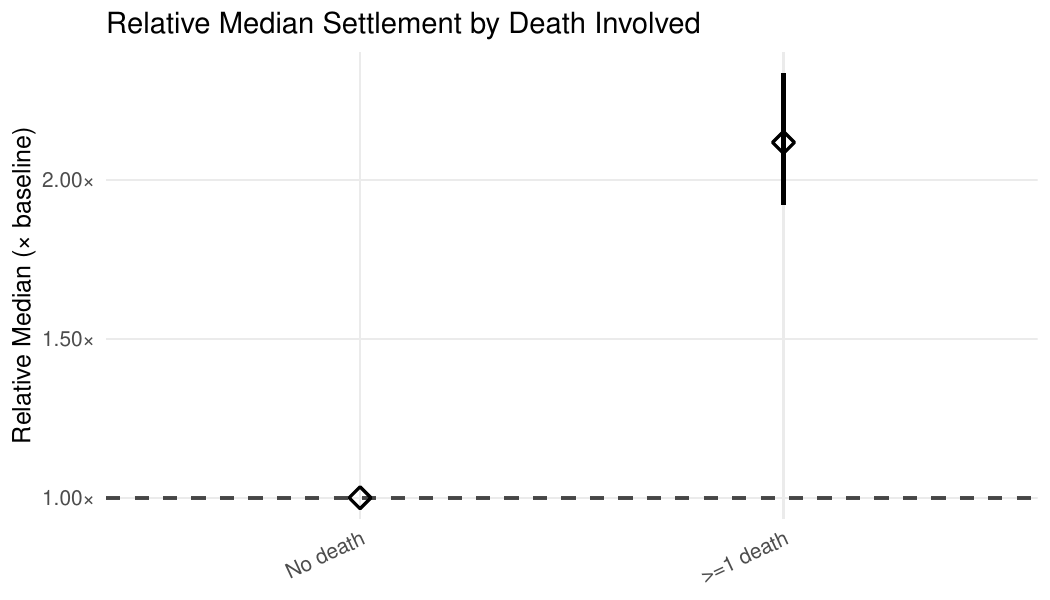}
  \end{subfigure}\hfill
  \begin{subfigure}[t]{0.48\textwidth}
    \centering
    \includegraphics[width=\linewidth]{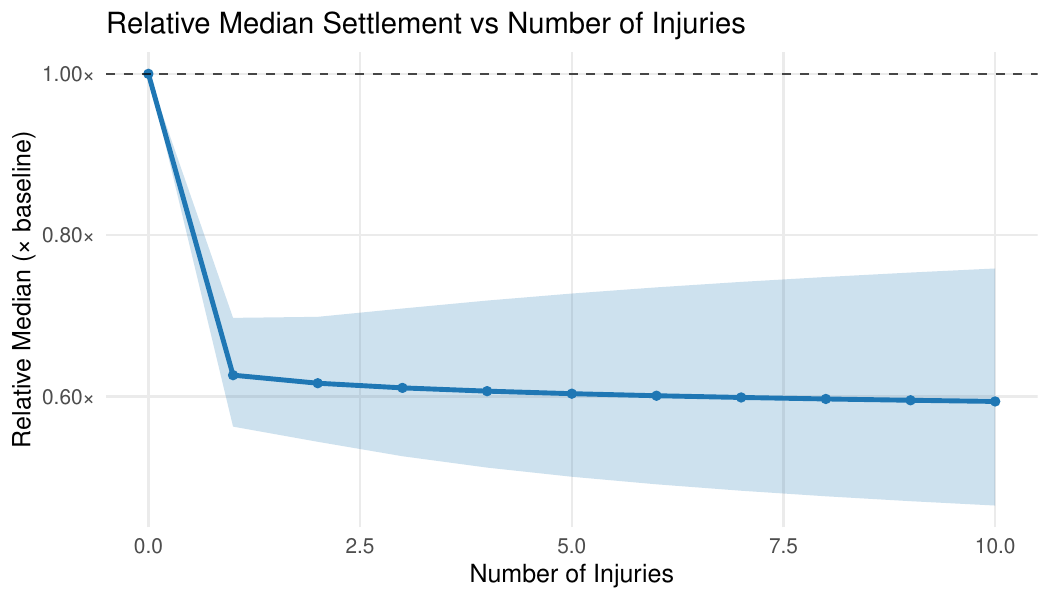}
  \end{subfigure}\hfill
  \begin{subfigure}[t]{0.48\textwidth}
    \centering
    \includegraphics[width=\linewidth]{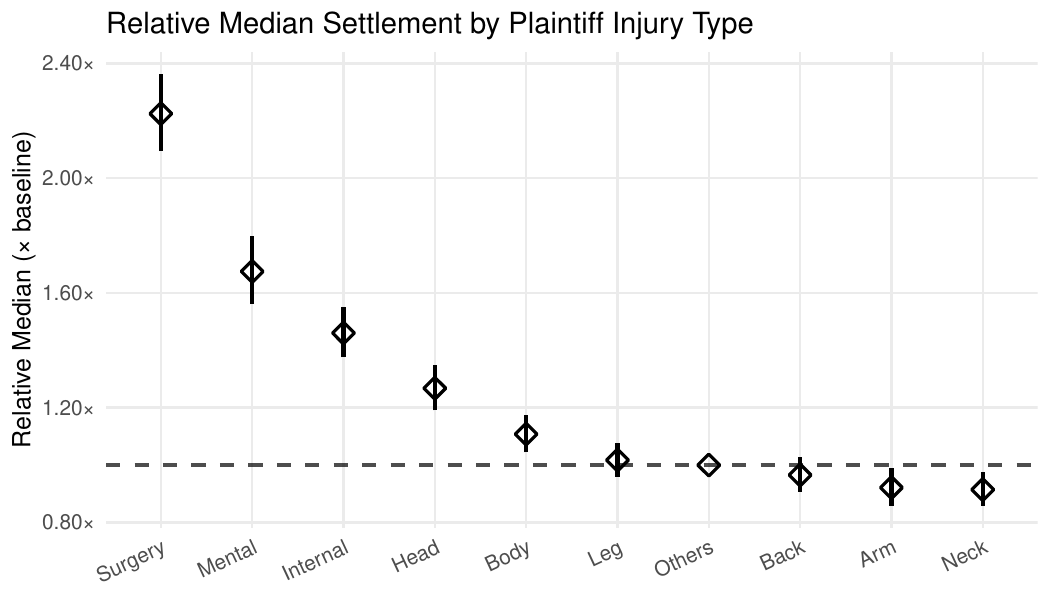}
  \end{subfigure}\hfill
  \begin{subfigure}[t]{0.48\textwidth}
    \centering
    \includegraphics[width=\linewidth]{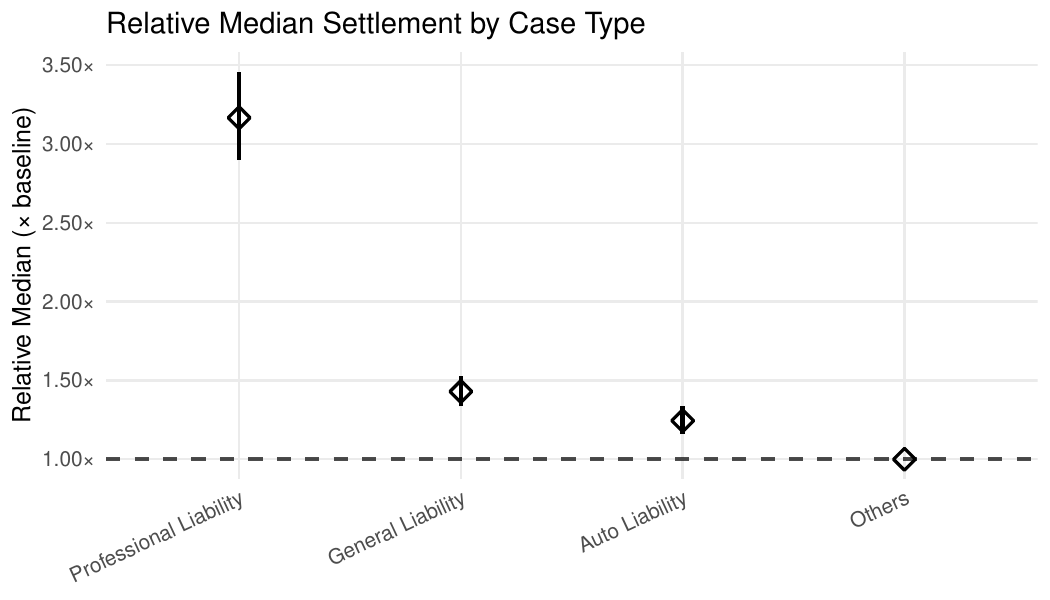}
  \end{subfigure}\hfill
  \begin{subfigure}[t]{0.48\textwidth}
    \centering
    \includegraphics[width=\linewidth]{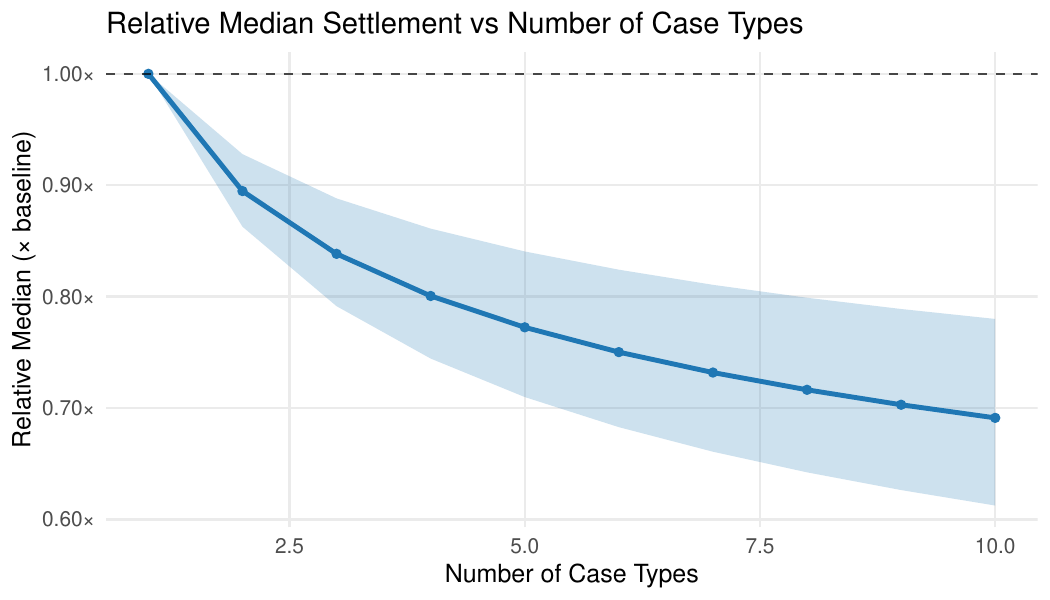}
  \end{subfigure}\hfill
  \caption{Relative median settlement amount vs. variables, with 95\% confidence intervals (part 1).}
  \label{fig:cov_sev_s_fig1}
\end{figure}

\begin{figure}[H]
  \centering
  \begin{subfigure}[t]{0.48\textwidth}
    \centering
    \includegraphics[width=\linewidth]{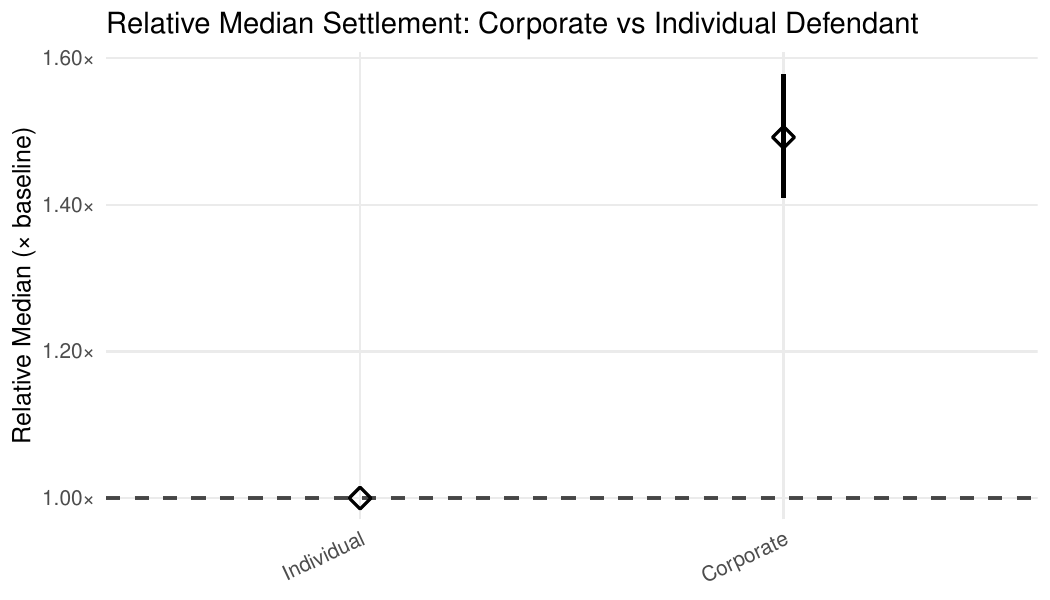}
  \end{subfigure}\hfill
  \begin{subfigure}[t]{0.48\textwidth}
    \centering
    \includegraphics[width=\linewidth]{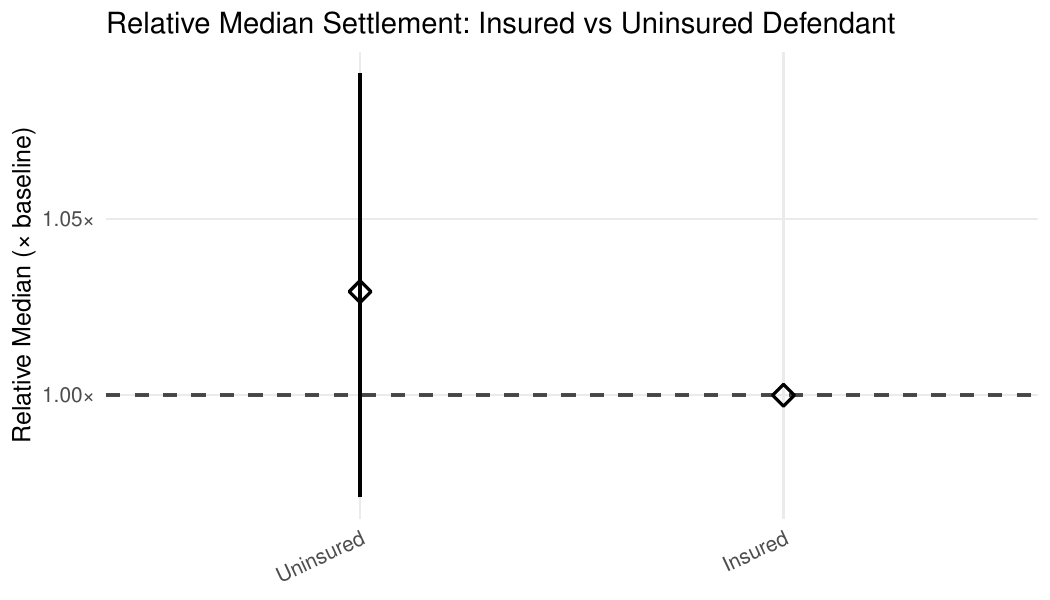}
  \end{subfigure}\hfill
    \begin{subfigure}[t]{0.48\textwidth}
    \centering
    \includegraphics[width=\linewidth]{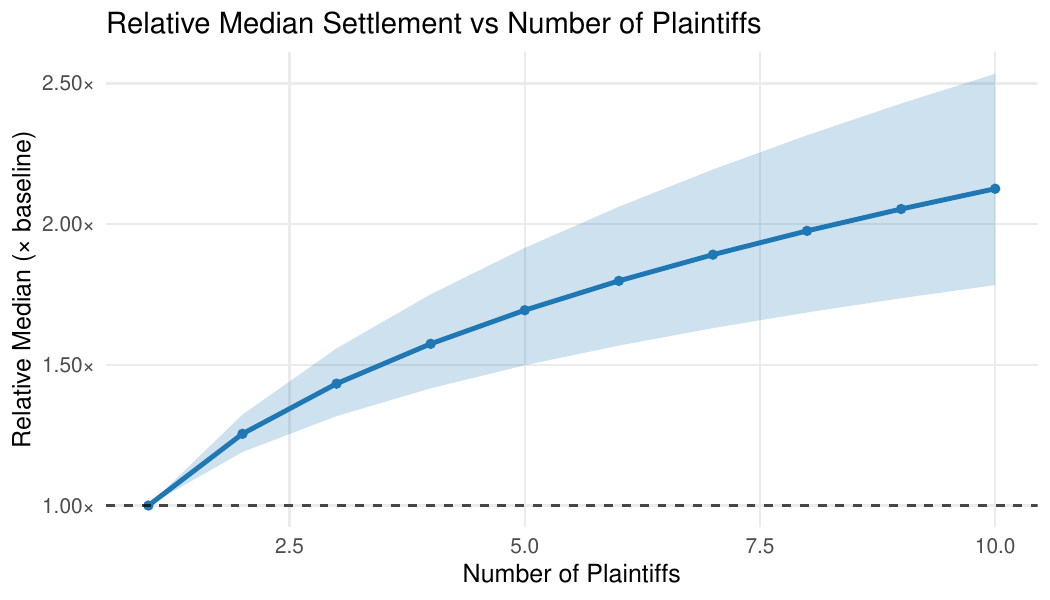}
  \end{subfigure}\hfill
  \begin{subfigure}[t]{0.48\textwidth}
    \centering
    \includegraphics[width=\linewidth]{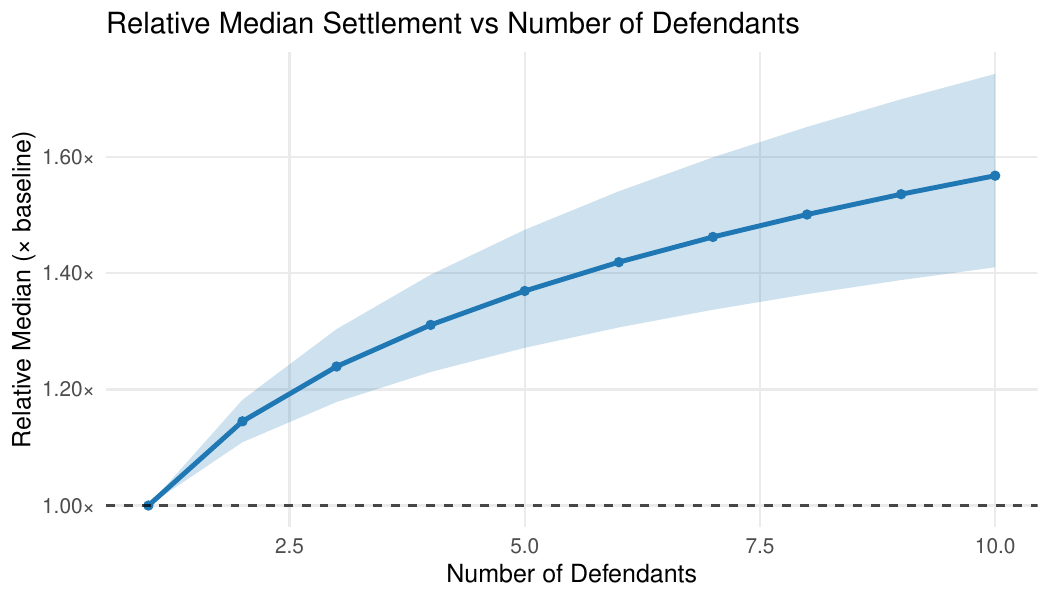}
  \end{subfigure}\hfill
  \begin{subfigure}[t]{0.48\textwidth}
    \centering
    \includegraphics[width=\linewidth]{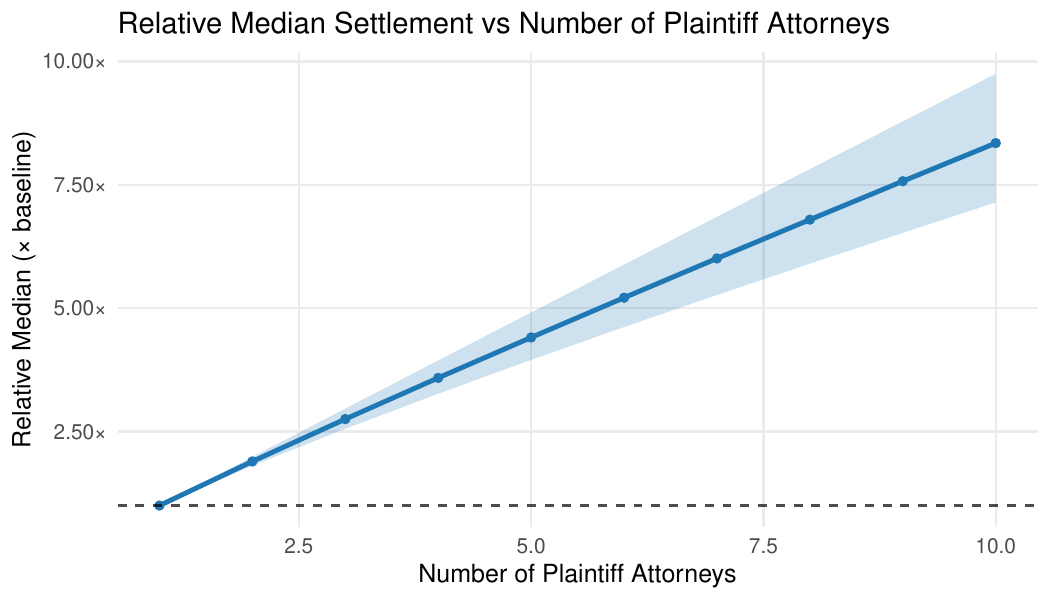}
  \end{subfigure}\hfill
  \begin{subfigure}[t]{0.48\textwidth}
    \centering
    \includegraphics[width=\linewidth]{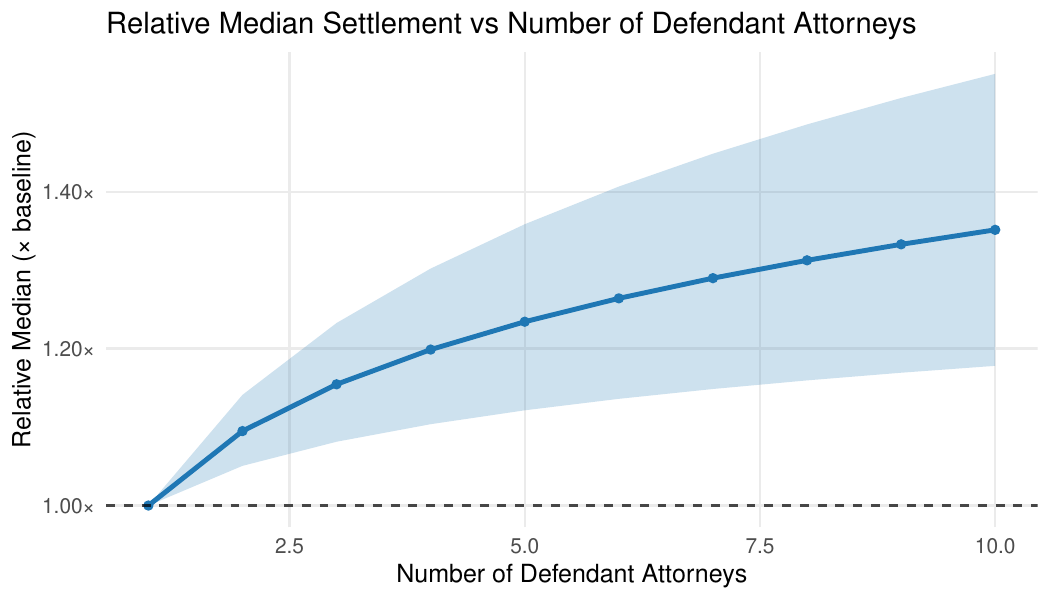}
  \end{subfigure}\hfill
  \begin{subfigure}[t]{0.48\textwidth}
    \centering
    \includegraphics[width=\linewidth]{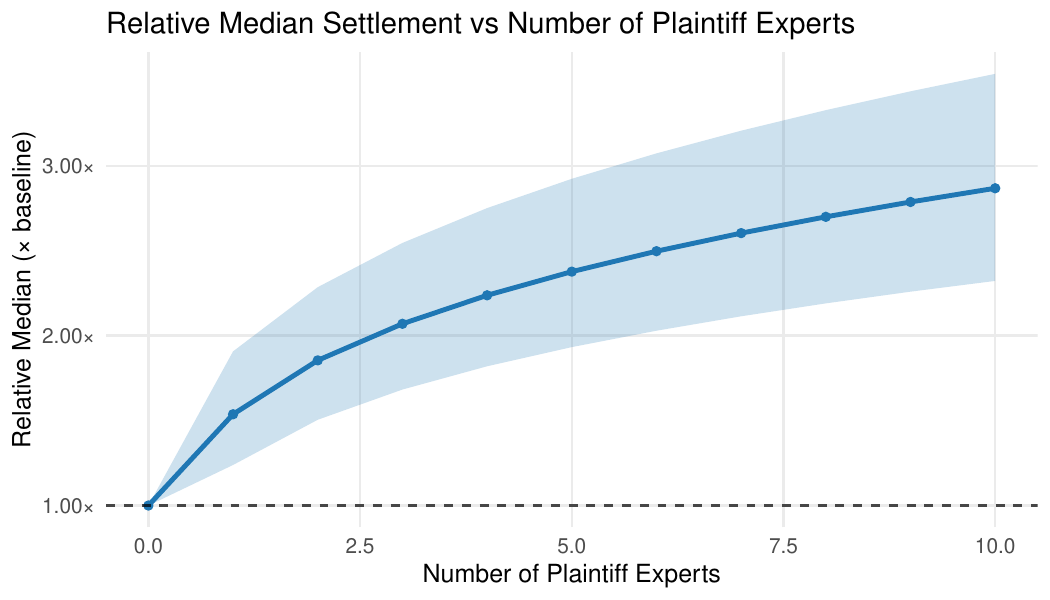}
  \end{subfigure}\hfill
  \begin{subfigure}[t]{0.48\textwidth}
    \centering
    \includegraphics[width=\linewidth]{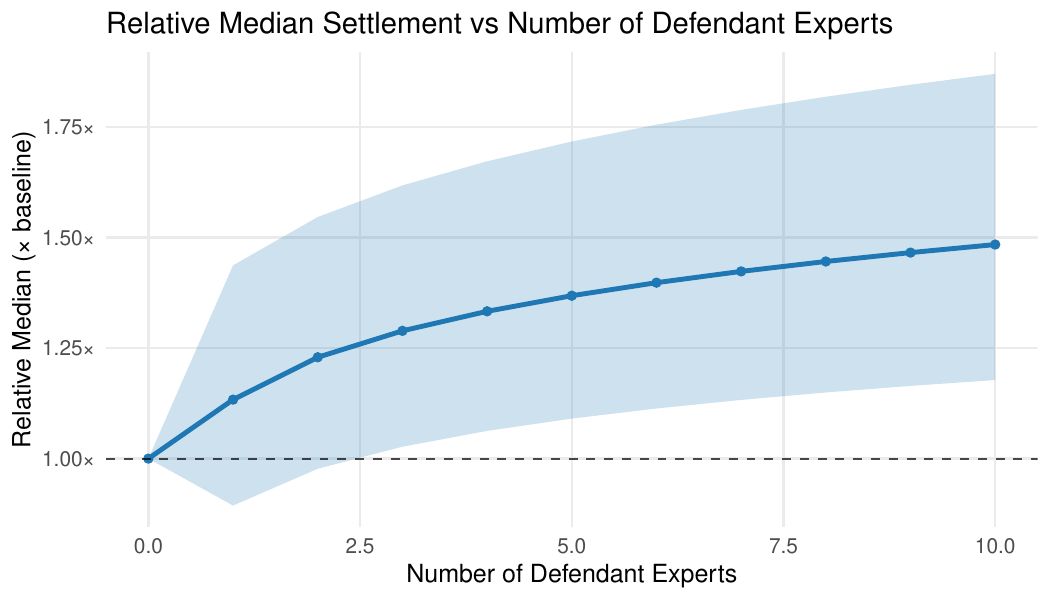}
  \end{subfigure}\hfill
  \begin{subfigure}[t]{0.48\textwidth}
    \centering
    \includegraphics[width=\linewidth]{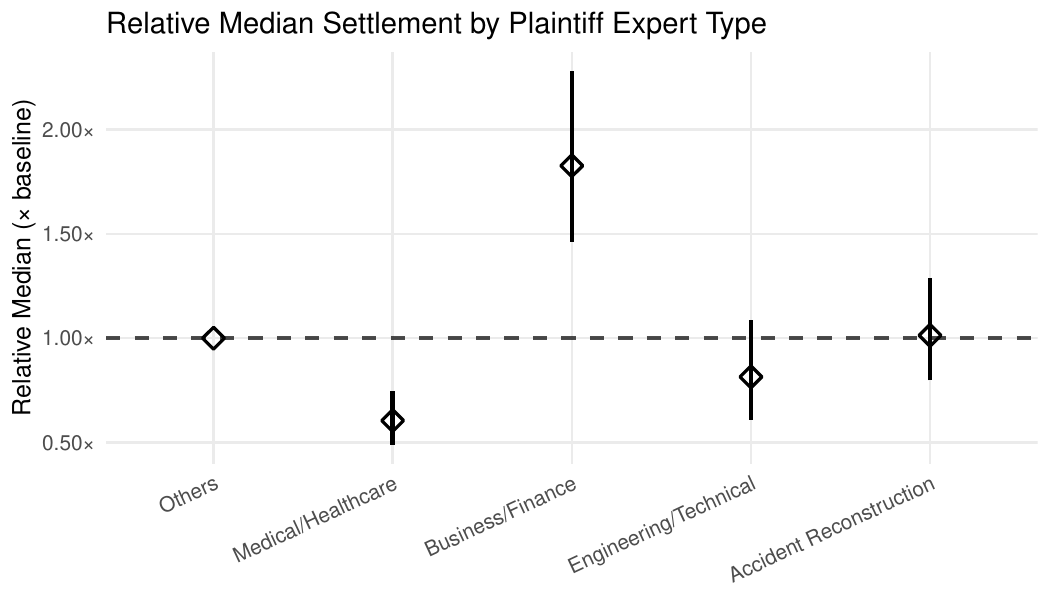}
  \end{subfigure}\hfill
  \begin{subfigure}[t]{0.48\textwidth}
    \centering
    \includegraphics[width=\linewidth]{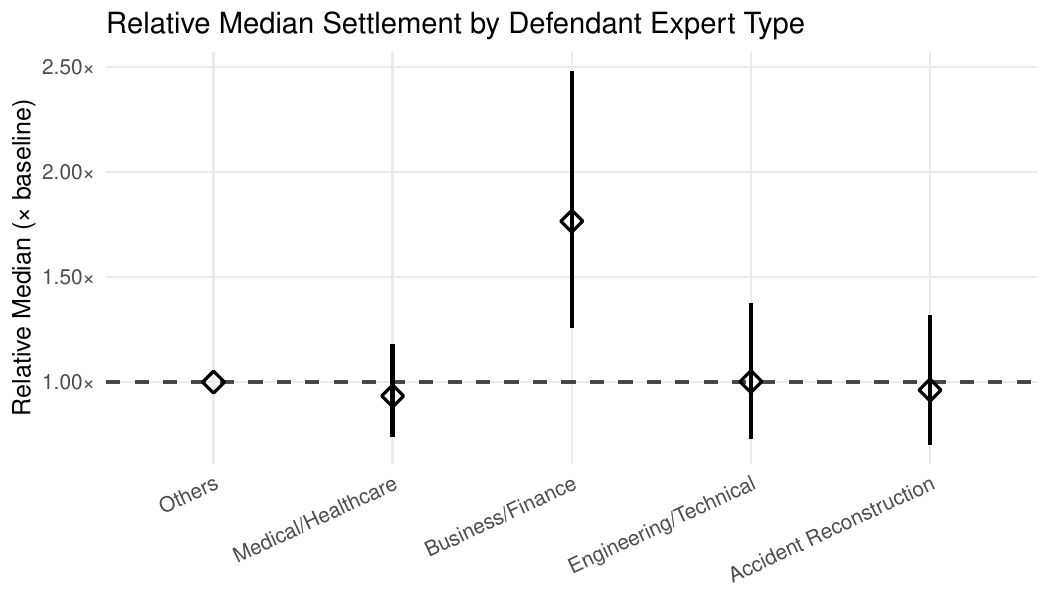}
  \end{subfigure}\hfill
  \caption{Relative median settlement amount vs. variables, with 95\% confidence intervals (part 2).}
  \label{fig:cov_sev_s_fig2}
\end{figure}

We conduct a parallel analysis for settlement amounts by fitting a median quantile regression to the log settlement amount, conditional on cases that resolve in settlement with positive payments. Implementation mirrors that described in Section \ref{sec:prelim:sev_p}: We use \texttt{rq()} to estimate the linear relationship between the median log settlement and the covariates, then exponentiate coefficients to obtain relative median settlement amounts. Figures \ref{fig:cov_sev_s_fig1} and \ref{fig:cov_sev_s_fig2} present these effects and their 95\% confidence intervals.

Geographic variation remains substantial. New England states have the highest median settlement amounts, followed by California, New York, and New Jersey. Settlements in these jurisdictions are roughly two to four times as large at the median as those in low-settlement states such as Texas and Ohio. These patterns echo the venue effects seen in verdict awards.

Plaintiff demographics influence settlement amounts in directions broadly consistent with the verdict results but with somewhat smaller magnitudes. Cases involving plaintiffs under 16 years of age have the highest median settlements among age groups, reflecting large projected future losses and sympathetic fact patterns. In contrast to the verdict model, gender, marital status, and parental status appear somewhat more salient for settlements. Cases involving male married plaintiffs with children tend to have higher median settlements than cases involving female single plaintiffs without children.

Injury and death variables are again key drivers. Settlements in cases involving death are more than two times larger at the median than in nonfatal cases. Among nonfatal injury types, surgery, mental injuries, and internal injuries are associated with the highest settlement medians, paralleling the verdict results. Consistent with the verdict model, cases involving no injuries are associated with the highest median settlement amounts. By contrast, cases with one or more injuries exhibit substantially lower relative median settlements, with insignificant changes as the number of injuries increases. This pattern may reflect that noninjury cases are more likely to involve higher property damage or other compensable losses that drive settlement values.

Case categories and party structure exhibit patterns somewhat similar to those seen for verdicts, though attenuated. Professional liability cases generate the largest median settlements, and cases involving corporate defendants have higher median settlements than those involving only individual defendants. Settlements in cases with insured defendants are slightly lower than those in cases with entirely uninsured defendants, though the difference here is insignificant. The number of case types per dispute, however, is negatively associated with the median settlement, suggesting that highly complex, multi-theory cases may be more likely to settle at discounts that reflect litigation risk on both sides if such cases do not proceed to verdict.

Finally, litigation intensity is strongly related to settlement size. The median settlement amount rises sharply with the number of plaintiff attorneys and plaintiff experts, and more moderately with the number of defense attorneys and experts. Expert specialization again matters. Similar to the verdict results, settlements are larger when plaintiffs retain business/finance experts.


\subsection{Summary of results}
Taken together, the logistic and quantile regression results presented in Sections \ref{sec:prelim:prob_p} to \ref{sec:prelim:sev_s} show that all four response variables of interest (plaintiff victory probability, settlement probability, verdict award amount, and settlement amount) are jointly shaped by a broad set of observable covariates. Geography, plaintiff demographics, injury presence and severity, case category, party structure (e.g., corporate and insured defendants), procedural features (trial and jury length), and the intensity and specialization of legal and expert resources all have economically and statistically significant effects. Moreover, the direction and magnitude of these effects can sometimes resonate across models. For example, cases involving death, young plaintiffs, and/or large numbers of attorneys/experts representing the plaintiffs increase both the likelihood and level of payments.

A central implication is that the social inflation phenomenon operates through multiple channels at once: Higher probabilities that plaintiffs win when cases reach a verdict, lower probabilities that disputes resolve via settlement, and higher conditional award and settlement amounts when plaintiffs do recover. At the same time, many of the covariates that drive these channels, e.g., case mix across states and liability types, the severity and composition of injuries, the prevalence of corporate and insured defendants, and the intensity of attorney and expert involvement, are themselves evolving over time, as documented in Section \ref{sec:data}. Raw time trends in plaintiff win rates or in case award amounts therefore conflate two components: (1) genuine temporal shifts in the judicial and litigation environment (the object we label ``social inflation''), and (2) mechanical effects induced by changes in the composition and complexity of cases that reach verdict or settlement.

The regression results in this section are thus not only of substantive interest but also foundational for determining how we should study and quantify social inflation in the subsequent analysis. They make clear that any credible social inflation index must be case-mix adjusted: It should reflect how outcomes would have changed over time if the underlying distribution of factual and strategic covariates had been held fixed, rather than tracking unadjusted averages driven by an increasingly selected and complex set of disputes. Logistic regressions provide the tools to construct such adjusted indices for plaintiff win and settlement probabilities by comparing predicted odds over time for a reference case profile or for a fixed baseline distribution of covariates. Quantile regressions play an analogous role for verdict and settlement amounts, allowing us to trace the evolution of conditional quantiles of awards for a given case mix while remaining robust to extreme nuclear verdicts that can destabilize mean-based measures.

In short, the preliminary analysis highlights both the strength and pervasiveness of covariate effects and the degree to which those covariates are shifting over time. The central message of this section is that social inflation cannot be inferred from univariate time series of awards or win rates alone. Instead, it must be recovered as an adjusted temporal effect after controlling for the case mix. The next section builds directly on these insights, using the logistic and quantile regression frameworks as building blocks for formal, case-mix-adjusted social inflation indices that separate genuine inflationary pressures in the civil justice system from evolving case composition.

\section{Methodology} \label{sec:method}
This section develops a family of social inflation indices based on the cross-sectional logistic and quantile regression frameworks introduced in Section \ref{sec:prelim}. The goal is to quantify social inflation separately for (1) plaintiff win probability given verdict, (2) settlement probability, (3) verdict award amount given a positive plaintiff award, (4) settlement amount given settlement, and (5) total loss payable to the plaintiff. Crucially, these indices are designed to control for the factual and strategic explanatory variables observed in the VerdictSearch data. By controlling for these variables, we disentangle changes in observed plaintiff win rates (which affect the conditional likelihood that a litigated dispute results in a positive payment---a frequency-type channel distinct from exposure-based claim incidence) and changes in verdict/settlement amounts (which affect loss severity, especially in reinsurance layers) into two components: (1) shifts in case composition over time, primarily reflected by factual covariates, and (2) genuine social inflation, reflected by time-varying outcome dynamics after holding case mix fixed. In addition, by incorporating strategic variables, the framework provides a quantitative lens for assessing how evolving litigation tactics may explain or amplify social inflation. Finally, we quantify uncertainty for each index using a random-weighted bootstrap procedure. 

In addition to the empirical application to VerdictSearch in Section \ref{sec:result}, Appendix \ref{apx:sec:sim} provides a controlled synthetic-data illustration that verifies that the proposed construction can recover the true (data-generating) social inflation indices and their uncertainty bands. The synthetic experiment also highlights the importance of controlling for an evolving case mix: Ignoring covariates leads to systematically overstated social inflation in both probability and severity channels.

We now introduce notation used throughout Section \ref{sec:method}. For each case $i\in\mathcal{N}$ in the VerdictSearch dataset (where $\mathcal{N}$ denotes the set of all case indices), let $P_i$, $D_i$, and $S_i$ be binary indicators for whether case $i$ ends in a plaintiff win, defendant win, or settlement, respectively. For example, $P_i=1$ if the plaintiff wins case $i$ and $P_i=0$ otherwise; note that $P_i+D_i+S_i=1$. Let $Y_i$ denote the verdict award or settlement amount payable to the plaintiff in case $i$, where $Y_i\ge 0$. Typically $Y_i=0$ when $D_i=1$, and $Y_i$ may also equal 0 even when $P_i=1$ in rare cases in which the award is not material. Let $\bm{X}_i$ be a vector of factual and/or strategic explanatory variables (as listed in Tables~2--3) for case $i$, excluding the calendar-year variable \texttt{YEAR}. Let $T_i$ denote the calendar year of the case end date, corresponding to \texttt{YEAR}. We define the following index sets:
\[
\mathcal{P}:=\{i\in\mathcal{N}:P_i=1\},\qquad
\mathcal{S}:=\{i\in\mathcal{N}:S_i=1\},\qquad
\mathcal{V}:=\{i\in\mathcal{N}:S_i=0\},
\]
where $\mathcal{V}$ is the set of cases resulting in a verdict outcome (a plaintiff or defendant win, but not settlement). For a positive integer $s$, we define
\[
\mathcal{A}_{t-s:t+1}:=\{i\in\mathcal{N}:t-s\le T_i\le t+1\},\qquad
\mathcal{A}_t:=\mathcal{A}_{t:t},
\]
as the sets of cases ending between years $t-s$ and $t$, and during year $t$, respectively. In our VerdictSearch data, we consider $t=2009,\ldots,2023$, i.e., social inflation is quantified up to year $t+1=2024$.

\subsection{Social inflation in plaintiff win probability} \label{sec:method:p_p}
We first quantify social inflation in plaintiff win rates by modeling, for each case $i$, the conditional probability that the plaintiff wins given that the case does not settle. Specifically, we define
\[
p_i^P := p^P(\bm{X}_i,T_i) \;=\; \Pr(P_i=1\mid S_i=0,\bm{X}_i,T_i).
\]
We model $p_i^P$ using a logistic regression with a logit link:
\begin{equation}
\label{eq:logit_model_P}
P_i \mid (S_i=0,\bm{X}_i,T_i)\sim \mathrm{Bernoulli}(p_i^P),\qquad
\operatorname{logit}(p_i^P)=\log\!\left(\frac{p_i^P}{1-p_i^P}\right)=\alpha_t^P+\bm{\beta}_t^{P}\bm{X}_i,
\end{equation}
for cases $i\in\mathcal{A}_t\cap\mathcal{V}$ (i.e., verdict cases ending in year $t$). Here, $\alpha_t^P$ is a year-$t$ intercept capturing the overall propensity for plaintiff wins in year $t$, while $\bm{\beta}_t^P$ is a vector of regression coefficients capturing how the explanatory variables affect the plaintiff win log-odds around year $t$.

To estimate $(\alpha_t^P,\bm{\beta}_t^P)$, we adopt a rolling-window maximum likelihood procedure that differs from the standard single-year logistic regression in two important ways. First, we pool verdict cases from the most recent five years (i.e., years $t-4$ to $t$) and the subsequent year (i.e., year $t+1$) to estimate the covariate effects, which stabilizes estimation as the annual number of VerdictSearch cases declines. Second, we allow the intercept to vary by calendar year within the pooled window, because the intercept is precisely the component intended to capture time-varying shifts in plaintiff win propensity (i.e., potential social inflation effects), and pooling intercepts would dampen and bias those shifts. Formally, for a given calendar year $t$, we maximize the log-likelihood function
\begin{align}
\label{eq:ll_P}
\mathcal{L}\!\left(\bm{\alpha}_{t-4:t+1}^P,\bm{\beta}^P_t\right)
= \sum_{i\in \mathcal{A}_{t-4:t+1}\cap\mathcal{V}}
\Big\{&
P_i\log\!\left(\operatorname{logit}^{-1}\!\left(\alpha_{T_i}^P+\bm{\beta}_t^{P}\bm{X}_i\right)\right)
\nonumber\\
&+(1-P_i)\log\!\left(1-\operatorname{logit}^{-1}\!\left(\alpha_{T_i}^P+\bm{\beta}_t^{P}\bm{X}_i\right)\right)
\Big\},
\end{align}
with respect to $\left(\bm{\alpha}_{t-4:t+1}^P,\bm{\beta}^P_t\right)$, where
$\bm{\alpha}_{t-4:t+1}^P := (\alpha_{t-4}^P,\ldots,\alpha_{t+1}^P)$.
Let the optimizer be denoted by
\[
\left(\hat{\bm{\alpha}}_{t-4:t+1}^P,\hat{\bm{\beta}}_t^P\right),
\qquad\text{and in particular}\qquad
\left(\hat{\alpha}_t^P,\hat{\bm{\beta}}_t^P\right)
\]
for the year-$t$ intercept and slope vector. The estimated year-$(t+1)$ intercept $\hat{\alpha}_{t+1}^P$ is also obtained.

For a verdict case $i\in\mathcal{A}_t\cap\mathcal{V}$, the fitted probability that the plaintiff wins in year $t$ is
\[
\hat{p}^P(\bm{X}_i,t)=\operatorname{logit}^{-1}\!\left(\hat{\alpha}_t^P+\hat{\bm{\beta}}_t^{P}\bm{X}_i\right),
\]
while the counterfactual predicted probability for the same case characteristics if the case were to end in year $t+1$ is
\[
\hat{p}^P(\bm{X}_i,t+1)=\operatorname{logit}^{-1}\!\left(\hat{\alpha}_{t+1}^P+\hat{\bm{\beta}}_{t}^{P}\bm{X}_i\right).
\]

We then construct the ASIR for plaintiff win probability from $t$ to $t+1$ by holding the case mix fixed at year $t$ and comparing the average predicted plaintiff win probability under year-$t$ versus year-$(t+1)$ parameters:
\begin{equation}
\label{eq:sir_prob_P}
ASIR^{\mathrm{prob},P}_{t+1}
=\frac{\sum_{i\in\mathcal{A}_t\cap\mathcal{V}}\hat{p}^P(\bm{X}_i,t+1)}
{\sum_{i\in\mathcal{A}_t\cap\mathcal{V}}\hat{p}^P(\bm{X}_i,t)}-1.
\end{equation}
This quantity measures the relative change in plaintiff win probability that is attributable to temporal shifts in the fitted model rather than to changes in the distribution of $\bm{X}_i$.

Let $t_0$ denote the base year (in our dataset, $t_0=2009$). The CSII at year $t+1$ is then defined as the chain product of annual growth factors:
\begin{equation}
\label{eq:sii_prob_P}
CSII^{\mathrm{prob},P}_{t+1}
=\prod_{j=t_0}^{t}\left(1+ASIR^{\mathrm{prob},P}_{j+1}\right),
\end{equation}
which represents the cumulative change in the case-mix-adjusted plaintiff win probability from year $t_0$ to year $t+1$.

To quantify uncertainty in Equations \eqref{eq:sir_prob_P} and \eqref{eq:sii_prob_P}, we use a random-weighted bootstrap with $B=500$ replicates. For each bootstrap replicate $b\in\{1,\ldots,B\}$ we perform the following steps:
\begin{itemize}
\item \textbf{Step 1.} For each case $i\in\mathcal{N}$, we draw an independent random weight $W_i^{(b)}$ from a distribution with mean $1$ and variance $1$, such as $\mathrm{Exp}(1)$, i.e., a standard exponential distribution.

\item \textbf{Step 2.} For each calendar year $t$, we maximize the weighted log-likelihood
\begin{align}
\label{eq:wll_P}
\mathcal{L}^{(b)}\!\left(\bm{\alpha}_{t-4:t+1}^{P(b)},\bm{\beta}_t^{P(b)}\right)
=\sum_{i\in \mathcal{A}_{t-4:t+1}\cap\mathcal{V}}W_i^{(b)}
\Big\{&
P_i\log\!\left(\operatorname{logit}^{-1}\!\left(\alpha_{T_i}^{P(b)}+\bm{\beta}_t^{P(b)}\bm{X}_i\right)\right)
\nonumber\\
&+(1-P_i)\log\!\left(1-\operatorname{logit}^{-1}\!\left(\alpha_{T_i}^{P(b)}+\bm{\beta}_t^{P(b)}\bm{X}_i\right)\right)
\Big\},
\end{align}
with respect to $\left(\bm{\alpha}_{t-4:t+1}^{P(b)},\bm{\beta}_t^{P(b)}\right)$, where
$\bm{\alpha}_{t-4:t+1}^{P(b)}=(\alpha_{t-4}^{P(b)},\ldots,\alpha_{t+1}^{P(b)})$.
Denote the resulting estimators by $(\hat{\alpha}_t^{P(b)},\hat{\bm{\beta}}_t^{P(b)})$ and $\hat{\alpha}_{t+1}^{P(b)}$.

\item \textbf{Step 3.} We compute the bootstrap fitted probabilities
\[
\hat{p}^{P(b)}(\bm{X}_i,t)=\operatorname{logit}^{-1}\!\left(\hat{\alpha}_t^{P(b)}+\hat{\bm{\beta}}_t^{P(b)}\bm{X}_i\right),
\]
and then compute
\begin{align}
\label{eq:boot_sir_sii_P}
ASIR^{\mathrm{prob},P(b)}_{t+1}
&=\frac{\sum_{i\in\mathcal{A}_t\cap\mathcal{V}}W_i^{(b)}\hat{p}^{P(b)}(\bm{X}_i,t+1)}
{\sum_{i\in\mathcal{A}_t\cap\mathcal{V}}W_i^{(b)}\hat{p}^{P(b)}(\bm{X}_i,t)}-1,
\nonumber\\
CSII^{\mathrm{prob},P(b)}_{t+1}
&=\prod_{j=t_0}^{t}\left(1+ASIR^{\mathrm{prob},P(b)}_{j+1}\right).
\end{align}

\item \textbf{Step 4.} We repeat Steps 1 to 3 for $b=1,\ldots,B$ to obtain $\{ASIR^{\mathrm{prob},P(b)}_{t+1}\}_{b=1}^B$ and $\{CSII^{\mathrm{prob},P(b)}_{t+1}\}_{b=1}^B$. Let
\[
\widehat{SE}\!\left(ASIR^{\mathrm{prob},P}_{t+1}\right)
=\operatorname{sd}\!\left(\{ASIR^{\mathrm{prob},P(b)}_{t+1}\}_{b=1}^B\right),
\qquad
\widehat{SE}\!\left(CSII^{\mathrm{prob},P}_{t+1}\right)
=\operatorname{sd}\!\left(\{CSII^{\mathrm{prob},P(b)}_{t+1}\}_{b=1}^B\right),
\]
where $\operatorname{sd}(\cdot)$ denotes the sample standard deviation.
A 95\% normal-approximation confidence interval for $ASIR^{\mathrm{prob},P}_{t+1}$ is
\[
\left[
ASIR^{\mathrm{prob},P}_{t+1}-1.96\,\widehat{SE}\!\left(ASIR^{\mathrm{prob},P}_{t+1}\right),\;
ASIR^{\mathrm{prob},P}_{t+1}+1.96\,\widehat{SE}\!\left(ASIR^{\mathrm{prob},P}_{t+1}\right)
\right],
\]
and similarly for $CSII^{\mathrm{prob},P}_{t+1}$:
\[
\left[
CSII^{\mathrm{prob},P}_{t+1}-1.96\,\widehat{SE}\!\left(CSII^{\mathrm{prob},P}_{t+1}\right),\;
CSII^{\mathrm{prob},P}_{t+1}+1.96\,\widehat{SE}\!\left(CSII^{\mathrm{prob},P}_{t+1}\right)
\right].
\]
\end{itemize}

{\begin{remark}
The random-weighted bootstrap bands reported throughout the paper quantify estimation uncertainty in the fitted logistic regressions (or quantile regressions in subsequent subsections) and the derived social inflation trends, conditional on the model specification and the observed covariates. They do not incorporate model/specification uncertainty (e.g., incorrect model class assumptions and omitted variables) or other unmodeled sources of variation, which are not something standard statistical procedures can quantify objectively. Accordingly, the reported confidence intervals should be interpreted as a lower bound on total uncertainty.
\end{remark}}

{ \begin{remark}
We pool cases from the most recent five-year (and the subsequent-year) window primarily to stabilize estimation of the covariate effects $\bm{\beta}_t^P$ in the rolling-window regressions, especially in later years with fewer observations. The social inflation indices, however, are driven mainly by the estimated year effects/intercepts (e.g., $\alpha_t^P$ and $\beta_t^P$) that encode time variation after controlling for covariates, rather than by the slopes $\bm{\beta}_t$. Consequently, the resulting ASIR and CSII series are not expected to be highly sensitive to modest changes in the window length. Empirically, re-estimating the key indices using a six-year (and the subsequent-year) window produces very similar or even virtually identical CSIIs (see Appendix \ref{apx:sec:robust}).
\end{remark}} 

\subsection{Social inflation in settlement probability} \label{sec:method:s_p}
We next quantify social inflation in settlement behavior by modeling the conditional probability that a case resolves through settlement, given observed case characteristics. Analogous to Section \ref{sec:method:p_p}, we define
\[
p_i^S := p^S(\bm{X}_i,T_i) \;=\; \Pr(S_i=1\mid \bm{X}_i,T_i).
\]
We model $p_i^S$ using a logistic regression with a logit link:
\begin{equation}
\label{eq:logit_model_S}
S_i \mid (\bm{X}_i,T_i)\sim \mathrm{Bernoulli}(p_i^S),\qquad
\operatorname{logit}(p_i^S)=\log\!\left(\frac{p_i^S}{1-p_i^S}\right)=\alpha_t^S+\bm{\beta}_t^{S}\bm{X}_i,
\end{equation}
for cases $i\in\mathcal{A}_t$ (i.e., all cases ending in calendar year $t$). Here, $\alpha_t^S$ captures the overall settlement propensity in year $t$, while $\bm{\beta}_t^S$ captures how the explanatory variables affect settlement log-odds around year $t$.

Estimation follows the same rolling-window maximum likelihood principle as in Section \ref{sec:method:p_p}: We pool cases from years $t-4$ through $t+1$ to stabilize estimation of the covariate effects while allowing year-specific intercepts to vary within the pooled window. Specifically, we maximize
\begin{align}
\label{eq:ll_S}
\mathcal{L}\!\left(\bm{\alpha}_{t-4:t+1}^S,\bm{\beta}^S_t\right)
= \sum_{i\in \mathcal{A}_{t-4:t+1}}
\Big\{&
S_i\log\!\left(\operatorname{logit}^{-1}\!\left(\alpha_{T_i}^S+\bm{\beta}_t^{S}\bm{X}_i\right)\right)
\nonumber\\
&+(1-S_i)\log\!\left(1-\operatorname{logit}^{-1}\!\left(\alpha_{T_i}^S+\bm{\beta}_t^{S}\bm{X}_i\right)\right)
\Big\},
\end{align}
with respect to $\left(\bm{\alpha}_{t-4:t+1}^S,\bm{\beta}^S_t\right)$, where
$\bm{\alpha}_{t-4:t+1}^S := (\alpha_{t-4}^S,\ldots,\alpha_{t+1}^S)$.
We denote the resulting estimators by $\left(\hat{\bm{\alpha}}_{t-4:t+1}^S,\hat{\bm{\beta}}_t^S\right)$. In particular, the year-$t$ parameters are $\left(\hat{\alpha}_t^S,\hat{\bm{\beta}}_t^S\right)$, and the year-$(t+1)$ estimated intercept is $\hat{\alpha}_{t+1}^S$.

For any case $i\in\mathcal{A}_t$, we define the fitted settlement probabilities under year-$t$ and counterfactual year-$(t+1)$ parameters as
\[
\hat{p}^S(\bm{X}_i,t)=\operatorname{logit}^{-1}\!\left(\hat{\alpha}_t^S+\hat{\bm{\beta}}_t^{S}\bm{X}_i\right),
\qquad
\hat{p}^S(\bm{X}_i,t+1)=\operatorname{logit}^{-1}\!\left(\hat{\alpha}_{t+1}^S+\hat{\bm{\beta}}_{t}^{S}\bm{X}_i\right).
\]
We then define the ASIR and the CSII for settlement probability as
\begin{equation}
\label{eq:sir_sii_prob_S}
ASIR^{\mathrm{prob},S}_{t+1}
=\frac{\sum_{i\in\mathcal{A}_t}\hat{p}^S(\bm{X}_i,t+1)}
{\sum_{i\in\mathcal{A}_t}\hat{p}^S(\bm{X}_i,t)}-1,
\qquad
CSII^{\mathrm{prob},S}_{t+1}
=\prod_{j=t_0}^{t}\left(1+ASIR^{\mathrm{prob},S}_{j+1}\right).
\end{equation}
Uncertainty quantification for $ASIR^{\mathrm{prob},S}_{t+1}$ and $CSII^{\mathrm{prob},S}_{t+1}$ can be carried out using the same random-weighted bootstrap procedure described in Section \ref{sec:method:p_p}; we omit the explicit steps here to avoid redundancy.

\subsection{Social inflation in verdict award amount}
\label{sec:method:p_s}

We now turn to social inflation in verdict severity. Because verdict awards are highly skewed and heavy-tailed (Section \ref{sec:data}), quantifying severity using sample averages can be unstable and overly sensitive to a small number of extreme awards. We therefore adopt a quantile-based approach and measure severity through conditional value-at-risk (VaR). For a probability level $\tau\in(0,1)$, we define the conditional VaR of the plaintiff award amount $Y_i$ given covariates $\bm{X}_i$ and conditional on a plaintiff verdict as
\begin{align}
\label{eq:var_def_P}
\mathrm{VaR}^P(\tau\mid \bm{X}_i,T_i)
:=\sup\Big\{y:\Pr(Y_i\le y\mid P_i=1,\bm{X}_i,T_i)\le \tau\Big\},
\end{align}
which is interpreted as the $100\tau$-th percentile of the payment to the plaintiff among verdict cases the plaintiff wins. We model $\mathrm{VaR}^P(\tau\mid \bm{X}_i,T_i)$ via a log-linear quantile regression:
\begin{equation}
\label{eq:qr_var_P}
\mathrm{VaR}^P(\tau\mid \bm{X}_i,T_i)=\exp\{\gamma_t^P+\bm{\eta}_t^P\bm{X}_i\},
\qquad i\in\mathcal{A}_t\cap\mathcal{P},
\end{equation}
where $\gamma_t^P$ is a year-$t$ intercept capturing the overall level of verdict severity in year $t$, and $\bm{\eta}_t^P$ captures how the explanatory variables shift the conditional $\tau$-quantile of verdict awards. The log link ensures positivity of fitted VaR levels. In practice, the quantile regression is applied on the log-scale and thus requires $Y_i>0$; this restriction is innocuous at the quantile levels used in our analysis because 0 or immaterial awards among plaintiff wins are very rare.

Estimation follows the same rolling-window principle as in Sections \ref{sec:method:p_p} and \ref{sec:method:s_p}. For each year $t$, we pool plaintiff-win verdict cases from the most recent five-year window to stabilize estimation of covariate effects, while allowing intercepts to vary by calendar year within the window. Specifically, we estimate $(\gamma_t^P,\bm{\eta}_t^P)$ by minimizing
\begin{align}
\label{eq:ll_sev_P}
\mathcal{L}\!\left(\bm{\gamma}_{t-4:t+1}^P,\bm{\eta}^P_t\right)
= \sum_{i\in \mathcal{A}_{t-4:t+1}\cap\mathcal{P}}
\rho_{\tau}\left(\log Y_i-(\gamma_{T_i}^P+\bm{\eta}_t^P\bm{X}_i)\right),
\end{align}
with respect to $\left(\bm{\gamma}_{t-4:t+1}^P,\bm{\eta}_t^P\right)$, where
$\bm{\gamma}_{t-4:t+1}^P=(\gamma_{t-4}^P,\ldots,\gamma_{t+1}^P)$ and
$\rho_\tau(u)=u\{\tau-\mathbb{I}(u<0)\}$ is the check function. We denote the resulting estimators by $(\hat{\bm{\gamma}}_{t-4:t+1}^P,\hat{\bm{\eta}}_t^P)$. In particular, the year-$t$ parameters are $(\hat{\gamma}_{t+1}^P,\hat{\bm{\eta}}_t^P)$, and the year-$(t+1)$ estimated intercept is $\hat{\gamma}_{t+1}^P$.

For any plaintiff-win verdict case $i\in\mathcal{A}_t\cap\mathcal{P}$, we define the fitted conditional VaR under year-$t$ and counterfactual year-$(t+1)$ parameters as
\[
\widehat{\mathrm{VaR}}^P(\tau\mid \bm{X}_i,t)=\exp\{\hat{\gamma}_{t}^P+\hat{\bm{\eta}}_{t}^P\bm{X}_i\},
\qquad
\widehat{\mathrm{VaR}}^P(\tau\mid \bm{X}_i,t+1)=\exp\{\hat{\gamma}_{t+1}^P+\hat{\bm{\eta}}_{t}^P\bm{X}_i\}.
\]
We then construct the ASIR for verdict award amount at quantile level $\tau$ as
\begin{equation}
\label{eq:sir_sev_P}
ASIR^{\mathrm{amt},P}_{t+1}
= \exp\left\{\frac{\sum\limits_{i\in\mathcal{A}_t\cap\mathcal{P}}
\left(\log\widehat{\mathrm{VaR}}^P(\tau\mid\bm{X}_i,t+1)-\log\widehat{\mathrm{VaR}}^P(\tau\mid\bm{X}_i,t)\right)}
{\left|\mathcal{A}_t\cap\mathcal{P}\right|}\right\}-1,
\end{equation}
and the corresponding CSII relative to the base year $t_0$ as
\begin{equation}
\label{eq:sii_sev_P}
CSII^{\mathrm{amt},P}_{t+1}
=\prod_{j=t_0}^{t}\left(1+ASIR^{\mathrm{amt},P}_{j+1}\right).
\end{equation}
The definition in Equation \eqref{eq:sir_sev_P} compares year-$(t+1)$ versus year-$t$ fitted VaR values while holding the year-$t$ case mix fixed. We use an exponential of the average log change (i.e., a geometric-mean growth rate) rather than a ratio of arithmetic averages because heavy-tailed award distributions can make arithmetic means unstable even after conditioning.

Uncertainty quantification for Equations \eqref{eq:sir_sev_P} and \eqref{eq:sii_sev_P} is obtained via the same random-weighted bootstrap principle as in Section \ref{sec:method:p_p}, using $B=500$ replicates. To avoid redundancy, we only highlight the key modifications relative to Section \ref{sec:method:p_p}. For bootstrap replicate $b$, the weighted rolling-window quantile regression objective becomes
\[
\mathcal{L}^{(b)}\!\left(\bm{\gamma}_{t-4:t+1}^{P(b)},\bm{\eta}^{P(b)}_t\right)
= \sum_{i\in \mathcal{A}_{t-4:t+1}\cap\mathcal{P}}
W_i^{(b)}\rho_{\tau}\left(\log Y_i-(\gamma_{T_i}^{P(b)}+\bm{\eta}_t^{P(b)}\bm{X}_i)\right),
\]
yielding bootstrapped estimators, given by $(\hat{\gamma}_t^{P(b)},\hat{\bm{\eta}}_t^{P(b)})$, and fitted VaR, given by $\widehat{\mathrm{VaR}}^{P(b)}(\tau\mid\bm{X}_i,t)=\exp\{\hat{\gamma}_t^{P(b)}+\hat{\bm{\eta}}_t^{P(b)}\bm{X}_i\}$.
The bootstrapped ASIR and CSII are computed as
\[
ASIR^{\mathrm{amt},P(b)}_{t+1}
= \exp\left\{\frac{\sum\limits_{i\in\mathcal{A}_t\cap\mathcal{P}}W_i^{(b)}
\left(\log\widehat{\mathrm{VaR}}^{P(b)}(\tau\mid\bm{X}_i,t+1)-\log\widehat{\mathrm{VaR}}^{P(b)}(\tau\mid\bm{X}_i,t)\right)}
{\sum\limits_{i\in\mathcal{A}_t\cap\mathcal{P}}W_i^{(b)}}\right\}-1,
\]
\[CSII^{\mathrm{amt},P(b)}_{t+1}
=\prod_{j=t_0}^{t}\left(1+ASIR^{\mathrm{amt},P(b)}_{j+1}\right),\]
from which standard errors and confidence intervals are obtained in the same manner as in Section \ref{sec:method:p_p}.

\subsection{Social inflation in settlement amount}
\label{sec:method:s_s}

We next quantify social inflation in settlement severity using the same quantile-based framework. For $\tau\in(0,1)$, we define the conditional VaR of the settlement payment as
\begin{align}
\label{eq:var_def_S}
\mathrm{VaR}^S(\tau\mid \bm{X}_i,T_i)
:=\sup\Big\{y:\Pr(Y_i\le y\mid S_i=1,\bm{X}_i,T_i)\le \tau\Big\},
\end{align}
interpreted as the $100\tau$-th percentile of settlement amounts conditional on settlement. We model
\begin{equation}
\label{eq:qr_var_S}
\mathrm{VaR}^S(\tau\mid \bm{X}_i,T_i)=\exp\{\gamma_t^S+\bm{\eta}_t^S\bm{X}_i\},
\qquad i\in\mathcal{A}_t\cap\mathcal{S},
\end{equation}
and estimate $(\gamma_t^S,\bm{\eta}_t^S)$ using the same rolling-window quantile regression strategy as in Equation \eqref{eq:ll_sev_P}, replacing $\mathcal{P}$ with $\mathcal{S}$ and $(\gamma_t^P,\bm{\eta}_t^P)$ with $(\gamma_t^S,\bm{\eta}_t^S)$. We denote the resulting estimators by $(\hat{\gamma}_t^S,\hat{\bm{\eta}}_t^S)$.

For $i\in\mathcal{A}_t\cap\mathcal{S}$, the fitted conditional VaR is
$\widehat{\mathrm{VaR}}^S(\tau\mid \bm{X}_i,t)=\exp\{\hat{\gamma}_{t}^S+\hat{\bm{\eta}}_t^S\bm{X}_i\}$.
Analogous to Equations \eqref{eq:sir_sev_P} and \eqref{eq:sii_sev_P}, we define the ASIR and CSII as
\begin{equation}
\label{eq:sir_sev_S}
ASIR^{\mathrm{amt},S}_{t+1}
= \exp\left\{\frac{\sum\limits_{i\in\mathcal{A}_t\cap\mathcal{S}}
\left(\log\widehat{\mathrm{VaR}}^S(\tau\mid\bm{X}_i,t+1)-\log\widehat{\mathrm{VaR}}^S(\tau\mid\bm{X}_i,t)\right)}
{\left|\mathcal{A}_t\cap\mathcal{S}\right|}\right\}-1,
\end{equation}
\begin{equation}
\label{eq:sii_sev_S}
CSII^{\mathrm{amt},S}_{t+1}
=\prod_{j=t_0}^{t}\left(1+ASIR^{\mathrm{amt},S}_{j+1}\right).
\end{equation}
Uncertainty quantification for $ASIR^{\mathrm{amt},S}_{t+1}$ and $CSII^{\mathrm{amt},S}_{t+1}$ is obtained using the same random-weighted bootstrap procedure as in Section \ref{sec:method:p_s}; details are omitted for conciseness.

\subsection{Social inflation in total amount payable to the plaintiff}
\label{sec:method:t_s}

Finally, we define an aggregate social inflation index for the total payment $Y_i$ payable to the plaintiff, which incorporates all channels jointly (plaintiff win probabilities, settlement behavior, and severity conditional on payment). For $\tau\in(0,1)$, we define the conditional VaR of total plaintiff payment given covariates as
\begin{align}
\label{eq:var_def_T}
\mathrm{VaR}^T(\tau\mid \bm{X}_i,T_i)
:=\sup\Big\{y:\Pr(Y_i\le y\mid \bm{X}_i,T_i)\le \tau\Big\}.
\end{align}
We model $\mathrm{VaR}^T(\tau\mid \bm{X}_i,T_i)$ through the same log-linear quantile regression form,
\begin{equation}
\label{eq:qr_var_T}
\mathrm{VaR}^T(\tau\mid \bm{X}_i,T_i)=\exp\{\gamma_t^T+\bm{\eta}_t^T\bm{X}_i\},
\qquad i\in\mathcal{A}_t,
\end{equation}
and estimate $(\gamma_t^T,\bm{\eta}_t^T)$ using the rolling-window quantile regression procedure analogous to Equation \eqref{eq:ll_sev_P}, now pooling all cases in $\mathcal{A}_{t-4:t+1}$. Although the distribution of $Y_i$ has a point mass at 0 due to defense wins, the quantile levels used in our analysis are chosen so that the target quantiles are strictly positive, making the log-scale modeling appropriate.

Let $\widehat{\mathrm{VaR}}^T(\tau\mid \bm{X}_i,t)=\exp\{\hat{\gamma}_{t}^T+\hat{\bm{\eta}}_t^T\bm{X}_i\}$ denote the fitted VaR under year-$t$ parameters. The annual and cumulative social inflation measures for total loss are then defined analogously to Equations \eqref{eq:sir_sev_P} and \eqref{eq:sii_sev_P}:
\begin{equation}
\label{eq:sir_sev_T}
ASIR^{\mathrm{amt},T}_{t+1}
= \exp\left\{\frac{\sum\limits_{i\in\mathcal{A}_t}
\left(\log\widehat{\mathrm{VaR}}^T(\tau\mid\bm{X}_i,t+1)-\log\widehat{\mathrm{VaR}}^T(\tau\mid\bm{X}_i,t)\right)}
{\left|\mathcal{A}_t\right|}\right\}-1,
\end{equation}
\begin{equation}
\label{eq:sii_sev_T}
CSII^{\mathrm{amt},T}_{t+1}
=\prod_{j=t_0}^{t}\left(1+ASIR^{\mathrm{amt},T}_{j+1}\right).
\end{equation}
The indices defined in Equations \eqref{eq:sir_sev_T} and \eqref{eq:sii_sev_T} provide a generic, case-mix-adjusted measure of social inflation in total plaintiff payment, absorbing inflationary effects through both frequency-type and severity-type channels. Uncertainty quantification is carried out using the random-weighted bootstrap approach described in Sections \ref{sec:method:p_s} and \ref{sec:method:s_s}; details are omitted.

\section{Data analysis results} \label{sec:result}
This section applies the methodology developed in Section \ref{sec:method} to the VerdictSearch ground-up loss dataset and provides quantitative evidence of how social inflation evolves over time through multiple channels. Specifically, we estimate ASIRs and CSIIs for (1) plaintiff win probability given verdict, (2) settlement probability, (3) verdict award amount conditional on a positive plaintiff award, (4) settlement amount conditional on settlement, and (5) the total amount payable to the plaintiff. These channels have direct implications for insurance and reinsurance risk: Changes in plaintiff win rates (conditional on verdict) and settlement propensities affect the likelihood that a litigated dispute results in a positive payment (a frequency-type component), while changes in verdict and settlement amounts affect loss severity, especially in the upper tail relevant to reinsurance attachments. Because VerdictSearch does not provide exposure-based claim counts or accident/claim arrival information, these probability-based indices do not fully parameterize the overall actuarial claim frequency trend.

Beyond documenting aggregate trends, this section has four analytical objectives. First, we evaluate whether the estimated social inflation measures are statistically significant and how they evolve over time across channels. Second, we assess the importance of controlling for factual case characteristics when measuring social inflation and quantify the potential overestimation that arises when evolution in the case mix is ignored. Third, we quantify the extent to which strategic variables (proxying litigation intensity and tactics) can explain the remaining case-mix-adjusted social inflation. Fourth, we study heterogeneity across case environments and jurisdictions to test several industry-motivated hypotheses, including whether social inflation differs for (1) cases involving corporate defendants versus only individual defendants, (2) cases involving insured versus uninsured defendants, (3) motor versus general versus professional liability cases, (4) states with versus without tort-cap laws, (5) states with versus without TPLF regulations, and (6) jury trials versus bench trials.

To implement items (4) and (5) above, we conducted a state-by-state search of the most recent laws and regulations governing tort caps and TPLF. As of December 31, 2024, we identify 29 of the 50 states with some form of cap on noneconomic and/or punitive damages in general tort litigation, covering 39,711 of the 74,188 cases from 2009 to 2024 (approximately 53.5\%); these states are listed in Table \ref{tab:tortcap_states}. Over the same period and as of December 31, 2024, we identify 16 of the 50 states with TPLF-related laws or regulations (e.g., transparency/disclosure requirements, cost and contract-term protections, and/or foreign-funding rules), covering 8,564 of the 74,188 cases (approximately 11.5\%); these states are listed in Table \ref{tab:tplf_states}. Among these 16 states, Kentucky effectively bans TPLF in the covered context (via champerty-related doctrines that void funding contracts), whereas the remaining states permit TPLF subject to varying regulatory constraints.

\begin{table}[!h]
\centering
\caption{States with tort-cap regulations (as of December 31, 2024).}
\label{tab:tortcap_states}
\setlength{\tabcolsep}{8pt}
\resizebox{\columnwidth}{!}{
\begin{tabular}{p{0.16\textwidth} p{0.16\textwidth} p{0.16\textwidth} p{0.16\textwidth} p{0.16\textwidth} p{0.18\textwidth}}
\toprule
\multicolumn{6}{c}{\textbf{Tort-cap-regulated states}}\\
\midrule
Alabama & Alaska & Colorado & Florida & Georgia & Hawaii\\
Idaho & Indiana & Kansas & Louisiana & Massachusetts & Michigan\\
Mississippi & Missouri & Montana & Nebraska & Nevada & New Hampshire\\
New Jersey & North Carolina & Ohio & Oklahoma & South Carolina & Tennessee\\
Texas & Virginia & Washington & West Virginia & Wisconsin & \\
\bottomrule
\end{tabular}}
\end{table}

\begin{table}[!h]
\centering
\caption{States with TPLF regulations (as of December 31, 2024).}
\label{tab:tplf_states}
\setlength{\tabcolsep}{8pt}
\resizebox{\columnwidth}{!}{
\begin{tabular}{p{0.16\textwidth} p{0.16\textwidth} p{0.16\textwidth} p{0.16\textwidth} p{0.16\textwidth} p{0.16\textwidth}}
\toprule
\multicolumn{6}{c}{\textbf{TPLF-regulated states}}\\
\midrule
Colorado & Illinois & Indiana & Kentucky & Louisiana & Maine\\
Montana & Nebraska & Nevada & Ohio & South Carolina & Tennessee\\
Utah & Vermont & West Virginia & Wisconsin & & \\
\bottomrule
\end{tabular}}
\end{table}

Operationally, Figures \ref{fig:idx_prob_p_0} to \ref{fig:idx_sev_t1_6} plot the estimated ASIRs (left panels) and CSIIs (right panels) against calendar year. Each plot reports three specifications: a baseline index without covariate controls (blue), a case-mix-adjusted index controlling for factual explanatory variables only (green), and a further-adjusted index controlling for both factual and strategic explanatory variables (red). \textbf{We advocate the factual-only (green) specification as the headline social inflation index}, because the factual variables (case characteristics largely determined at the outset of litigation and difficult to alter through litigation tactics once the dispute is in court) are the appropriate controls for isolating loss growth beyond general economic inflation net of evolving case mix. 

Consistent with this interpretation, the blue-green gap in each figure quantifies how much of the observed temporal change can be mechanically attributed to shifts in case composition (and thus the potential overstatement of social inflation in unadjusted trend analysis). The red specification is therefore best viewed as a ``residual'' index: By additionally conditioning on endogenous strategic variables (trial and deliberation length, and attorney/expert involvement) that can be shaped by litigation decisions, it shows how much of the headline (green) social inflation trend is explained by evolving litigation intensity and related strategic factors. Accordingly, such endogenous strategic variables should be regarded as a channel of or contributing factor to social inflation, rather than as controls that should be netted out when quantifying headline social inflation. The green-red gap provides an interpretable decomposition of the headline social inflation index into the portion explained by observed strategic evolution versus the portion that remains after conditioning on those strategic variables. Shaded bands provide 95\% confidence intervals based on the random-weighted bootstrap described in Section \ref{sec:method}. 

We present results both for the full sample and for stratified subsamples, including (1) all cases, (2) corporate versus individual defendants, (3) insured versus uninsured defendants, (4) liability type (i.e., the proxy insurance LoB: motor vehicle cases (auto liability), general liability, professional liability, and “others”), (5) tort-cap states versus noncap states, (6) TPLF-regulated states versus nonregulated states, and (7) jury versus bench trials.


\subsection{Social inflation in plaintiff win probability} \label{sec:result:p_p}
\begin{figure}[H]
  \centering
  \begin{subfigure}[t]{0.48\textwidth}
    \centering
    \includegraphics[width=\linewidth]{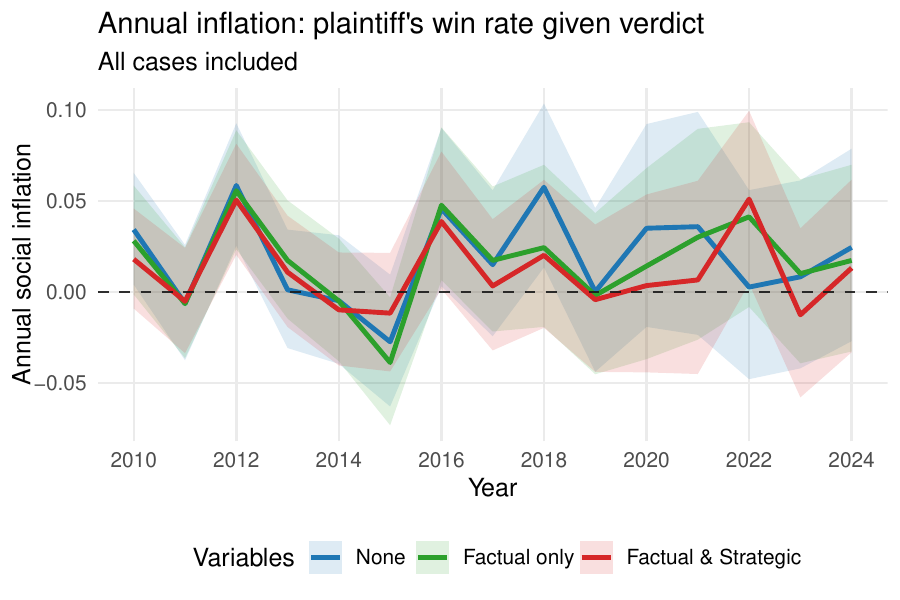}
  \end{subfigure}\hfill
  \begin{subfigure}[t]{0.48\textwidth}
    \centering
    \includegraphics[width=\linewidth]{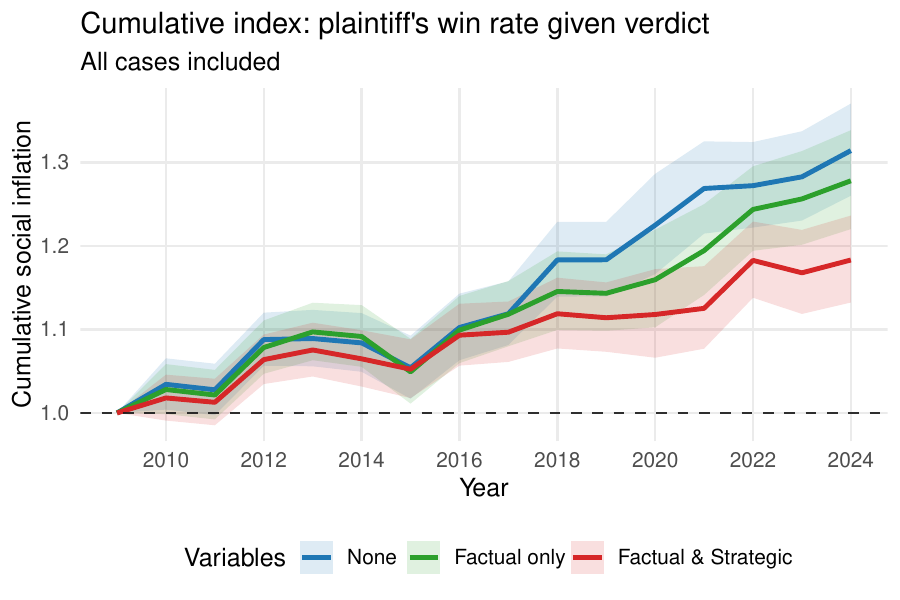}
  \end{subfigure}\hfill
  \caption{Annual (\textit{left panel}) and cumulative (\textit{right panel}) social inflation in plaintiff's win rate given verdict. All cases are included.}
  \label{fig:idx_prob_p_0}
\end{figure}

\begin{figure}[H]
  \centering
  \begin{subfigure}[t]{0.48\textwidth}
    \centering
    \includegraphics[width=\linewidth]{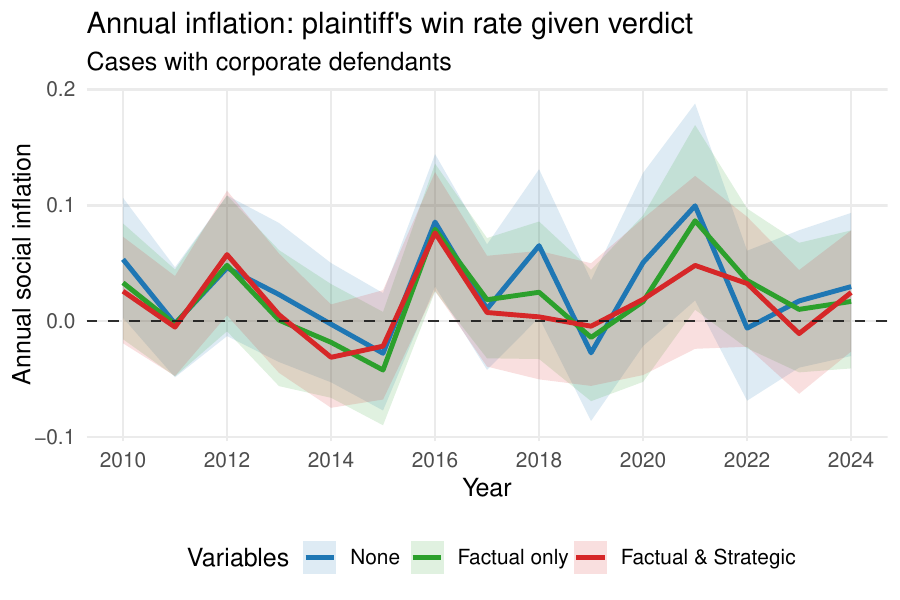}
  \end{subfigure}\hfill
  \begin{subfigure}[t]{0.48\textwidth}
    \centering
    \includegraphics[width=\linewidth]{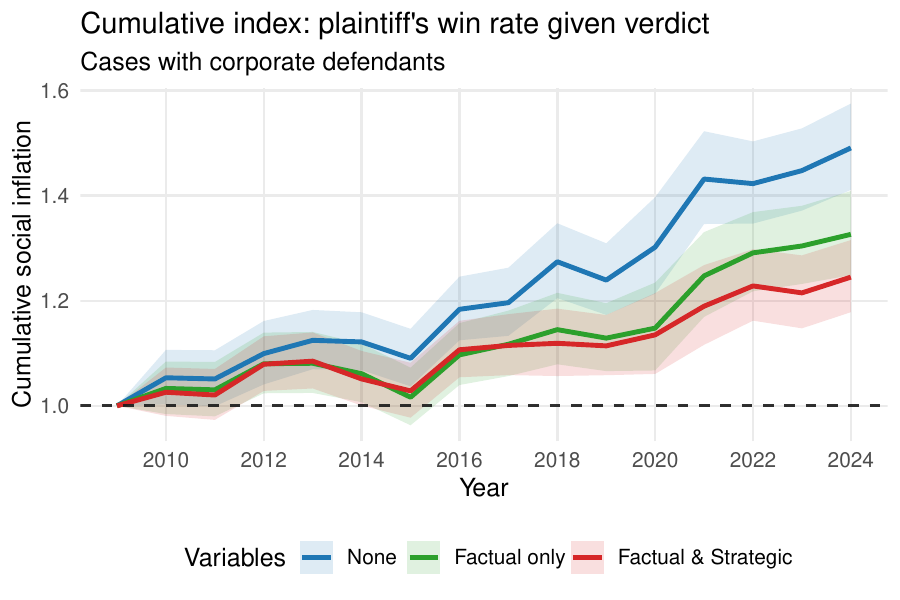}
  \end{subfigure}\hfill
    \begin{subfigure}[t]{0.48\textwidth}
    \centering
    \includegraphics[width=\linewidth]{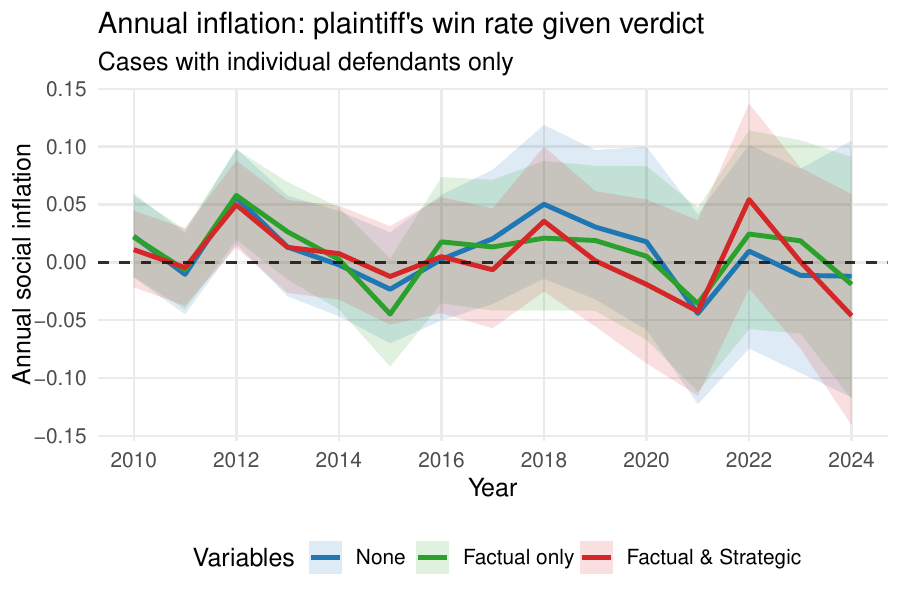}
  \end{subfigure}\hfill
  \begin{subfigure}[t]{0.48\textwidth}
    \centering
    \includegraphics[width=\linewidth]{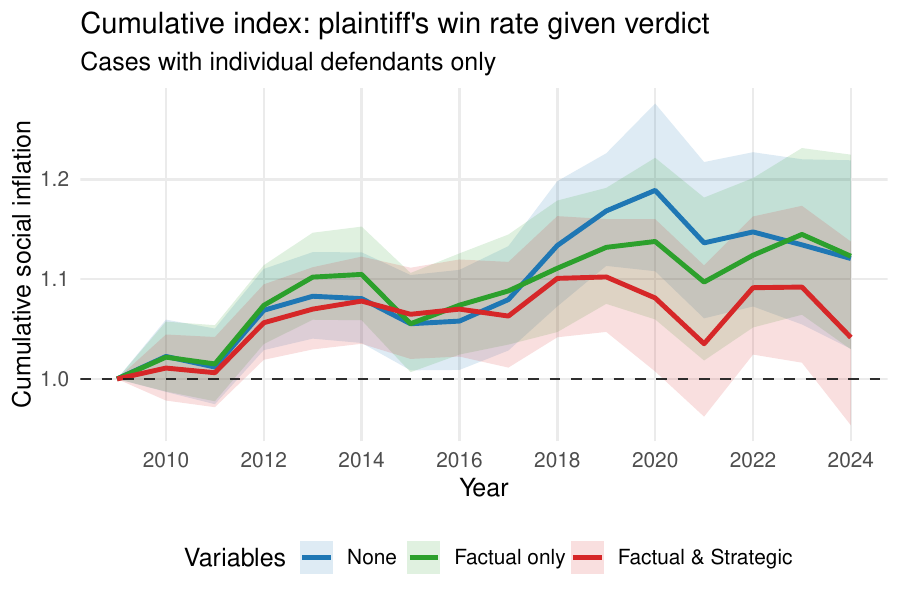}
  \end{subfigure}\hfill
  \caption{Annual (\textit{left panels}) and cumulative (\textit{right panels}) social inflation in plaintiff's win rate given verdict. \textit{Top panels}: cases with corporate defendants; \textit{bottom panels}: cases with individual defendants only.}
  \label{fig:idx_prob_p_1}
\end{figure}

\begin{figure}[H]
  \centering
  \begin{subfigure}[t]{0.48\textwidth}
    \centering
    \includegraphics[width=\linewidth]{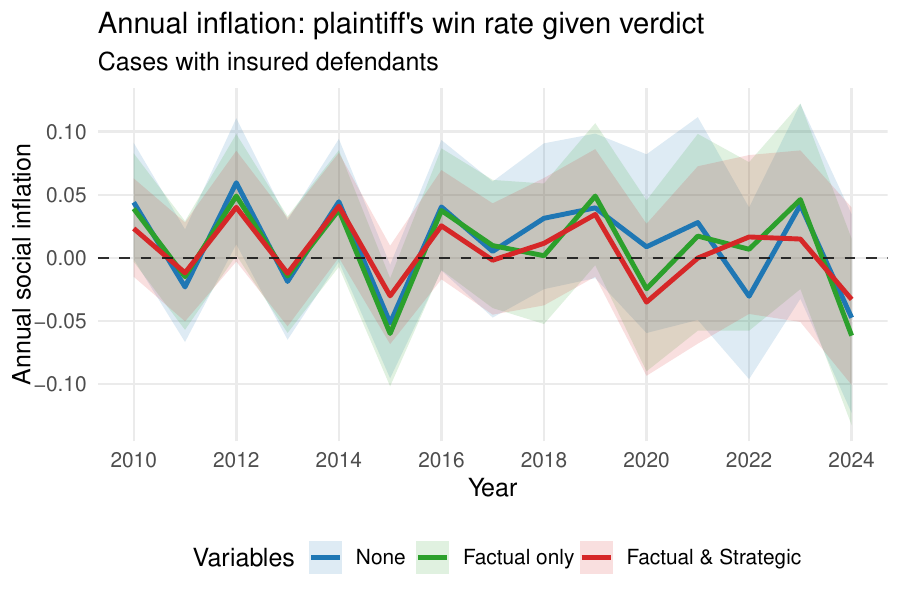}
  \end{subfigure}\hfill
  \begin{subfigure}[t]{0.48\textwidth}
    \centering
    \includegraphics[width=\linewidth]{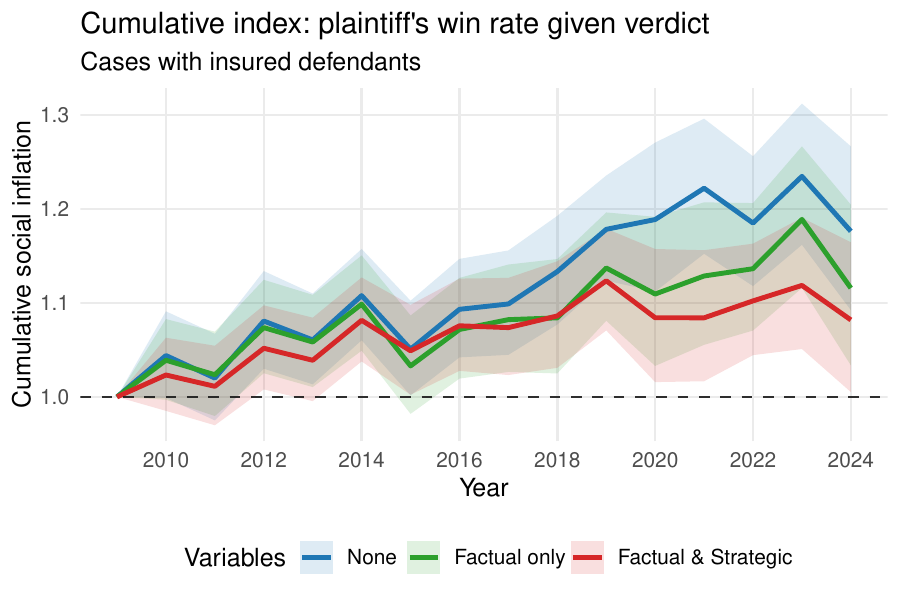}
  \end{subfigure}\hfill
    \begin{subfigure}[t]{0.48\textwidth}
    \centering
    \includegraphics[width=\linewidth]{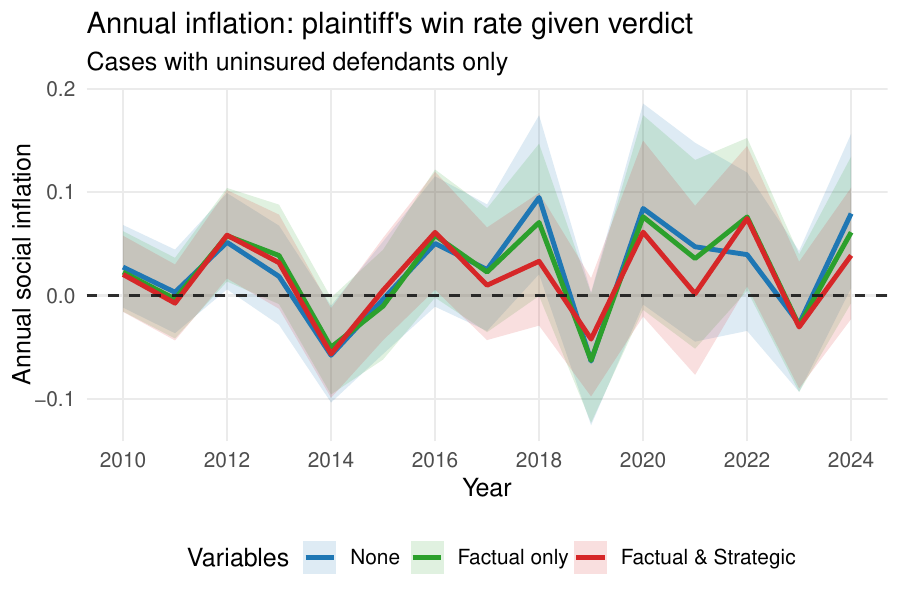}
  \end{subfigure}\hfill
  \begin{subfigure}[t]{0.48\textwidth}
    \centering
    \includegraphics[width=\linewidth]{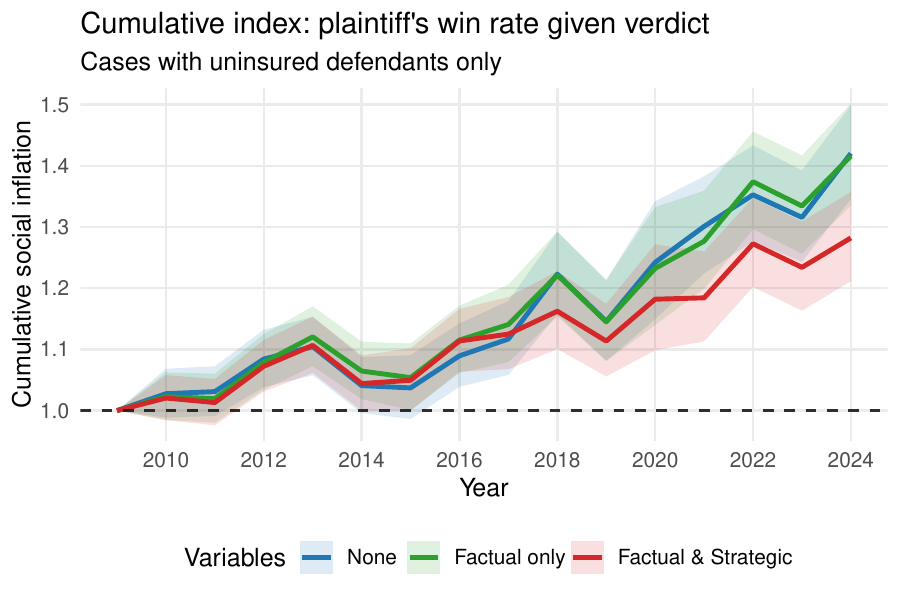}
  \end{subfigure}\hfill
  \caption{Annual (\textit{left panels}) and cumulative (\textit{right panels}) social inflation in plaintiff's win rate given verdict. \textit{Top panels}: cases with insured defendants; \textit{bottom panels}: cases with uninsured defendants only.}
  \label{fig:idx_prob_p_2}
\end{figure}

\begin{figure}[H]
  \centering
  \begin{subfigure}[t]{0.48\textwidth}
    \centering
    \includegraphics[width=\linewidth]{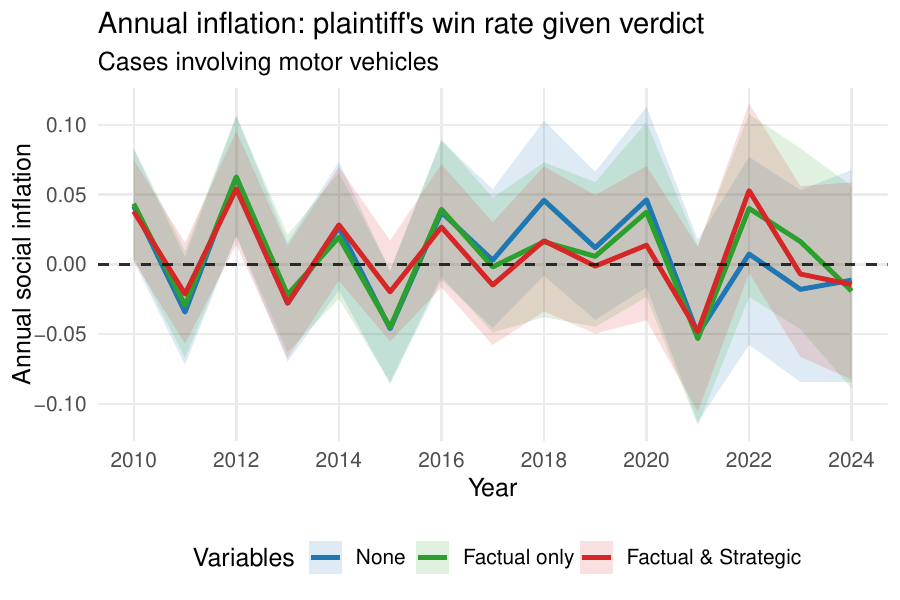}
  \end{subfigure}\hfill
  \begin{subfigure}[t]{0.48\textwidth}
    \centering
    \includegraphics[width=\linewidth]{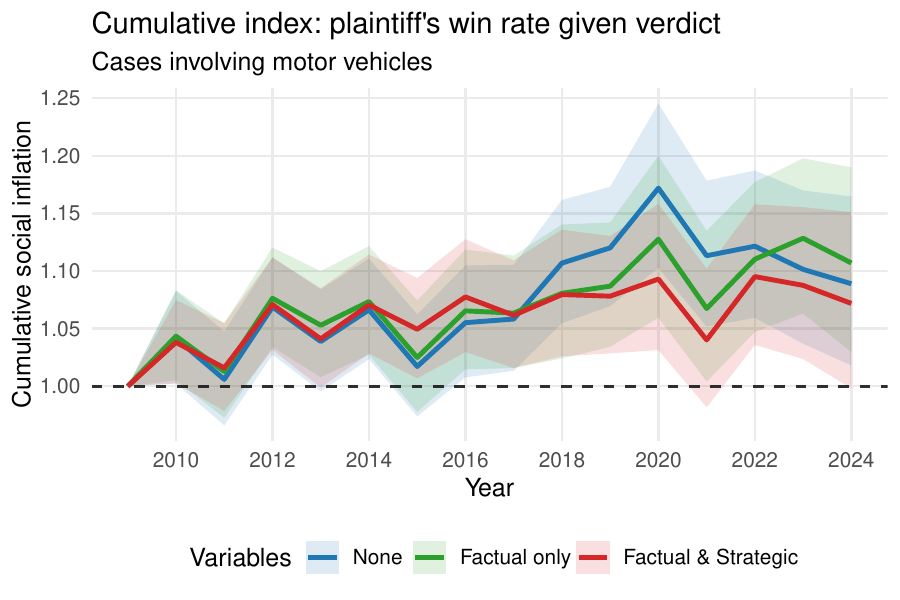}
  \end{subfigure}\hfill
    \begin{subfigure}[t]{0.48\textwidth}
    \centering
    \includegraphics[width=\linewidth]{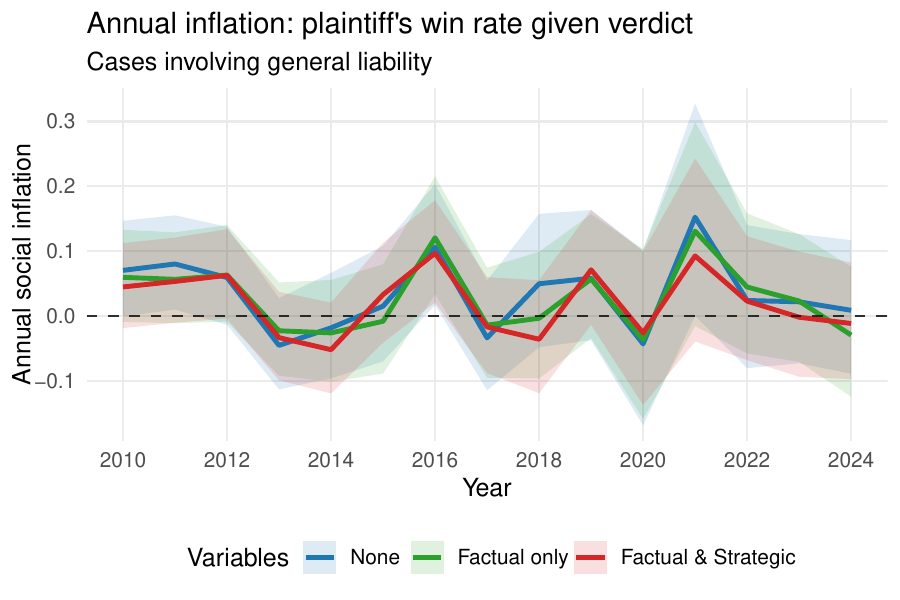}
  \end{subfigure}\hfill
  \begin{subfigure}[t]{0.48\textwidth}
    \centering
    \includegraphics[width=\linewidth]{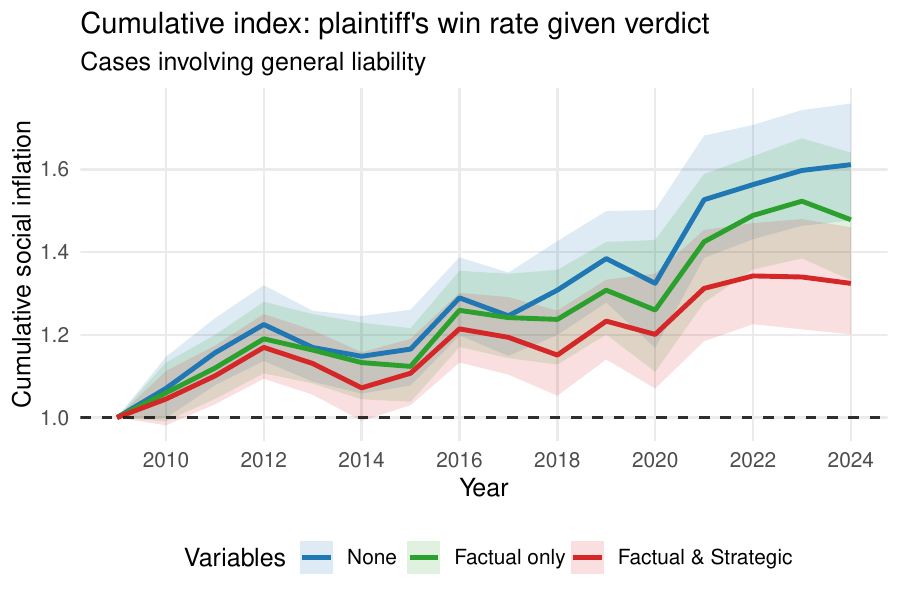}
  \end{subfigure}\hfill
  \begin{subfigure}[t]{0.48\textwidth}
    \centering
    \includegraphics[width=\linewidth]{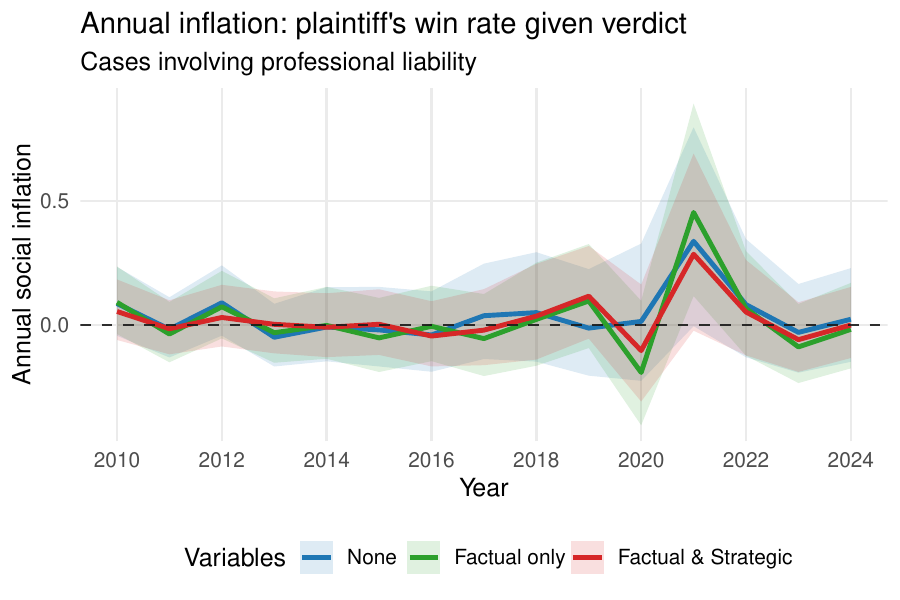}
  \end{subfigure}\hfill
  \begin{subfigure}[t]{0.48\textwidth}
    \centering
    \includegraphics[width=\linewidth]{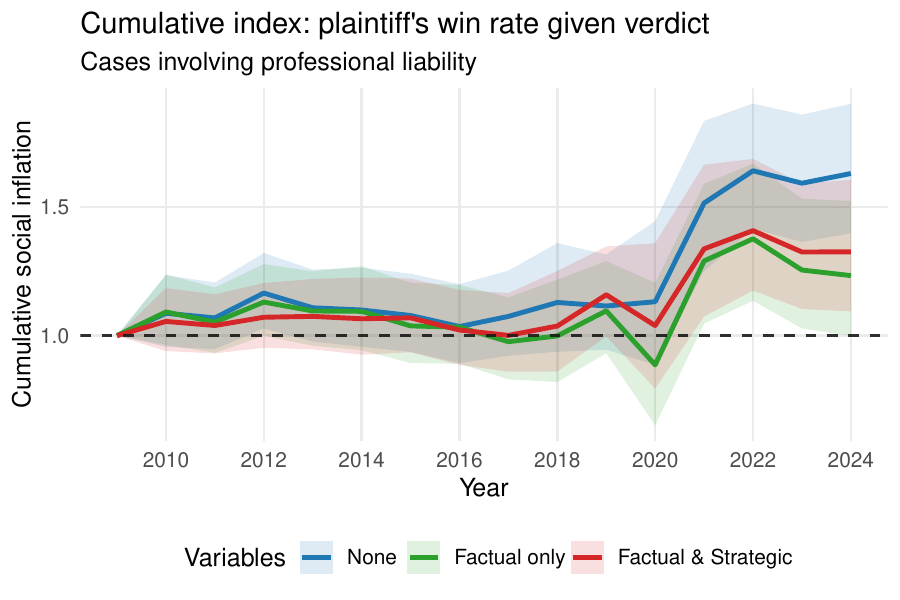}
  \end{subfigure}\hfill
  \begin{subfigure}[t]{0.48\textwidth}
    \centering
    \includegraphics[width=\linewidth]{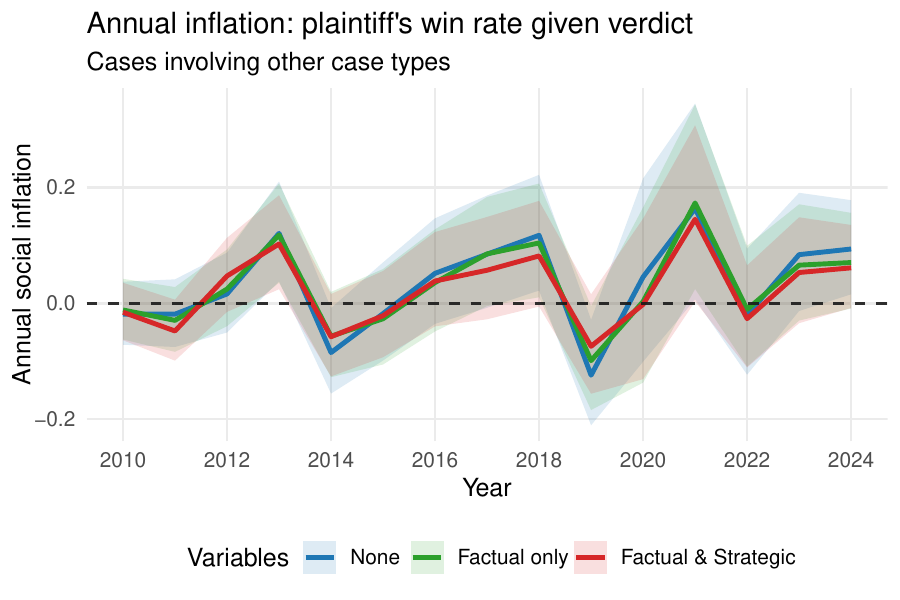}
  \end{subfigure}\hfill
  \begin{subfigure}[t]{0.48\textwidth}
    \centering
    \includegraphics[width=\linewidth]{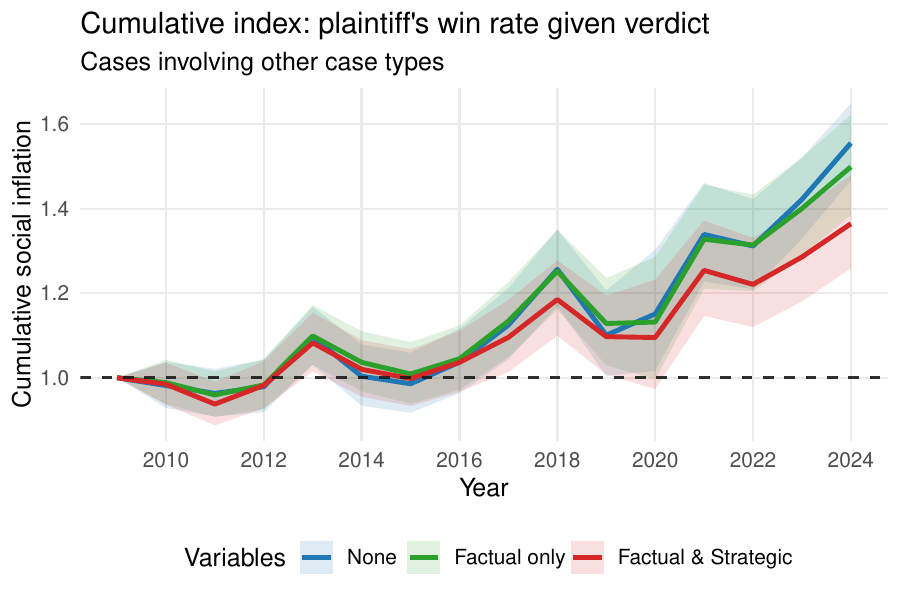}
  \end{subfigure}\hfill
  \caption{Annual (\textit{left panels}) and cumulative (\textit{right panels}) social inflation in plaintiff's win rate given verdict. \textit{First row}: motor liability cases; \textit{second row}: general liability cases;\textit{ third row}: professional liability cases; \textit{fourth row}: other cases.}
  \label{fig:idx_prob_p_3}
\end{figure}

\begin{figure}[H]
  \centering
  \begin{subfigure}[t]{0.48\textwidth}
    \centering
    \includegraphics[width=\linewidth]{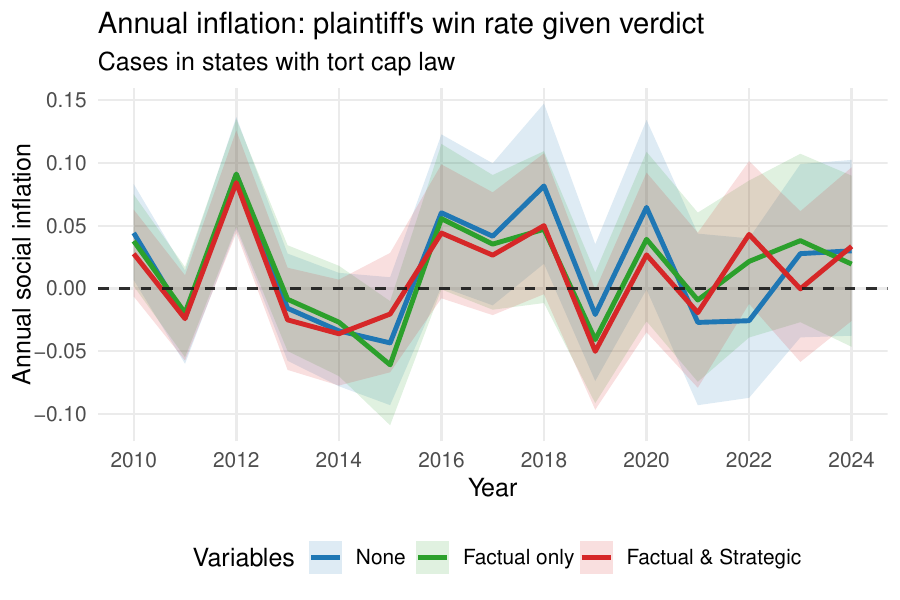}
  \end{subfigure}\hfill
  \begin{subfigure}[t]{0.48\textwidth}
    \centering
    \includegraphics[width=\linewidth]{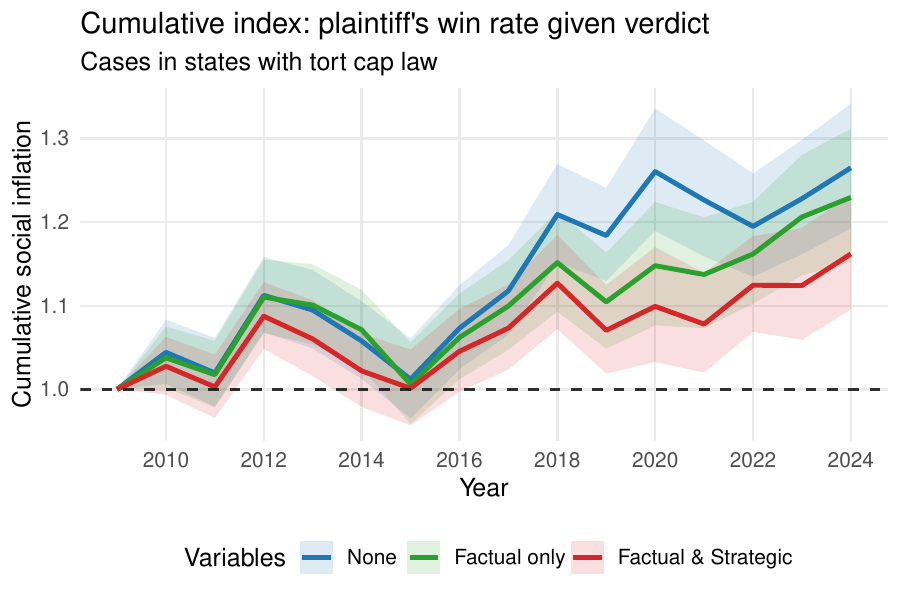}
  \end{subfigure}\hfill
    \begin{subfigure}[t]{0.48\textwidth}
    \centering
    \includegraphics[width=\linewidth]{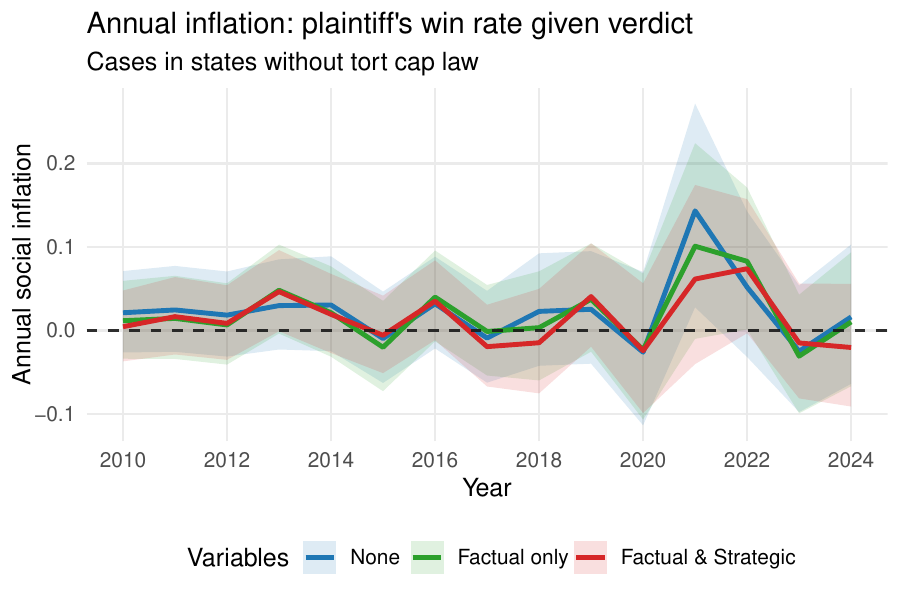}
  \end{subfigure}\hfill
  \begin{subfigure}[t]{0.48\textwidth}
    \centering
    \includegraphics[width=\linewidth]{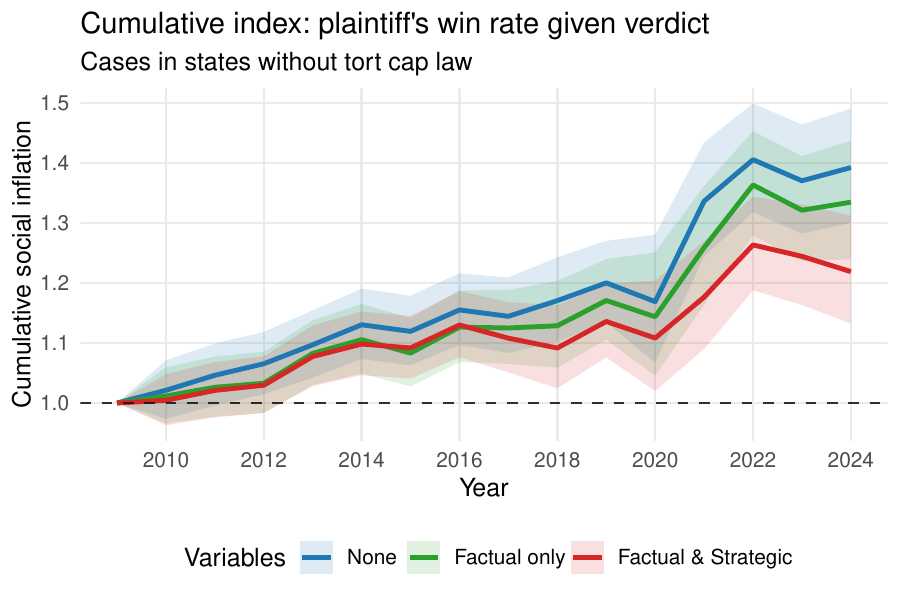}
  \end{subfigure}\hfill
  \caption{Annual (\textit{left panels}) and cumulative (\textit{right panels}) social inflation in plaintiff's win rate given verdict. \textit{Top panels}: cases in states with tort-cap laws; \textit{bottom panels}: cases in states without tort-cap laws.}
  \label{fig:idx_prob_p_4}
\end{figure}

\begin{figure}[H]
  \centering
  \begin{subfigure}[t]{0.48\textwidth}
    \centering
    \includegraphics[width=\linewidth]{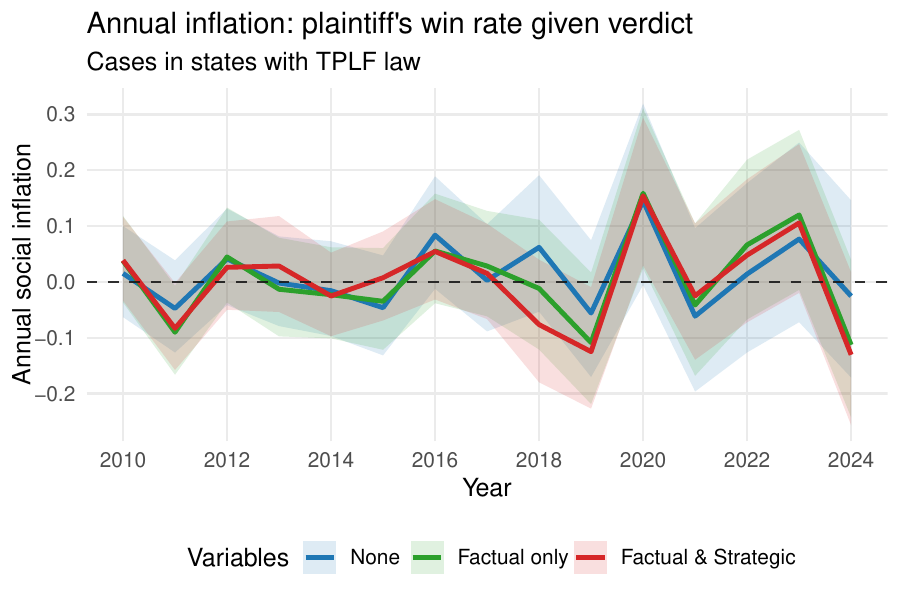}
  \end{subfigure}\hfill
  \begin{subfigure}[t]{0.48\textwidth}
    \centering
    \includegraphics[width=\linewidth]{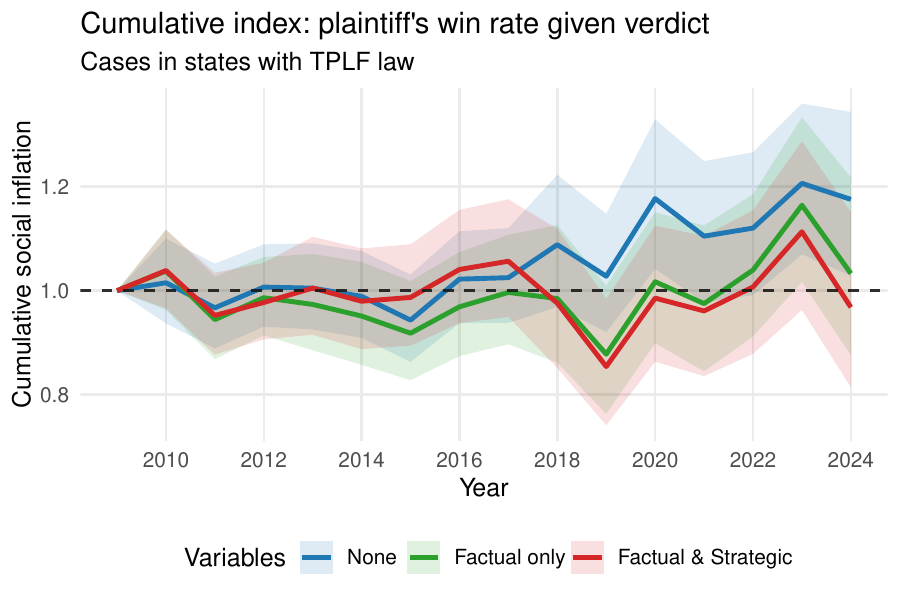}
  \end{subfigure}\hfill
    \begin{subfigure}[t]{0.48\textwidth}
    \centering
    \includegraphics[width=\linewidth]{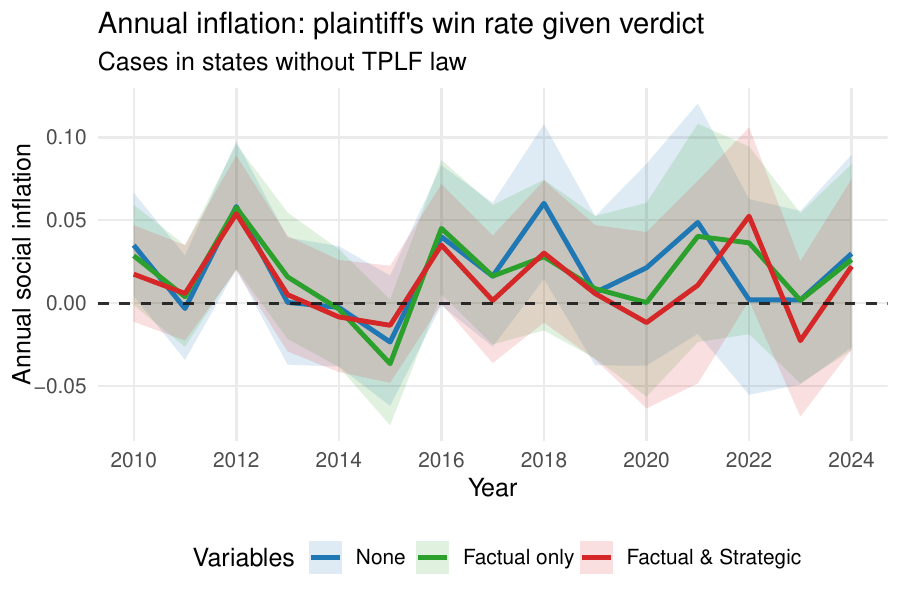}
  \end{subfigure}\hfill
  \begin{subfigure}[t]{0.48\textwidth}
    \centering
    \includegraphics[width=\linewidth]{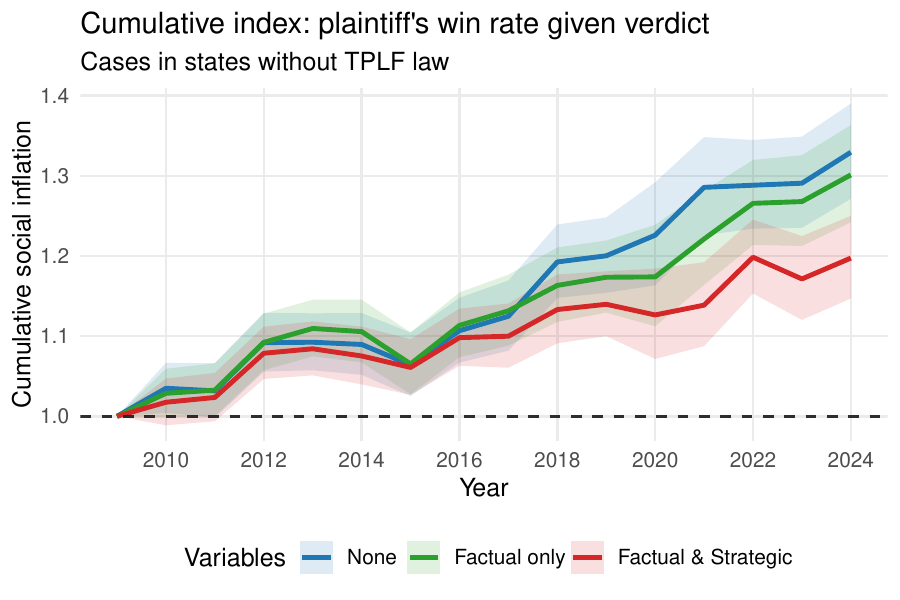}
  \end{subfigure}\hfill
  \caption{Annual (\textit{left panels}) and cumulative (\textit{right panels}) social inflation in plaintiff's win rate given verdict. \textit{Top panels}: cases in states with TPLF laws; \textit{bottom panels}: cases in states without TPLF laws.}
  \label{fig:idx_prob_p_5}
\end{figure}

\begin{figure}[H]
  \centering
  \begin{subfigure}[t]{0.48\textwidth}
    \centering
    \includegraphics[width=\linewidth]{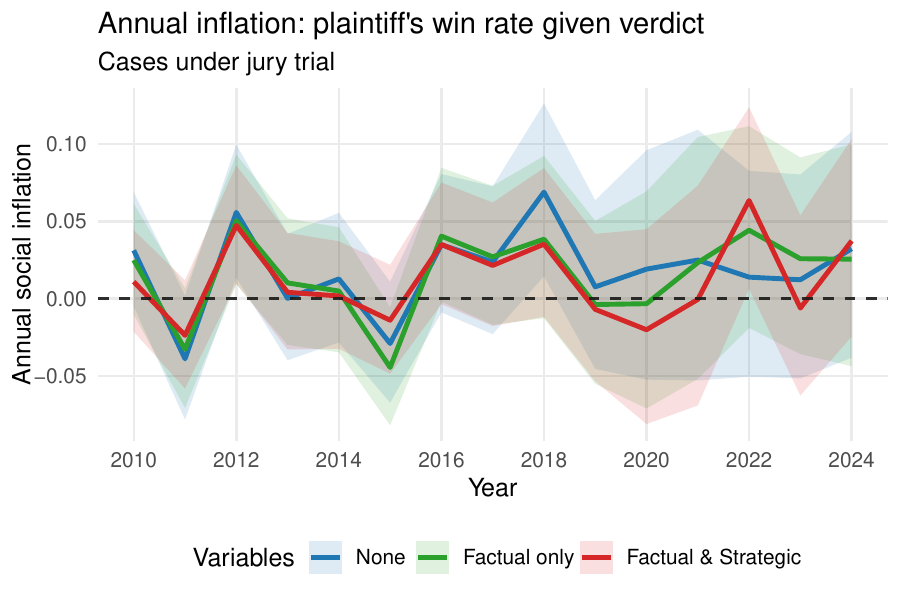}
  \end{subfigure}\hfill
  \begin{subfigure}[t]{0.48\textwidth}
    \centering
    \includegraphics[width=\linewidth]{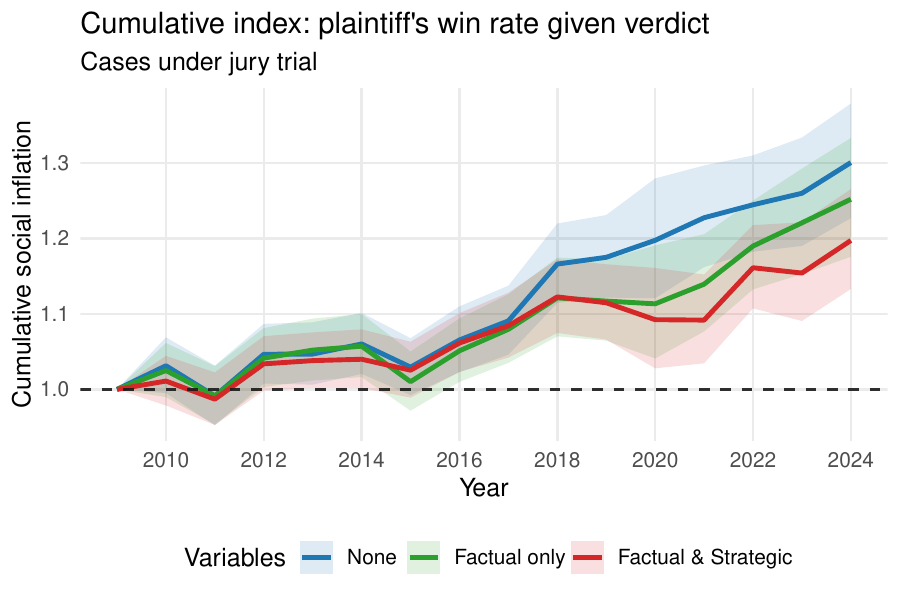}
  \end{subfigure}\hfill
    \begin{subfigure}[t]{0.48\textwidth}
    \centering
    \includegraphics[width=\linewidth]{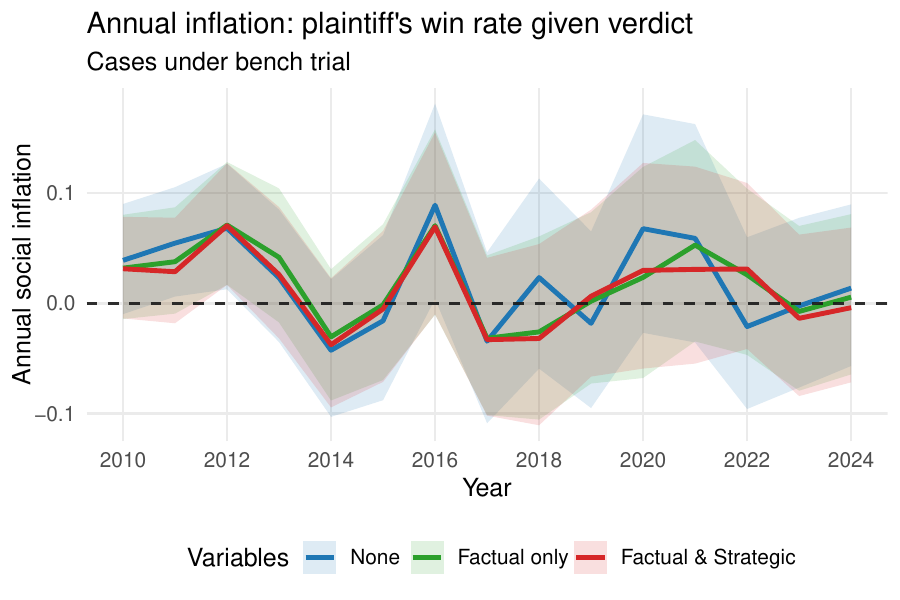}
  \end{subfigure}\hfill
  \begin{subfigure}[t]{0.48\textwidth}
    \centering
    \includegraphics[width=\linewidth]{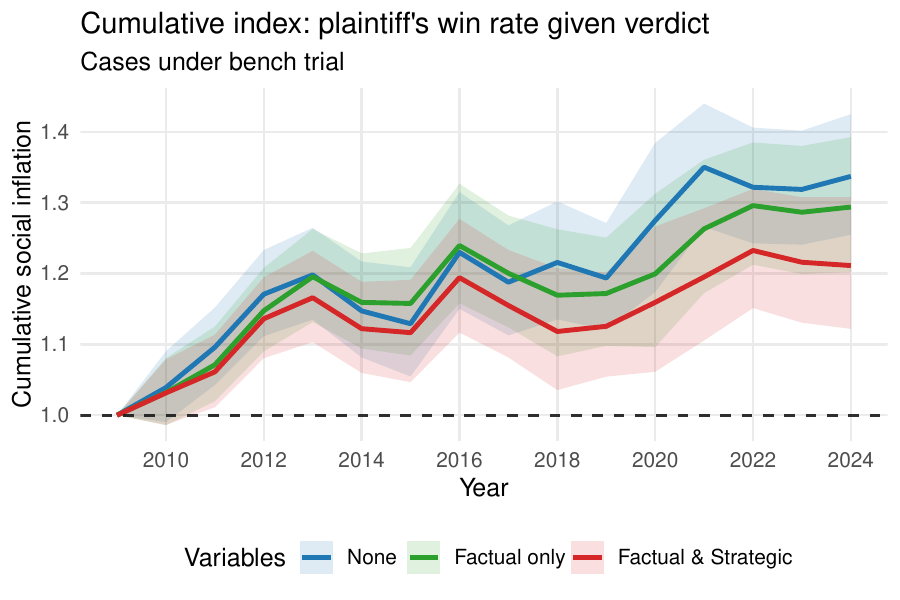}
  \end{subfigure}\hfill
  \caption{Annual (\textit{left panels}) and cumulative (\textit{right panels}) social inflation in plaintiff's win rate given verdict. \textit{Top panels}: cases under jury trial; \textit{bottom panels}: cases under bench trial.}
  \label{fig:idx_prob_p_6}
\end{figure}

This subsection reports social inflation results for plaintiff win probability conditional on reaching a verdict (i.e., excluding settlements), using the indices defined in Section \ref{sec:method:p_p}. Because the plaintiff win probability directly affects the likelihood that a verdict produces a positive payment, social inflation in this channel can be interpreted as a frequency-type component of loss inflation, in the sense of changing conditional payment probabilities for litigated cases, rather than exposure-based claim incidence. Figures \ref{fig:idx_prob_p_0} to \ref{fig:idx_prob_p_6} summarize how the plaintiff-win index evolves for the full sample and for key subsamples of interest.

Figure \ref{fig:idx_prob_p_0} (right panel) shows that plaintiff win probability exhibits a sustained upward trend from 2009 to 2024 and that the increase remains statistically significant under all model specifications. In the unadjusted specification (blue), the cumulative index rises to above 1.30 by 2024, implying an overall increase in plaintiff win probability of more than 30\% relative to 2009. When factual covariates are included (green), the cumulative increase is smaller (below 30\%), indicating that part of the observed rise reflects a changing case mix over time. When both factual and strategic covariates are included (red), the cumulative increase is further reduced (below 20\%), implying that evolving litigation intensity and tactics explain an additional portion of the upward trend. Importantly, however, the red curve still increases meaningfully over time, suggesting that a nontrivial component of social inflation in plaintiff win probability is not fully explained by the strategic variables observed in VerdictSearch and may reflect broader, exogenous shifts in legal outcomes (e.g., changing jury norms, judicial standards, or other unobserved institutional factors). The confidence bands for the CSII in Figure \ref{fig:idx_prob_p_0} lie well above the 2009 baseline of 1 for most of the period, confirming strong statistical significance of the cumulative effect.

The year-by-year ASIR estimates in Figure \ref{fig:idx_prob_p_0} (left panel) are generally positive across many years, but they are more volatile and less frequently statistically significant than the cumulative indices. This is expected: Annual rates rely on a single-year comparison and therefore contain substantially more sampling variation, especially in the later years when the number of verdict observations declines sharply. The widening confidence intervals after 2020 highlight the greater uncertainty in year-specific estimates, whereas the cumulative index aggregates information over time and therefore tends to deliver more stable inference.

Figures \ref{fig:idx_prob_p_1} and \ref{fig:idx_prob_p_2} highlight substantial heterogeneity in social inflation in plaintiff win probability across case environments that are central to reinsurance exposure. Figure \ref{fig:idx_prob_p_1} shows that cases involving corporate defendants experience markedly stronger social inflation than cases involving only individual defendants. When controlling for factual covariates, the cumulative increase from 2009 to 2024 exceeds 30\% for corporate-defendant cases but is below 15\% for individual-only cases. When both factual and strategic variables are controlled for, the corporate-defendant CSII remains statistically significant from the 2009 baseline of 1, whereas the individual-only CSII becomes statistically indistinguishable from the 2009 baseline. This pattern is consistent with the industry concern that corporate defendants may increasingly be viewed as ``deep pockets,'' potentially amplifying plaintiff success probabilities over time; it also reinforces the importance of accounting for the rising proportion of corporate-defendant cases documented in Section \ref{sec:data}, as that compositional shift can mechanically inflate unadjusted trends.

In contrast, Figure \ref{fig:idx_prob_p_2} shows that uninsured-defendant cases exhibit substantially more social inflation in plaintiff win probability than insured-defendant cases. After controlling for factual variables, the cumulative increase from 2009 to 2024 exceeds 40\% for uninsured-only cases, while it is only slightly above 10\% for insured cases. This finding runs counter to a simple ``insured deep-pocket'' narrative and suggests that insurer involvement may dampen plaintiff-win social inflation through more sophisticated claims handling, defense resources, and litigation management.

Figure \ref{fig:idx_prob_p_3} demonstrates that social inflation in plaintiff win probability also varies materially by liability type. General liability cases exhibit the most pronounced upward trajectory, with CSIIs increasing more rapidly and reaching higher levels than those for motor and professional liability cases. This heterogeneity is important for insurers and reinsurers because it suggests that the frequency-type component of social inflation is not uniform across business lines; in particular, portfolios concentrated in general liability exposures may face stronger upward pressure on plaintiff win rates than portfolios dominated by motor or professional liability claims.

Figures \ref{fig:idx_prob_p_4} and \ref{fig:idx_prob_p_5} examine whether legal and regulatory environments are associated with differential social inflation in plaintiff win probability. Figure \ref{fig:idx_prob_p_4} shows that cases in states without tort-cap laws experience higher cumulative social inflation than cases in tort-cap states; the difference in cumulative increase from 2009 to 2024 is on the order of 10 percentage points. This is consistent with the role of caps as a structural constraint on potential trial outcomes, which may indirectly affect bargaining tactics and litigation incentives. Figure \ref{fig:idx_prob_p_5} reports an even sharper contrast related to TPLF regulation: Cases in states with TPLF laws exhibit little to no upward cumulative trend, with the case-mix-adjusted and strategy-adjusted index (red) even trending slightly below 1 by 2024 (though the decrease is not statistically significant). By comparison, states without TPLF laws show a steady and statistically significant increase, with the factual-adjusted cumulative increase reaching roughly 30\% by 2024. The results are consistent with the hypothesis that regulatory environments governing litigation finance may influence the evolution of plaintiff success probabilities over time.

Finally, Figure \ref{fig:idx_prob_p_6} compares jury trials and bench trials. The estimated ASIR and CSII trajectories are broadly similar across the two settings, and the differences are not significant. This suggests that the upward shift in plaintiff win probability conditional on verdict is not driven solely by juries; instead, the forces behind plaintiff-win social inflation appear to affect both judge-decided and jury-decided cases in a broadly comparable manner.

Overall, Figures \ref{fig:idx_prob_p_0} to \ref{fig:idx_prob_p_6} indicate that plaintiff win probability conditional on verdict has risen significantly since 2009 but that unadjusted trends overstate the increase. Case-mix adjustment via factual covariates reduces the cumulative growth, and adding strategic covariates reduces it further, implying that both shifting composition and intensified litigation practices explain part (though not all) of the observed social inflation. The effect is heterogeneous. It is stronger for corporate-defendant and uninsured-defendant cases, more pronounced in general liability than in motor or professional liability cases, and larger in states without tort caps or TPLF regulation, while jury versus bench trials show no meaningful difference. These findings motivate the next subsections, which examine whether comparable patterns arise in settlement probability and in severity-based channels, and how the channels combine to drive social inflation in total plaintiff payments.

\subsection{Social inflation in settlement probability} \label{sec:result:s_p}
\begin{figure}[!h]
  \centering
  \begin{subfigure}[t]{0.48\textwidth}
    \centering
    \includegraphics[width=\linewidth]{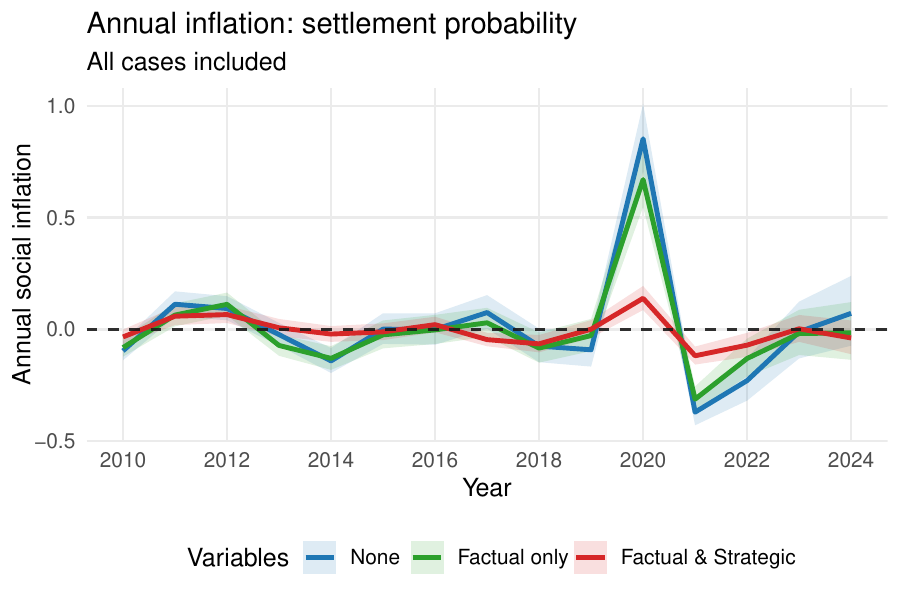}
  \end{subfigure}\hfill
  \begin{subfigure}[t]{0.48\textwidth}
    \centering
    \includegraphics[width=\linewidth]{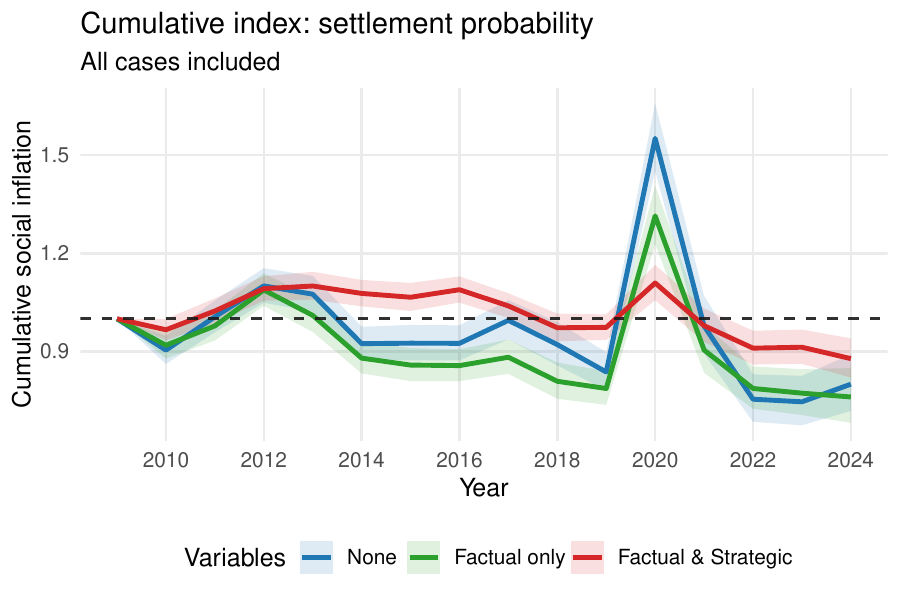}
  \end{subfigure}\hfill
  \caption{Annual (\textit{left panel}) and cumulative (\textit{right panel}) social inflation in settlement probability. All cases are included.}
  \label{fig:idx_prob_s_0}
\end{figure}

\begin{figure}[H]
  \centering
  \begin{subfigure}[t]{0.48\textwidth}
    \centering
    \includegraphics[width=\linewidth]{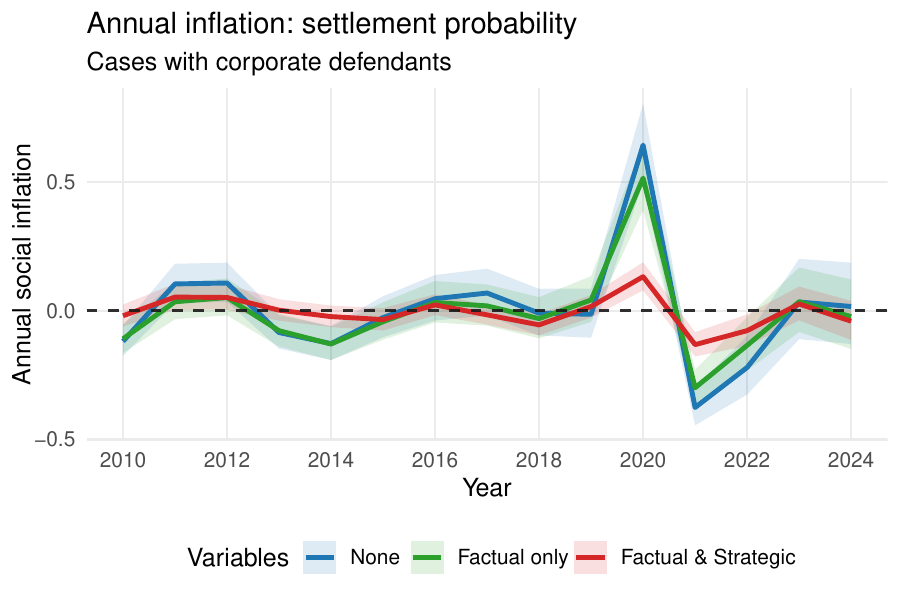}
  \end{subfigure}\hfill
  \begin{subfigure}[t]{0.48\textwidth}
    \centering
    \includegraphics[width=\linewidth]{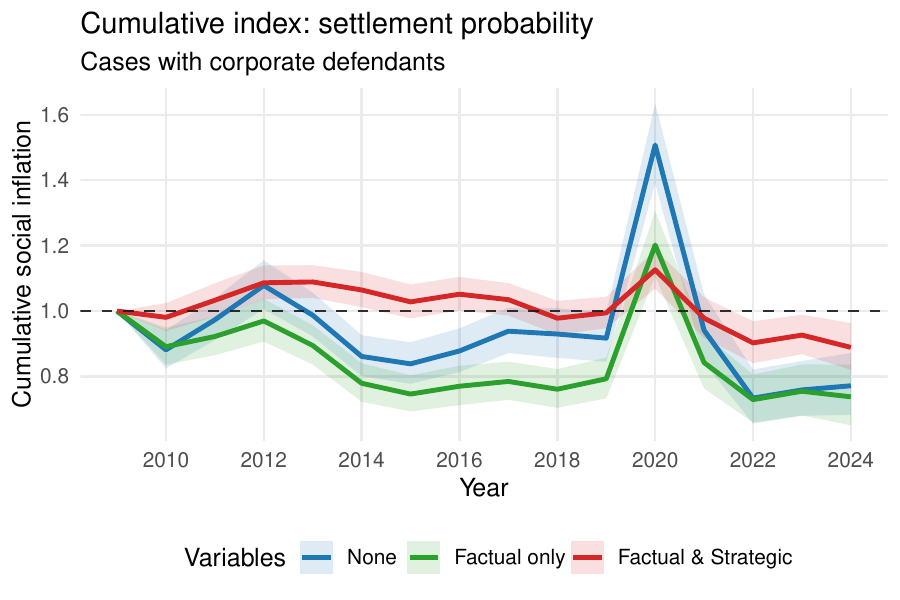}
  \end{subfigure}\hfill
    \begin{subfigure}[t]{0.48\textwidth}
    \centering
    \includegraphics[width=\linewidth]{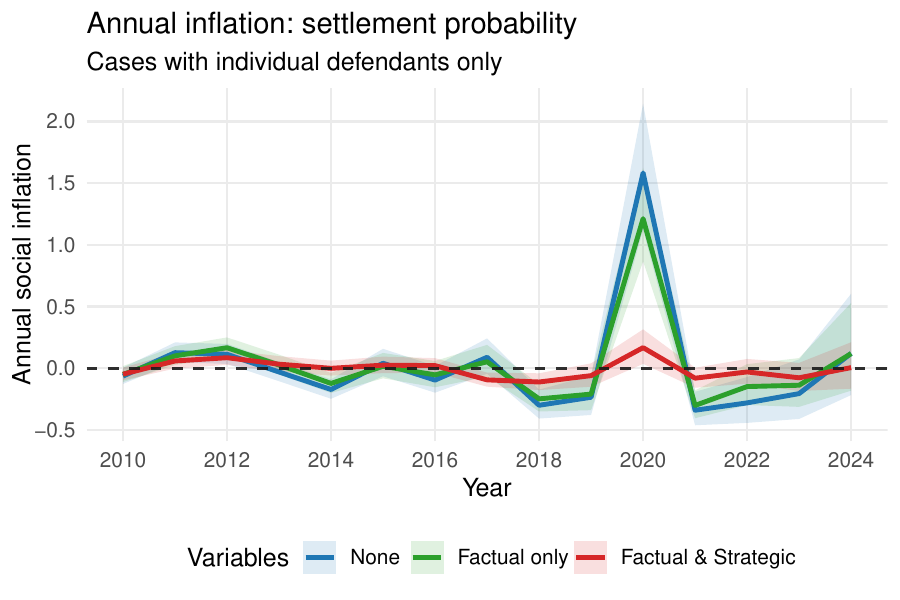}
  \end{subfigure}\hfill
  \begin{subfigure}[t]{0.48\textwidth}
    \centering
    \includegraphics[width=\linewidth]{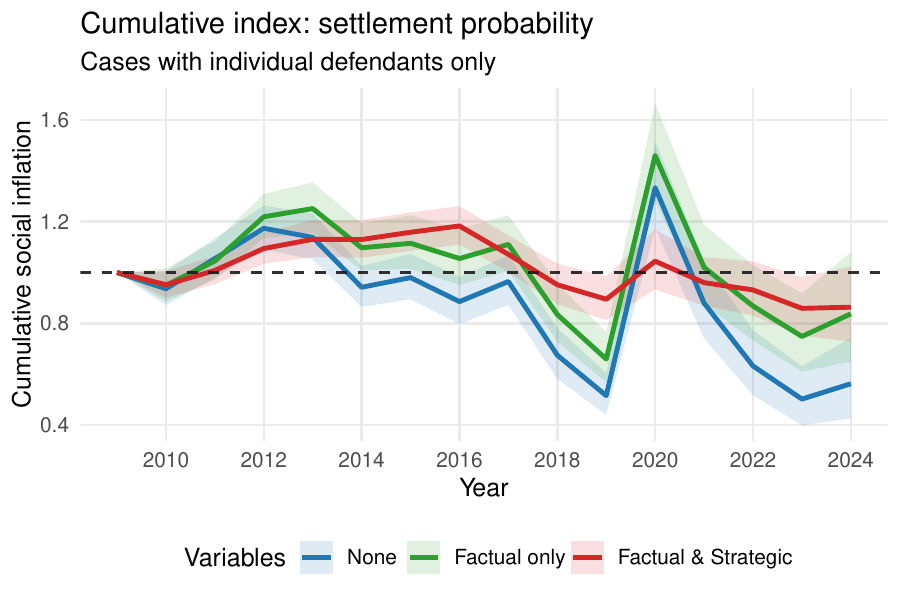}
  \end{subfigure}\hfill
  \caption{Annual (\textit{left panels}) and cumulative (\textit{right panels}) social inflation in settlement probability. \textit{Top panels}: cases with corporate defendants; \textit{bottom panels}: cases with individual defendants only.}
  \label{fig:idx_prob_s_1}
\end{figure}

\begin{figure}[H]
  \centering
  \begin{subfigure}[t]{0.48\textwidth}
    \centering
    \includegraphics[width=\linewidth]{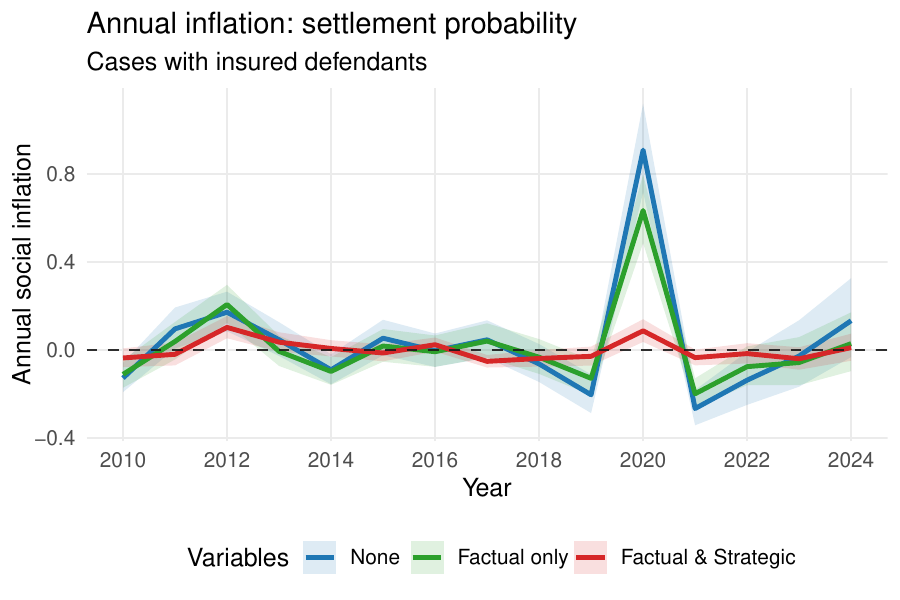}
  \end{subfigure}\hfill
  \begin{subfigure}[t]{0.48\textwidth}
    \centering
    \includegraphics[width=\linewidth]{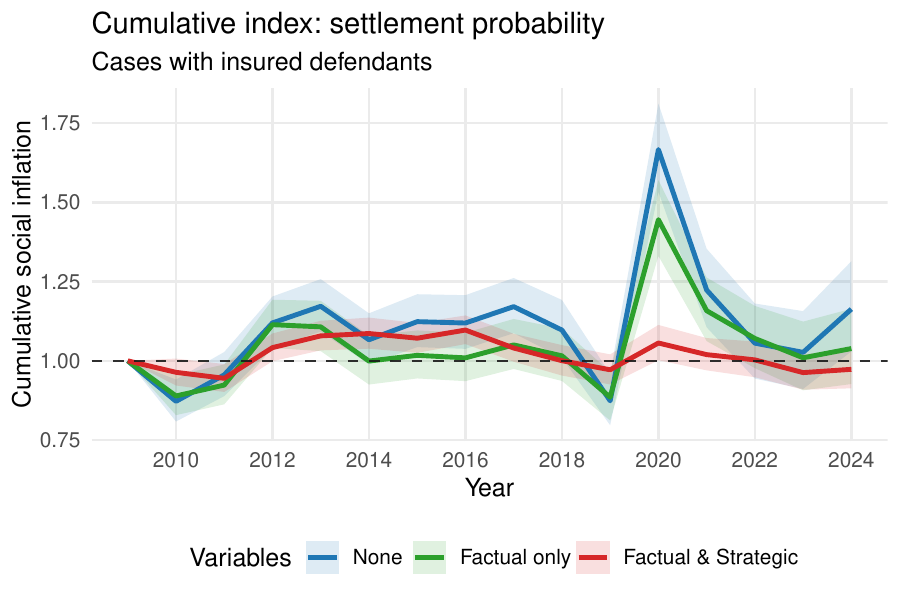}
  \end{subfigure}\hfill
    \begin{subfigure}[t]{0.48\textwidth}
    \centering
    \includegraphics[width=\linewidth]{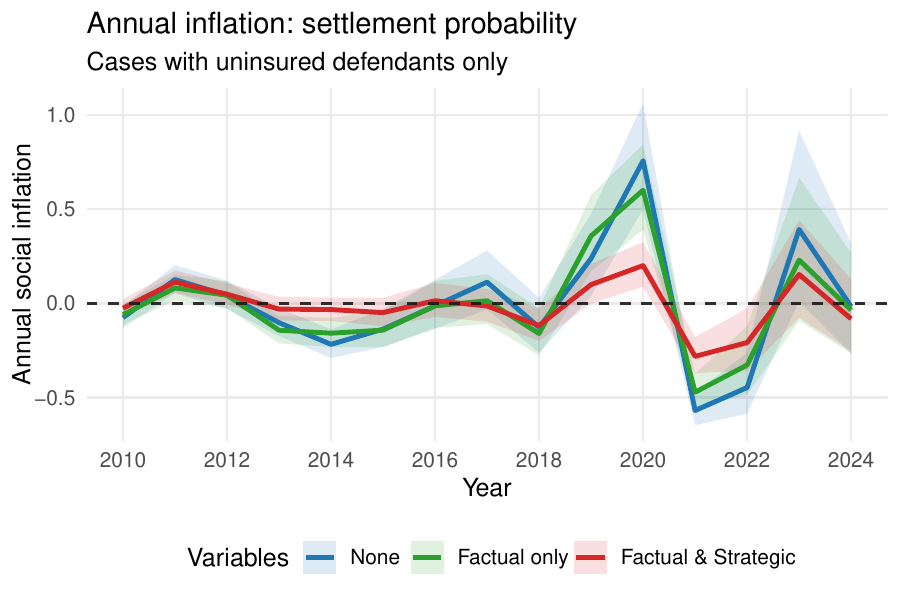}
  \end{subfigure}\hfill
  \begin{subfigure}[t]{0.48\textwidth}
    \centering
    \includegraphics[width=\linewidth]{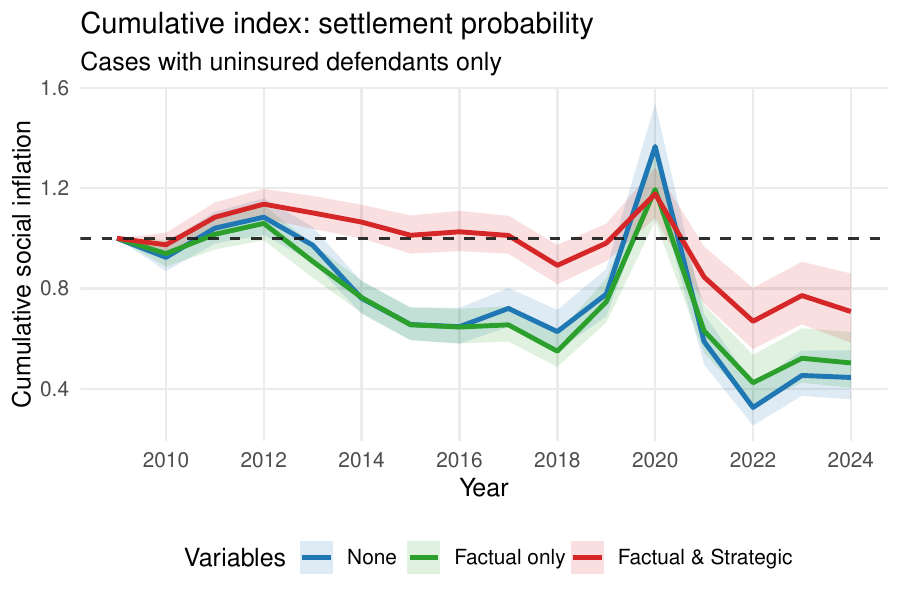}
  \end{subfigure}\hfill
  \caption{Annual (\textit{left panels}) and cumulative (\textit{right panels}) social inflation in settlement probability. \textit{Top panels}: cases with insured defendants; \textit{bottom panels}: cases with uninsured defendants only.}
  \label{fig:idx_prob_s_2}
\end{figure}

\begin{figure}[H]
  \centering
  \begin{subfigure}[t]{0.48\textwidth}
    \centering
    \includegraphics[width=\linewidth]{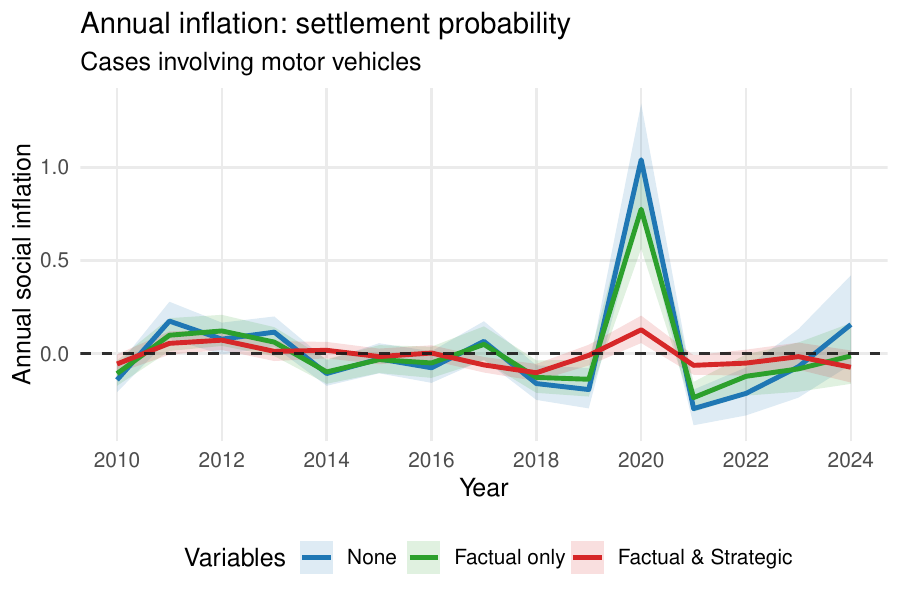}
  \end{subfigure}\hfill
  \begin{subfigure}[t]{0.48\textwidth}
    \centering
    \includegraphics[width=\linewidth]{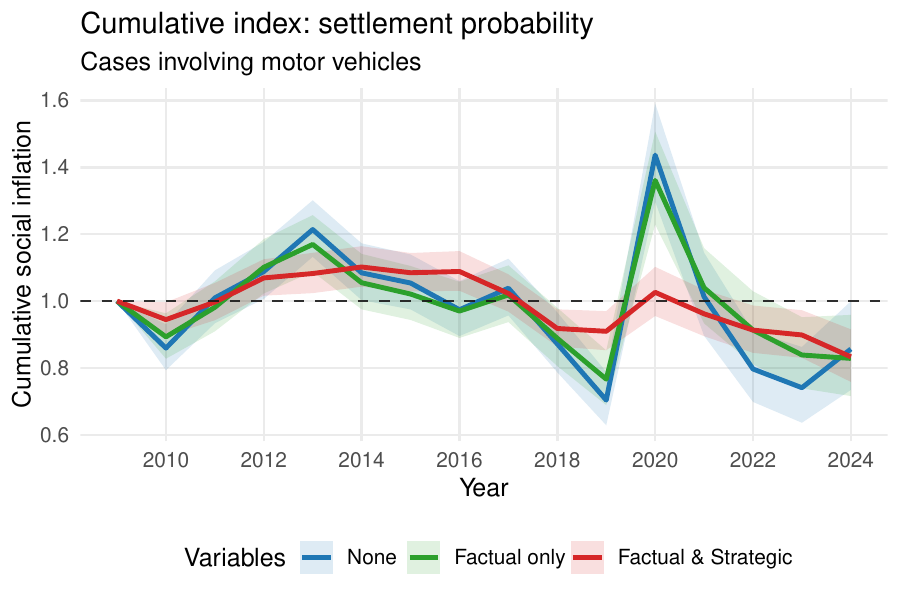}
  \end{subfigure}\hfill
    \begin{subfigure}[t]{0.48\textwidth}
    \centering
    \includegraphics[width=\linewidth]{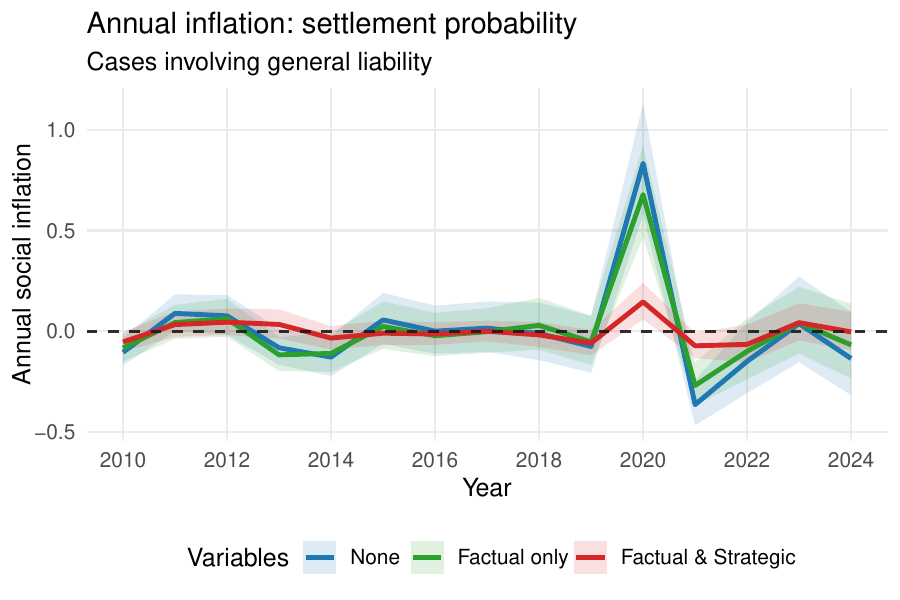}
  \end{subfigure}\hfill
  \begin{subfigure}[t]{0.48\textwidth}
    \centering
    \includegraphics[width=\linewidth]{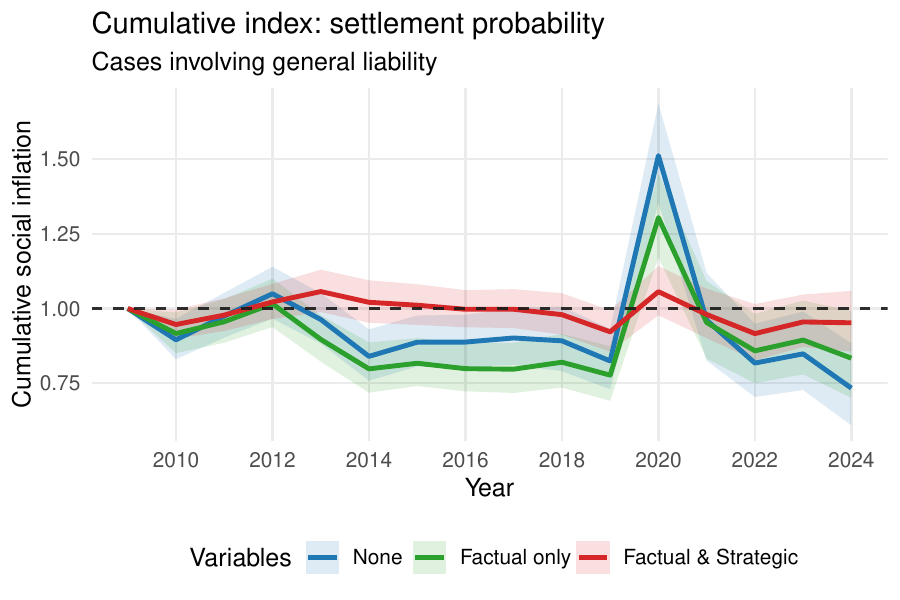}
  \end{subfigure}\hfill
  \begin{subfigure}[t]{0.48\textwidth}
    \centering
    \includegraphics[width=\linewidth]{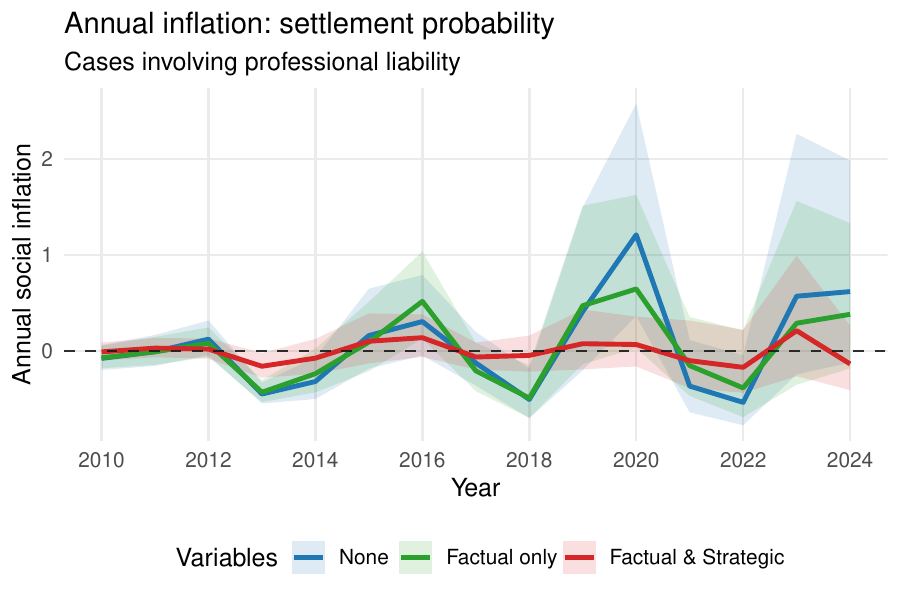}
  \end{subfigure}\hfill
  \begin{subfigure}[t]{0.48\textwidth}
    \centering
    \includegraphics[width=\linewidth]{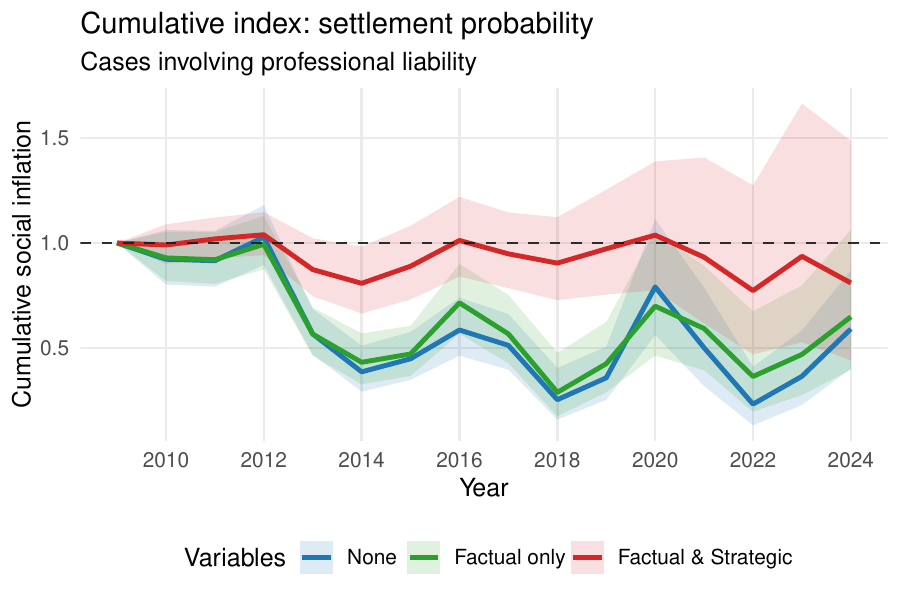}
  \end{subfigure}\hfill
  \begin{subfigure}[t]{0.48\textwidth}
    \centering
    \includegraphics[width=\linewidth]{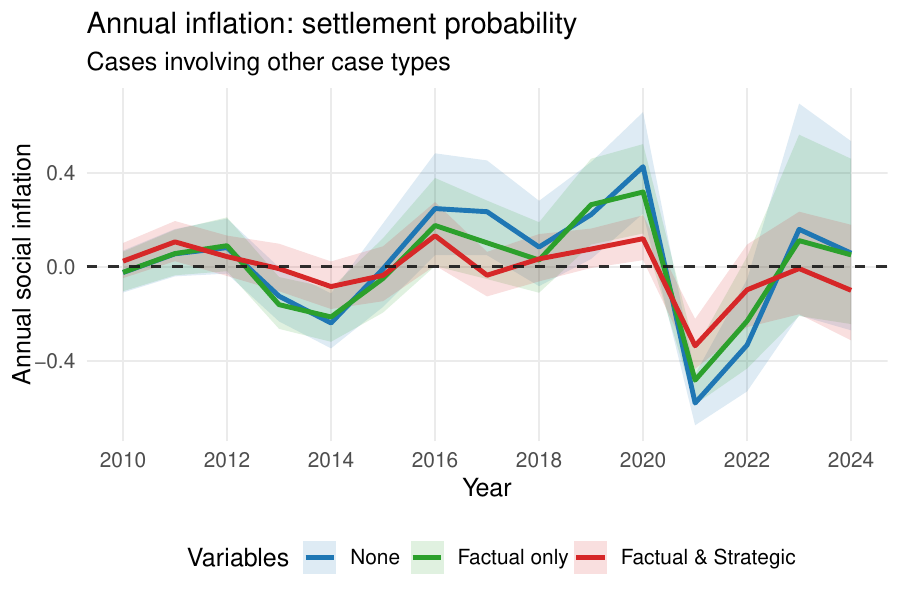}
  \end{subfigure}\hfill
  \begin{subfigure}[t]{0.48\textwidth}
    \centering
    \includegraphics[width=\linewidth]{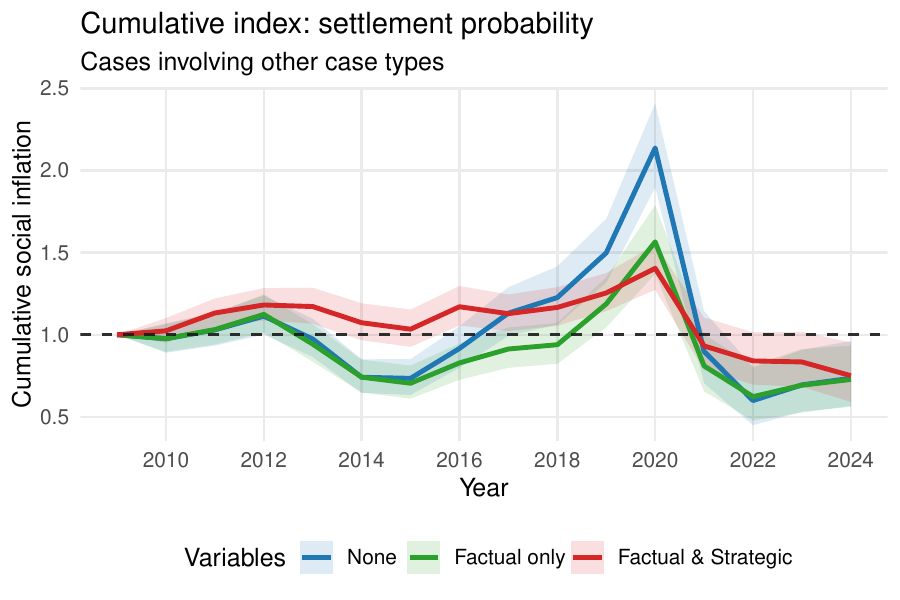}
  \end{subfigure}\hfill
  \caption{Annual (\textit{left panels}) and cumulative (\textit{right panels}) social inflation in settlement probability. \textit{First row}: motor liability cases; \textit{second row}: general liability cases; \textit{third row}: professional liability cases; \textit{fourth row}: other cases.}
  \label{fig:idx_prob_s_3}
\end{figure}

\begin{figure}[H]
  \centering
  \begin{subfigure}[t]{0.48\textwidth}
    \centering
    \includegraphics[width=\linewidth]{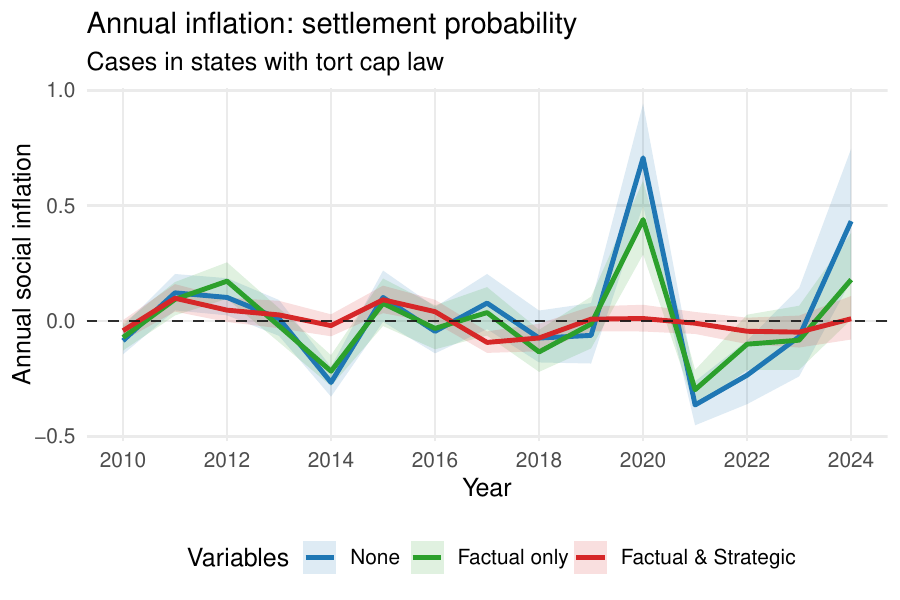}
  \end{subfigure}\hfill
  \begin{subfigure}[t]{0.48\textwidth}
    \centering
    \includegraphics[width=\linewidth]{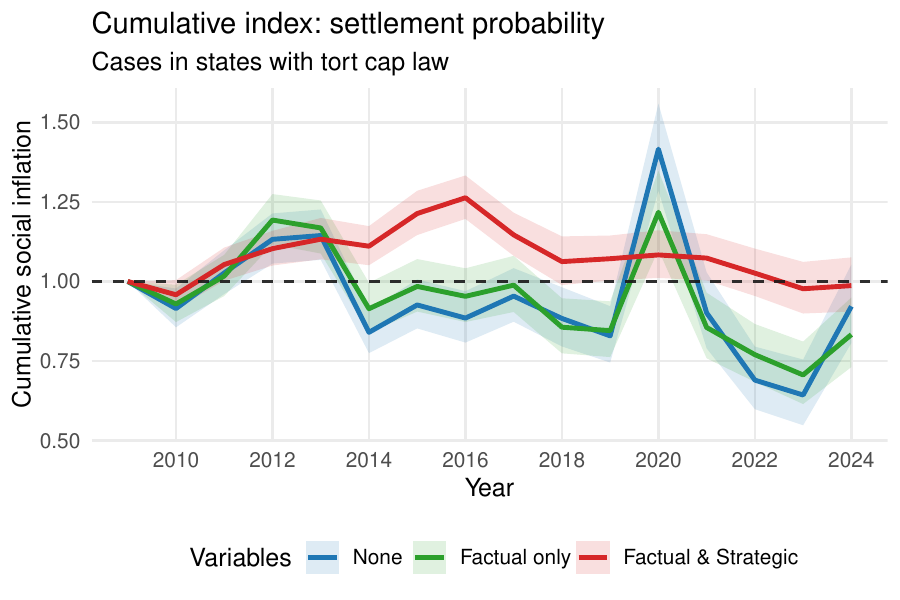}
  \end{subfigure}\hfill
    \begin{subfigure}[t]{0.48\textwidth}
    \centering
    \includegraphics[width=\linewidth]{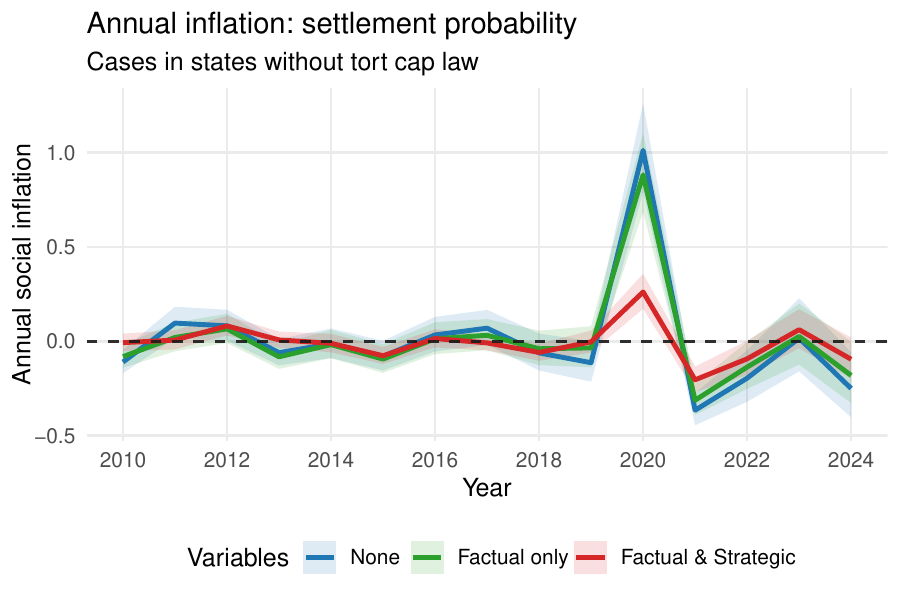}
  \end{subfigure}\hfill
  \begin{subfigure}[t]{0.48\textwidth}
    \centering
    \includegraphics[width=\linewidth]{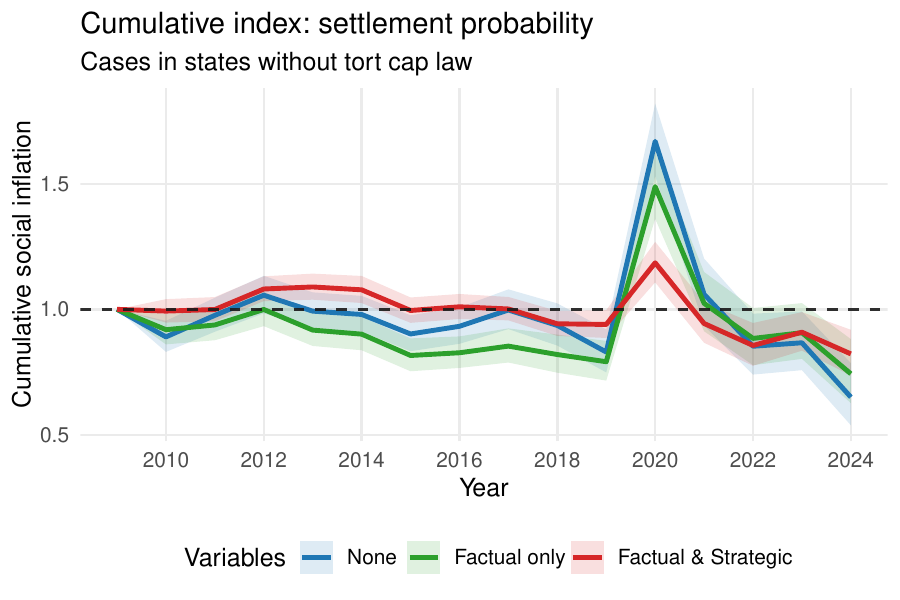}
  \end{subfigure}\hfill
  \caption{Annual (\textit{left panels}) and cumulative (\textit{right panels}) social inflation in settlement probability. \textit{Top panels}: cases in states with tort-cap laws; \textit{bottom panels}: cases in states without tort-cap laws.}
  \label{fig:idx_prob_s_4}
\end{figure}

\begin{figure}[H]
  \centering
  \begin{subfigure}[t]{0.48\textwidth}
    \centering
    \includegraphics[width=\linewidth]{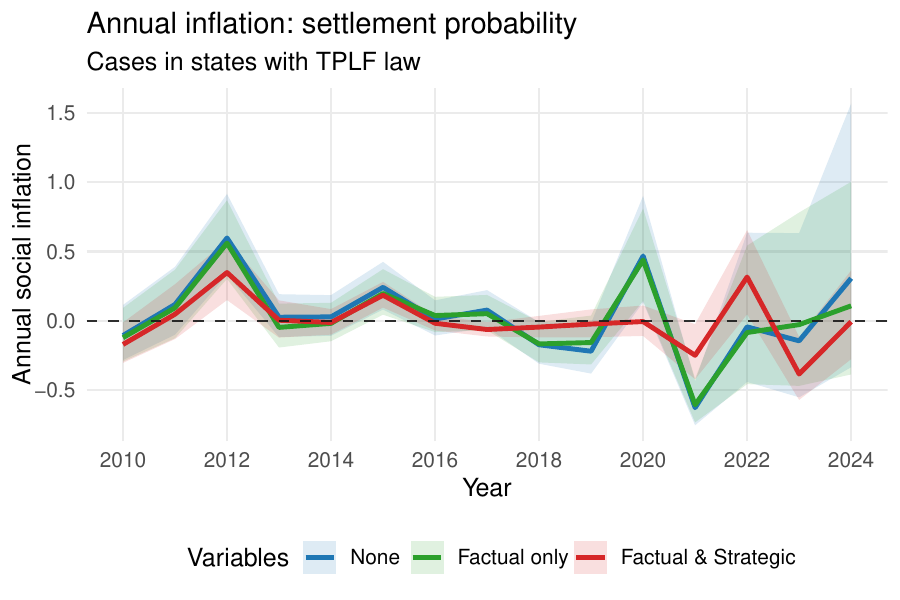}
  \end{subfigure}\hfill
  \begin{subfigure}[t]{0.48\textwidth}
    \centering
    \includegraphics[width=\linewidth]{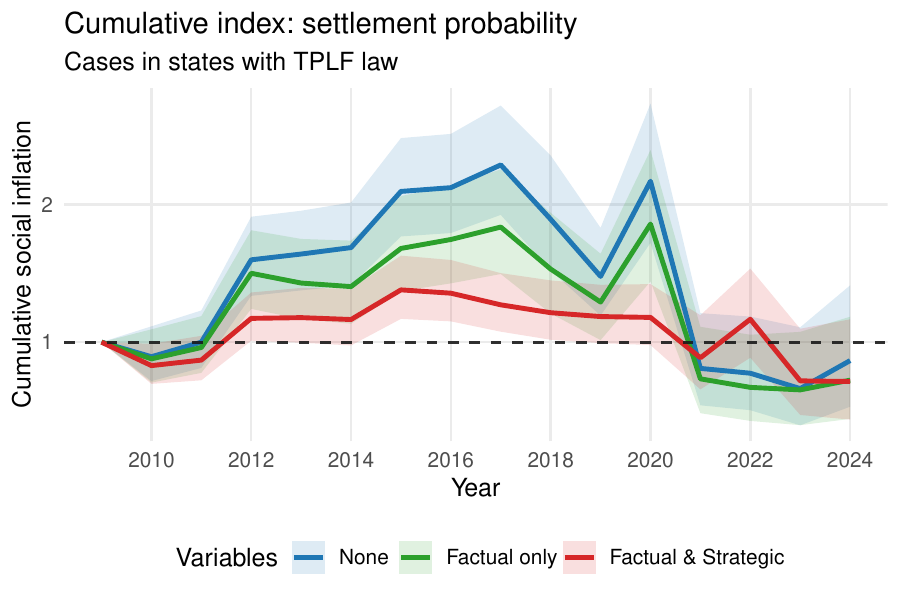}
  \end{subfigure}\hfill
    \begin{subfigure}[t]{0.48\textwidth}
    \centering
    \includegraphics[width=\linewidth]{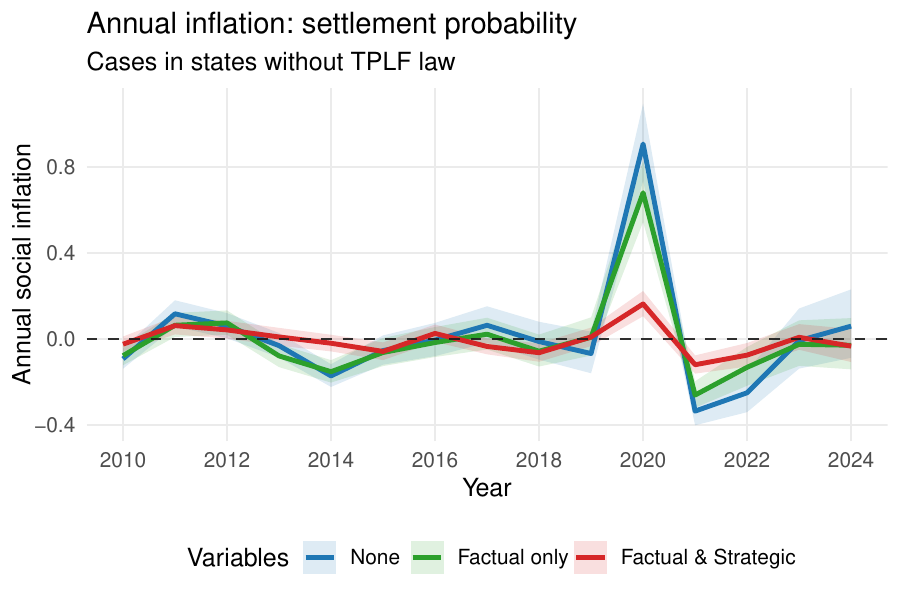}
  \end{subfigure}\hfill
  \begin{subfigure}[t]{0.48\textwidth}
    \centering
    \includegraphics[width=\linewidth]{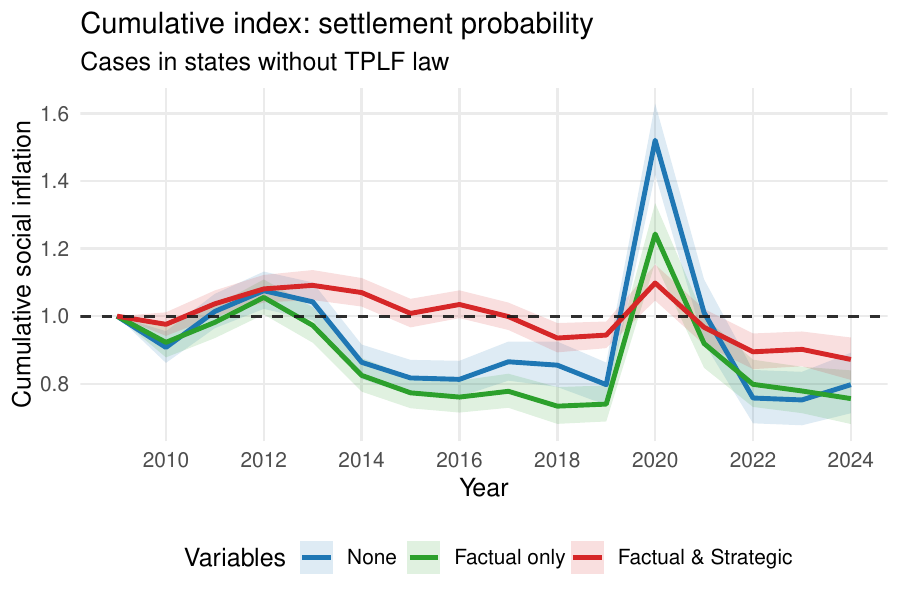}
  \end{subfigure}\hfill
  \caption{Annual (\textit{left panels}) and cumulative (\textit{right panels}) social inflation in settlement probability. \textit{Top panels}: cases in states with TPLF laws; \textit{bottom panels}: cases in states without TPLF laws.}
  \label{fig:idx_prob_s_5}
\end{figure}

This subsection reports social inflation results for settlement probability, based on the indices defined in Section \ref{sec:method:s_p} and summarized in Figures \ref{fig:idx_prob_s_0} to \ref{fig:idx_prob_s_5}. In contrast to plaintiff win probability (Section \ref{sec:result:p_p}), settlement probability exhibits a clear downward trend over time. Figure \ref{fig:idx_prob_s_0} shows that the cumulative settlement-probability index declines steadily from 2009 to 2024 under all specifications. After adjusting for factual covariates, the CSII is approximately 0.8 in 2024, which is significantly below the 2009 baseline, implying a highly significant relative decline of roughly 20\% in settlement probability over the sample period. This finding is consistent with the descriptive evidence presented in Section \ref{sec:data} and supports the interpretation that disputes have become, on net, less likely to resolve through settlement and more likely to proceed to verdict over time.

A notable exception occurs in 2020, when settlement probability spikes sharply, as reflected by a positive ASIR in that year and a visible deviation from the otherwise decreasing trajectory (Figure \ref{fig:idx_prob_s_0}). This pattern is plausibly attributable to COVID-19 disruptions: Court closures, suspension of jury trials, and the resulting backlog likely increased incentives for parties to settle, while courts may have prioritized only the most urgent or high-profile matters for limited trial capacity. Importantly, the regression-based adjustments indicate that a substantial portion of the 2020 spike can be explained by observed factual and strategic covariates (e.g., shifts in case composition and litigation intensity during the pandemic period), though not fully, suggesting that COVID-19 introduced additional unobserved institutional shocks affecting settlement behavior.

Figures \ref{fig:idx_prob_s_1} and \ref{fig:idx_prob_s_2} compare settlement-probability social inflation across defendant types and insurance statuses. Figure \ref{fig:idx_prob_s_1} illustrates that, without covariate controls, cases involving only individual defendants appear to experience a larger decline in settlement probability than cases involving corporate defendants. However, once factual and/or strategic covariates are incorporated, the difference between corporate-defendant and individual-defendant cases diminishes materially. This offers a central methodological message: Cross-group comparisons based on unadjusted indices can be misleading when the underlying case mix differs systematically across groups and evolves over time. By contrast, Figure \ref{fig:idx_prob_s_2} shows a more persistent divergence by insurance status. Cases involving only uninsured defendants exhibit a substantially stronger and statistically significant decline in settlement probability, with the CSII in 2024 significantly below 1 regardless of whether covariates are included. For cases involving insured defendants, the settlement-probability CSII is much closer to the baseline. This pattern is consistent with insurer involvement and incentives stabilizing settlement behavior over time for better risk management.

Figure \ref{fig:idx_prob_s_3} shows that the declining settlement-probability trend is broadly present across major liability lines (motor, general, professional, and other), with the 2020 spike appearing across categories as well. The strength and statistical significance of the decline varies by line, indicating that the settlement channel of social inflation is not perfectly uniform across business segments.

Finally, Figures \ref{fig:idx_prob_s_4} and \ref{fig:idx_prob_s_5} suggest that legal environments related to tort caps and TPLF regulation are associated with differential settlement-probability trends. Cases in states without tort-cap laws and/or without TPLF regulations display a more pronounced decline in settlement probability over time than cases in states with such laws. A plausible interpretation is that tort caps and TPLF regulation may reduce the expected upside of proceeding to trial, either by limiting recoverable damages or by restricting litigation-finance mechanisms that can support prolonged trial strategies, thereby weakening plaintiffs' incentives to forgo settlement in pursuit of large verdicts. 

Overall, while our analysis is not causal, we observe a statistically significant decline in settlement propensity since 2009, interrupted by a COVID-driven spike in 2020, with the downward trend most pronounced for uninsured-defendant cases and for jurisdictions without tort caps or TPLF regulation.

\subsection{Social inflation in verdict award amount} \label{sec:result:p_s}
\begin{figure}[H]
  \centering
  \begin{subfigure}[t]{0.48\textwidth}
    \centering
    \includegraphics[width=\linewidth]{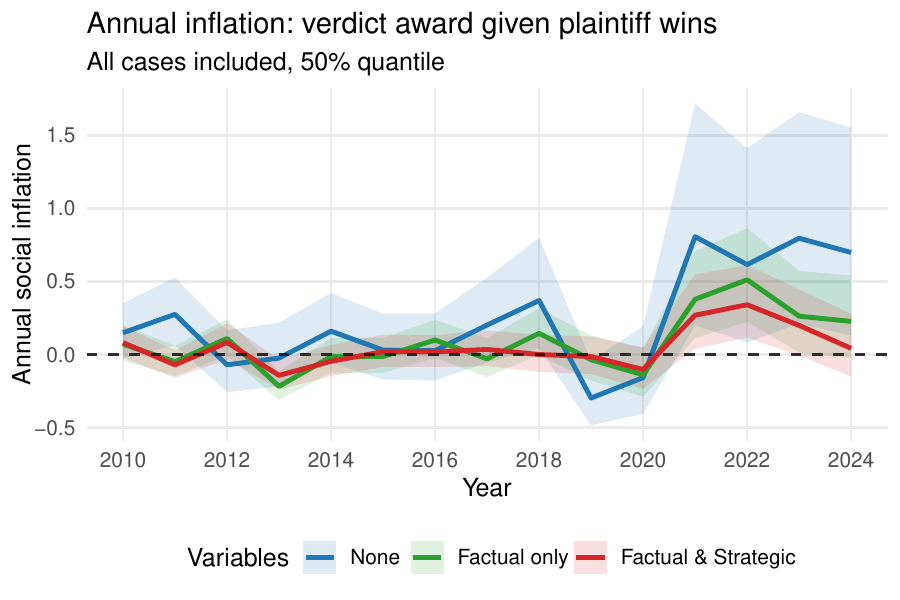}
  \end{subfigure}\hfill
  \begin{subfigure}[t]{0.48\textwidth}
    \centering
    \includegraphics[width=\linewidth]{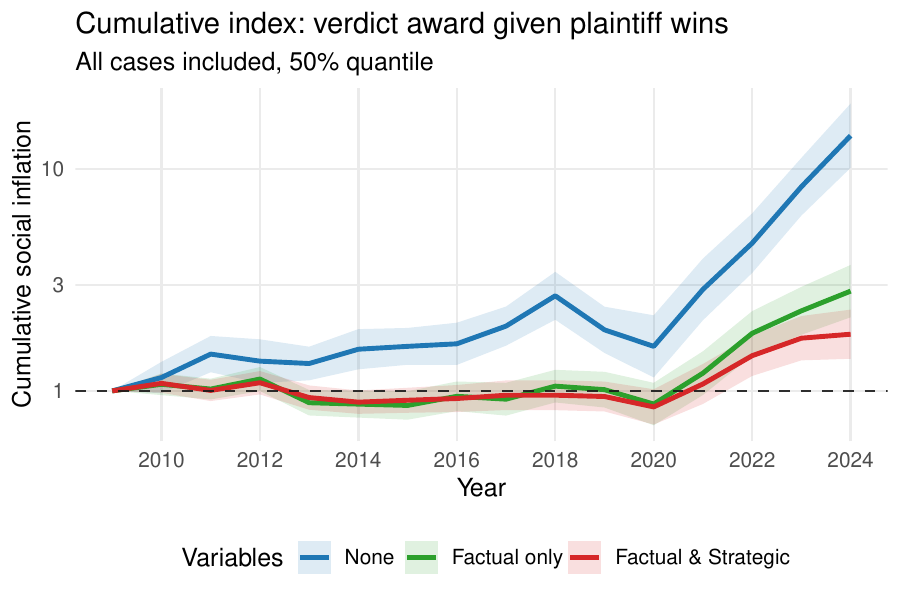}
  \end{subfigure}\hfill
  \caption{Annual (\textit{left panel}) and cumulative (\textit{right panel}) social inflation in verdict award amount (at 50\% quantile level) given plaintiff wins. All cases are included.}
  \label{fig:idx_sev_p1_0}
\end{figure}

\begin{figure}[H]
  \centering
  \begin{subfigure}[t]{0.48\textwidth}
    \centering
    \includegraphics[width=\linewidth]{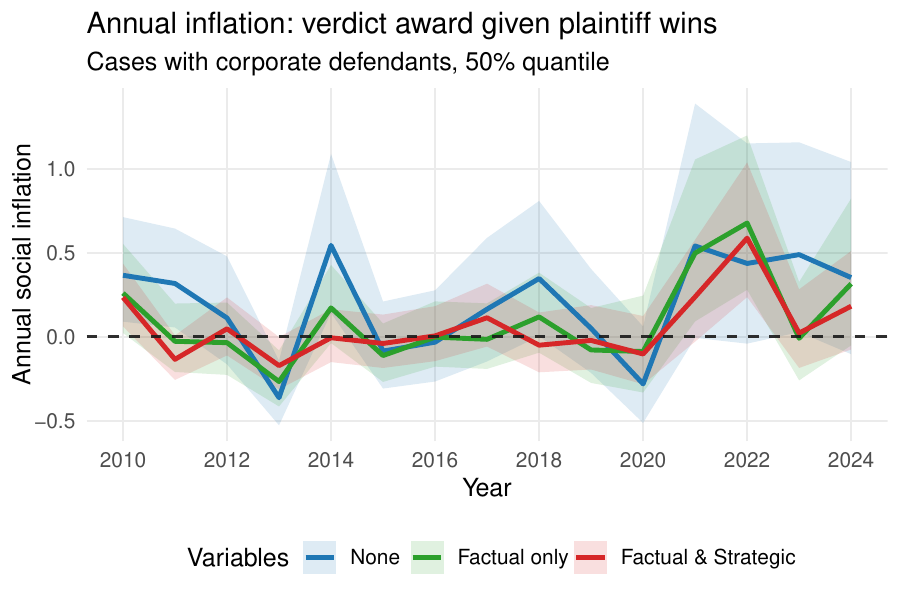}
  \end{subfigure}\hfill
  \begin{subfigure}[t]{0.48\textwidth}
    \centering
    \includegraphics[width=\linewidth]{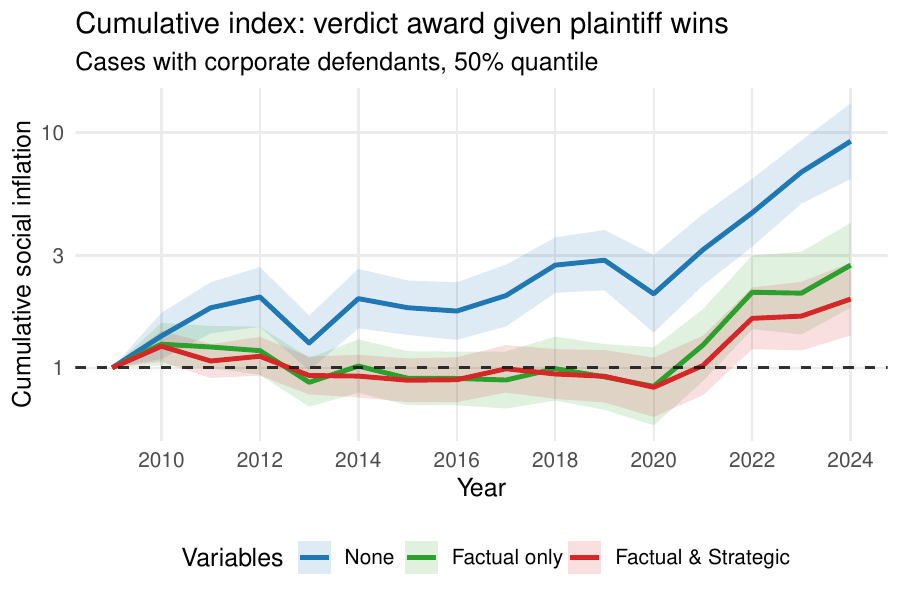}
  \end{subfigure}\hfill
    \begin{subfigure}[t]{0.48\textwidth}
    \centering
    \includegraphics[width=\linewidth]{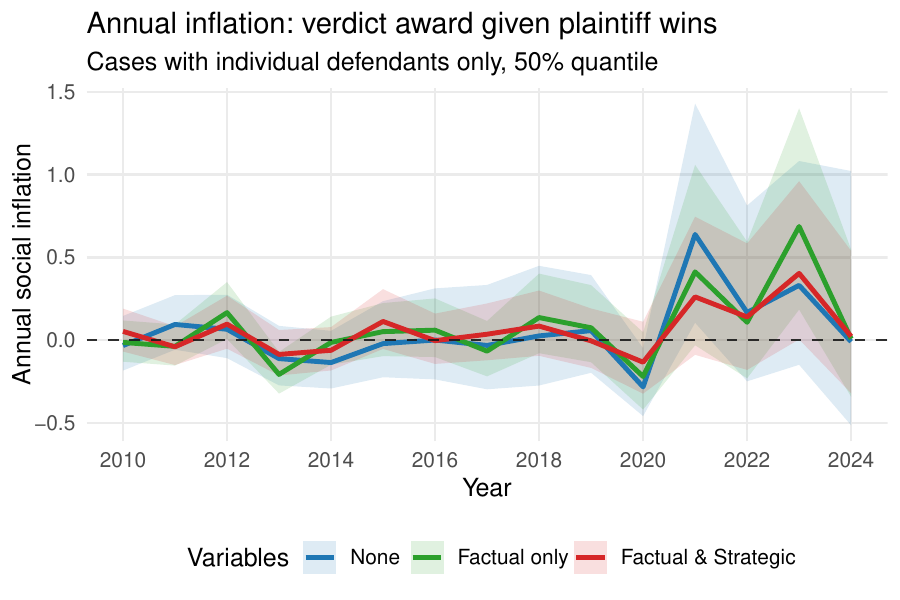}
  \end{subfigure}\hfill
  \begin{subfigure}[t]{0.48\textwidth}
    \centering
    \includegraphics[width=\linewidth]{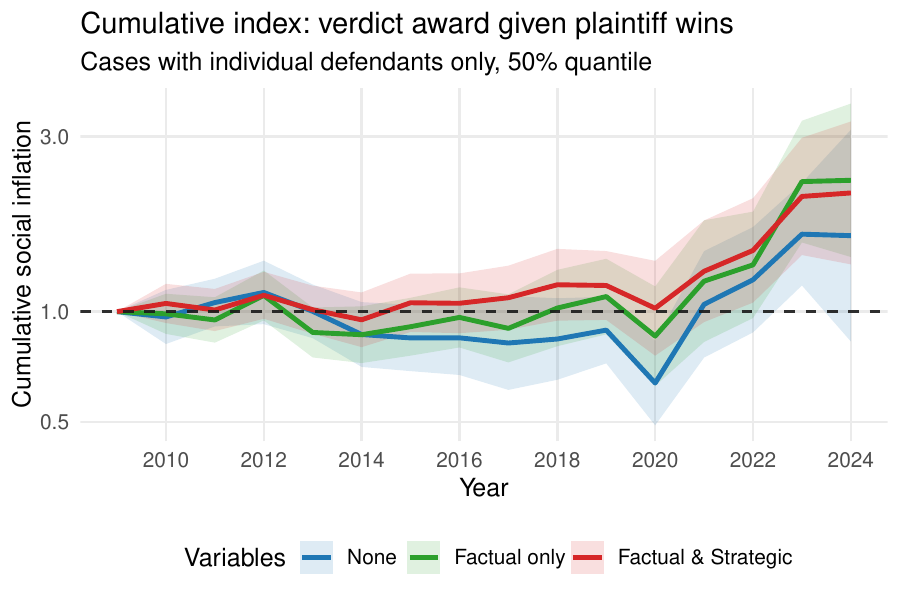}
  \end{subfigure}\hfill
  \caption{Annual (\textit{left panels}) and cumulative (\textit{right panels}) social inflation in verdict award amount (at 50\% quantile level) given plaintiff wins. \textit{Top panels}: cases with corporate defendants; \textit{bottom panels}: cases with individual defendants only.}
  \label{fig:idx_sev_p1_1}
\end{figure}

\begin{figure}[H]
  \centering
  \begin{subfigure}[t]{0.48\textwidth}
    \centering
    \includegraphics[width=\linewidth]{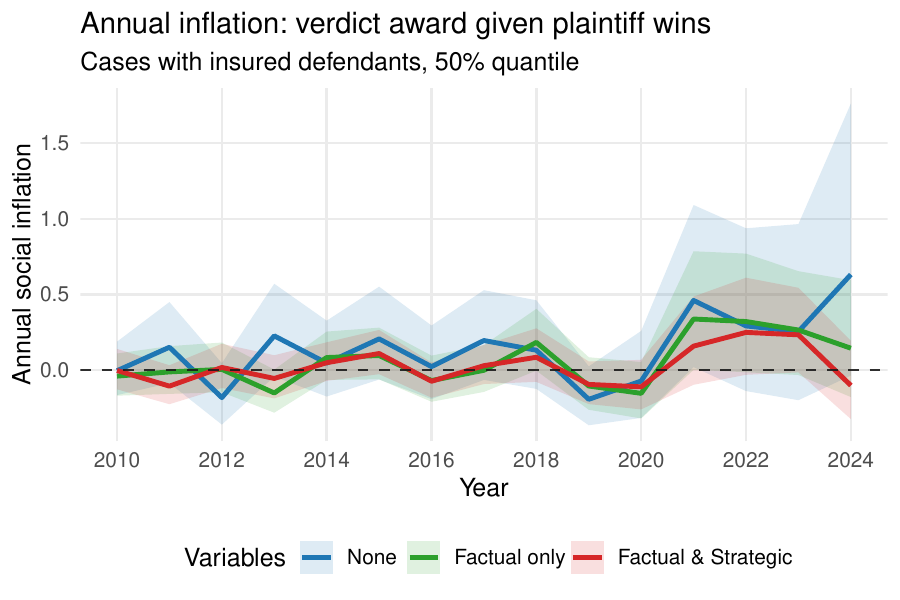}
  \end{subfigure}\hfill
  \begin{subfigure}[t]{0.48\textwidth}
    \centering
    \includegraphics[width=\linewidth]{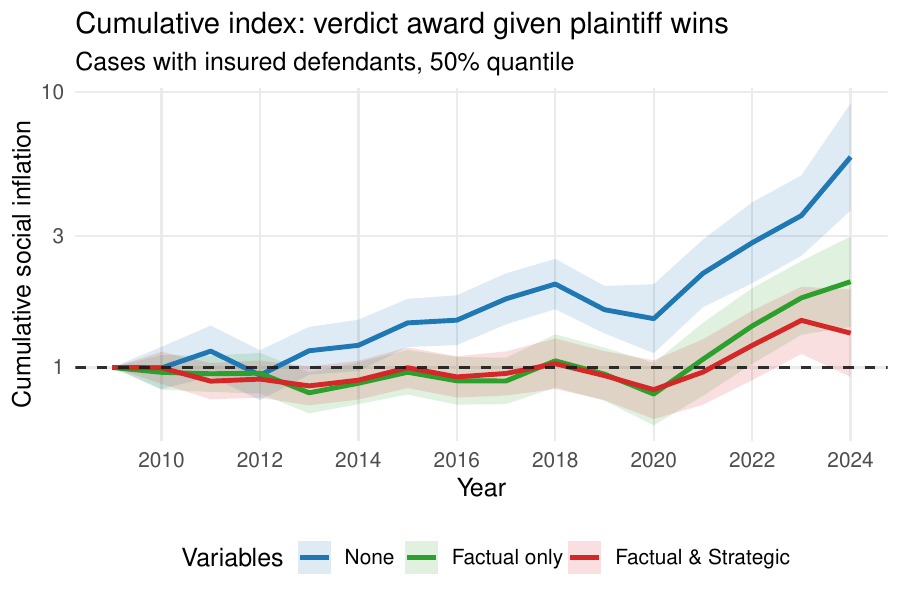}
  \end{subfigure}\hfill
    \begin{subfigure}[t]{0.48\textwidth}
    \centering
    \includegraphics[width=\linewidth]{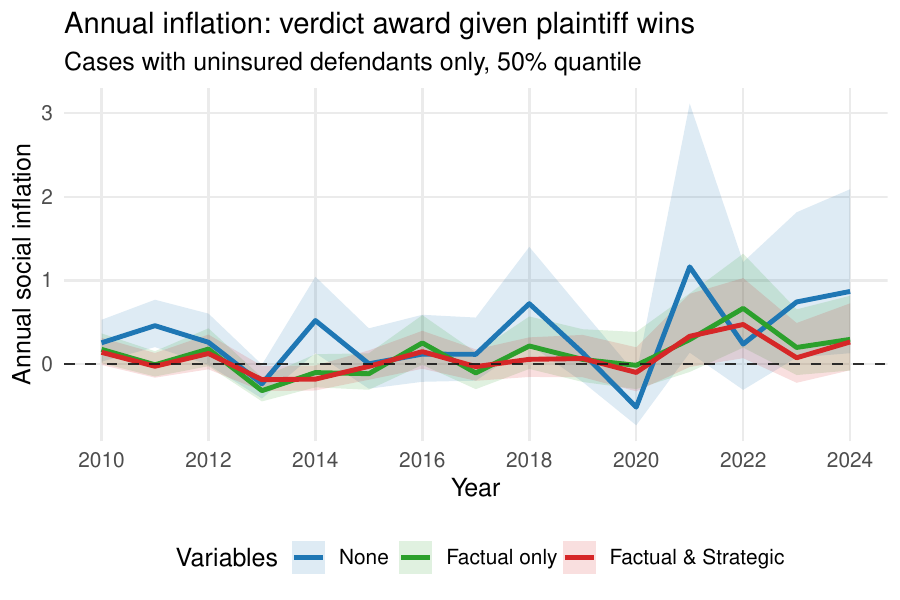}
  \end{subfigure}\hfill
  \begin{subfigure}[t]{0.48\textwidth}
    \centering
    \includegraphics[width=\linewidth]{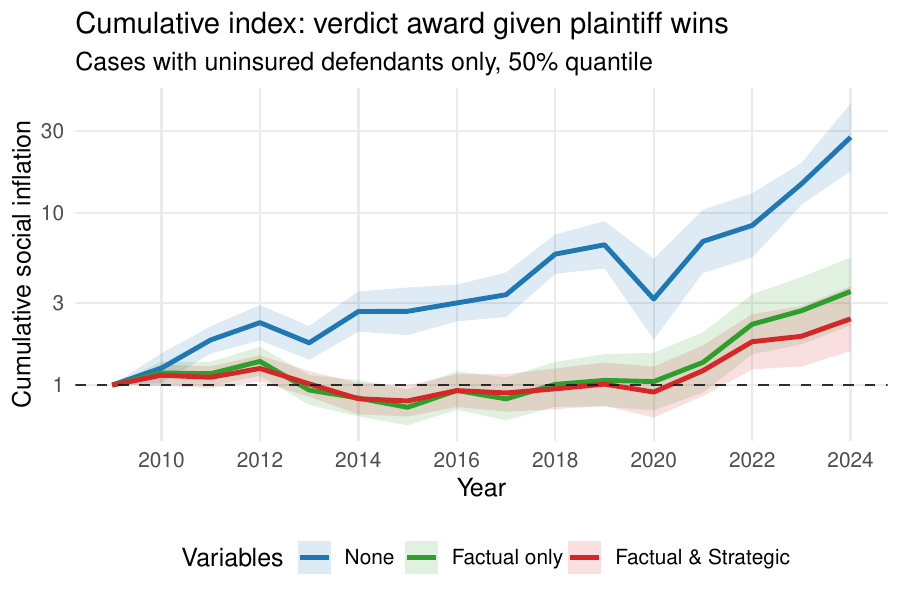}
  \end{subfigure}\hfill
  \caption{Annual (\textit{left panels}) and cumulative (\textit{right panels}) social inflation in verdict award amount (at 50\% quantile level) given plaintiff wins. \textit{Top panels}: cases with insured defendants; \textit{bottom panels}: cases with uninsured defendants only.}
  \label{fig:idx_sev_p1_2}
\end{figure}

\begin{figure}[H]
  \centering
  \begin{subfigure}[t]{0.48\textwidth}
    \centering
    \includegraphics[width=\linewidth]{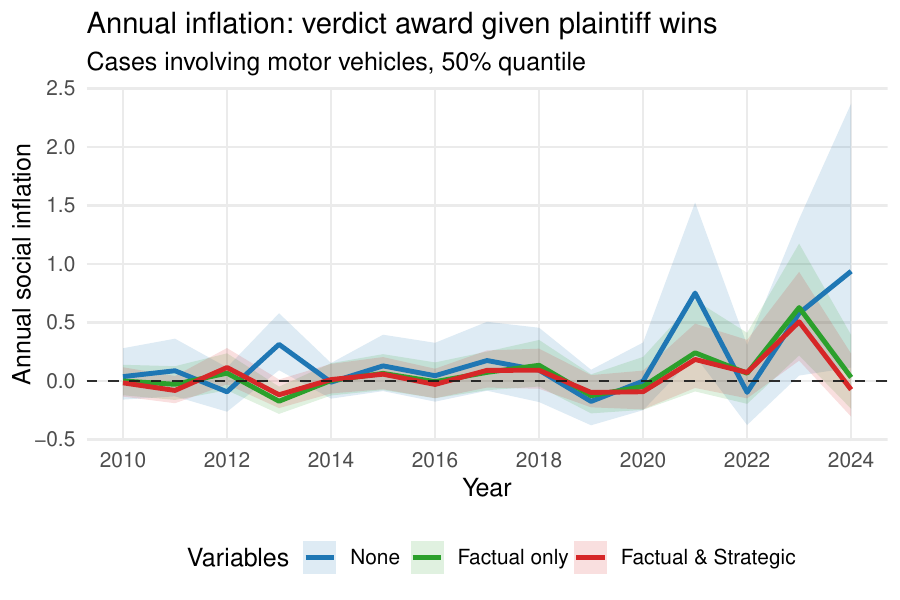}
  \end{subfigure}\hfill
  \begin{subfigure}[t]{0.48\textwidth}
    \centering
    \includegraphics[width=\linewidth]{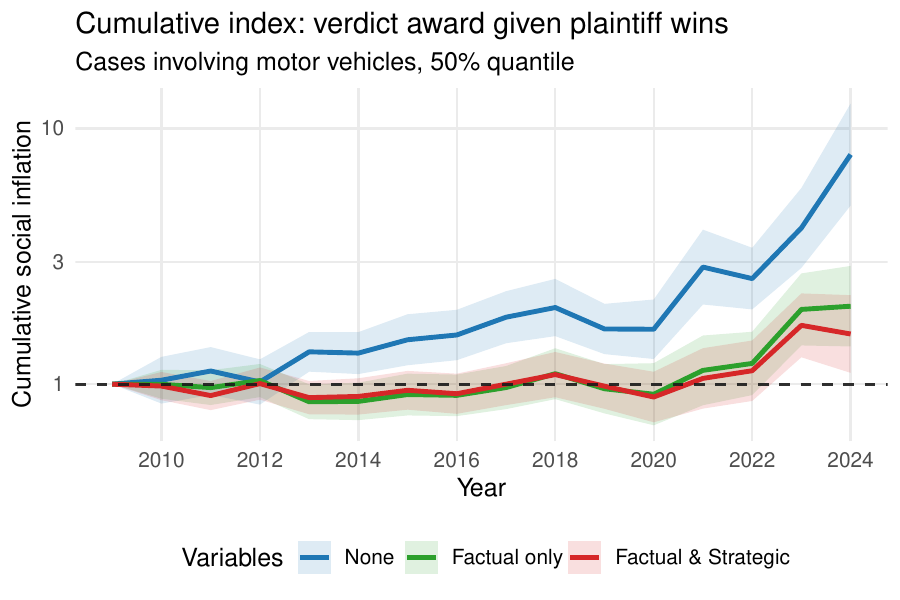}
  \end{subfigure}\hfill
    \begin{subfigure}[t]{0.48\textwidth}
    \centering
    \includegraphics[width=\linewidth]{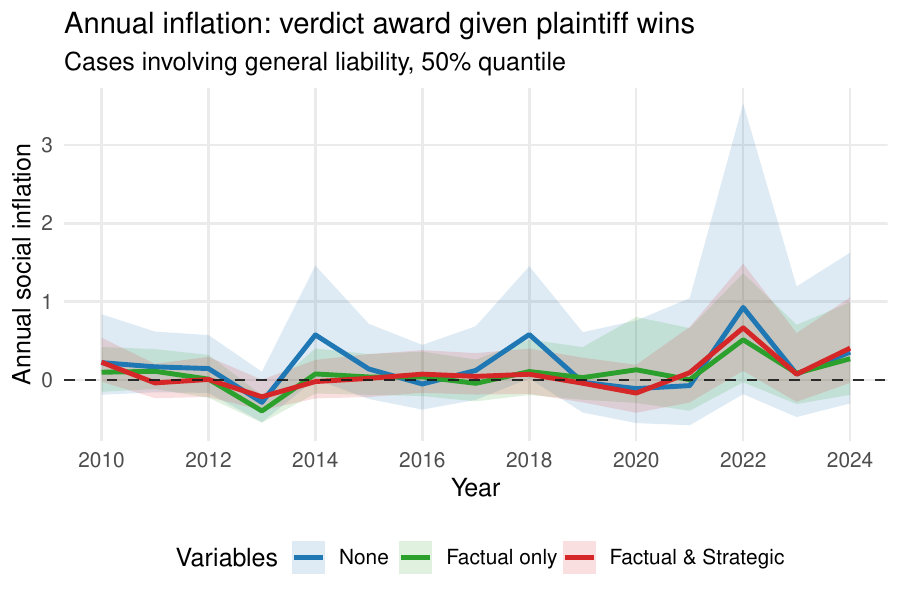}
  \end{subfigure}\hfill
  \begin{subfigure}[t]{0.48\textwidth}
    \centering
    \includegraphics[width=\linewidth]{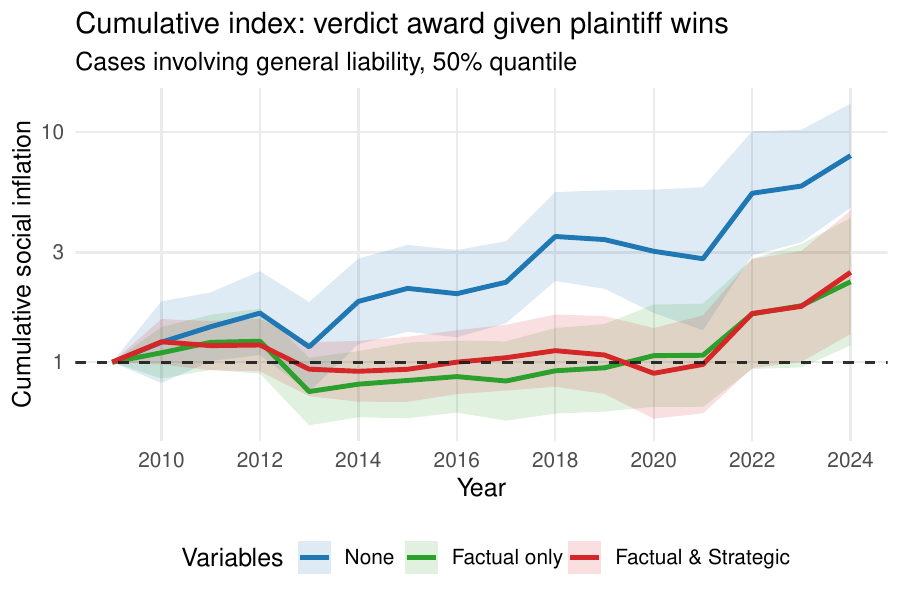}
  \end{subfigure}\hfill
    \begin{subfigure}[t]{0.48\textwidth}
    \centering
    \includegraphics[width=\linewidth]{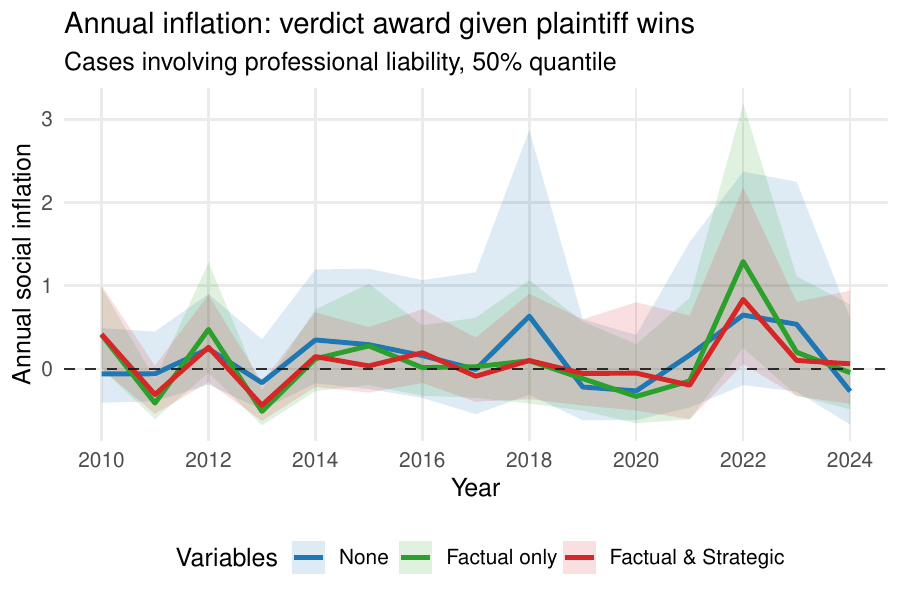}
  \end{subfigure}\hfill
  \begin{subfigure}[t]{0.48\textwidth}
    \centering
    \includegraphics[width=\linewidth]{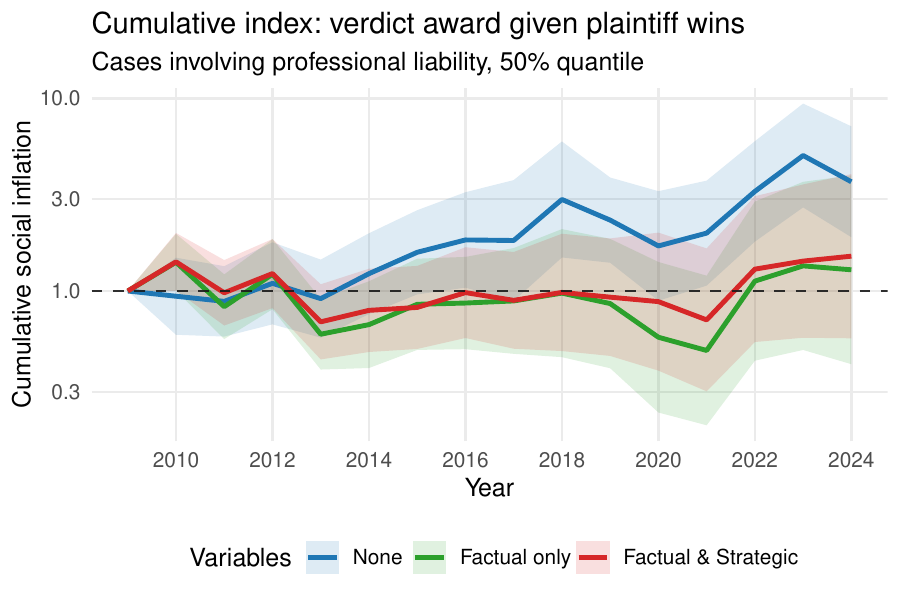}
  \end{subfigure}\hfill
    \begin{subfigure}[t]{0.48\textwidth}
    \centering
    \includegraphics[width=\linewidth]{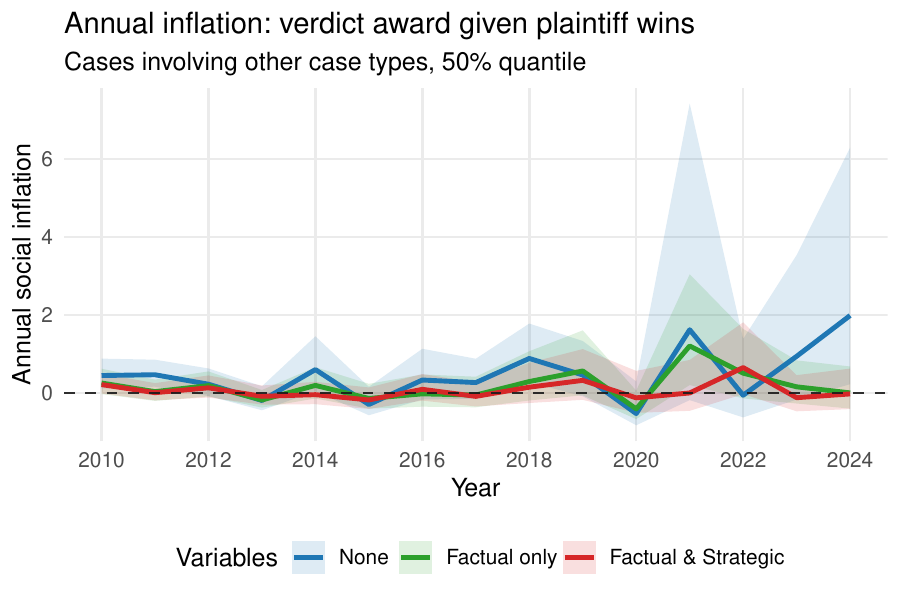}
  \end{subfigure}\hfill
  \begin{subfigure}[t]{0.48\textwidth}
    \centering
    \includegraphics[width=\linewidth]{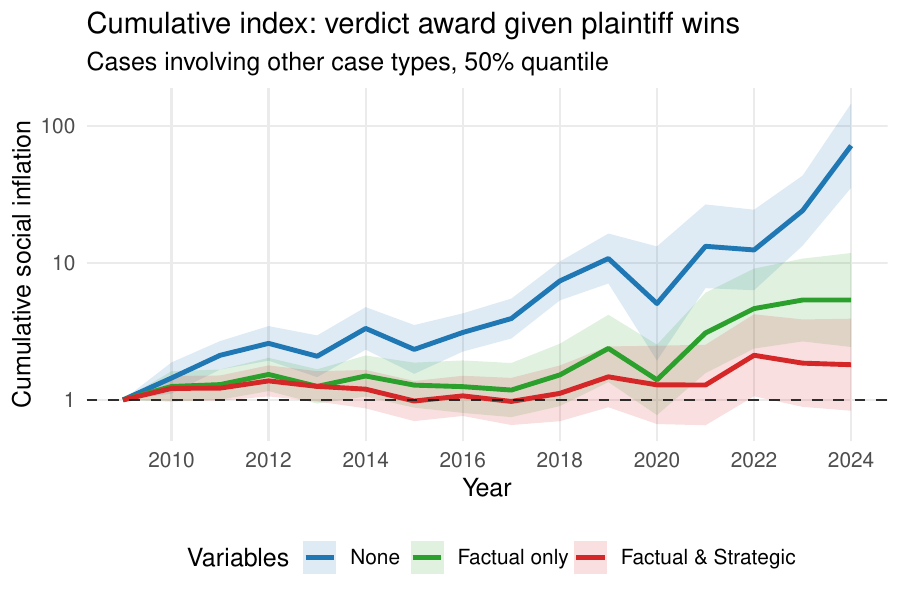}
  \end{subfigure}\hfill
  \caption{Annual (\textit{left panels}) and cumulative (\textit{right panels}) social inflation in verdict award amount (at 50\% quantile level) given plaintiff wins. \textit{First row}: motor liability cases; \textit{second row}: general liability cases; \textit{third row}: professional liability cases; \textit{fourth row}: other cases.}
  \label{fig:idx_sev_p1_3}
\end{figure}

\begin{figure}[H]
  \centering
  \begin{subfigure}[t]{0.48\textwidth}
    \centering
    \includegraphics[width=\linewidth]{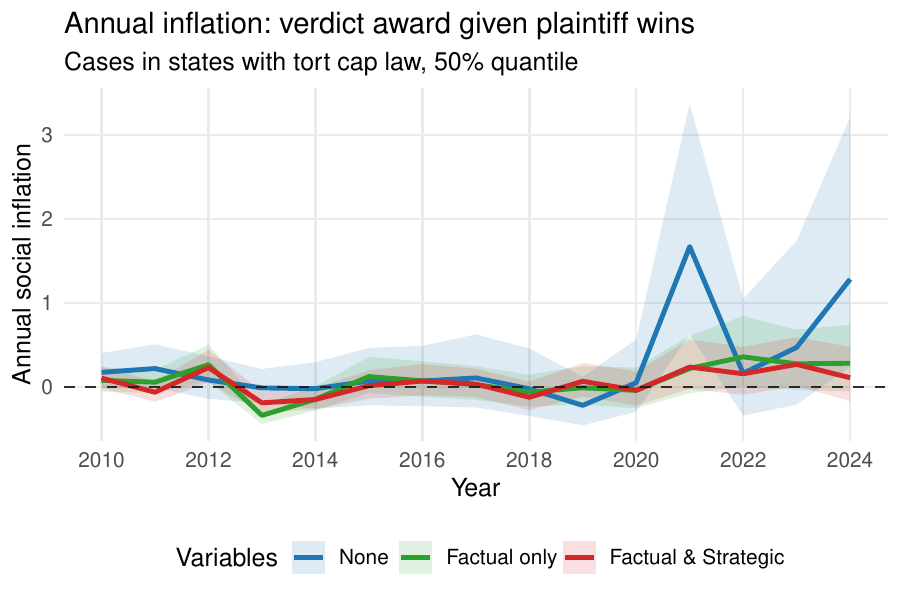}
  \end{subfigure}\hfill
  \begin{subfigure}[t]{0.48\textwidth}
    \centering
    \includegraphics[width=\linewidth]{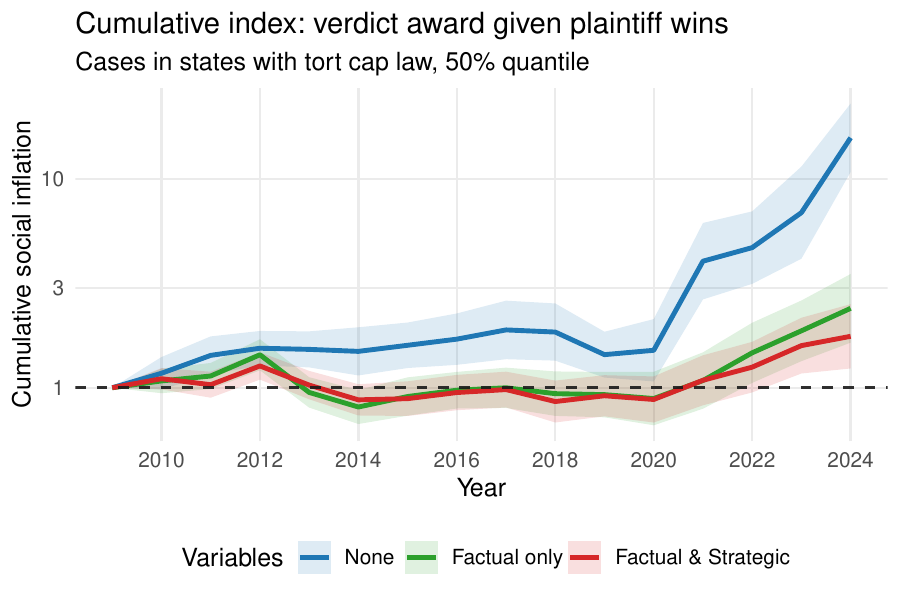}
  \end{subfigure}\hfill
    \begin{subfigure}[t]{0.48\textwidth}
    \centering
    \includegraphics[width=\linewidth]{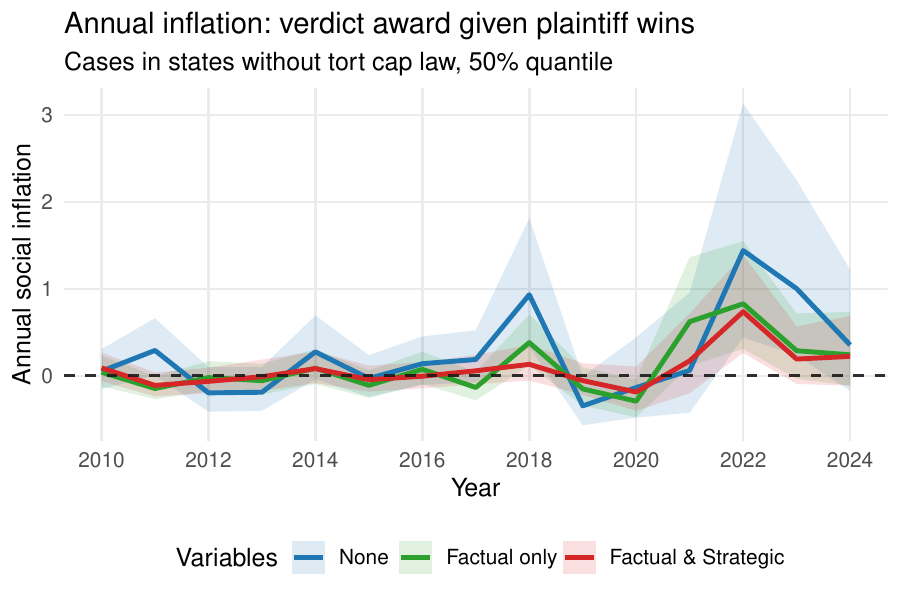}
  \end{subfigure}\hfill
  \begin{subfigure}[t]{0.48\textwidth}
    \centering
    \includegraphics[width=\linewidth]{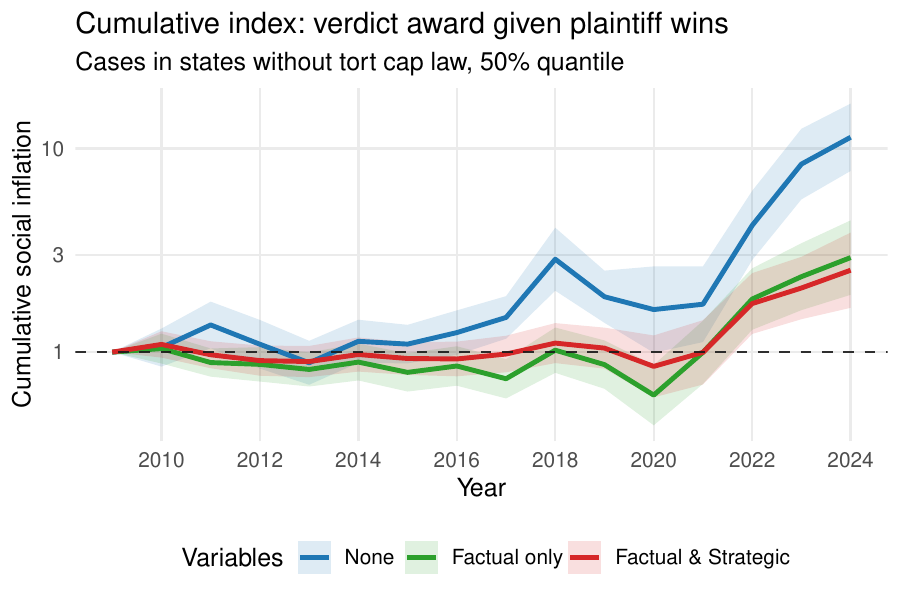}
  \end{subfigure}\hfill
  \caption{Annual (\textit{left panels}) and cumulative (\textit{right panels}) social inflation in verdict award amount (at 50\% quantile level) given plaintiff wins. \textit{Top panels}: cases in states with tort-cap laws; \textit{bottom panels}: cases in states without tort-cap laws.}
  \label{fig:idx_sev_p1_4}
\end{figure}

\begin{figure}[H]
  \centering
  \begin{subfigure}[t]{0.48\textwidth}
    \centering
    \includegraphics[width=\linewidth]{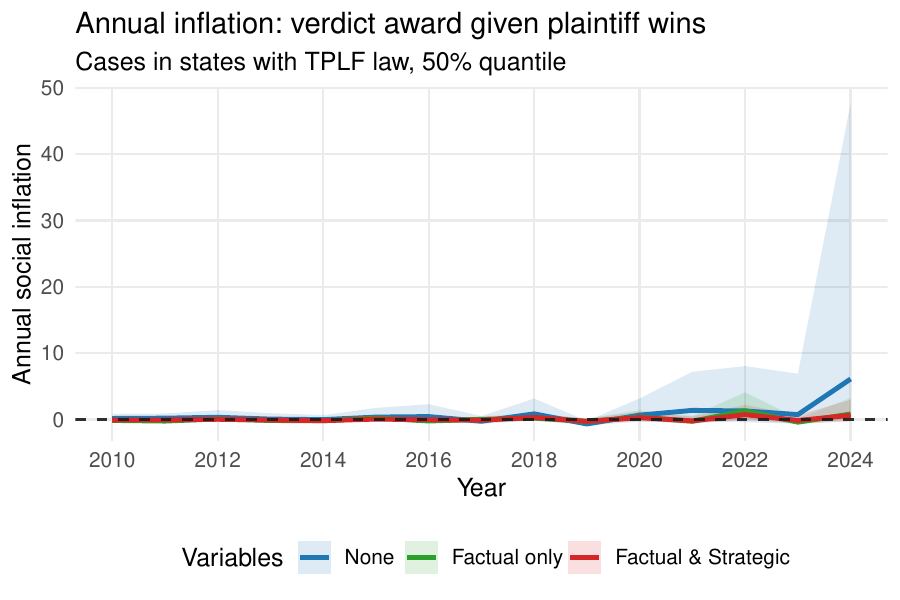}
  \end{subfigure}\hfill
  \begin{subfigure}[t]{0.48\textwidth}
    \centering
    \includegraphics[width=\linewidth]{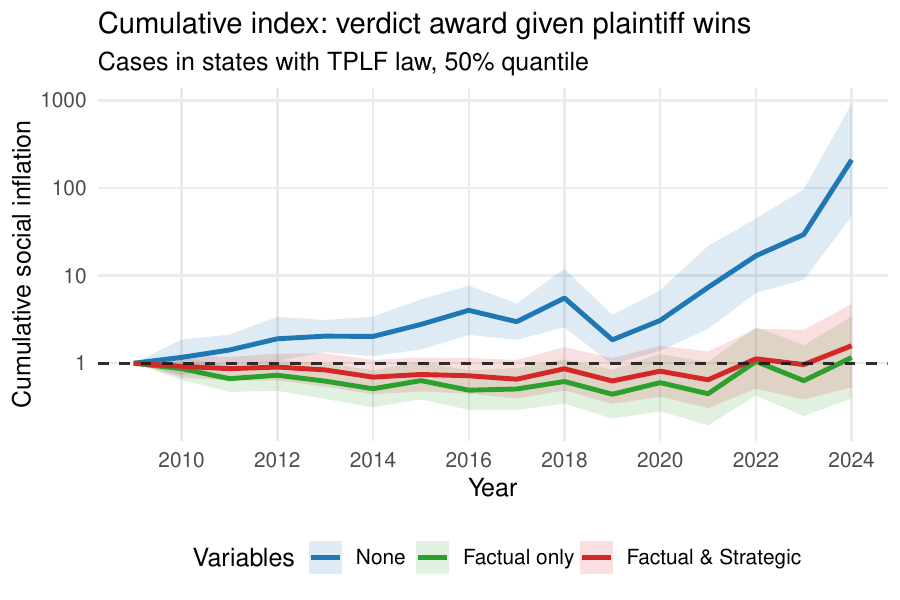}
  \end{subfigure}\hfill
    \begin{subfigure}[t]{0.48\textwidth}
    \centering
    \includegraphics[width=\linewidth]{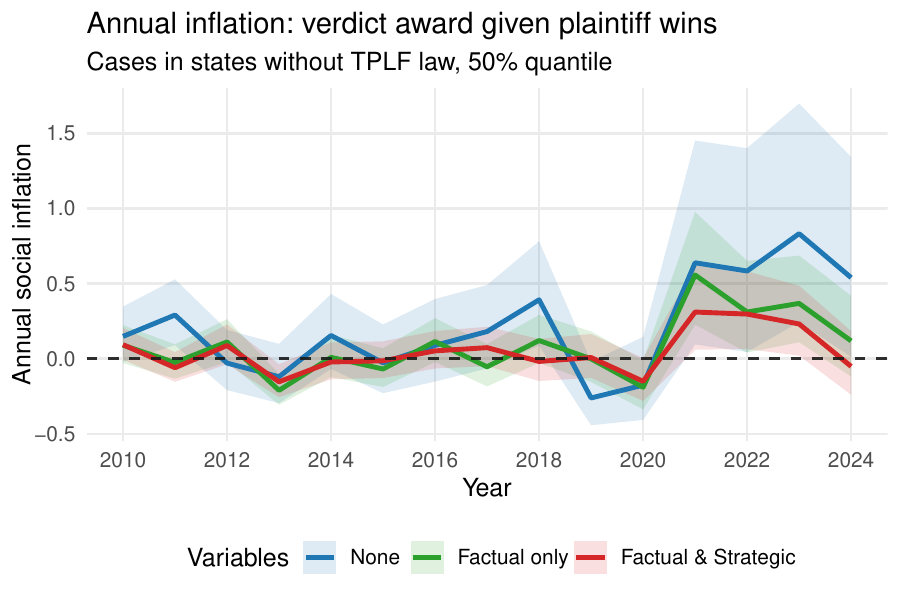}
  \end{subfigure}\hfill
  \begin{subfigure}[t]{0.48\textwidth}
    \centering
    \includegraphics[width=\linewidth]{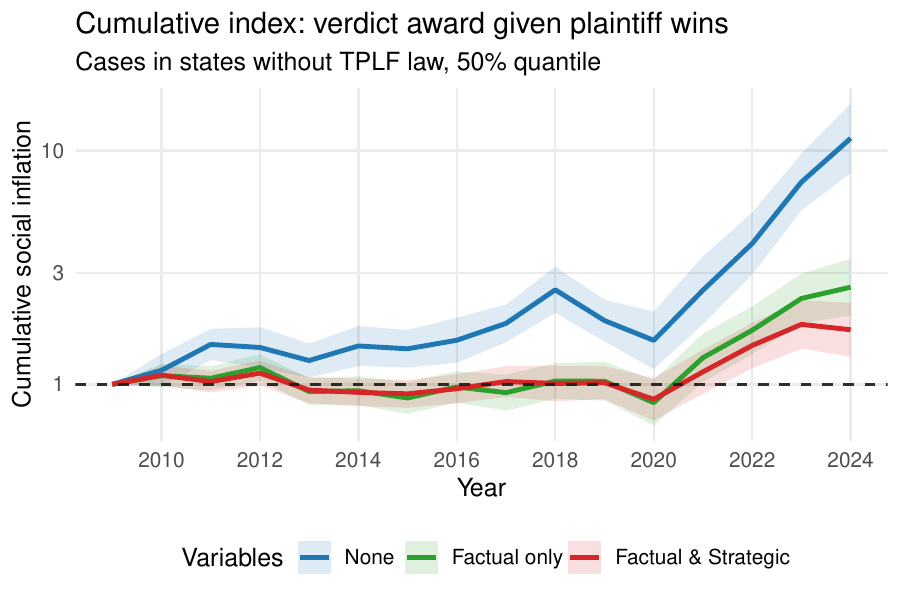}
  \end{subfigure}\hfill
  \caption{Annual (\textit{left panels}) and cumulative (\textit{right panels}) social inflation in verdict award amount (at 50\% quantile level) given plaintiff wins. \textit{Top panels}: cases in states with TPLF laws; \textit{bottom panels}: cases in states without TPLF  laws.}
  \label{fig:idx_sev_p1_5}
\end{figure}

\begin{figure}[H]
  \centering
  \begin{subfigure}[t]{0.48\textwidth}
    \centering
    \includegraphics[width=\linewidth]{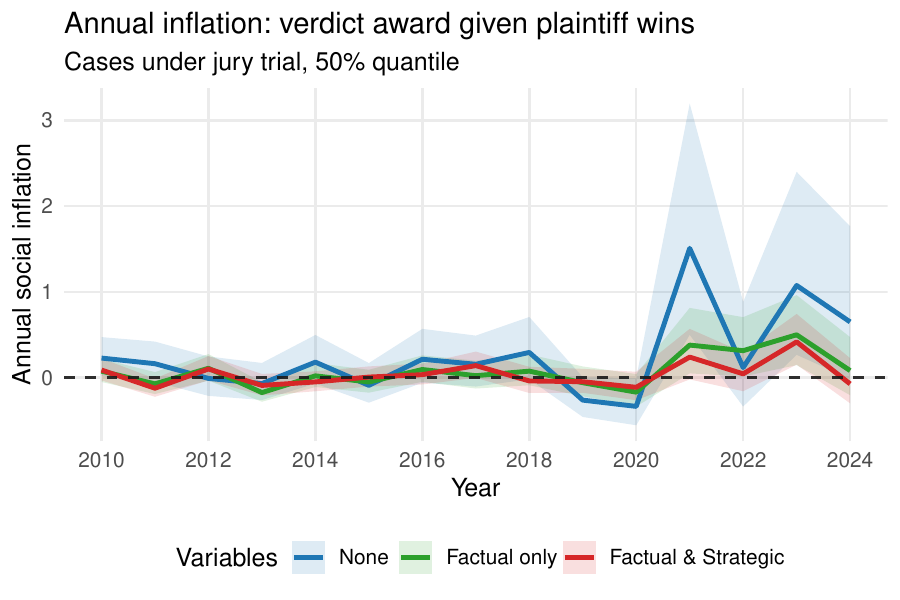}
  \end{subfigure}\hfill
  \begin{subfigure}[t]{0.48\textwidth}
    \centering
    \includegraphics[width=\linewidth]{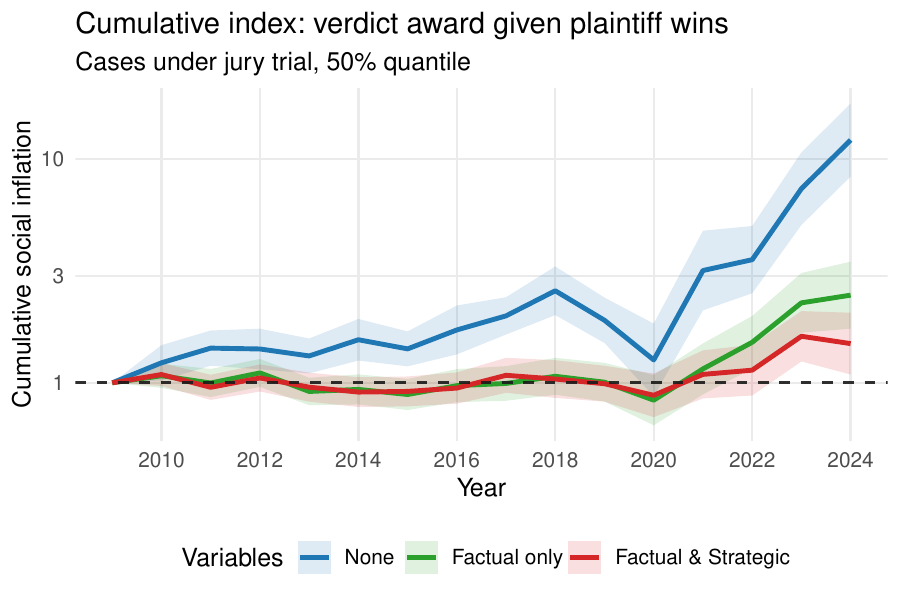}
  \end{subfigure}\hfill
    \begin{subfigure}[t]{0.48\textwidth}
    \centering
    \includegraphics[width=\linewidth]{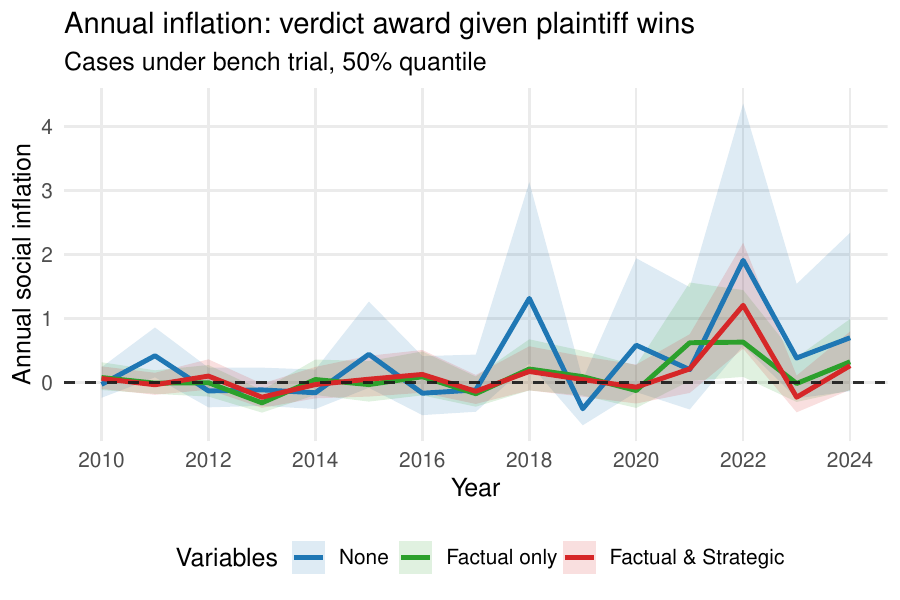}
  \end{subfigure}\hfill
  \begin{subfigure}[t]{0.48\textwidth}
    \centering
    \includegraphics[width=\linewidth]{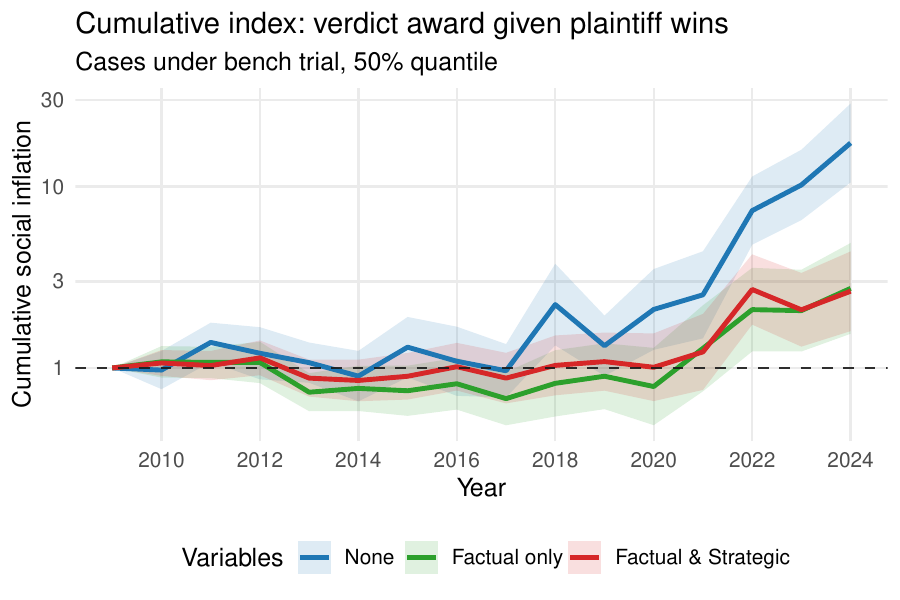}
  \end{subfigure}\hfill
  \caption{Annual (\textit{left panels}) and cumulative (\textit{right panels}) social inflation in verdict award amount (at 50\% quantile level) given plaintiff wins. \textit{Top panels}: cases under jury trial; \textit{bottom panels}: cases under bench trial.}
  \label{fig:idx_sev_p1_6}
\end{figure}

This subsection reports social inflation results for verdict award severity, using the quantile-based indices presented in Section \ref{sec:method:p_s} with $\tau=0.5$ (i.e., the conditional median verdict award among plaintiff-win verdicts). For robustness, we also repeat the same analysis at a higher quantile level ($\tau=0.9$) to probe the upper tail of verdict severity. The qualitative conclusions and interpretations are highly consistent with those reported for $\tau=0.5$, so we omit the $\tau=0.9$ plots to avoid redundancy. Figures \ref{fig:idx_sev_p1_0} to \ref{fig:idx_sev_p1_6} plot the corresponding ASIRs and CSIIs (with $\tau=0.5$) under three specifications: unadjusted (no covariates), factual-only adjustment, and factual-plus-strategic adjustment. Because verdict awards are strongly heavy-tailed and represent the most salient ``severity'' channel of social inflation for reinsurance layers, these results are central to the overall interpretation of social inflation in the dataset.

Figure \ref{fig:idx_sev_p1_0} provides the key baseline finding. Without incorporating any explanatory variables, the median verdict award CSII rises gradually from 2009 through 2020 to roughly the 2--3 range and then accelerates sharply after the COVID-19 pandemic, increasing from around 2 in 2020 to above 10 by 2024. This unadjusted pattern is consistent with the common industry narrative that social inflation intensified materially in the post-pandemic period. However, once we control for factual covariates, the pre-pandemic growth largely disappears: From 2009 to 2019, the factual-adjusted CSII is close to 1, implying that inflation-adjusted verdict severity exhibits little systematic increase prior to the pandemic, after case-mix adjustment. In contrast, the post-2020 period still shows prominent and statistically significant growth even after controlling for evolving case mix. By 2024, the factual-adjusted median verdict CSII reaches roughly 2 to 3, and the confidence intervals are well above 1, indicating robust evidence of post-pandemic social inflation in verdict severity beyond both economic inflation and inflation driven by observable shifts in case composition.

Comparing the three specifications in Figure \ref{fig:idx_sev_p1_0} further clarifies the roles of evolving case mix and litigation strategy. The green line (factual-only adjustment) lies below the blue line (unadjusted), indicating that part of the raw increase in verdict severity is mechanically attributable to composition shifts toward more severe and complex cases over time. The red line (factual plus strategic adjustment) lies below the green line, indicating that observed strategic factors, such as expanded attorney and expert involvement and longer trials, explain an additional portion of the case-mix-adjusted increase. Nonetheless, a substantial residual increase remains after controlling for both sets of covariates, especially after 2020, suggesting that broader shifts in the litigation environment (e.g., changing jury norms or other factors not directly captured by our strategic proxies) contribute materially to post-pandemic verdict severity inflation.

The magnitude of verdict-severity social inflation also dominates the frequency-type channel documented in Section \ref{sec:result:p_p}. Whereas the plaintiff-win CSIIs rise to roughly 1.2--1.3 by 2024 (depending on covariate controls), the verdict-award CSIIs in Figure \ref{fig:idx_sev_p1_0} range from about 2 to 3 (after full adjustment) to above 10 (without adjustment). This large contrast indicates that, in this dataset, social inflation operates far more strongly through severity than through plaintiff win probability. For reinsurance portfolios, this implies that changes in ground-up loss severity, and especially the tail behavior proxied by quantile-based measures, are likely to be the dominant driver of social inflation impacts on reinsurance losses.

Figures \ref{fig:idx_sev_p1_1} and \ref{fig:idx_sev_p1_2} examine heterogeneity across defendant types and insurance statuses. Figure \ref{fig:idx_sev_p1_1} shows that, without covariate controls, cases involving corporate defendants exhibit extremely large cumulative social inflation in median verdict awards (CSII around 10 by 2024), while cases involving only individual defendants show much smaller growth (CSII below 2). At face value, the unadjusted comparison would suggest that most verdict-severity social inflation is concentrated in corporate-defendant litigation. However, after controlling for factual and/or strategic variables, the gap narrows substantially and the patterns become more nuanced. For corporate-defendant cases, the adjusted indices (green/red) fall below the unadjusted index (blue), indicating that part of the extreme unadjusted growth reflects composition and strategy changes within the corporate-defendant subsample. In contrast, for individual-only cases, the adjusted indices (green/red) rise above the unadjusted index, implying that the unadjusted trend understates the case-mix-adjusted social inflation for that subsample. Overall, while corporate-defendant cases still exhibit higher adjusted CSIIs than individual-only cases, the results indicate that both groups contribute to verdict-severity social inflation once compositional differences are accounted for. This again reinforces the methodological point that naive stratified trend comparisons can misattribute social inflation to particular groups when group-specific case mixes evolve differently over time.

Figure \ref{fig:idx_sev_p1_2} shows a clearer and more robust pattern: Uninsured-defendant cases exhibit substantially higher social inflation in median verdict awards than insured-defendant cases, and this ordering persists across all specifications. This finding parallels and supports the evidence presented in Section \ref{sec:result:p_p} that uninsured-defendant cases experience stronger plaintiff-win social inflation as well.

Figure \ref{fig:idx_sev_p1_3} shows that positive verdict-severity social inflation is present across all major liability lines (motor, general, professional, and other), though the magnitude and statistical significance differ by line. Similar to the conclusion drawn in Section \ref{sec:result:p_p}, this heterogeneity is important for LoB pricing and for reinsurance portfolio construction: The severity component of social inflation is not uniform across liability segments.

Figures \ref{fig:idx_sev_p1_4} and \ref{fig:idx_sev_p1_5} evaluate whether statutory environments related to tort caps and TPLF regulation are associated with differential verdict-severity social inflation. Consistent with the plaintiff-win results in Section \ref{sec:result:p_p}, cases in states without tort caps and/or without TPLF regulation exhibit higher adjusted CSIIs than cases in states with such legal constraints. These patterns are directionally consistent with the notion that state regulations may limit the upside from trial and reduce incentives to pursue exceptionally large awards. Figure \ref{fig:idx_sev_p1_5} also highlights an important practical benefit of the proposed regression-based approach. Without covariate controls, the CSII for cases in TPLF-regulated states becomes extremely large (exceeding 100 by 2024) and is accompanied by very wide confidence intervals, indicating instability driven by a small number of extreme observations and/or a thin sample in recent years. After controlling for factual and strategic variables, the corresponding indices become far more stable and interpretable, with substantially narrower uncertainty bands. This illustrates that case-mix adjustment is not only conceptually necessary for isolating social inflation, but also empirically critical for producing reliable severity indices.

Finally, Figure \ref{fig:idx_sev_p1_6} compares jury trials versus bench trials. As in Section \ref{sec:result:p_p}, we do not observe meaningful differences in the evolution of the adjusted verdict-severity indices between jury and bench trial cases. While jury behavior is often highlighted in discussions of nuclear verdicts, the data suggest that the post-2009 upward pressure in median verdict awards is not exclusively a jury-trial phenomenon.

In summary, verdict award severity exhibits the largest and most consequential social inflation signal in the VerdictSearch data, particularly after 2020. Much of the apparent pre-pandemic growth in unadjusted indices can be explained by evolving case composition, but a sizable, statistically significant post-pandemic increase remains after controlling for both factual and strategic covariates. The results also reveal meaningful heterogeneity across defendant types, insurance statuses, liability lines, and legal environments and highlight that regression-based case-mix adjustment is essential for both accurate attribution and statistical stability when quantifying verdict-severity social inflation.

\subsection{Social inflation in settlement amount} \label{sec:result:s_s}
\begin{figure}[H]
  \centering
  \begin{subfigure}[t]{0.48\textwidth}
    \centering
    \includegraphics[width=\linewidth]{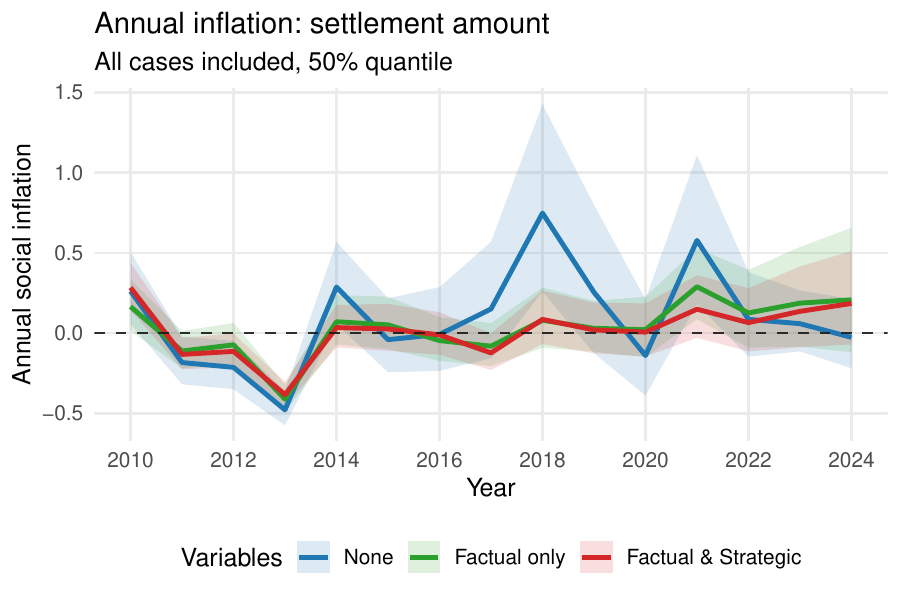}
  \end{subfigure}\hfill
  \begin{subfigure}[t]{0.48\textwidth}
    \centering
    \includegraphics[width=\linewidth]{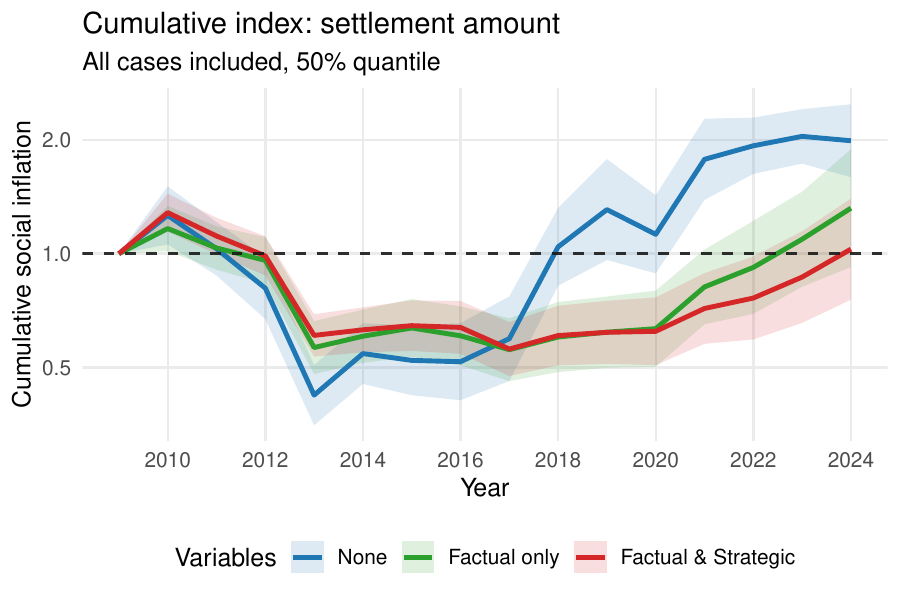}
  \end{subfigure}\hfill
  \caption{Annual (\textit{left panels}) and cumulative (\textit{right panels}) social inflation in settlement amount (at 50\% quantile level). All cases are included.}
  \label{fig:idx_sev_s1_0}
\end{figure}

\begin{figure}[H]
  \centering
  \begin{subfigure}[t]{0.48\textwidth}
    \centering
    \includegraphics[width=\linewidth]{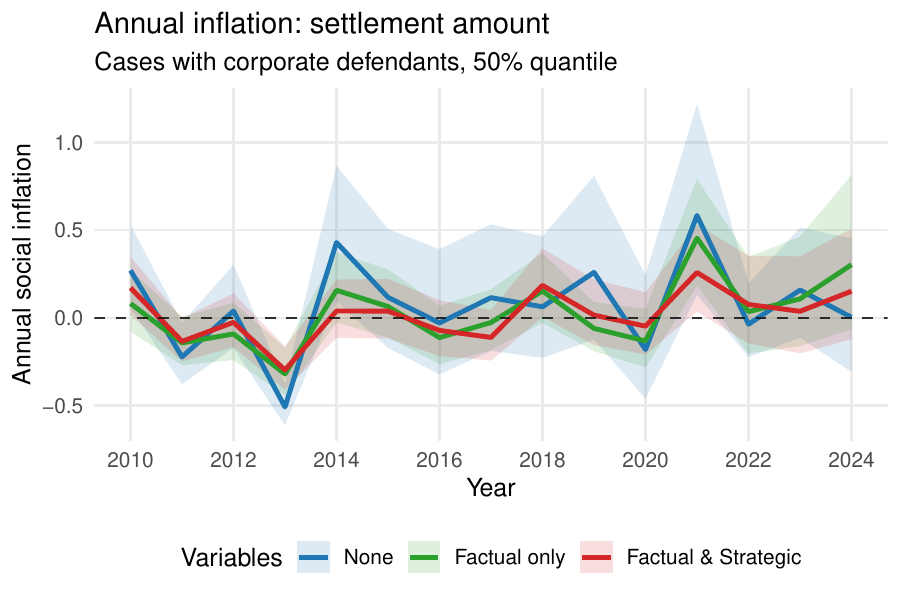}
  \end{subfigure}\hfill
  \begin{subfigure}[t]{0.48\textwidth}
    \centering
    \includegraphics[width=\linewidth]{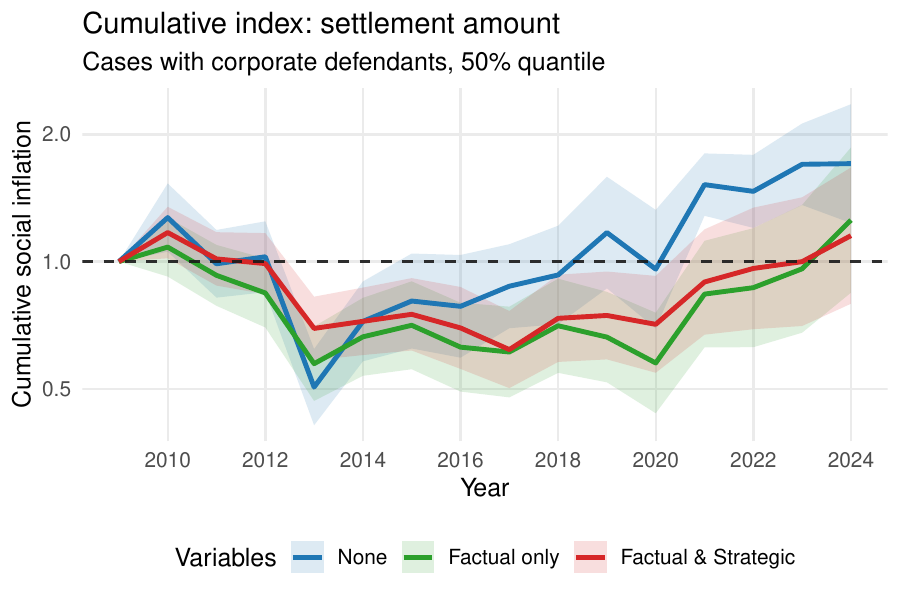}
  \end{subfigure}\hfill
    \begin{subfigure}[t]{0.48\textwidth}
    \centering
    \includegraphics[width=\linewidth]{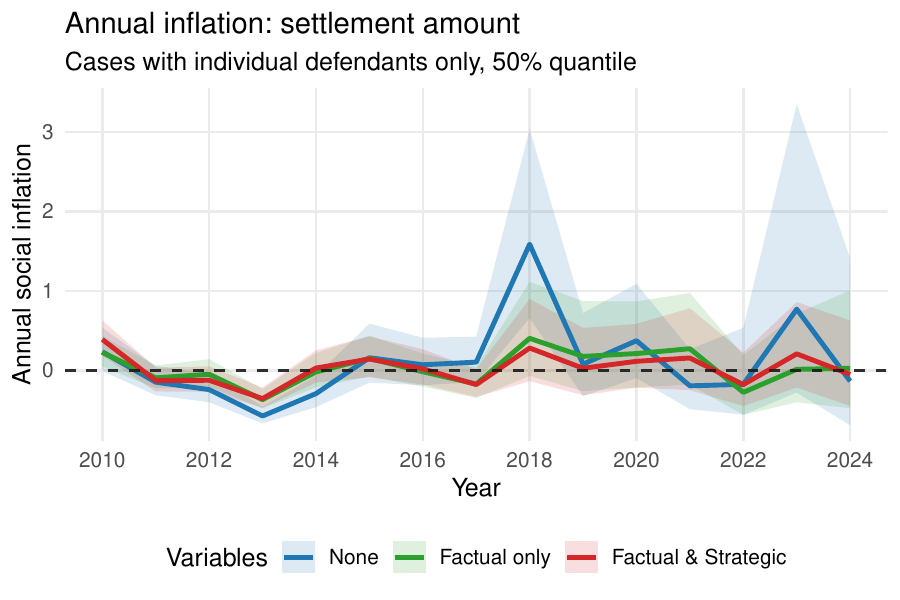}
  \end{subfigure}\hfill
  \begin{subfigure}[t]{0.48\textwidth}
    \centering
    \includegraphics[width=\linewidth]{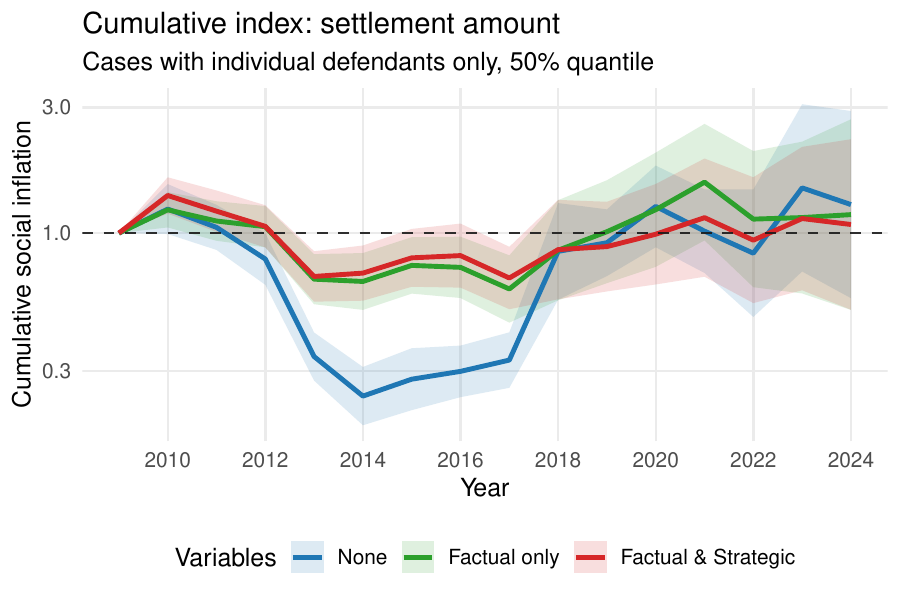}
  \end{subfigure}\hfill
  \caption{Annual (\textit{left panels}) and cumulative (\textit{right panels}) social inflation in settlement amount (at 50\% quantile level). \textit{Top panels}: cases with corporate defendants; \textit{bottom panels}: cases with individual defendants only.}
  \label{fig:idx_sev_s1_1}
\end{figure}

\begin{figure}[H]
  \centering
  \begin{subfigure}[t]{0.48\textwidth}
    \centering
    \includegraphics[width=\linewidth]{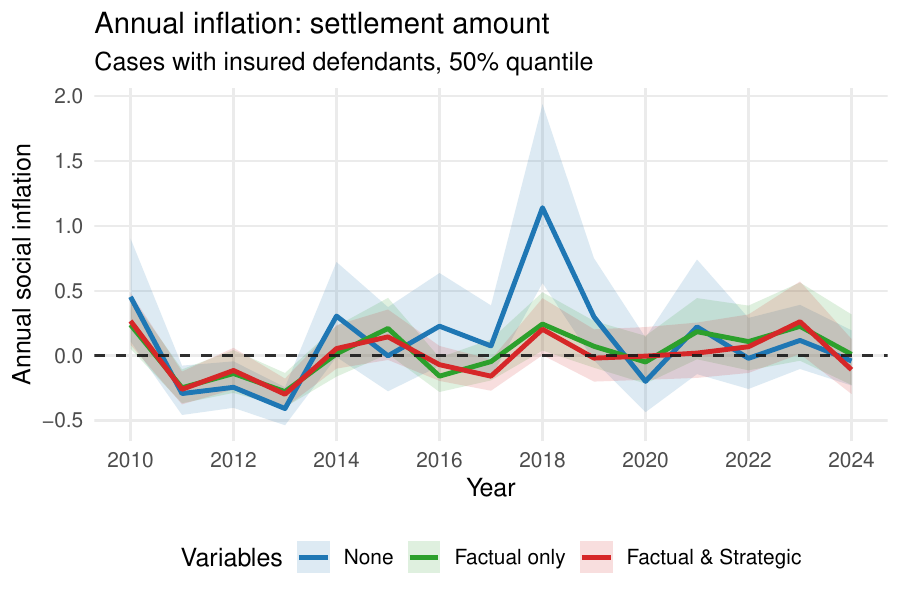}
  \end{subfigure}\hfill
  \begin{subfigure}[t]{0.48\textwidth}
    \centering
    \includegraphics[width=\linewidth]{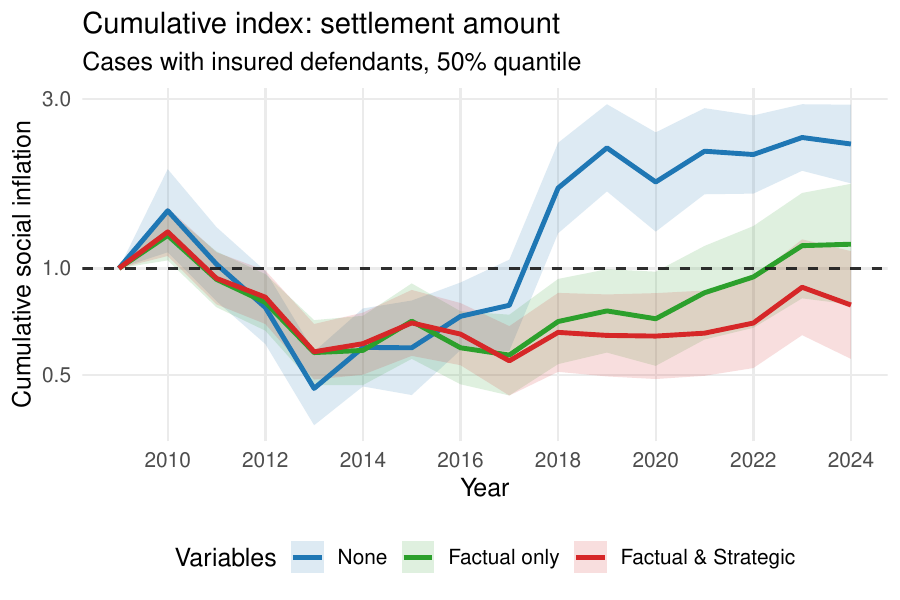}
  \end{subfigure}\hfill
    \begin{subfigure}[t]{0.48\textwidth}
    \centering
    \includegraphics[width=\linewidth]{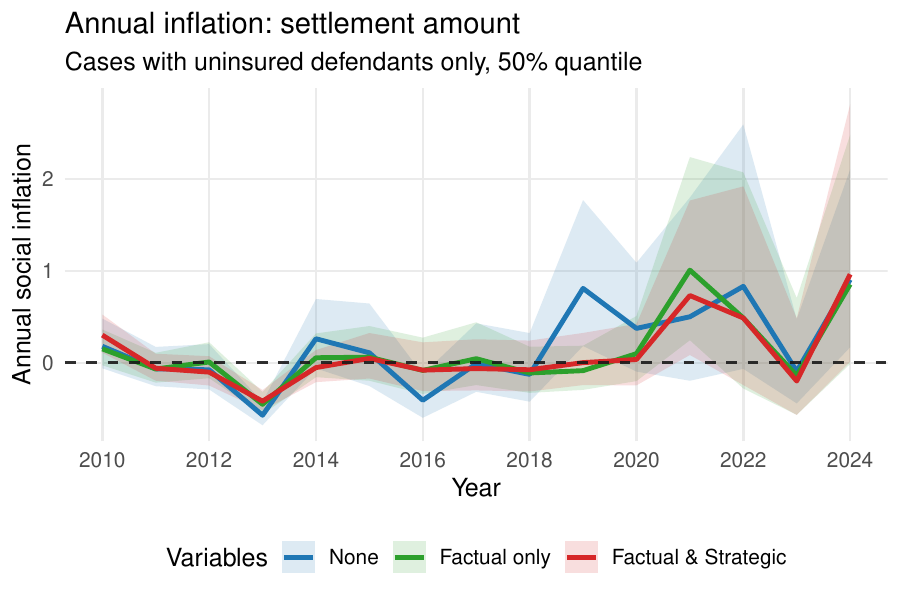}
  \end{subfigure}\hfill
  \begin{subfigure}[t]{0.48\textwidth}
    \centering
    \includegraphics[width=\linewidth]{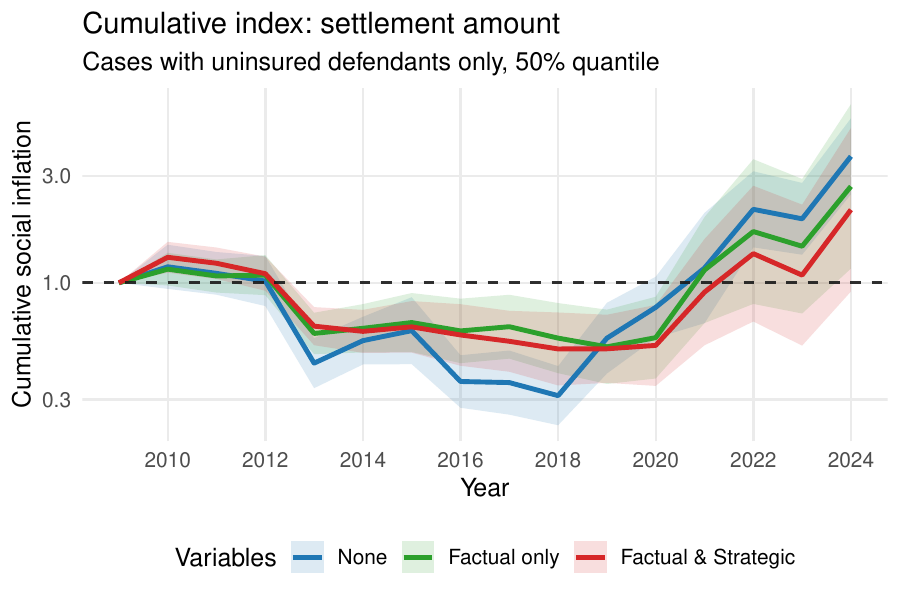}
  \end{subfigure}\hfill
  \caption{Annual (\textit{left panels}) and cumulative (\textit{right panels}) social inflation in settlement amount (at 50\% quantile level). \textit{Top panels}: cases with insured defendants; \textit{bottom panels}: cases with uninsured defendants only.}
  \label{fig:idx_sev_s1_2}
\end{figure}

\begin{figure}[H]
  \centering
  \begin{subfigure}[t]{0.48\textwidth}
    \centering
    \includegraphics[width=\linewidth]{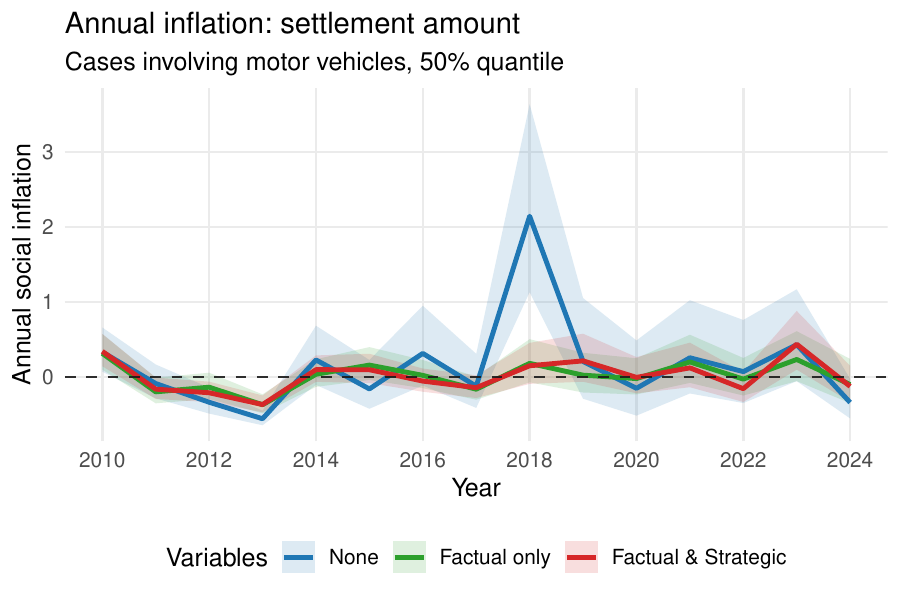}
  \end{subfigure}\hfill
  \begin{subfigure}[t]{0.48\textwidth}
    \centering
    \includegraphics[width=\linewidth]{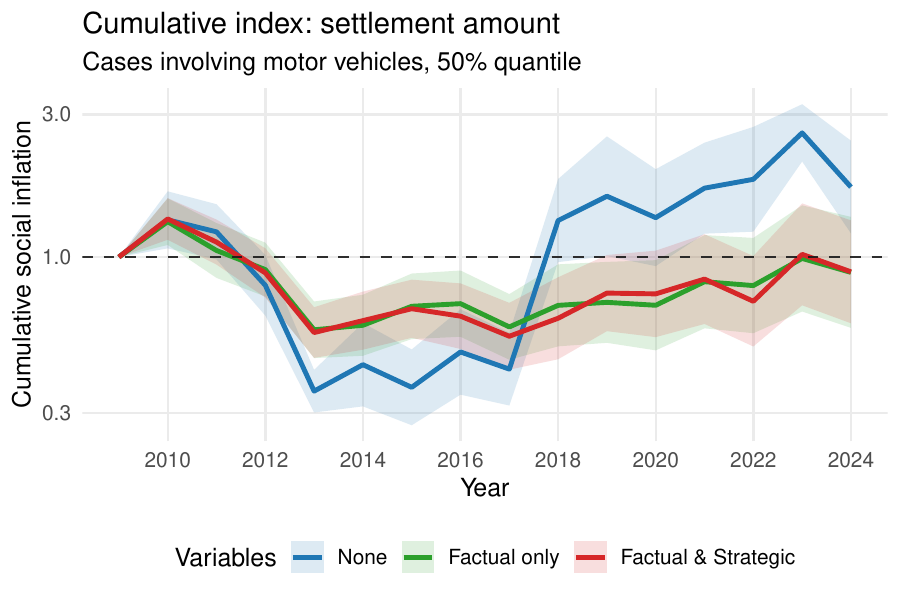}
  \end{subfigure}\hfill
    \begin{subfigure}[t]{0.48\textwidth}
    \centering
    \includegraphics[width=\linewidth]{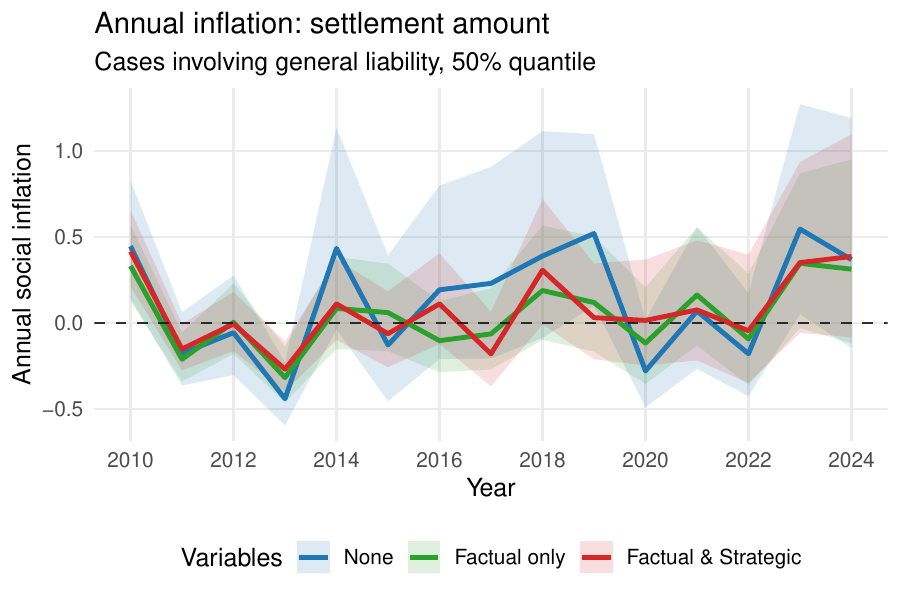}
  \end{subfigure}\hfill
  \begin{subfigure}[t]{0.48\textwidth}
    \centering
    \includegraphics[width=\linewidth]{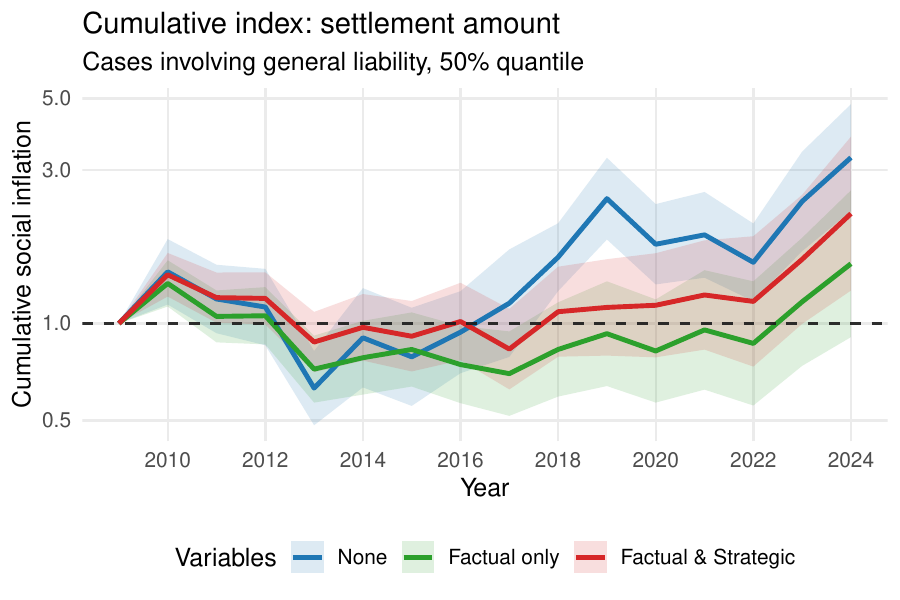}
  \end{subfigure}\hfill
    \begin{subfigure}[t]{0.48\textwidth}
    \centering
    \includegraphics[width=\linewidth]{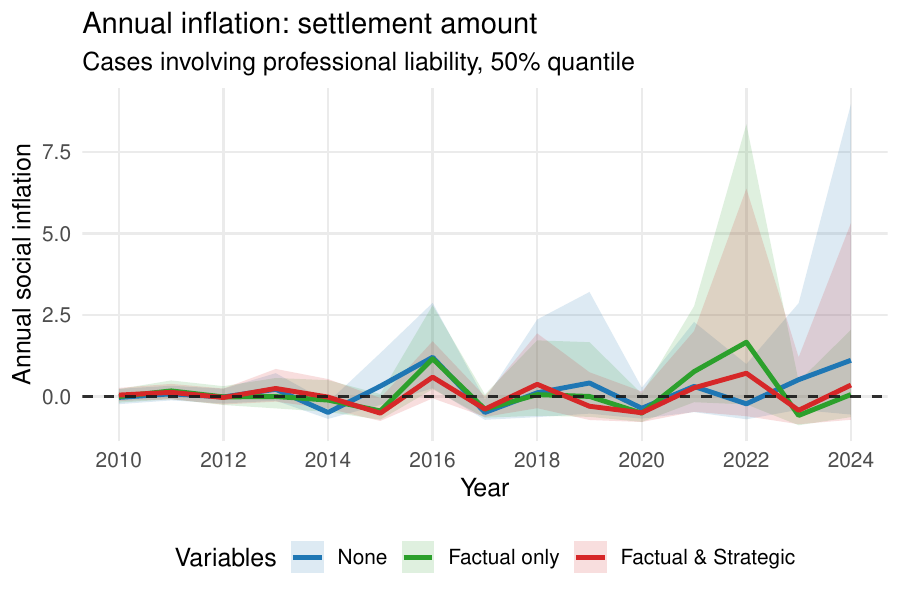}
  \end{subfigure}\hfill
  \begin{subfigure}[t]{0.48\textwidth}
    \centering
    \includegraphics[width=\linewidth]{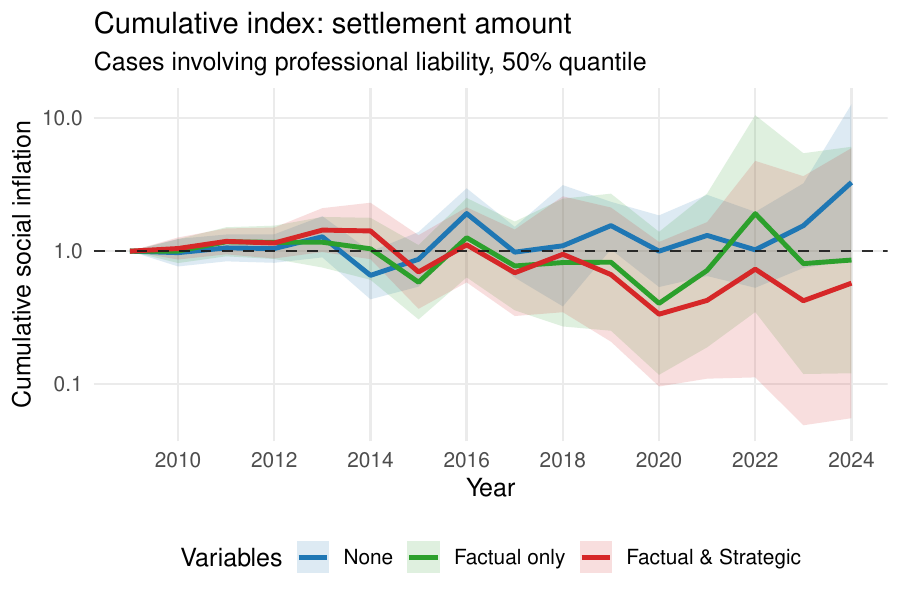}
  \end{subfigure}\hfill
    \begin{subfigure}[t]{0.48\textwidth}
    \centering
    \includegraphics[width=\linewidth]{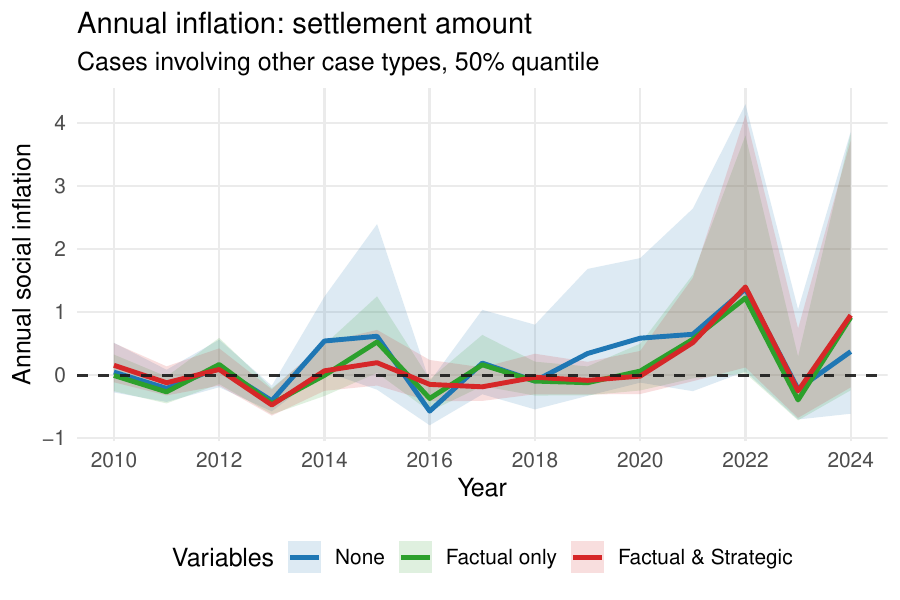}
  \end{subfigure}\hfill
  \begin{subfigure}[t]{0.48\textwidth}
    \centering
    \includegraphics[width=\linewidth]{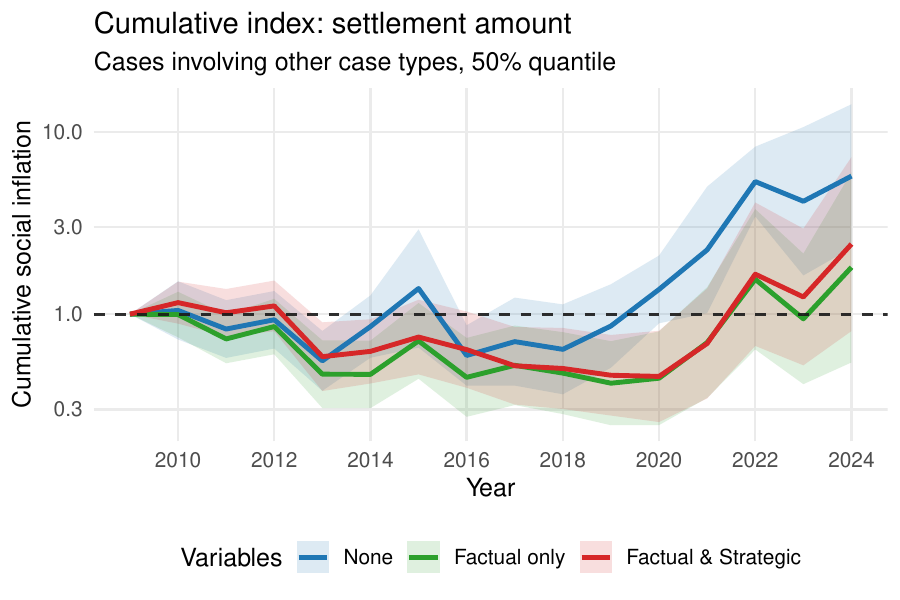}
  \end{subfigure}\hfill
  \caption{Annual (\textit{left panels}) and cumulative (\textit{right panels}) social inflation in settlement amount (at 50\% quantile level). \textit{First row}: motor liability cases; \textit{second row}: general liability cases; \textit{third row}: professional liability cases; \textit{fourth row}: other cases.}
  \label{fig:idx_sev_s1_3}
\end{figure}

\begin{figure}[H]
  \centering
  \begin{subfigure}[t]{0.48\textwidth}
    \centering
    \includegraphics[width=\linewidth]{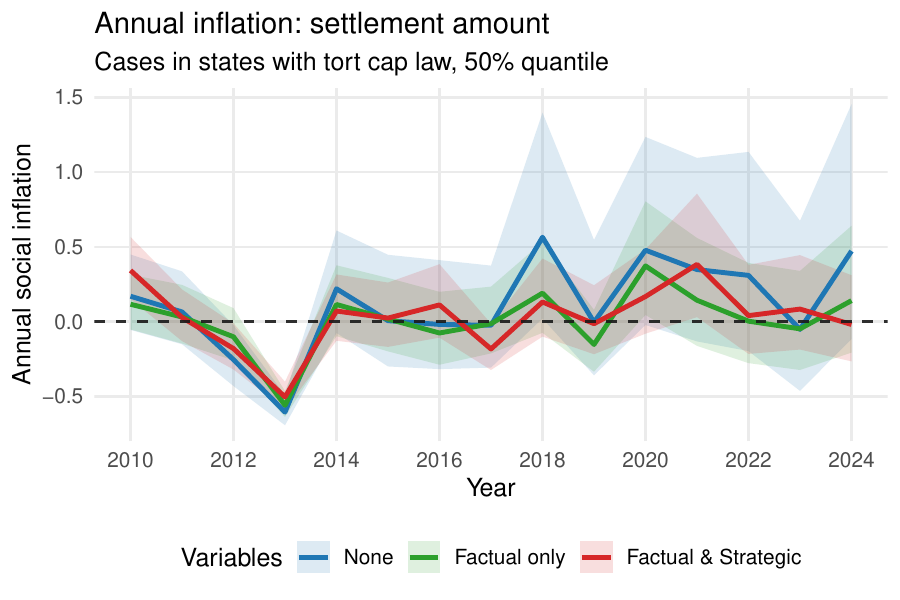}
  \end{subfigure}\hfill
  \begin{subfigure}[t]{0.48\textwidth}
    \centering
    \includegraphics[width=\linewidth]{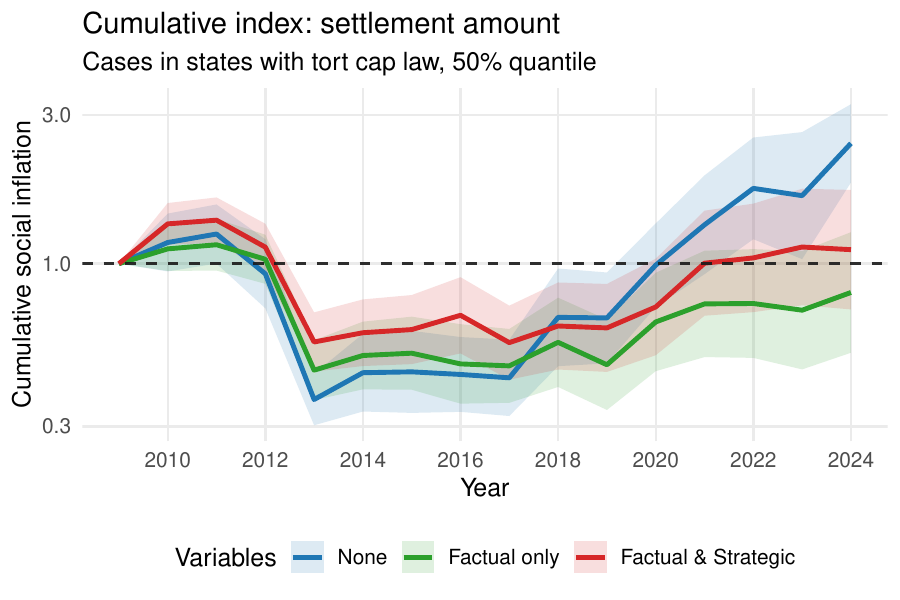}
  \end{subfigure}\hfill
    \begin{subfigure}[t]{0.48\textwidth}
    \centering
    \includegraphics[width=\linewidth]{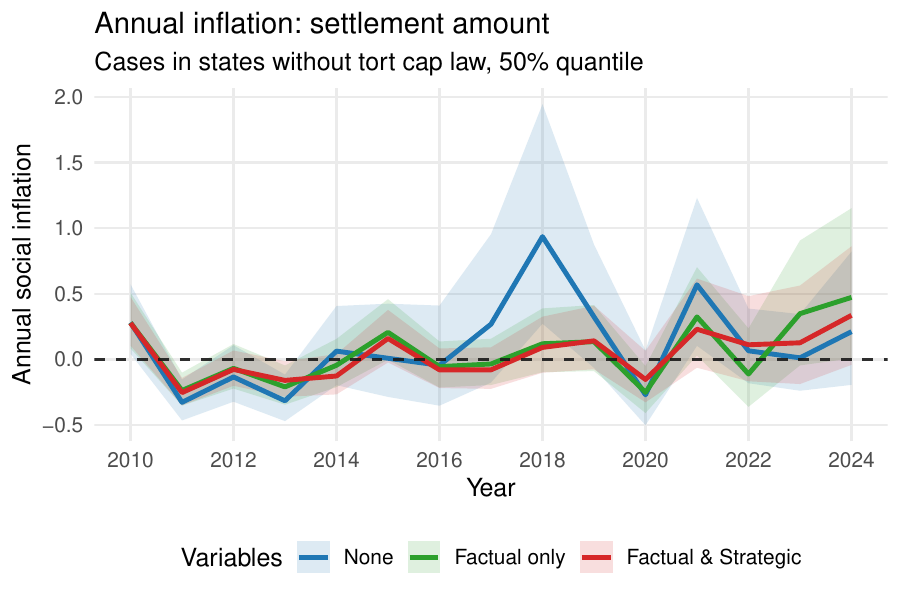}
  \end{subfigure}\hfill
  \begin{subfigure}[t]{0.48\textwidth}
    \centering
    \includegraphics[width=\linewidth]{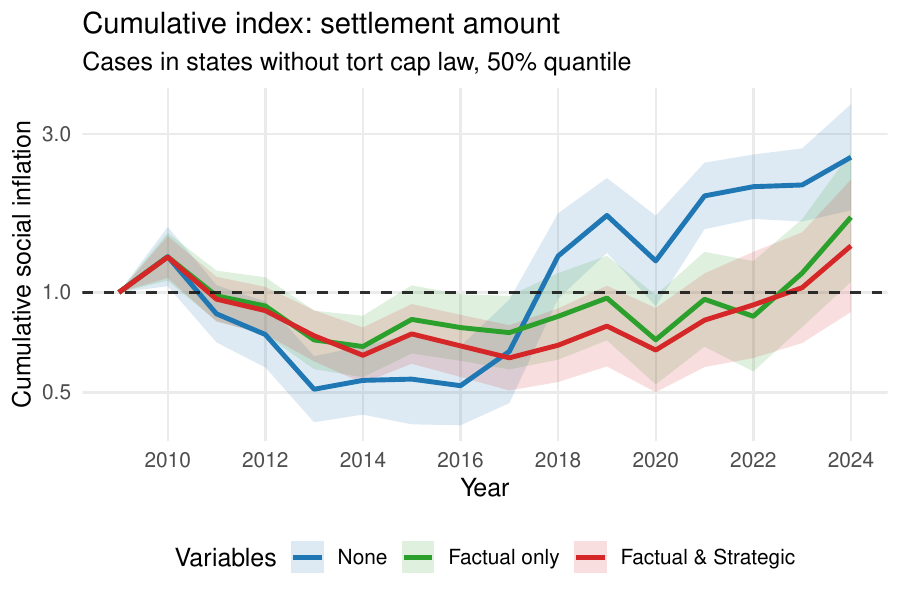}
  \end{subfigure}\hfill
  \caption{Annual (\textit{left panels}) and cumulative (\textit{right panels}) social inflation in settlement amount (at 50\% quantile level). \textit{Top panels}: cases in states with tort-cap laws; \textit{bottom panels}: cases in states without tort-cap laws.}
  \label{fig:idx_sev_s1_4}
\end{figure}

\begin{figure}[H]
  \centering
  \begin{subfigure}[t]{0.48\textwidth}
    \centering
    \includegraphics[width=\linewidth]{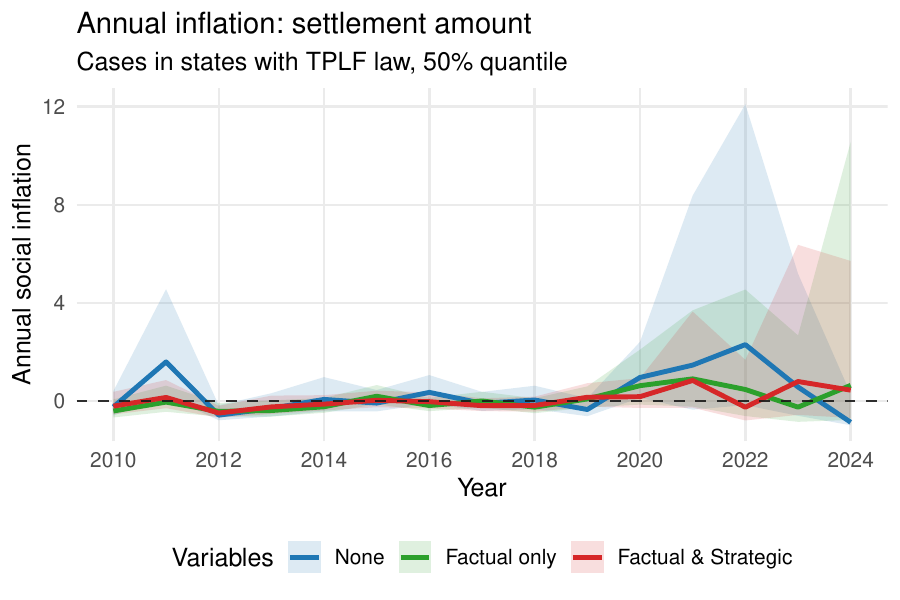}
  \end{subfigure}\hfill
  \begin{subfigure}[t]{0.48\textwidth}
    \centering
    \includegraphics[width=\linewidth]{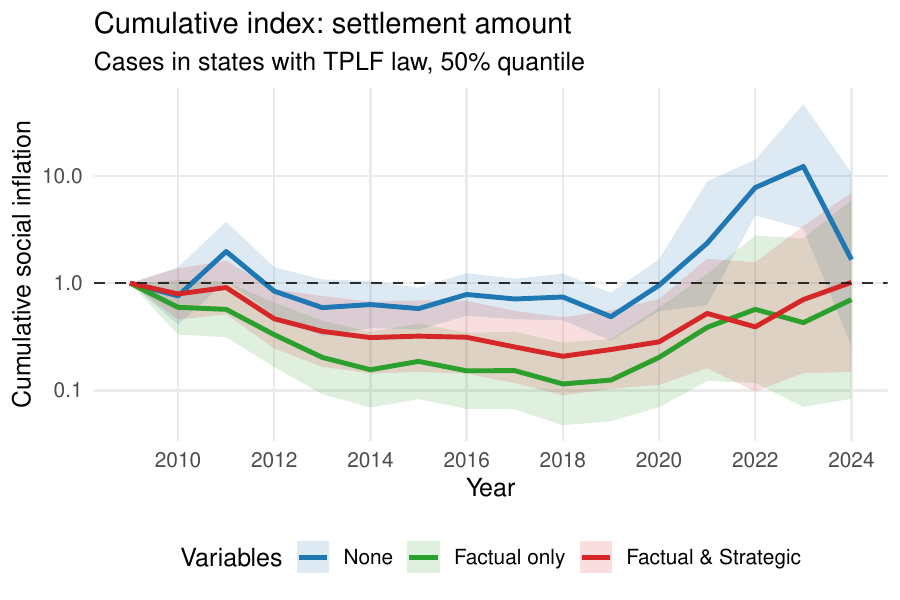}
  \end{subfigure}\hfill
    \begin{subfigure}[t]{0.48\textwidth}
    \centering
    \includegraphics[width=\linewidth]{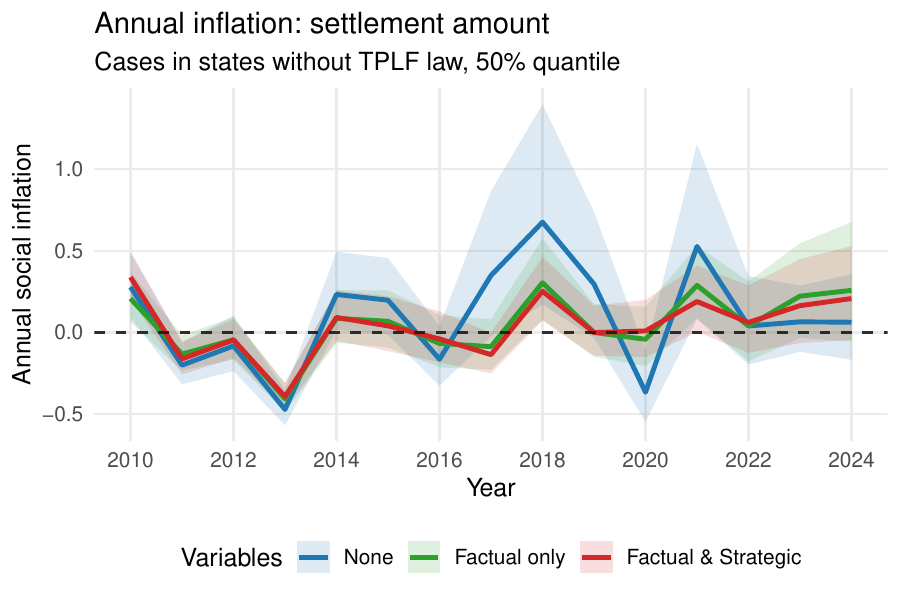}
  \end{subfigure}\hfill
  \begin{subfigure}[t]{0.48\textwidth}
    \centering
    \includegraphics[width=\linewidth]{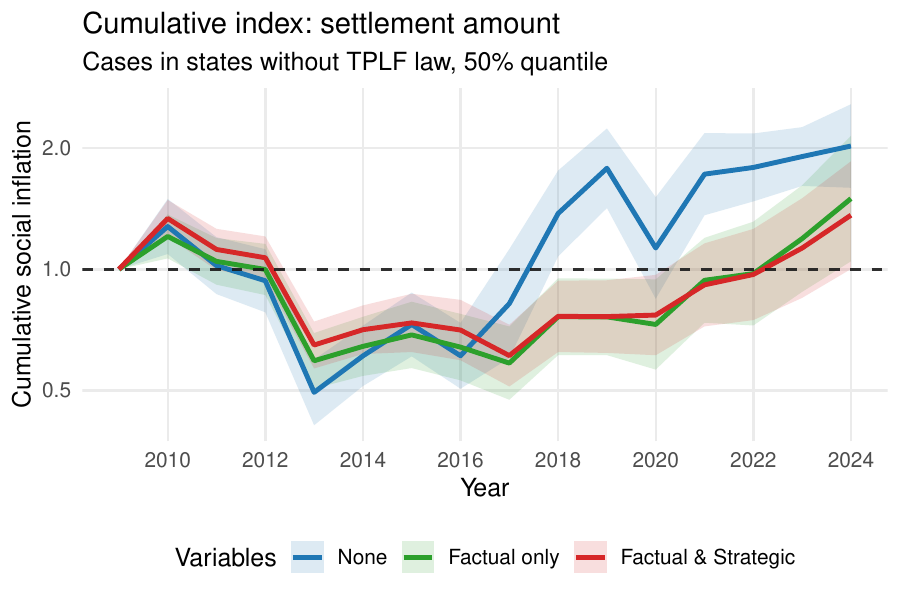}
  \end{subfigure}\hfill
  \caption{Annual (\textit{left panels}) and cumulative (\textit{right panels}) social inflation in settlement amount (at 50\% quantile level). \textit{Top panels}: cases in states with TPLF laws; \textit{bottom panels}: cases in states without TPLF laws.}
  \label{fig:idx_sev_s1_5}
\end{figure}

This subsection reports social inflation results for settlement severity, using the quantile-based indices presented in Section \ref{sec:method:s_s} with $\tau=0.5$ (median settlement amount) and summarized in Figures \ref{fig:idx_sev_s1_0} to \ref{fig:idx_sev_s1_5}; we obtain very similar qualitative conclusions at $\tau=0.9$ and therefore omit those plots for conciseness. Overall, once we properly control for evolving case mix and litigation strategies, social inflation in inflation-adjusted settlement amounts is generally small and mostly statistically insignificant across essentially all settings considered, including the full sample, corporate versus individual defendants, liability lines, and states with versus without tort caps or TPLF regulation. This contrasts sharply with the strong and statistically significant post-2020 social inflation observed for verdict award severity discussed in Section \ref{sec:result:p_s}. Taken together, these patterns suggest that, in the VerdictSearch data, the dominant severity channel of social inflation is primarily a verdict-driven phenomenon rather than a broad-based escalation of negotiated settlement values, which is consistent with insurance companies' increasing focus on the role of juries, trial dynamics, and the increasing incidence of extreme verdict outcomes rather than on how the cases are settled.

\subsection{Social inflation in total amount payable to the plaintiff} \label{sec:result:t_s}
\begin{figure}[H]
  \centering
  \begin{subfigure}[t]{0.48\textwidth}
    \centering
    \includegraphics[width=\linewidth]{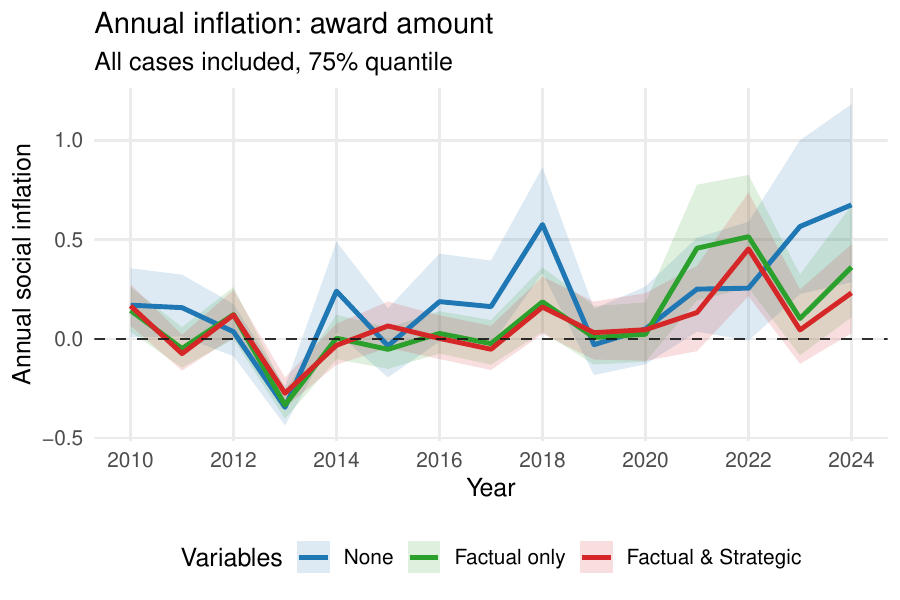}
  \end{subfigure}\hfill
  \begin{subfigure}[t]{0.48\textwidth}
    \centering
    \includegraphics[width=\linewidth]{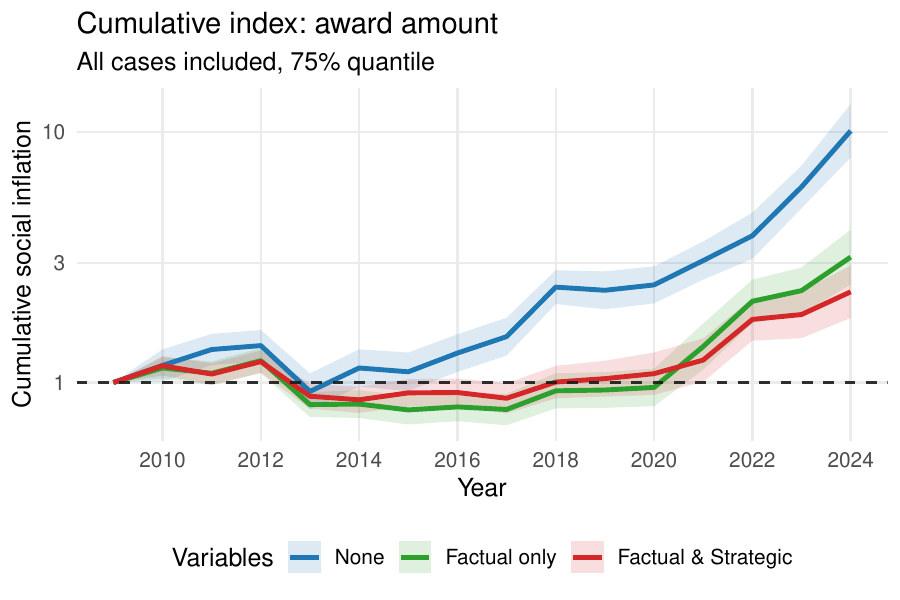}
  \end{subfigure}\hfill
  \caption{Annual (\textit{left panels}) and cumulative (\textit{right panels}) social inflation in total amount payable to plaintiff (at 75\% quantile level). All cases are included.}
  \label{fig:idx_sev_t1_0}
\end{figure}

\begin{figure}[H]
  \centering
  \begin{subfigure}[t]{0.48\textwidth}
    \centering
    \includegraphics[width=\linewidth]{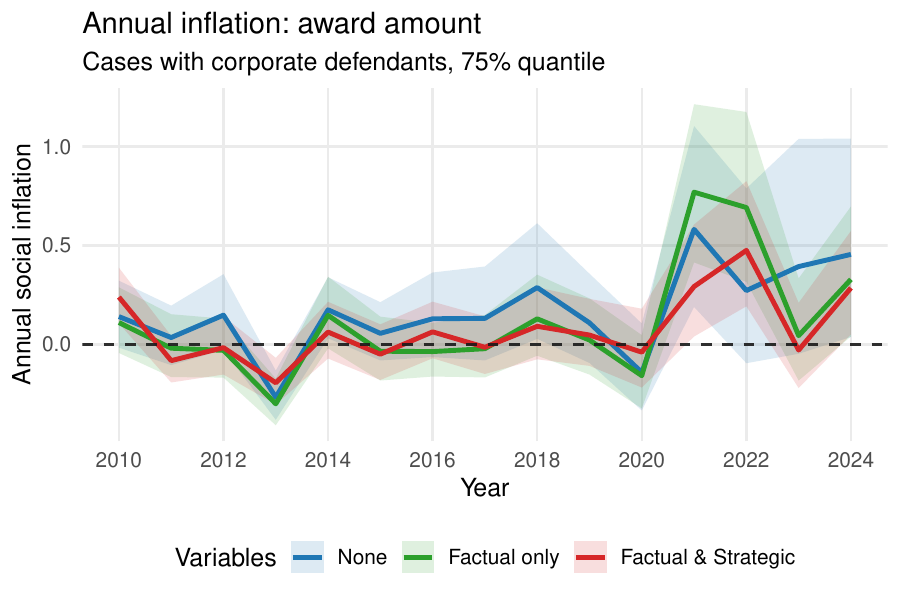}
  \end{subfigure}\hfill
  \begin{subfigure}[t]{0.48\textwidth}
    \centering
    \includegraphics[width=\linewidth]{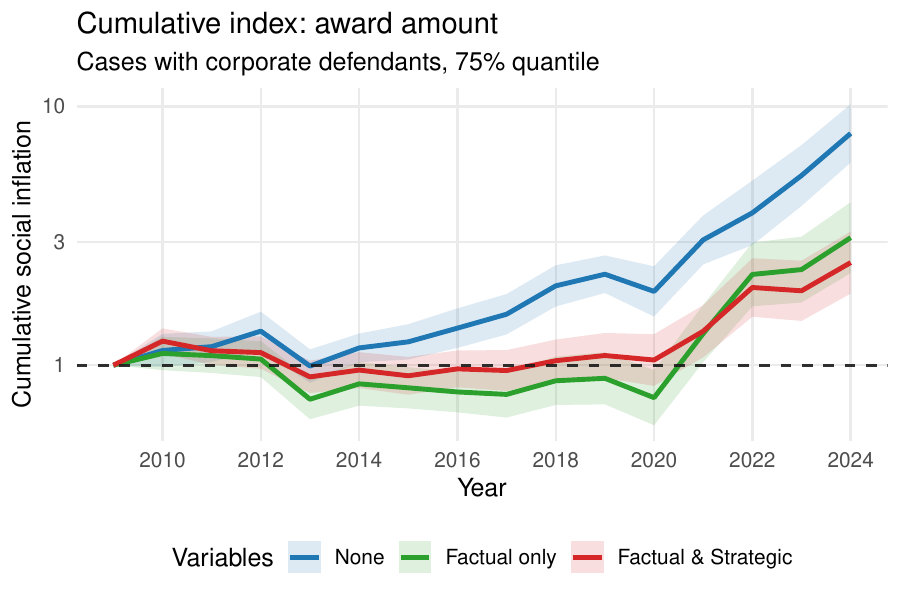}
  \end{subfigure}\hfill
    \begin{subfigure}[t]{0.48\textwidth}
    \centering
    \includegraphics[width=\linewidth]{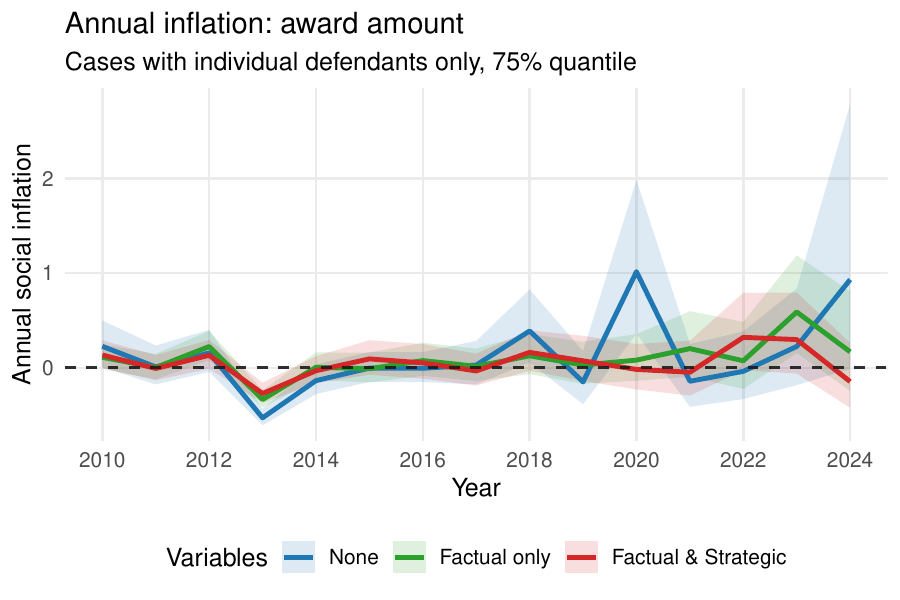}
  \end{subfigure}\hfill
  \begin{subfigure}[t]{0.48\textwidth}
    \centering
    \includegraphics[width=\linewidth]{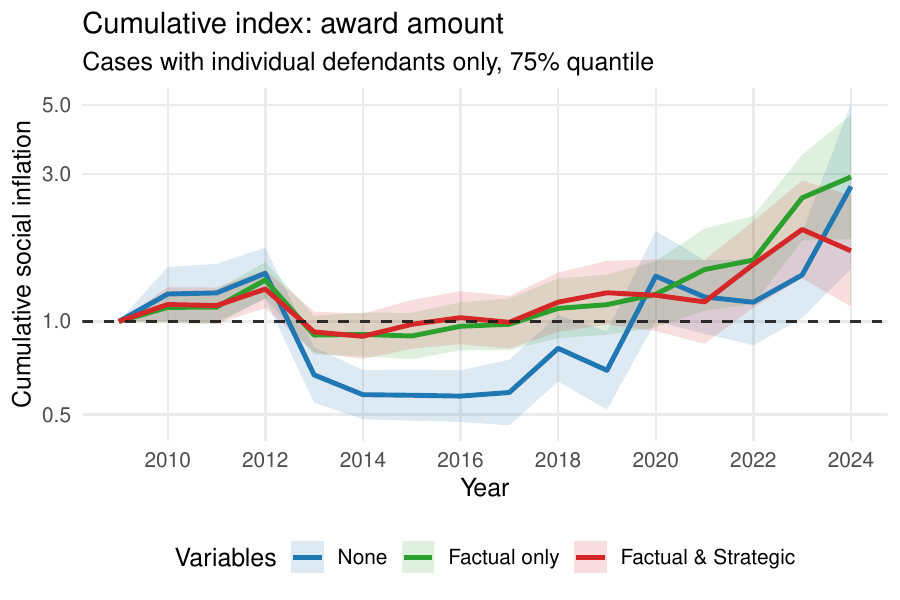}
  \end{subfigure}\hfill
  \caption{Annual (\textit{left panels}) and cumulative (\textit{right panels}) social inflation in total amount payable to plaintiff (at 75\% quantile level). \textit{Top panels}: cases with corporate defendants; \textit{bottom panels}: cases with individual defendants only.}
  \label{fig:idx_sev_t1_1}
\end{figure}

\begin{figure}[H]
  \centering
  \begin{subfigure}[t]{0.48\textwidth}
    \centering
    \includegraphics[width=\linewidth]{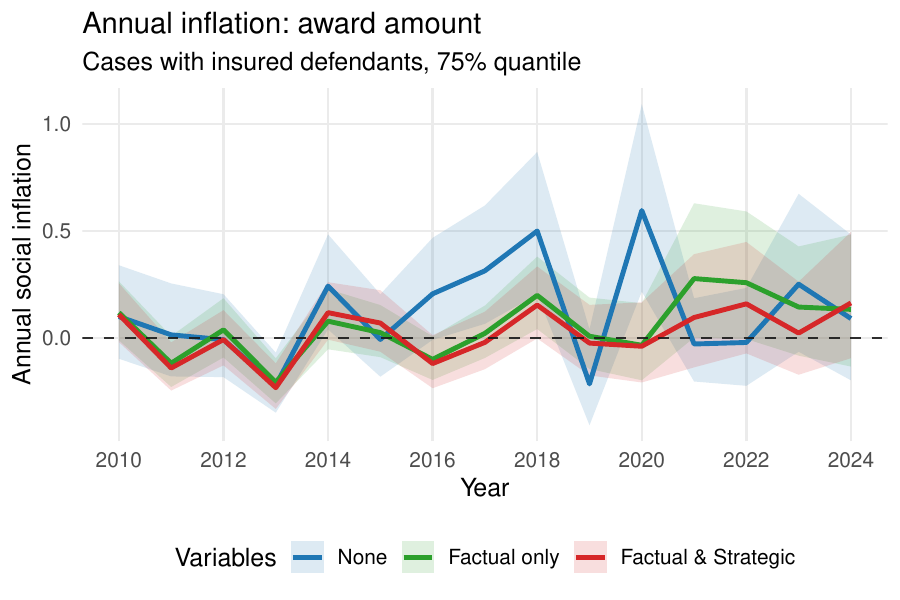}
  \end{subfigure}\hfill
  \begin{subfigure}[t]{0.48\textwidth}
    \centering
    \includegraphics[width=\linewidth]{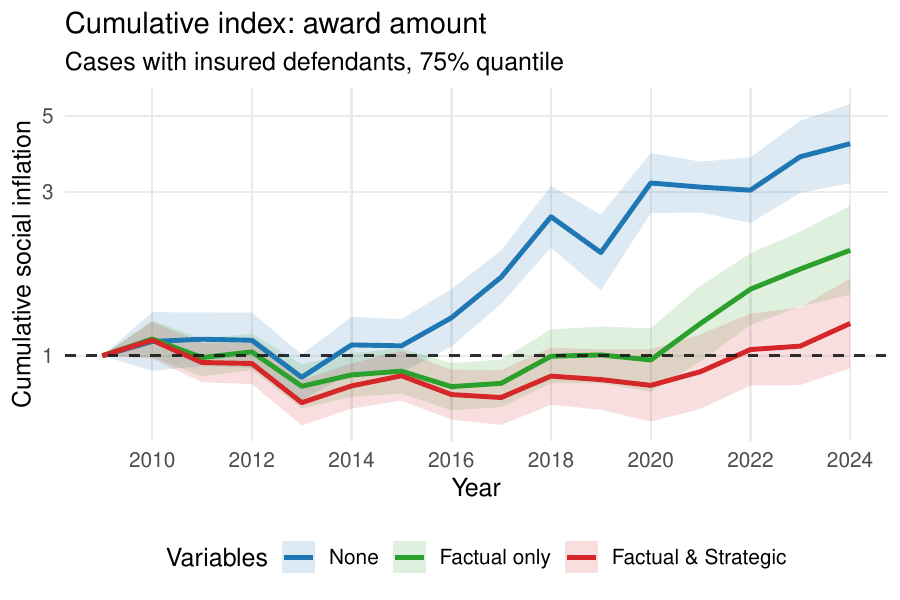}
  \end{subfigure}\hfill
    \begin{subfigure}[t]{0.48\textwidth}
    \centering
    \includegraphics[width=\linewidth]{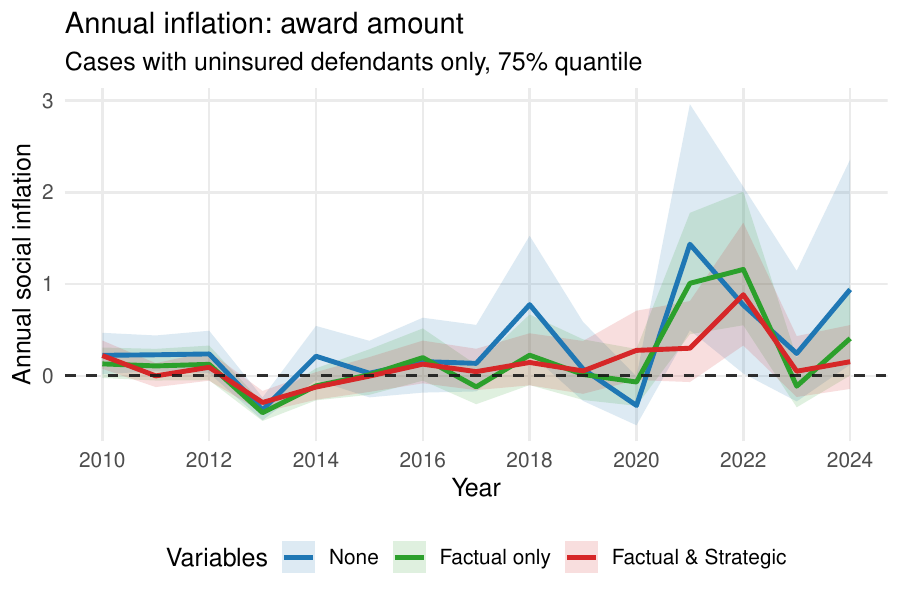}
  \end{subfigure}\hfill
  \begin{subfigure}[t]{0.48\textwidth}
    \centering
    \includegraphics[width=\linewidth]{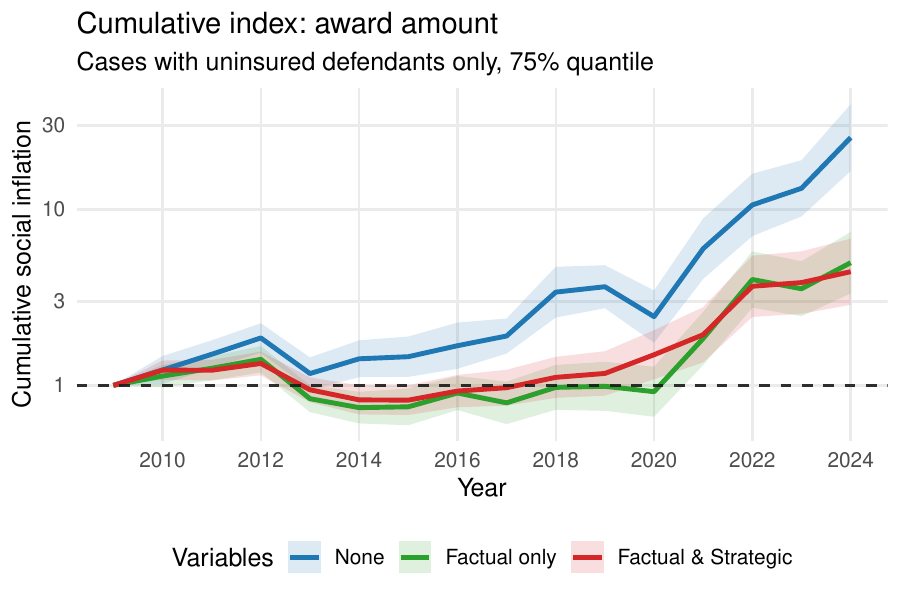}
  \end{subfigure}\hfill
  \caption{Annual (\textit{left panels}) and cumulative (\textit{right panels}) social inflation in total amount payable to plaintiff (at 75\% quantile level). \textit{Top panels}: cases with insured defendants; \textit{bottom panels}: cases with uninsured defendants only.}
  \label{fig:idx_sev_t1_2}
\end{figure}

\begin{figure}[H]
  \centering
  \begin{subfigure}[t]{0.48\textwidth}
    \centering
    \includegraphics[width=\linewidth]{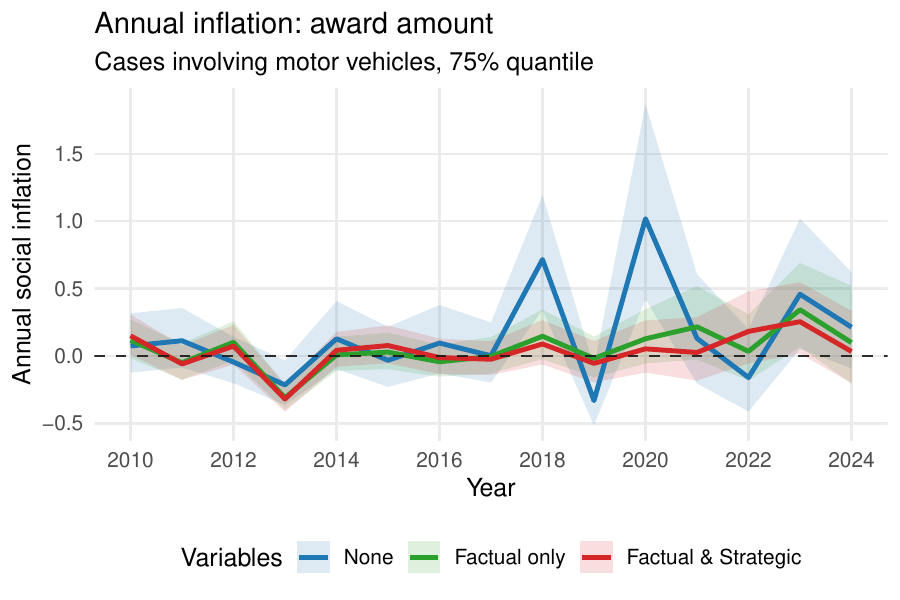}
  \end{subfigure}\hfill
  \begin{subfigure}[t]{0.48\textwidth}
    \centering
    \includegraphics[width=\linewidth]{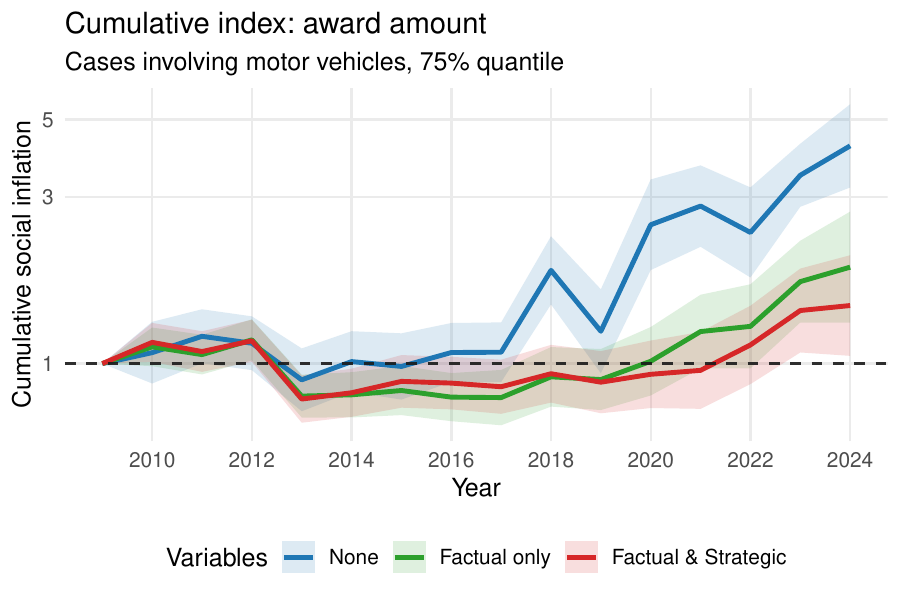}
  \end{subfigure}\hfill
    \begin{subfigure}[t]{0.48\textwidth}
    \centering
    \includegraphics[width=\linewidth]{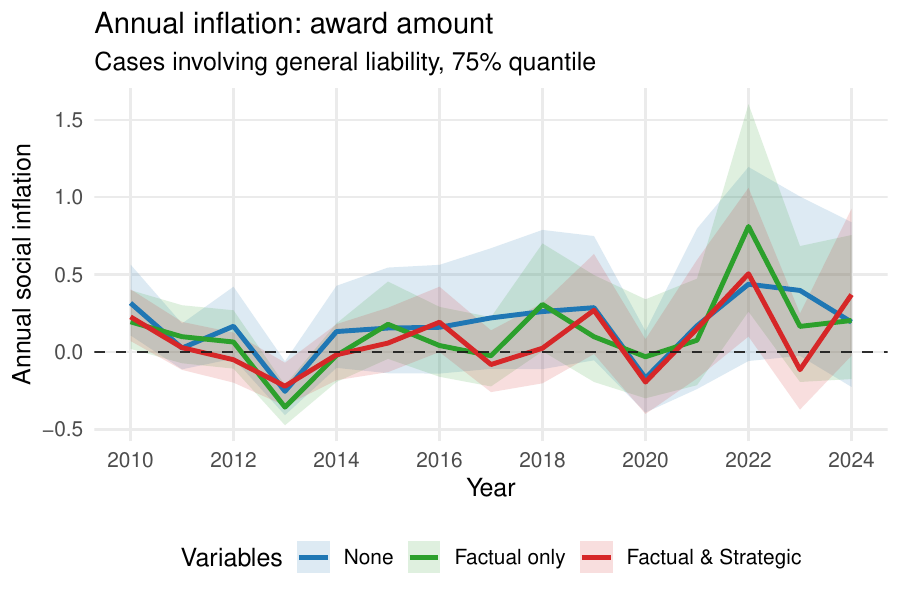}
  \end{subfigure}\hfill
  \begin{subfigure}[t]{0.48\textwidth}
    \centering
    \includegraphics[width=\linewidth]{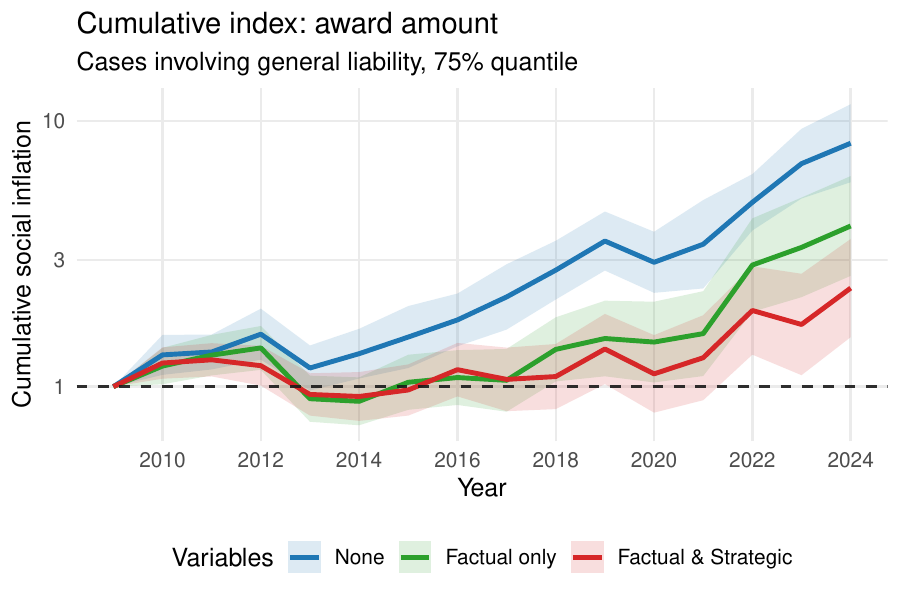}
  \end{subfigure}\hfill
    \begin{subfigure}[t]{0.48\textwidth}
    \centering
    \includegraphics[width=\linewidth]{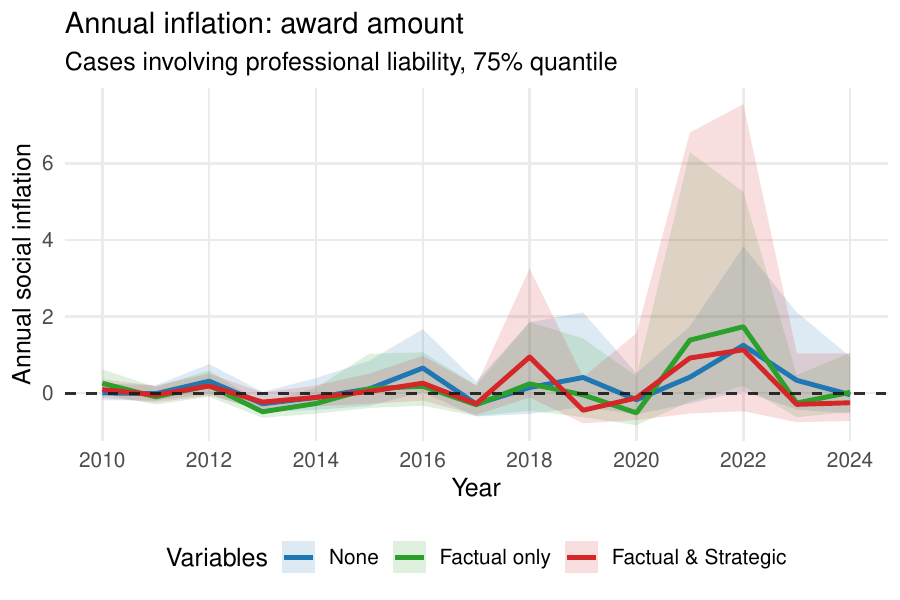}
  \end{subfigure}\hfill
  \begin{subfigure}[t]{0.48\textwidth}
    \centering
    \includegraphics[width=\linewidth]{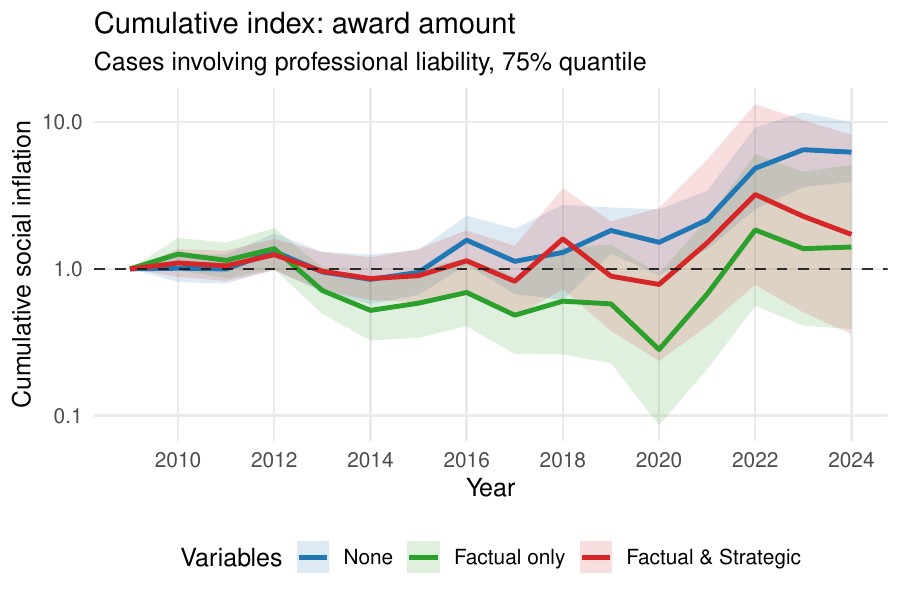}
  \end{subfigure}\hfill
    \begin{subfigure}[t]{0.48\textwidth}
    \centering
    \includegraphics[width=\linewidth]{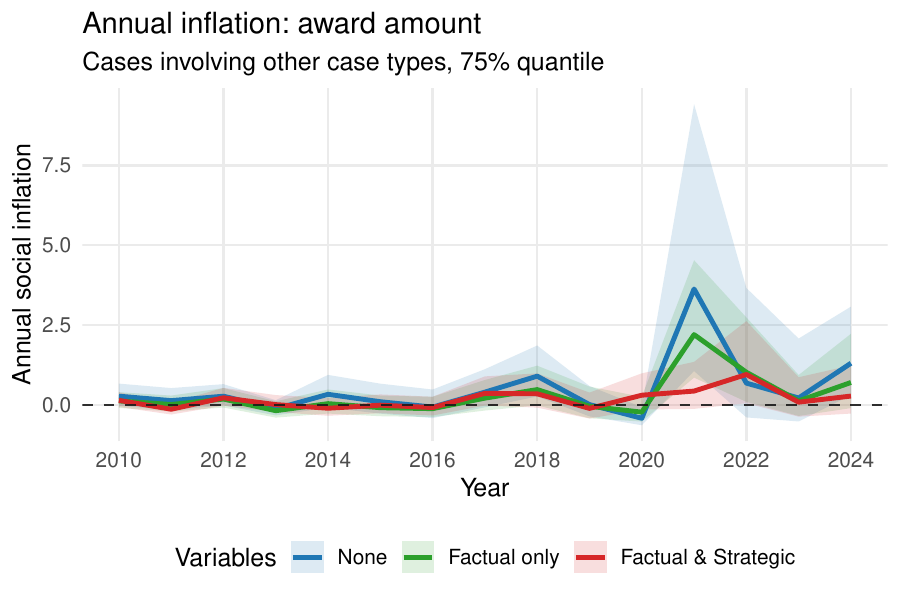}
  \end{subfigure}\hfill
  \begin{subfigure}[t]{0.48\textwidth}
    \centering
    \includegraphics[width=\linewidth]{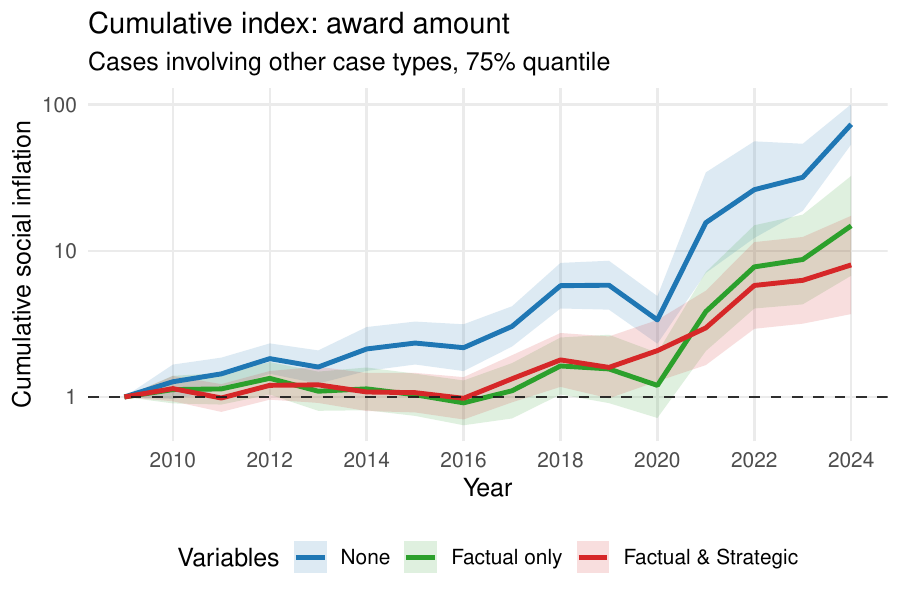}
  \end{subfigure}\hfill
  \caption{Annual (\textit{left panels}) and cumulative (\textit{right panels}) social inflation in total amount payable to plaintiff (at 75\% quantile level). \textit{First row}: motor liability cases; \textit{second row}: general liability cases; \textit{third row}: professional liability cases; \textit{fourth row}: other cases.}
  \label{fig:idx_sev_t1_3}
\end{figure}

\begin{figure}[H]
  \centering
  \begin{subfigure}[t]{0.48\textwidth}
    \centering
    \includegraphics[width=\linewidth]{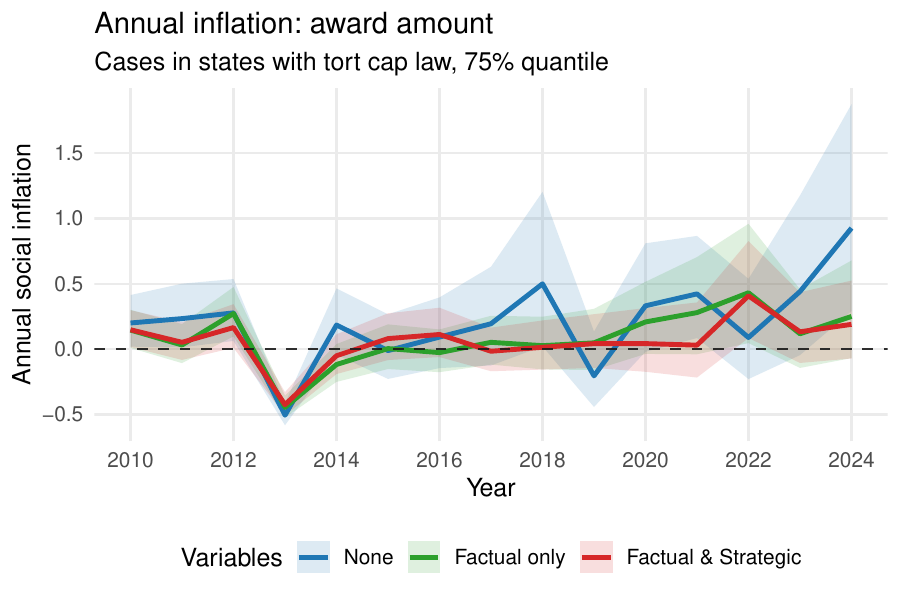}
  \end{subfigure}\hfill
  \begin{subfigure}[t]{0.48\textwidth}
    \centering
    \includegraphics[width=\linewidth]{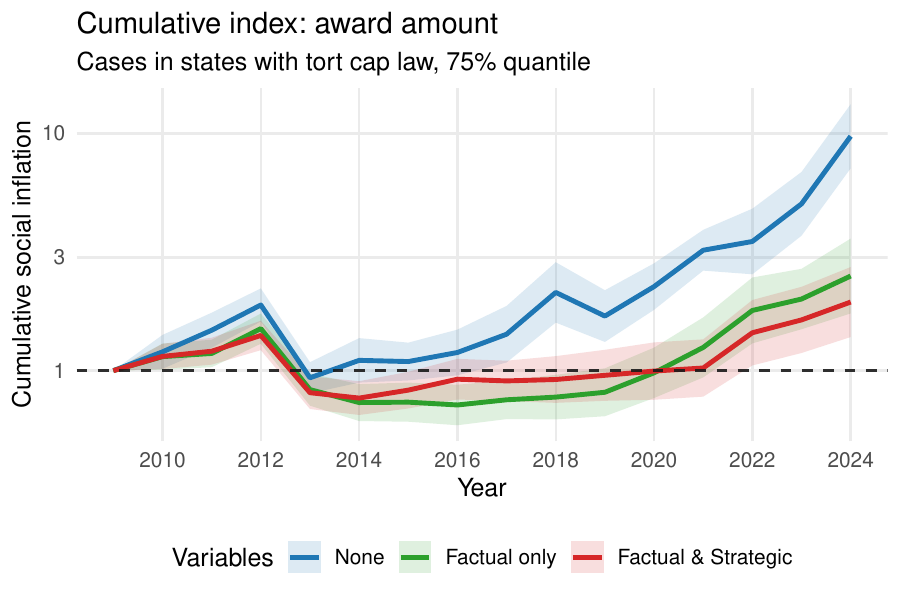}
  \end{subfigure}\hfill
    \begin{subfigure}[t]{0.48\textwidth}
    \centering
    \includegraphics[width=\linewidth]{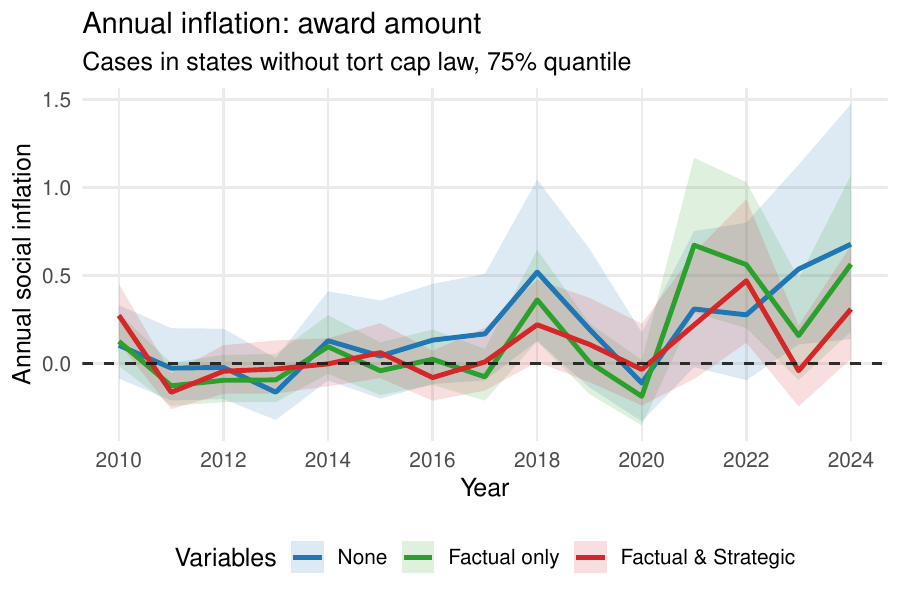}
  \end{subfigure}\hfill
  \begin{subfigure}[t]{0.48\textwidth}
    \centering
    \includegraphics[width=\linewidth]{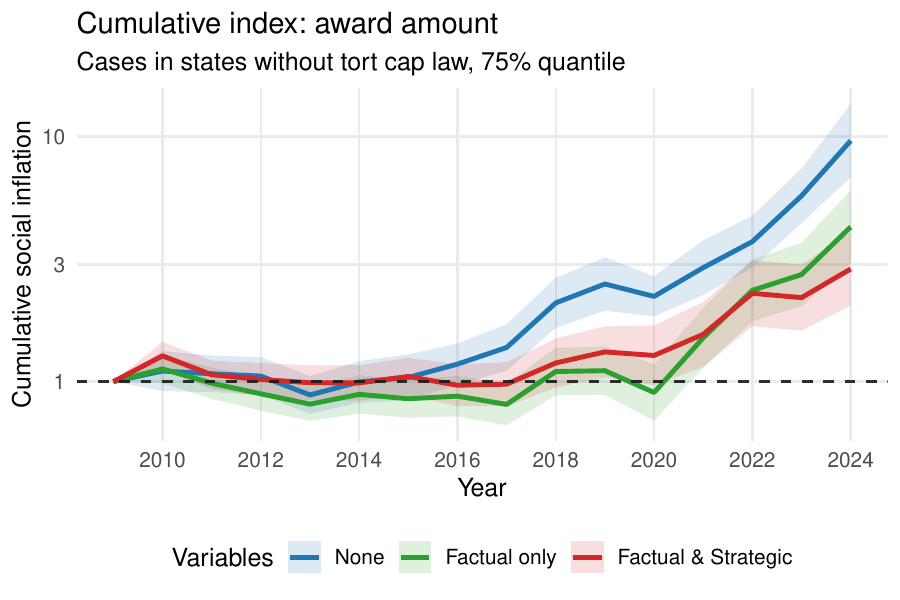}
  \end{subfigure}\hfill
  \caption{Annual (\textit{left panels}) and cumulative (\textit{right panels}) social inflation in total amount payable to plaintiff (at 75\% quantile level). \textit{Top panels}: cases in states with tort-cap laws; \textit{bottom panels}: cases in states without tort-cap laws.}
  \label{fig:idx_sev_t1_4}
\end{figure}

\begin{figure}[H]
  \centering
  \begin{subfigure}[t]{0.48\textwidth}
    \centering
    \includegraphics[width=\linewidth]{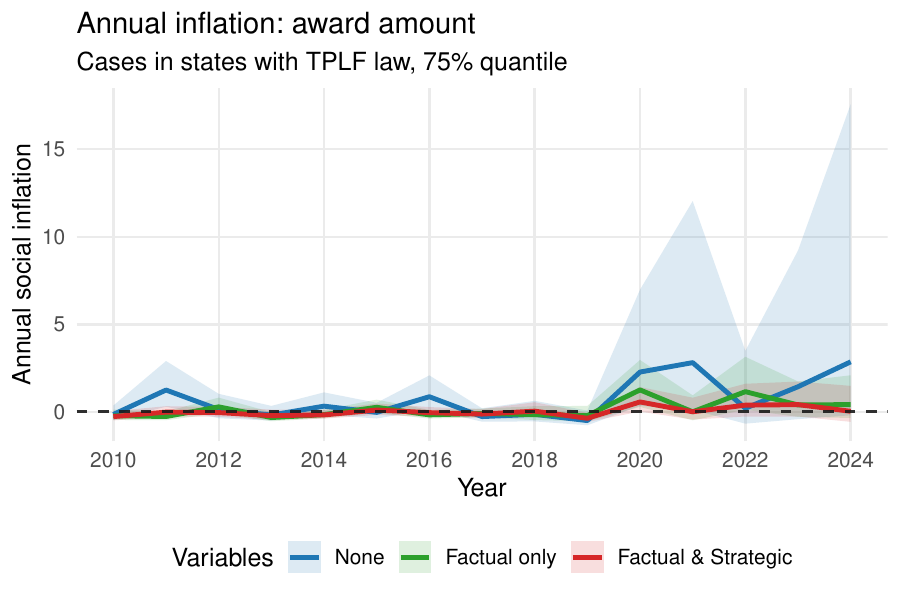}
  \end{subfigure}\hfill
  \begin{subfigure}[t]{0.48\textwidth}
    \centering
    \includegraphics[width=\linewidth]{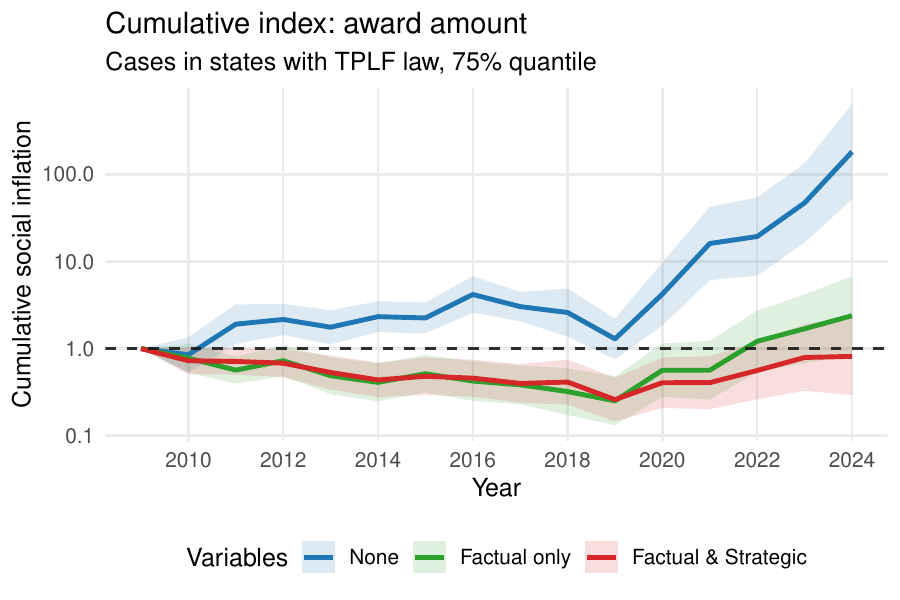}
  \end{subfigure}\hfill
    \begin{subfigure}[t]{0.48\textwidth}
    \centering
    \includegraphics[width=\linewidth]{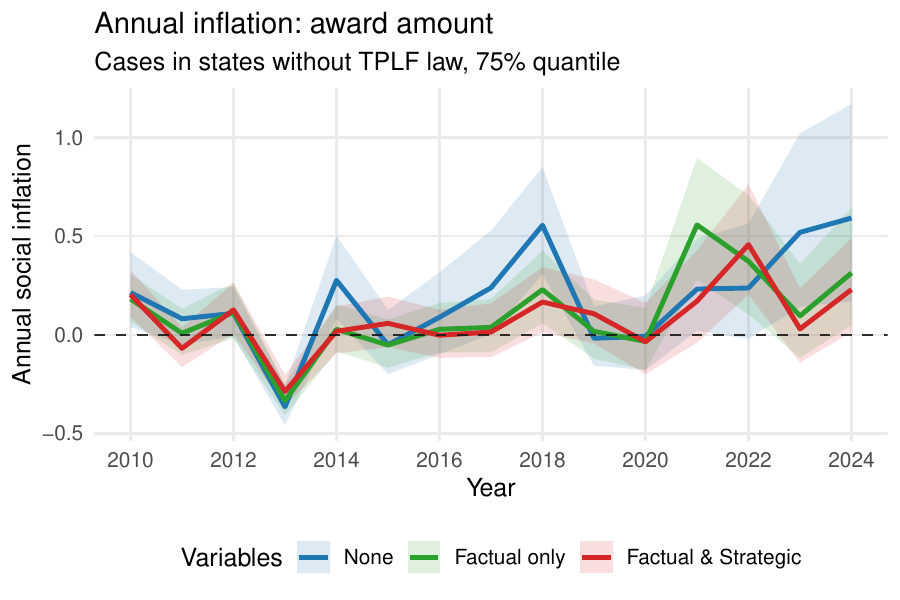}
  \end{subfigure}\hfill
  \begin{subfigure}[t]{0.48\textwidth}
    \centering
    \includegraphics[width=\linewidth]{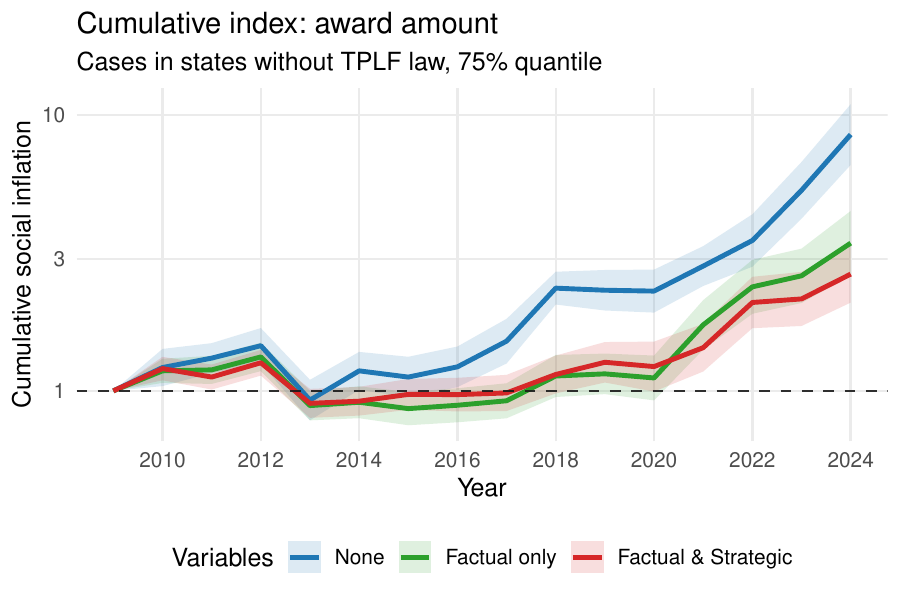}
  \end{subfigure}\hfill
  \caption{Annual (\textit{left panels}) and cumulative (\textit{right panels}) social inflation in total amount payable to plaintiff (at 75\% quantile level). \textit{Top panels}: cases in states with TPLF laws; \textit{bottom panels}: cases in states without TPLF  laws.}
  \label{fig:idx_sev_t1_5}
\end{figure}

\begin{figure}[H]
  \centering
  \begin{subfigure}[t]{0.48\textwidth}
    \centering
    \includegraphics[width=\linewidth]{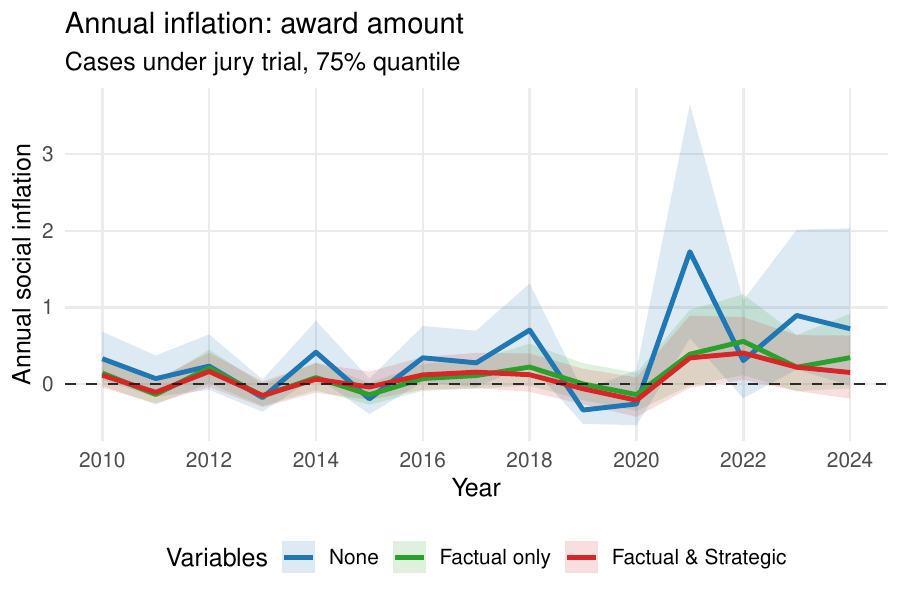}
  \end{subfigure}\hfill
  \begin{subfigure}[t]{0.48\textwidth}
    \centering
    \includegraphics[width=\linewidth]{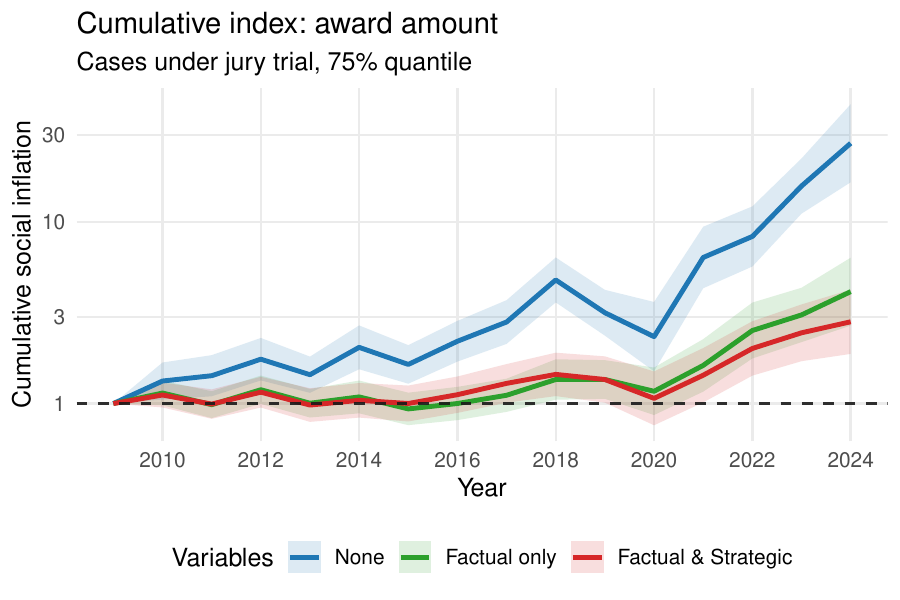}
  \end{subfigure}\hfill
    \begin{subfigure}[t]{0.48\textwidth}
    \centering
    \includegraphics[width=\linewidth]{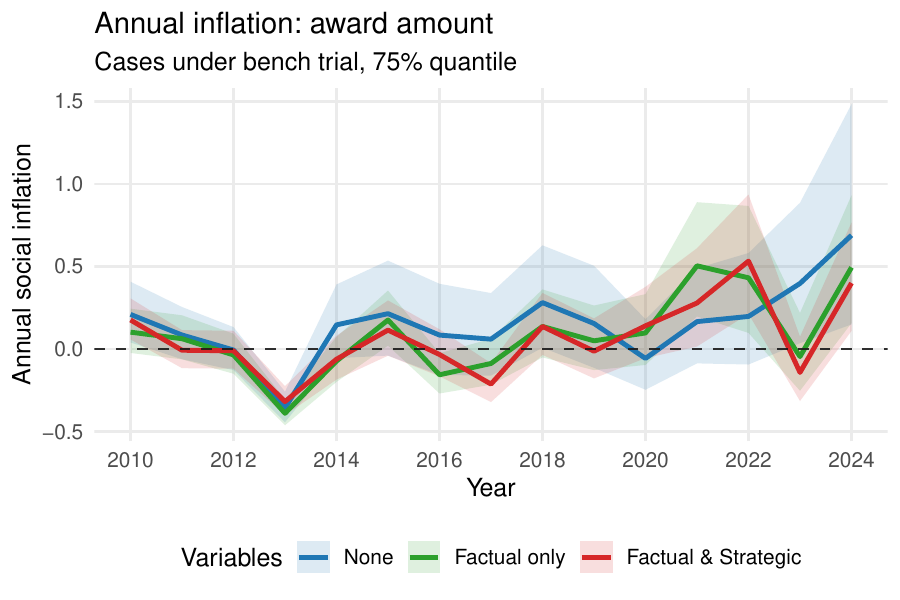}
  \end{subfigure}\hfill
  \begin{subfigure}[t]{0.48\textwidth}
    \centering
    \includegraphics[width=\linewidth]{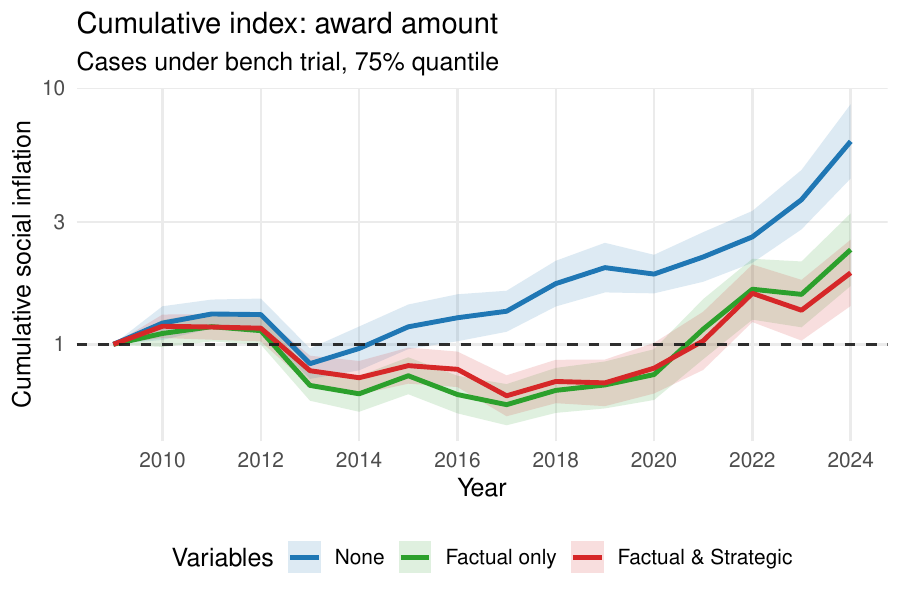}
  \end{subfigure}\hfill
  \caption{Annual (\textit{left panels}) and cumulative (\textit{right panels}) social inflation in total amount payable to plaintiff (at 75\% quantile level). \textit{Top panels}: cases under jury trial; \textit{bottom panels}: cases under bench trial.}
  \label{fig:idx_sev_t1_6}
\end{figure}

This subsection reports social inflation results for the total amount payable to the plaintiff, using the aggregate quantile-based indices presented in Section \ref{sec:method:t_s} with $\tau=0.75$ (Figures \ref{fig:idx_sev_t1_0} to \ref{fig:idx_sev_t1_6}). We adopt a higher quantile than in Sections \ref{sec:result:p_s} and \ref{sec:result:s_s} because the total-payment distribution contains a nontrivial mass at 0 due to defense verdicts; we also repeat the analysis at $\tau=0.95$ and obtain qualitatively identical conclusions, so we omit those additional plots for brevity. Across all stratifications considered, the estimated ASIR/CSII trajectories for total payments closely mirror those for verdict award severity in Figures \ref{fig:idx_sev_p1_0} to \ref{fig:idx_sev_p1_6}, both in shape and in the direction of covariate-adjustment effects. In particular, the post-2020 acceleration remains pronounced and statistically significant after case-mix and strategy adjustment, while a substantial portion of the pre-pandemic growth in unadjusted indices is attenuated once factual covariates are included. The near one-to-one correspondence between Figures \ref{fig:idx_sev_t1_0} to \ref{fig:idx_sev_t1_6} and the results on verdict award amounts reinforces the central conclusion of Sections \ref{sec:result:p_p} to \ref{sec:result:s_s}: The overall evolution of social inflation in total plaintiff payments is driven predominantly by the severity channel associated with verdict awards, whereas changes in plaintiff win probabilities, settlement probabilities, and settlement amounts play lesser roles in explaining aggregate payment inflation over time.

\section{Quantifying differential social inflation}
\label{sec:diff}

A common perception in practice attributes social inflation primarily to nuclear verdicts, where extreme awards often reach tens (or hundreds) of millions of dollars, and it is frequently hypothesized that the upper tail of the loss distribution has inflated faster than more moderate verdict or settlement sizes. While this hypothesis is intuitively appealing and widely discussed, to our best knowledge the literature has not provided a rigorous, index-based comparison of social inflation dynamics across distinct segments of the severity distribution.

The goal of this section is to quantify how social inflation in severity varies across different parts of the conditional severity distribution. We focus on three severity channels: verdict award amount (conditional on a plaintiff verdict), settlement amount (conditional on settlement), and total amount payable to the plaintiff (unconditional). The analysis is designed to answer a simple question: After controlling for case mix and litigation strategies, do higher quantiles inflate faster than central quantiles?

To avoid ambiguity, we make explicit the dependence of the severity-based indices on the quantile level $\tau$. Recall from Sections \ref{sec:method:p_s} to \ref{sec:method:t_s} that the ASIRs and cumulative CSIIs for verdict award amount, settlement amount, and total plaintiff payment are constructed using quantile regression VaR models at probability level $\tau$. We therefore write
$ASIR^{\mathrm{amt},P}_{t+1}(\tau)$ and $CSII^{\mathrm{amt},P}_{t+1}(\tau)$
for the ASIR and CSII in Equations \eqref{eq:sir_sev_P} and \eqref{eq:sii_sev_P} evaluated at quantile level $\tau$ and analogously denote
$ASIR^{\mathrm{amt},S}_{t+1}(\tau)$, $CSII^{\mathrm{amt},S}_{t+1}(\tau)$,
$ASIR^{\mathrm{amt},T}_{t+1}(\tau)$, and $CSII^{\mathrm{amt},T}_{t+1}(\tau)$
for the corresponding quantities in Equations \eqref{eq:sir_sev_S}, \eqref{eq:sii_sev_S}, \eqref{eq:sir_sev_T}, and \eqref{eq:sii_sev_T}.

We measure differential social inflation by comparing severity inflation at two risk levels, $\tau_1$ and $\tau_2$, with $\tau_1>\tau_2$ (e.g., an upper-tail quantile versus the median). The differential social inflation rate (DSIR) is defined as the difference between the two ASIRs. For verdict award amounts,
\begin{equation}
\label{eq:dsir_P}
DSIR^{\mathrm{amt},P}_{t+1}(\tau_1,\tau_2)
:= ASIR^{\mathrm{amt},P}_{t+1}(\tau_1)-ASIR^{\mathrm{amt},P}_{t+1}(\tau_2),
\end{equation}
and we define $DSIR^{\mathrm{amt},S}_{t+1}(\tau_1,\tau_2)$ and $DSIR^{\mathrm{amt},T}_{t+1}(\tau_1,\tau_2)$ analogously for settlement and total amounts. A positive DSIR indicates that the upper tail inflates faster than the lower quantile in that year, while a negative DSIR indicates the opposite. We also define a corresponding differential social inflation index (DSII) relative to the base year $t_0$ by compounding annual differential growth:
\begin{equation}
\label{eq:dsii_P}
DSII^{\mathrm{amt},P}_{t+1}(\tau_1,\tau_2)
:= \prod_{j=t_0}^{t}\left(1+DSIR^{\mathrm{amt},P}_{j+1}(\tau_1,\tau_2)\right),
\end{equation}
with analogous definitions for $DSII^{\mathrm{amt},S}_{t+1}(\tau_1,\tau_2)$ and $DSII^{\mathrm{amt},T}_{t+1}(\tau_1,\tau_2)$. In implementation, DSIR and DSII are computed under the same three specifications used throughout Section \ref{sec:result} (no covariates, factual-only, and factual-plus-strategic), so that any apparent tail-versus-center differences can be assessed both with and without case-mix adjustment.

\begin{figure}[H]
  \centering
  \begin{subfigure}[t]{0.48\textwidth}
    \centering
    \includegraphics[width=\linewidth]{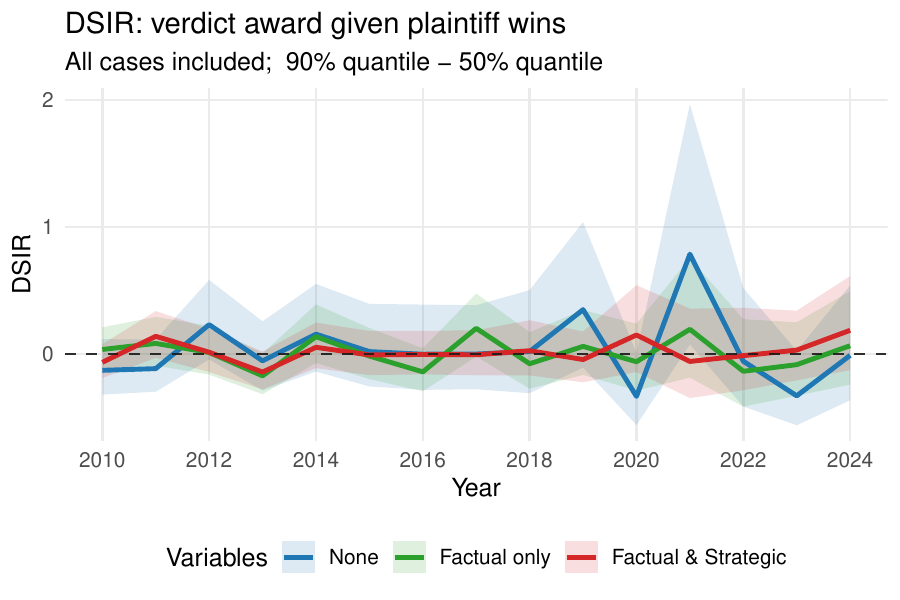}
  \end{subfigure}\hfill
  \begin{subfigure}[t]{0.48\textwidth}
    \centering
    \includegraphics[width=\linewidth]{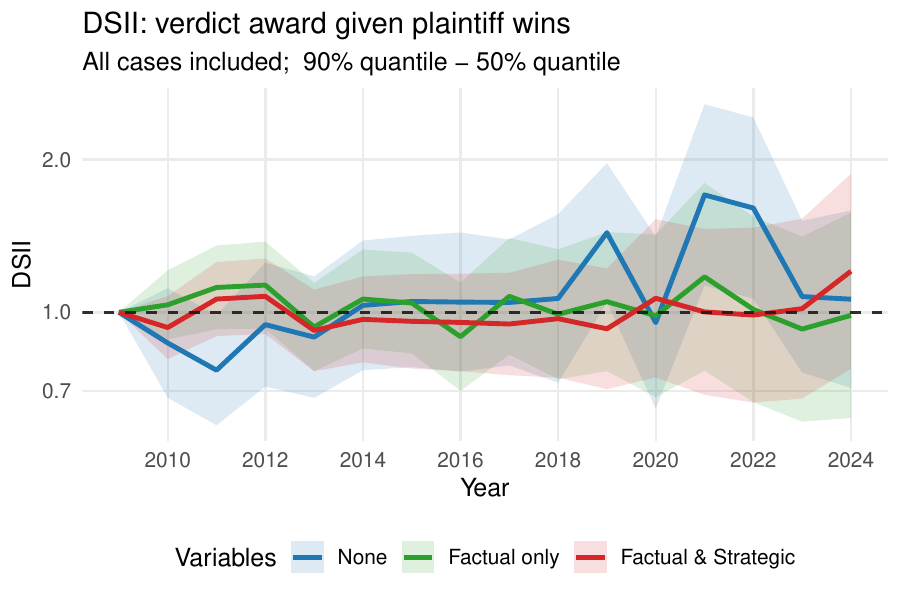}
  \end{subfigure}\hfill
    \begin{subfigure}[t]{0.48\textwidth}
    \centering
    \includegraphics[width=\linewidth]{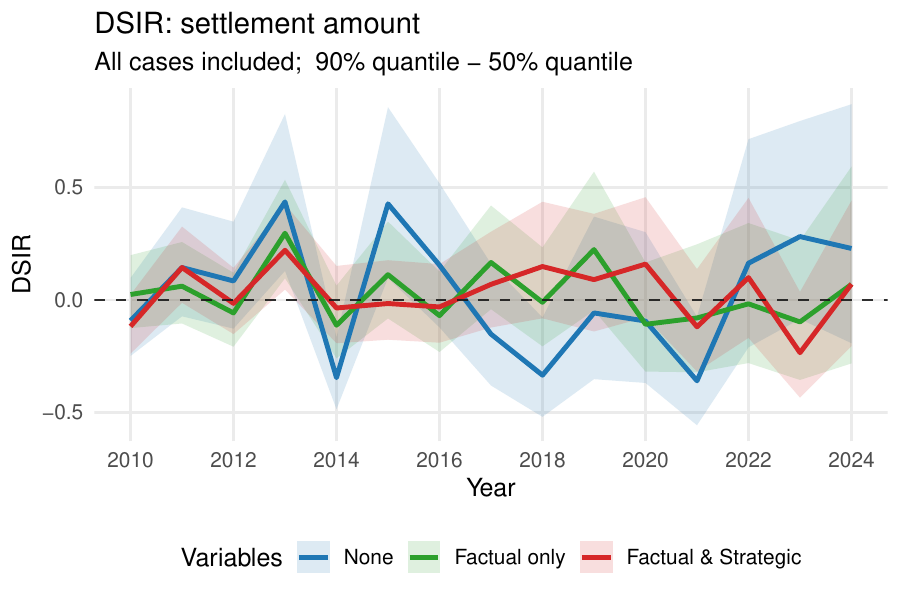}
  \end{subfigure}\hfill
  \begin{subfigure}[t]{0.48\textwidth}
    \centering
    \includegraphics[width=\linewidth]{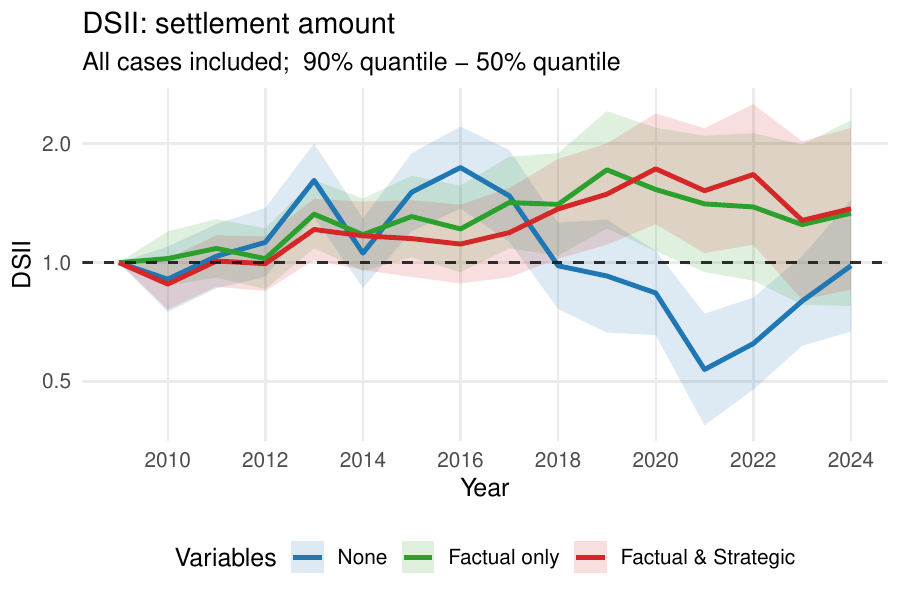}
  \end{subfigure}\hfill
    \begin{subfigure}[t]{0.48\textwidth}
    \centering
    \includegraphics[width=\linewidth]{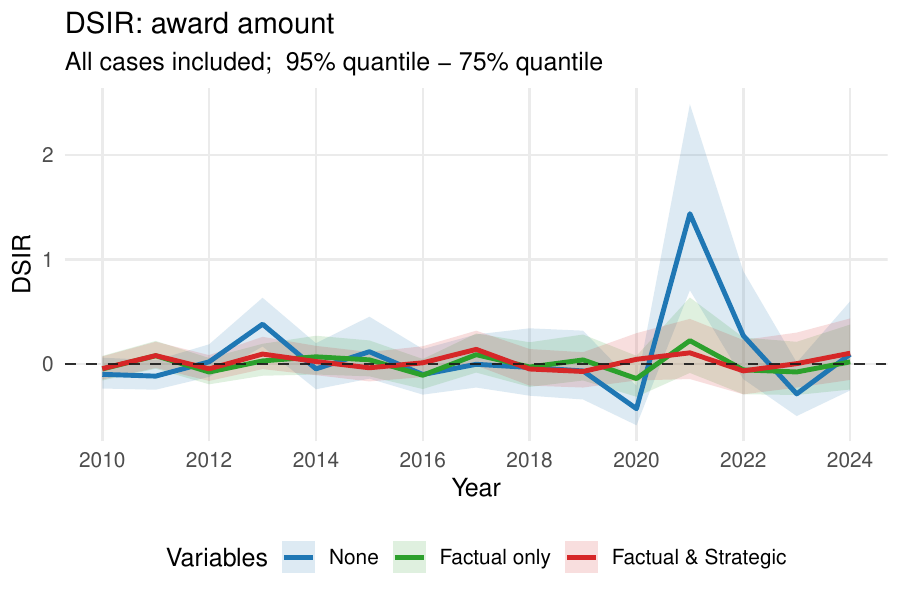}
  \end{subfigure}\hfill
  \begin{subfigure}[t]{0.48\textwidth}
    \centering
    \includegraphics[width=\linewidth]{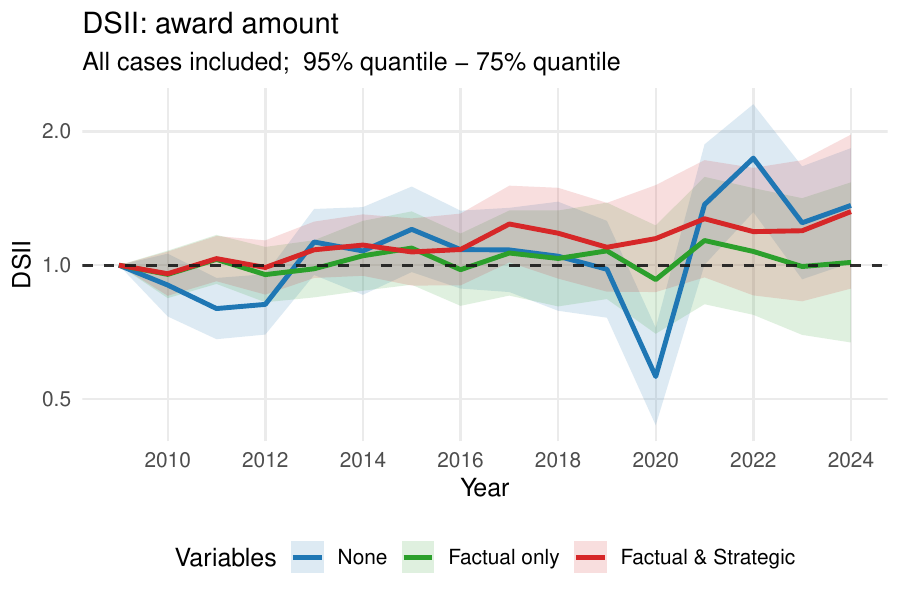}
  \end{subfigure}\hfill
  \caption{Differential annual (\textit{left panels}) and cumulative (\textit{right panels}) social inflation in verdict award amount (\textit{top panels}, with $\tau_1=0.9$ and $\tau_2=0.5$), settlement amount (\textit{middle panels}, with $\tau_1=0.9$ and $\tau_2=0.5$), and total amount payable to plaintiff (\textit{bottom panels}, with $\tau_1=0.95$ and $\tau_2=0.75$). All cases are included.}
  \label{fig:diff_sev}
\end{figure}

Figure \ref{fig:diff_sev} summarizes the evolution of DSIRs (left panels) and DSIIs (right panels) over time for all cases. We set $(\tau_1,\tau_2)=(0.9,0.5)$ for verdict award and settlement amounts and $(\tau_1,\tau_2)=(0.95,0.75)$ for total plaintiff payments. The key message is that differential social inflation is not prevalent once case mix and litigation intensity are controlled for. While the unadjusted specification occasionally produces statistically significant DSIRs/DSIIs in one or two years for settlement and total amounts, these differential effects largely disappear after adjusting for factual and/or strategic covariates: The DSIRs are statistically insignificant from 0 and the DSIIs are statistically insignificant from 1 in almost all years and across all three severity channels. Substantively, this suggests that social inflation operates more like a broad upward shift of the conditional severity distribution rather than a phenomenon concentrated exclusively in the extreme tail. In other words, the data provide little support for the hypothesis that nuclear outcomes have inflated systematically faster than moderate awards, once changes in case composition and litigation behavior are appropriately accounted for.

\section{Discussion} \label{sec:discuss}
This paper develops case-mix-adjusted social inflation indices using 74,188 US civil verdict and settlement records from VerdictSearch. By combining rolling-window logistic models for outcome probabilities with quantile regressions for severities, and by quantifying uncertainty via a random-weighted bootstrap, we separate genuine temporal effects from shifts in case composition and observable litigation intensity. We find that plaintiff win probability rises while settlement probability declines. The dominant inflation channel is verdict severity, which increases sharply after 2020 even after adjustment, whereas the increase in settlement severity is generally not as substantial. Total plaintiff-payment inflation closely tracks verdict severity. Social inflation is more pronounced in corporate-defendant and uninsured-defendant cases and in states without tort caps or TPLF regulation, and we find little evidence that the upper-tailed nuclear verdicts inflate faster than moderate awards.

A key limitation is that our analysis is grounded in ground-up litigation outcomes (verdict/settlement probabilities and amounts) rather than the claim-level payments ultimately borne by insurers after accounting for contractual terms such as deductibles, limits, attachment points, and reinsurance structures. {A related limitation is that VerdictSearch is not organized by insurers’ statutory lines of business. While motor vehicle cases provide a fairly clean proxy for auto liability, our “general liability” and “professional liability” results are based on simple keyword rules applied to case-type descriptions and should be viewed as rough LoB proxies rather than exact statutory line assignments.} Ground-up outcomes are a critical leading indicator of loss-cost pressure, but translating them into insurer-paid loss inflation requires additional information on policy design and claims handling. A promising direction is to quantify social inflation directly from claims data. One data candidate is Verisk Insurance Services Office (ISO) claims and policy data, which contain broad coverage of insurer-reported policy and claim records and have been used in recent work (e.g., Dixon et al. 2024) to examine trends in claim payments. However, access constraints often imply analyses at aggregated levels, which limits statistical power and makes tail-focused severity modeling difficult. In addition, claims data may not capture ``near-misses,'' i.e., losses below deductibles or insured retentions that are not reported to insurers. If retentions, reporting thresholds, or claims-handling practices evolve over time, the resulting selection can mask important dynamics and bias inference about claim-level social inflation. Another candidate is the S\&P Global SNL statutory filings database derived from National Association of Insurance Commissioners financial statements, which provides structured insurer state-level loss and premium measures and rich controls for economic and financial conditions. Its main limitation for our purposes is that deductibles and policy limits (and how they evolve over time by insurer and line) are not directly observed, and omitting these terms can bias estimates of claim-level social inflation. Moreover, because statutory data are inherently insurer-paid, they are also subject to the same below-retention visibility issue. Future work could seek direct policy-feature data; incorporate claim count information across reporting layers where available; or develop credible proxies for evolving deductibles, limits, and reporting thresholds.

Several follow-on research questions are particularly relevant for reinsurance pricing and reserving. First, it would be valuable to connect ground-up social inflation (from VerdictSearch) to claim-level social inflation (from claims databases or statutory filings) to quantify how changes in verdict behavior propagate into insurer-paid losses and, ultimately, reinsurance recoveries. Second, measuring and modeling case backlog---the lag from accident date to case deliberation date---and its evolution during and after the COVID-19 pandemic could clarify how court delays shift cases between settlement and verdict, with direct implications for reported but not settled (RBNS) and incurred but not reported (IBNR) dynamics and developments. Third, future work should quantify insurers' behavioral responses to social inflation (premium increases, studied by Oh (2022), underwriting tightening, and changes in deductibles and/or limits), leveraging the premium and exposure information in statutory filings and evaluating how these responses interact with observed litigation outcomes over time.

\section*{Acknowledgments}
This research was funded by the Casualty Actuarial Society (CAS) through its research program on quantifying the reinsurance costs of social inflation. The authors gratefully acknowledge this support.

\clearpage
\section*{References}
\begin{singlespace}
\begin{description}

\item Bergen, Tod. 2020. ``Social Inflation: What Is It? What Causes It? Why Should You Care?" \textit{McConkey Insurance \& Benefits}, November 9.  \href{https://www.ekmcconkey.com/blog/social-inflation-what-is-it-what-causes-it-why-should-you-care/}{https://www.ekmcconkey.com/blog/social-inflation-what-is-it-what-causes-it-why-should-you-care/}.

\item Boerlin, Martin, James Finucane, Thomas Holzheu, et al. 2024. ``Social Inflation: Litigation Costs Drive Claims Inflation."  Swiss Re Institute. \href{https://www.swissre.com/institute/research/sigma-research/sigma-2024-04-social-inflation.html}{https://www.swissre.com/institute/research/sigma-research/sigma-2024-04-social-inflation.html}.

\item Buffett, Warren. 1978. ``Chairman’s letter---1977." Berkshire Hathaway. \href{https://www.berkshirehathaway.com/letters/1977.html}{https://www.berkshirehathaway.com/letters/1977.html}.

\item Dickinson, Andrea, and Meg Sutton. 2020. ``The Ripple Effect of Social Inflation and Nuclear Verdicts on the Insurance Industry." \textit{Amwins}, December 8. 
\href{https://www.amwins.com/resources-and-insights/market-insights/article/the-ripple-effect-of-social-inflation-and-nuclear-verdicts-on-the-insurance-industry}{https://www.amwins.com/resources-and-insights/market-insights/article/the-ripple-effect-of-social-inflation-and-nuclear-verdicts-on-the-insurance-industry}.

\item Dixon, Lloyd, Nicholas M. Pace, James Davidson, and Jamie Morikawa. 2024. ``What Is the Evidence for Social Inflation? Trends in Trial Awards and Insurance Claim Payments." \textit{RAND Corporation}, July 9. \href{https://www.rand.org/pubs/research_reports/RRA2645-1.html}{https://www.rand.org/pubs/research\_reports/RRA2645-1.html}.

\item Djazayeri, Alexander. 2020. ``Social Inflation: An Emerging Risk for Corporations." \textit{HDI Global}, June 24.  
\href{https://www.hdi.global/infocenter/insights/2020/social-inflation/}{https://www.hdi.global/infocenter/insights/2020/social-inflation/}.

\item Francis, Louise. 2024. ``Understanding and Coping with the “New Normal” in Insurance Inflation." \textit{CAS E-Forum} Summer (July). \href{https://eforum.casact.org/article/121372-understanding-and-coping-with-the-new-normal-in-insurance-inflation}{https://eforum.casact.org/article/121372-understanding-and-coping-with-the-new-normal-in-insurance-inflation}.

\item Frees, Edward W., Richard A. Derrig, and Glenn Meyers. 2014. \textit{Predictive Modeling Applications in Actuarial Science.} Cambridge University Press. \href{https://doi.org/10.1017/CBO9781139342674}{https://doi.org/10.1017/CBO9781139342674}.

\item Insurance Information Institute. 2022. ``Social Inflation: What It Is and Why It Matters." \textit{Trends and Insights}, February. \href{https://www.iii.org/sites/default/files/docs/pdf/triple-i_state_of_the_risk_social_inflation_02082022.pdf}{https://www.iii.org/sites/default/files/docs/pdf/triple-i\_state\_of\_the\_risk\_social\_inflation\_02082022.pdf}.

\item Lynch, Jim, and Dave Moore. 2022. ``Social Inflation and Loss Development---An Update." Casualty Actuarial Society, Insurance Information Institute. \href{https://www.casact.org/sites/default/files/2023-03/RP_Social_Inflation_Update.pdf}{https://www.casact.org/sites/default/files/2023-03/RP\_Social\_Inflation\_Update.pdf}.

\item Mackeprang, Christopher. 2020. ``Quantifying Social Inflation---Jury Awards, Income Inequality, and the Bronx Jury Hypothesis." \textit{Gen Re}, September 24. \href{https://www.genre.com/us/knowledge/publications/2020/september/quantifying-social-inflation-jury-awards-income-inequality-and-the-bronx-jury-hypothesis-en}{https://www.genre.com/us/knowledge/publications/2020/september/quantifying-social-inflation-jury-awards-income-inequality-and-the-bronx-jury-hypothesis-en}.

\item McKnight, David, and Paul Hinton. 2024. \textit{Tort Costs in America: An Empirical Analysis of Costs and Compensation in the U.S. Tort System}. 3rd ed. U.S. Chamber of Commerce, Institute for Legal Reform. 
\href{https://instituteforlegalreform.com/research/tort-costs-in-america-an-empirical-analysis-of-costs-and-compensation-in-the-u-s-tort-system}{https://instituteforlegalreform.com/research/tort-costs-in-america-an-empirical-analysis-of-costs-and-compensation-in-the-u-s-tort-system}. 
\item Moorcraft, Bethan. 2020. ``What Is Social Inflation, and Why Is It Hurting Insurance?' \textit{Insurance Business America}, January 3. \href{https://www.insurancebusinessmag.com/us/news/breaking-news/what-is-social-inflation-and-why-is-it-hurting-insurance-195626.aspx}{https://www.insurancebusinessmag.com/us/news/breaking-news/what-is-social-inflation-and-why-is-it-hurting-insurance-195626.aspx}.

\item NAIC (National Association of Insurance Commissioners). 2025. ``Social Inflation." Last updated December 19. \href{https://content.naic.org/insurance-topics/social-inflation}{https://content.naic.org/insurance-topics/social-inflation}.

\item Oh, Sangmin. 2022. ``\textit{Social Inflation}." SSRN. \href{https://doi.org/10.2139/ssrn.3685667}{https://doi.org/10.2139/ssrn.3685667}.

\item Travelers. 2023. ``What’s Driving Huge Jury Awards? Navigating Legal Liability in the Era of the Nuclear Verdict."  \href{https://www.travelers.com/resources/business-topics/top-100-verdicts/whats-driving-huge-jury-awards}{https://www.travelers.com/resources/business-topics/top-100-verdicts/whats-driving-huge-jury-awards}.

\item Wellington, Elizabeth. 2023. ``A Trend Model for Social Inflation in Medical Professional Liability." \textit{CAS E-Forum} Spring (May). \href{https://eforum.casact.org/article/74847-a-trend-model-for-social-inflation-in-medical-professional-liability}{https://eforum.casact.org/article/74847-a-trend-model-for-social-inflation-in-medical-professional-liability}.  

\end{description}
\end{singlespace}

\clearpage
\newpage
\begin{appendices}
\renewcommand{\thetable}{\thesection.\arabic{table}}
\renewcommand{\thefigure}{\thesection.\arabic{figure}}
\section{Additional tables: Social inflation by case characteristics} \label{apx:table}
\setcounter{table}{0}
\setcounter{figure}{0}
This appendix provides additional detailed summary tables of social inflation for total verdict/settlement amounts, reported separately for key case characteristics: (1) corporate versus individual defendants (Table \ref{apx:tab:summary_corp}); (2) insured versus uninsured defendants (Table \ref{apx:tab:summary_ins}); (3) liability category (motor vehicle, general liability, professional liability, and others) (Table \ref{apx:tab:summary_liab}); (4) tort-cap versus non-tort-cap states (Table \ref{apx:tab:summary_tort}); (5) states with versus without TPLF regulation (Table \ref{apx:tab:summary_tplf}); and (6) jury versus bench trials (Table \ref{apx:tab:summary_trial}).

\begin{table}[!htbp]
\centering
\small
\caption{Headline (case-mix-adjusted) social inflation summary for total verdict/settlement amounts for cases with corporate versus individual defendants. Values are extracted from the green lines in Figure \ref{fig:idx_sev_t1_1}. CSII is normalized to 100 in 2009.}
\label{apx:tab:summary_corp}
\begin{tabular}{lrrrr}
\toprule
& \multicolumn{2}{c}{Corporate defendants} & \multicolumn{2}{c}{Individual defendants} \\
\cmidrule(lr){2-3}\cmidrule(lr){4-5}
Year & ASIR (\%) & CSII & ASIR (\%) & CSII \\
\midrule
2010 &  11.0 & 111.0 &  11.0 & 111.0 \\
2011 &  --1.9& 108.9 &   0.2 & 111.2 \\
2012 &  --3.0& 105.6 &  22.4 & 136.2 \\
2013 & --30.1&  73.9 & --33.7&  90.3 \\
2014 &  14.7 &  84.7 &   0.4 &  90.7 \\
2015 &  --3.5&  81.8 &  --1.1&  89.7 \\
2016 &  --3.6&  78.8 &   7.5 &  96.5 \\
2017 &  --2.2&  77.1 &   1.5 &  97.9 \\
2018 &  12.9 &  87.1 &  12.5 & 110.2 \\
2019 &   2.2 &  89.0 &   2.8 & 113.3 \\
2020 & --15.8&  74.9 &   8.0 & 122.4 \\
2021 &  77.0 & 132.6 &  20.2 & 147.1 \\
2022 &  69.2 & 224.3 &   7.3 & 157.9 \\
2023 &   4.4 & 234.2 &  59.0 & 251.0 \\
2024 &  32.9 & 311.2 &  16.8 & 293.2 \\
\bottomrule
\end{tabular}
\end{table}

\begin{table}[!htbp]
\centering
\small
\caption{Headline (case-mix-adjusted) social inflation summary for total verdict/settlement amounts for cases with insured versus uninsured defendants. Values are extracted from the green lines in Figure \ref{fig:idx_sev_t1_2}. CSII is normalized to 100 in 2009.}
\label{apx:tab:summary_ins}
\begin{tabular}{lrrrr}
\toprule
& \multicolumn{2}{c}{Insured defendants} & \multicolumn{2}{c}{Uninsured defendants} \\
\cmidrule(lr){2-3}\cmidrule(lr){4-5}
Year & ASIR (\%) & CSII & ASIR (\%) & CSII \\
\midrule
2010 &  11.9 & 111.9 &  13.0 & 113.0 \\
2011 & --11.7&  98.8 &  10.7 & 125.1 \\
2012 &   3.9 & 102.6 &  12.5 & 140.7 \\
2013 & --20.6&  81.5 & --40.3&  84.0 \\
2014 &   7.9 &  87.9 & --11.0&  74.7 \\
2015 &   2.6 &  90.2 &   1.1 &  75.6 \\
2016 &  --9.9&  81.3 &  19.8 &  90.5 \\
2017 &   2.2 &  83.1 & --12.1&  79.6 \\
2018 &  20.0 &  99.7 &  22.4 &  97.5 \\
2019 &   0.9 & 100.6 &   1.4 &  98.8 \\
2020 &  --3.3&  97.2 &  --6.8&  92.1 \\
2021 &  27.8 & 124.2 & 100.8 & 184.9 \\
2022 &  25.9 & 156.4 & 115.9 & 399.2 \\
2023 &  14.5 & 179.1 & --11.4& 353.9 \\
2024 &  13.4 & 203.2 &  40.7 & 497.8 \\
\bottomrule
\end{tabular}
\end{table}

\begin{table}[!htbp]
\centering
\small
\caption{Headline (case-mix-adjusted) social inflation summary for total verdict/settlement amounts by liability category. Values are extracted from the green lines in Figure \ref{fig:idx_sev_t1_3}. CSII is normalized to 100 in 2009.}
\label{apx:tab:summary_liab}
\begin{tabular}{lrrrrrrrr}
\toprule
& \multicolumn{2}{c}{Motor vehicle} & \multicolumn{2}{c}{General liability} & \multicolumn{2}{c}{Professional liability} & \multicolumn{2}{c}{Others} \\
\cmidrule(lr){2-3}\cmidrule(lr){4-5}\cmidrule(lr){6-7}\cmidrule(lr){8-9}
Year & ASIR (\%) & CSII & ASIR (\%) & CSII & ASIR (\%) & CSII & ASIR (\%) & CSII \\
\midrule
2010 &  11.7 & 111.7 &  19.5 & 119.5 &  26.2 & 126.2 &  12.3 & 112.3 \\
2011 &  --5.0& 106.1 &   9.9 & 131.4 &  --9.4& 114.3 &   1.1 & 113.5 \\
2012 &  10.2 & 117.0 &   6.5 & 139.9 &  20.3 & 137.5 &  18.1 & 134.1 \\
2013 & --31.0&  80.7 & --35.6&  90.1 & --48.2&  71.2 & --18.4& 109.5 \\
2014 &   0.8 &  81.4 &  --2.2&  88.1 & --26.7&  52.1 &   3.8 & 113.6 \\
2015 &   2.9 &  83.7 &  17.9 & 103.9 &  12.0 &  58.4 &  --8.9& 103.5 \\
2016 &  --4.3&  80.2 &   4.2 & 108.2 &  18.3 &  69.1 & --12.0&  91.1 \\
2017 &  --0.3&  79.9 &  --2.4& 105.6 & --30.0&  48.4 &  21.0 & 110.2 \\
2018 &  14.7 &  91.6 &  30.8 & 138.1 &  24.6 &  60.2 &  47.8 & 162.9 \\
2019 &  --1.8&  90.0 &   9.9 & 151.8 &  --4.1&  57.8 &  --4.5& 155.5 \\
2020 &  12.8 & 101.5 &  --3.1& 147.1 & --51.1&  28.2 & --22.9& 120.0 \\
2021 &  21.7 & 123.5 &   7.6 & 158.3 & 138.7 &  67.3 & 220.0 & 384.0 \\
2022 &   3.5 & 127.8 &  81.1 & 286.8 & 173.6 & 184.2 & 102.0 & 775.8 \\
2023 &  34.3 & 171.7 &  16.6 & 334.3 & --25.4& 137.4 &  12.3 & 871.4 \\
2024 &  10.0 & 188.9 &  20.4 & 402.4 &   2.6 & 141.0 &  70.1 & 1,482.1\\
\bottomrule
\end{tabular}
\end{table}

\begin{table}[!htbp]
\centering
\small
\caption{Headline (case-mix-adjusted) social inflation summary for total verdict/settlement amounts for tort-cap versus non-tort-cap states. Values are extracted from the green lines in Figure \ref{fig:idx_sev_t1_4}. CSII is normalized to 100 in 2009.}
\label{apx:tab:summary_tort}
\begin{tabular}{lrrrr}
\toprule
& \multicolumn{2}{c}{Tort-cap states} & \multicolumn{2}{c}{Non-tort-cap states} \\
\cmidrule(lr){2-3}\cmidrule(lr){4-5}
Year & ASIR (\%) & CSII & ASIR (\%) & CSII \\
\midrule
2010 &  14.6 & 114.6 &  12.6 & 112.6 \\
2011 &   3.1 & 118.1 & --12.6&  98.5 \\
2012 &  27.2 & 150.3 &  --9.5&  89.1 \\
2013 & --44.6&  83.2 &  --9.2&  80.9 \\
2014 & --11.8&  73.4 &   9.6 &  88.7 \\
2015 &   0.4 &  73.6 &  --4.1&  85.1 \\
2016 &  --2.7&  71.7 &   2.5 &  87.2 \\
2017 &   5.2 &  75.4 &  --7.5&  80.7 \\
2018 &   2.6 &  77.4 &  36.2 & 109.9 \\
2019 &   4.8 &  81.1 &   0.8 & 110.8 \\
2020 &  20.8 &  97.9 & --18.6&  90.2 \\
2021 &  27.9 & 125.3 &  67.3 & 150.9 \\
2022 &  43.1 & 179.3 &  56.2 & 235.8 \\
2023 &  11.7 & 200.3 &  15.9 & 273.4 \\
2024 &  25.0 & 250.4 &  56.5 & 427.9 \\
\bottomrule
\end{tabular}
\end{table}

\begin{table}[!htbp]
\centering
\small
\caption{Headline (case-mix-adjusted) social inflation summary for total verdict/settlement amounts for states with versus without TPLF regulation. Values are extracted from the green lines in Figure \ref{fig:idx_sev_t1_5}. CSII is normalized to 100 in 2009.}
\label{apx:tab:summary_tplf}
\begin{tabular}{lrrrr}
\toprule
& \multicolumn{2}{c}{States with TPLF regulation} & \multicolumn{2}{c}{States with no TPLF regulation} \\
\cmidrule(lr){2-3}\cmidrule(lr){4-5}
Year & ASIR (\%) & CSII & ASIR (\%) & CSII \\
\midrule
2010 & --22.4&  77.6 &  18.2 & 118.2 \\
2011 & --27.0&  56.7 &   0.8 & 119.1 \\
2012 &  28.2 &  72.6 &  11.4 & 132.7 \\
2013 & --32.8&  48.8 & --33.4&  88.4 \\
2014 & --16.1&  41.0 &   2.8 &  90.9 \\
2015 &  25.7 &  51.5 &  --5.2&  86.2 \\
2016 & --17.6&  42.4 &   2.8 &  88.7 \\
2017 &  --9.5&  38.4 &   3.8 &  92.1 \\
2018 & --16.4&  32.1 &  23.0 & 113.2 \\
2019 & --21.8&  25.1 &   1.8 & 115.2 \\
2020 & 125.1 &  56.5 &  --3.4& 111.4 \\
2021 &   0.0 &  56.5 &  55.8 & 173.5 \\
2022 & 115.0 & 121.5 &  37.3 & 238.1 \\
2023 &  39.2 & 169.1 &   9.6 & 261.1 \\
2024 &  41.3 & 238.9 &  31.4 & 343.1 \\
\bottomrule
\end{tabular}
\end{table}

\begin{table}[!htbp]
\centering
\small
\caption{Headline (case-mix-adjusted) social inflation summary for total verdict/settlement amounts for jury versus bench trials. Values are extracted from the green lines in Figure \ref{fig:idx_sev_t1_6}. CSII is normalized to 100 in 2009.}
\label{apx:tab:summary_trial}
\begin{tabular}{lrrrr}
\toprule
& \multicolumn{2}{c}{Jury trials} & \multicolumn{2}{c}{Bench trials} \\
\cmidrule(lr){2-3}\cmidrule(lr){4-5}
Year & ASIR (\%) & CSII & ASIR (\%) & CSII \\
\midrule
2010 &  14.2 & 114.2 &  10.3 & 110.3 \\
2011 & --13.7&  98.5 &   6.3 & 117.2 \\
2012 &  20.9 & 119.1 &  --3.5& 113.1 \\
2013 & --15.7& 100.4 & --38.9&  69.1 \\
2014 &   8.3 & 108.7 &  --7.3&  64.1 \\
2015 & --14.1&  93.4 &  17.6 &  75.4 \\
2016 &   7.1 & 100.1 & --15.5&  63.7 \\
2017 &  11.0 & 111.0 &  --8.7&  58.1 \\
2018 &  22.1 & 135.6 &  13.7 &  66.1 \\
2019 &   0.0 & 135.6 &   5.0 &  69.4 \\
2020 & --13.9& 116.8 &   9.8 &  76.2 \\
2021 &  38.8 & 162.1 &  50.4 & 114.6 \\
2022 &  55.6 & 252.2 &  43.2 & 164.1 \\
2023 &  21.8 & 307.2 &  --4.5& 156.7 \\
2024 &  34.3 & 412.7 &  49.5 & 234.2 \\
\bottomrule
\end{tabular}
\end{table}

\clearpage

\section{Synthetic dataset illustration}\label{apx:sec:sim}
\setcounter{table}{0}
\setcounter{figure}{0}
This appendix empirically demonstrates, using a controlled synthetic experiment, that the statistical framework proposed in Section \ref{sec:method} can (1) recover social inflation indices over a 15-year horizon and (2) quantify the associated estimation uncertainty via 95\% confidence intervals. The synthetic setting is constructed so that the true (data-generating) social inflation is known. This allows us to verify that the proposed case-mix-adjusted indices track the ground truth closely and to illustrate why adjusting for evolving case mix (covariates) is necessary in order to avoid overestimated inflation in naive, unadjusted trend measures.

We generate a synthetic dataset over $15$ years, with $5,000$ cases in each calendar year $t=1,\ldots,15$. For each case $i\in\mathcal{A}_t$ (i.e., case $i$ observed in year $t$), the covariate $\bm{X}_i$ is univariate and follows a normal distribution with mean $0.05(t-1)$ and standard deviation $0.5$. This construction creates an explicitly time-varying case mix: The distribution of $\bm{X}_i$ shifts upward over time. Intuitively, $\bm{X}_i$ represents an observable case characteristic whose prevalence changes across years.

Plaintiff winning outcomes are generated from a logistic regression governed by Equation~\ref{eq:logit_model_P}, with intercept $\alpha_t^P=-0.8+0.1t$ and slope coefficient $\bm{\beta}_t^P=1$. Because $\alpha_t^P$ increases with $t$, the data-generating plaintiff winning odds rise over time, implying true social inflation in plaintiff win probability. The resulting ground-truth ASIRs and CSIIs for plaintiff win probability are shown in the last two columns of Table \ref{apx:tab:sim_prob}. For simplicity, we assume there are no settlement cases in this illustration, so each case resolves by verdict and either the plaintiff or the defendant wins.

Conditional on a plaintiff win for case $i\in\mathcal{A}_t$, we generate the corresponding (economically inflation-adjusted) verdict award amount $Y_i$ from a Pareto Type II (Lomax) distribution with density $f_{Y_i}(y;\theta_i,\gamma)=\gamma\theta_i^{\gamma}/(y+\theta_i)^{\gamma+1}$ for $y>0$. We set the shape parameter to $\gamma=5$ and specify the scale parameter as $\theta_i=4\exp(\bm{X}_i)\times CSII_{t}^{\text{amt},P}$, where $CSII_{t}^{\text{amt},P}$ is the ground-truth cumulative index governing verdict award severity and is reported as the last column of Table \ref{apx:tab:sim_sev}. The corresponding ground-truth ASIR for verdict award severity is reported in the second-to-last column of Table \ref{apx:tab:sim_sev}. This construction induces a heavy-tailed severity distribution and embeds a known time trend in the conditional severity level while also allowing the covariate $\bm{X}_i$ to shift the conditional severity in a multiplicative manner.

\begin{figure}[H]
  \centering
  \begin{subfigure}[t]{0.48\textwidth}
    \centering
    \includegraphics[width=\linewidth]{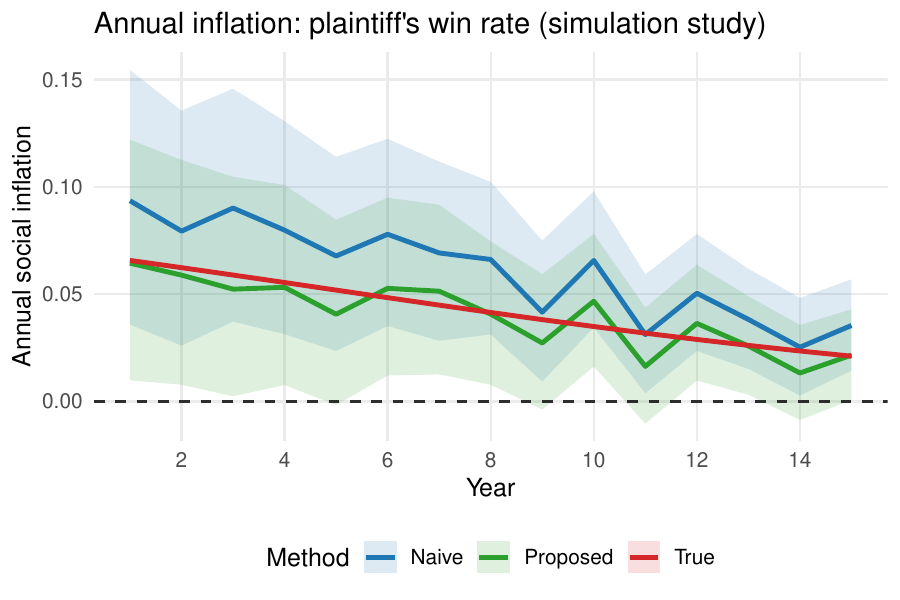}
  \end{subfigure}\hfill
  \begin{subfigure}[t]{0.48\textwidth}
    \centering
    \includegraphics[width=\linewidth]{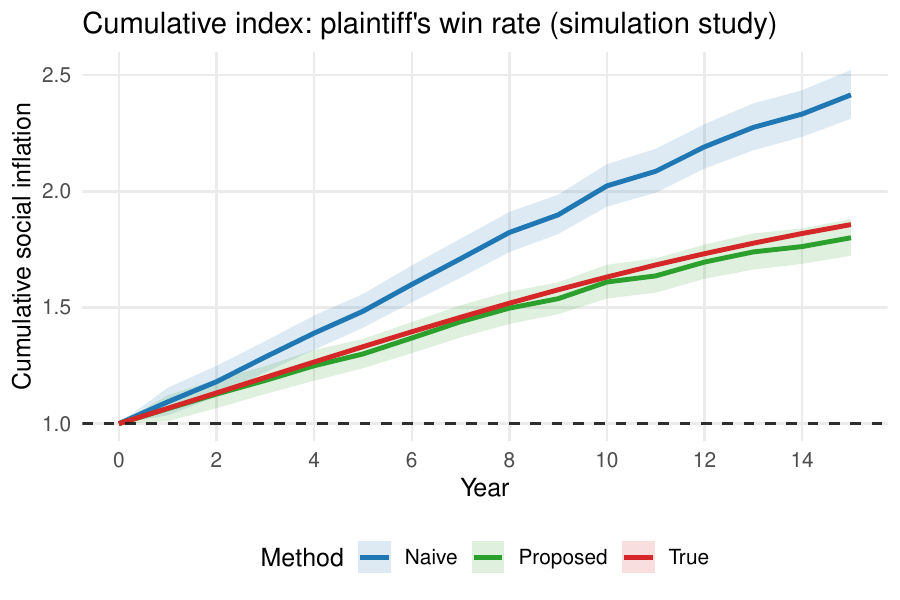}
  \end{subfigure}\hfill
  \caption{Synthetic dataset illustration: Annual (\textit{left panel}) and cumulative (\textit{right panel}) social inflation in plaintiff win probability.}
  \label{apx:fig:sim_prob}
\end{figure}

\begin{figure}[H]
  \centering
  \begin{subfigure}[t]{0.48\textwidth}
    \centering
    \includegraphics[width=\linewidth]{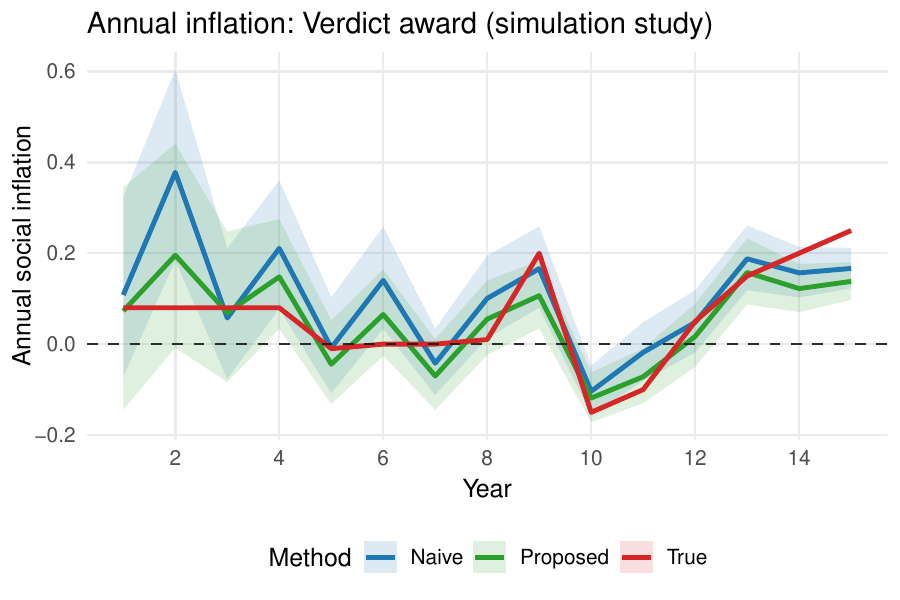}
  \end{subfigure}\hfill
  \begin{subfigure}[t]{0.48\textwidth}
    \centering
    \includegraphics[width=\linewidth]{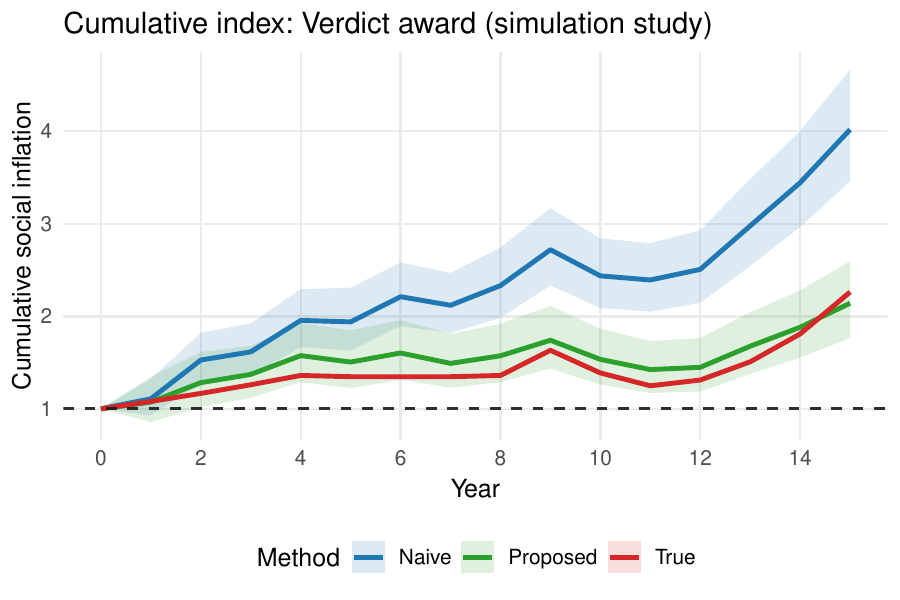}
  \end{subfigure}\hfill
  \caption{Synthetic dataset illustration: Annual (\textit{left panel}) and cumulative (\textit{right panel}) social inflation in verdict award severity (at 75\% quantile level).}
  \label{apx:fig:sim_sev}
\end{figure}

We then apply the estimation procedure proposed in Section \ref{sec:method} to estimate ASIRs and CSIIs for both plaintiff win probability and verdict award severity for each year $t=1,\ldots,15$, and we compute 95\% confidence intervals via the random-weighted bootstrap. We consider two estimation settings. In the naive setting, we omit $\bm{X}_i$ from the fitted models and therefore do not adjust for the evolving case mix. In the proposed setting, we include $\bm{X}_i$ in the fitted models, yielding case-mix-adjusted social inflation indices. Figures \ref{apx:fig:sim_prob} and \ref{apx:fig:sim_sev} compare the estimated ASIRs and CSIIs under the naive approach (blue) and the proposed approach (green) against the ground-truth indices implied by the data-generating process (red). The shaded regions display the bootstrap-based 95\% confidence intervals for the estimated ASIR/CSII series.

Two main patterns emerge. First, when the covariate is ignored, the naive indices substantially overstate social inflation for both plaintiff win probability and verdict award severity. This overstatement occurs because the upward drift in case mix (increasing $\bm{X}_i$ over time) is mechanically absorbed into the estimated time trend, and $\bm{X}_i$ is positively associated with both the probability of plaintiff success and the amount of awards. In other words, the naive approach conflates a change in what cases look like with a change in outcome dynamics over time. Second, once we adjust for case mix by including $\bm{X}_i$, the proposed indices track the ground truth closely: The estimated green curves in Figures \ref{apx:fig:sim_prob} and \ref{apx:fig:sim_sev} align well with the red curves, and the ground-truth series generally falls within the 95\% confidence intervals. Overall, this synthetic study supports the accuracy of the proposed methodology under a controlled setting and reinforces the importance of case-mix adjustment when quantifying social inflation from litigation outcome data.

\begin{table}[!htbp]
\centering
\small
\caption{Synthetic dataset illustration: Social inflation summary for plaintiff win probability. The table compares (1) a naive index without covariate control, (2) the proposed case-mix-adjusted index that controls for the covariate, and (3) the ground-truth social inflation implied by the data-generating process. CSII is normalized to 100 at year 0 (baseline).}
\label{apx:tab:sim_prob}
\begin{tabular}{lrrrrrr}
\toprule
& \multicolumn{2}{c}{Naive} & \multicolumn{2}{c}{Proposed} & \multicolumn{2}{c}{Ground-truth} \\
\cmidrule(lr){2-3}\cmidrule(lr){4-5}\cmidrule(lr){6-7}
Year & ASIR (\%) & CSII & ASIR (\%) & CSII & ASIR (\%) & CSII \\
\midrule
1  &  9.4 & 109.4 &  6.4 & 106.4 &  6.6 & 106.6 \\
2  &  7.9 & 118.0 &  5.9 & 112.7 &  6.2 & 113.2 \\
3  &  9.0 & 128.7 &  5.2 & 118.6 &  5.9 & 119.9 \\
4  &  8.0 & 138.9 &  5.3 & 124.9 &  5.5 & 126.5 \\
5  &  6.8 & 148.3 &  4.1 & 130.0 &  5.2 & 133.1 \\
6  &  7.8 & 159.9 &  5.3 & 136.8 &  4.8 & 139.5 \\
7  &  6.9 & 171.0 &  5.1 & 143.9 &  4.5 & 145.8 \\
8  &  6.6 & 182.3 &  4.1 & 149.7 &  4.1 & 151.8 \\
9  &  4.2 & 189.9 &  2.7 & 153.8 &  3.8 & 157.6 \\
10 &  6.6 & 202.3 &  4.7 & 161.0 &  3.5 & 163.1 \\
11 &  3.1 & 208.6 &  1.6 & 163.6 &  3.2 & 168.3 \\
12 &  5.0 & 219.1 &  3.6 & 169.5 &  2.9 & 173.1 \\
13 &  3.8 & 227.5 &  2.6 & 173.9 &  2.6 & 177.7 \\
14 &  2.5 & 233.3 &  1.3 & 176.2 &  2.3 & 181.8 \\
15 &  3.5 & 241.5 &  2.2 & 180.0 &  2.1 & 185.7 \\
\bottomrule
\end{tabular}
\end{table}

\begin{table}[!htbp]
\centering
\small
\caption{Synthetic dataset illustration: Social inflation summary for verdict award severity (award amount conditional on a positive plaintiff award). The table compares (1) a naive index without covariate control, (2) the proposed case-mix-adjusted index that controls for the covariate, and (3) the ground-truth social inflation implied by the data-generating process. CSII is normalized to 100 at year 0 (baseline).}
\label{apx:tab:sim_sev}
\begin{tabular}{lrrrrrr}
\toprule
& \multicolumn{2}{c}{Naive} & \multicolumn{2}{c}{Proposed} & \multicolumn{2}{c}{Ground-truth} \\
\cmidrule(lr){2-3}\cmidrule(lr){4-5}\cmidrule(lr){6-7}
Year & ASIR (\%) & CSII & ASIR (\%) & CSII & ASIR (\%) & CSII \\
\midrule
1  &  10.8 & 110.8 &   7.3 & 107.3 &   8.0 & 108.0 \\
2  &  37.8 & 152.7 &  19.6 & 128.3 &   8.0 & 116.6 \\
3  &   5.7 & 161.5 &   6.9 & 137.2 &   8.0 & 126.0 \\
4  &  21.1 & 195.5 &  14.8 & 157.5 &   8.0 & 136.0 \\
5  &  --0.8& 193.9 &  --4.4& 150.5 &  --1.0& 134.7 \\
6  &  14.0 & 221.1 &   6.5 & 160.3 &   0.0 & 134.7 \\
7  &  --4.2& 211.9 &  --7.0& 149.1 &   0.0 & 134.7 \\
8  &  10.1 & 233.2 &   5.5 & 157.3 &   1.0 & 136.0 \\
9  &  16.6 & 271.9 &  10.7 & 174.1 &  20.0 & 163.2 \\
10 & --10.4& 243.7 & --11.9& 153.4 & --15.0& 138.8 \\
11 &  --1.8& 239.2 &  --7.2& 142.4 & --10.0& 124.9 \\
12 &   4.8 & 250.7 &   1.7 & 144.8 &   5.0 & 131.1 \\
13 &  18.8 & 297.7 &  15.8 & 167.6 &  15.0 & 150.8 \\
14 &  15.7 & 344.4 &  12.2 & 188.1 &  20.0 & 181.0 \\
15 &  16.7 & 401.8 &  13.8 & 214.1 &  25.0 & 226.2 \\
\bottomrule
\end{tabular}
\end{table}

\section{Robustness check of rolling-window approach} \label{apx:sec:robust}
\setcounter{table}{0}
\setcounter{figure}{0}
Our methodology described in Section \ref{sec:method} employs a rolling-window design to improve the stability of the estimated social inflation indices, particularly when annual sample sizes are smaller in the later part of the sample. In the main VerdictSearch results presented in Section \ref{sec:result}, we use a five-year window. This appendix evaluates the sensitivity of the key findings to that choice by repeating the analysis using a six-year window.

\begin{figure}[H]
  \centering
  \begin{subfigure}[t]{0.48\textwidth}
    \centering
    \includegraphics[width=\linewidth]{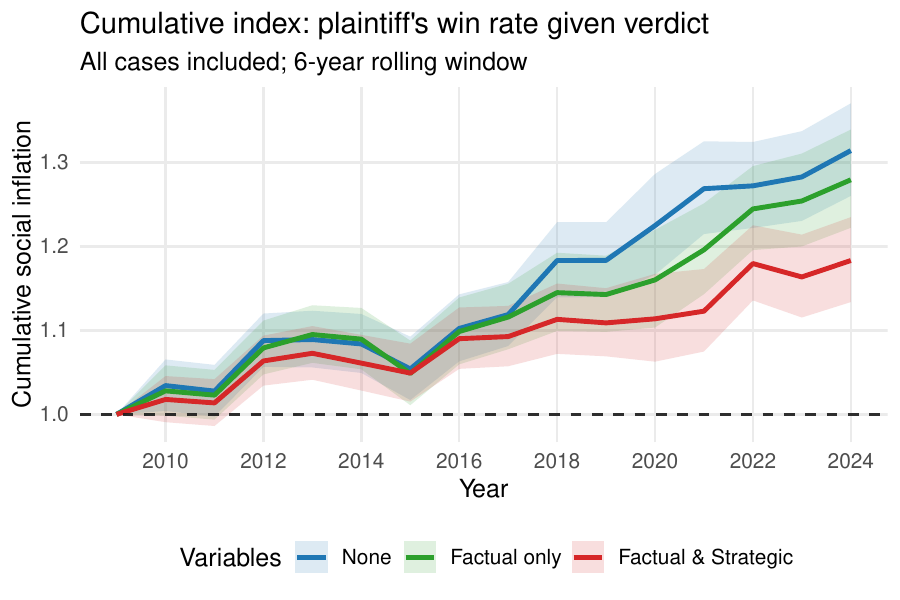}
  \end{subfigure}\hfill
  \begin{subfigure}[t]{0.48\textwidth}
    \centering
    \includegraphics[width=\linewidth]{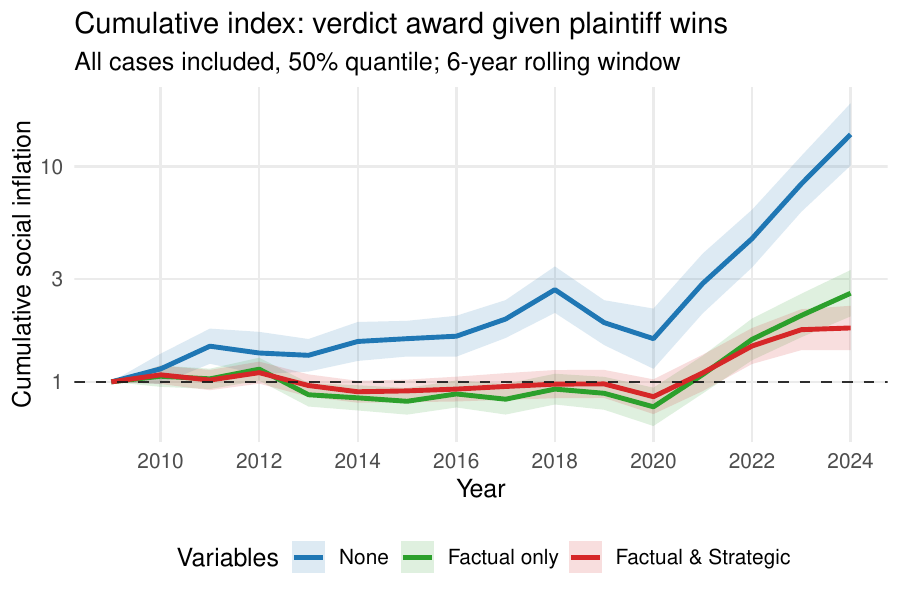}
  \end{subfigure}\hfill
  \caption{Cumulative CSIIs for plaintiff win probability (\textit{left panel}) and verdict award amount (\textit{right panel}; at 50\% quantile level) under a 6-year window length. All cases are included.}
  \label{apx:fig:idx_robust}
\end{figure}

Figure \ref{apx:fig:idx_robust} reports the resulting CSIIs for plaintiff win probability (left panel) and verdict award severity (right panel; at the 50\% quantile). These panels correspond to the cumulative-index panels of Figures \ref{fig:idx_prob_p_0} and \ref{fig:idx_sev_p1_0}, respectively, with the only change being the window length (six years instead of five).

The CSIIs are extremely similar across the two window lengths: The six-year-window curves closely track the five-year-window curves throughout the sample, with any visible differences being minimal (well inside the confidence intervals). This robustness check indicates that our main conclusions are not driven by the specific five-year choice and that the proposed social inflation measures are insensitive to modest changes in the rolling-window length.

\end{appendices}

\end{document}